\newif\ifshowcomments
\showcommentstrue 

\documentclass[prd,superscriptaddress,floatfix,notitlepage,nofootinbib]{revtex4-1} 

\pdfoutput=1
\usepackage{natbib,ifthen}
\usepackage{enumerate}
\usepackage{empheq}
\usepackage{feynmp}
\usepackage[english]{babel}
\usepackage{amsmath,amssymb,amsbsy,amstext, amsthm, simplewick,stackrel}
\usepackage{hyperref}
\usepackage{graphicx}
\usepackage{amsfonts}
\usepackage{amssymb}
\usepackage{upgreek}
 \usepackage{exscale,relsize}
 \usepackage[makeroom]{cancel}
\usepackage{float}
\usepackage{framed}
\usepackage{array}
\newcolumntype{C}[1]{>{\centering\arraybackslash}p{#1}}

\usepackage{placeins}
\newcommand{\MyFloatBarrier}{\FloatBarrier}  
\usepackage{color}
\usepackage{epsfig}
\usepackage{bm}
\usepackage{wasysym}
\usepackage{mathrsfs}

\usepackage{enumitem}

\newcommand{\appropto}{\mathrel{\vcenter{
  \offinterlineskip\halign{\hfil$##$\cr
    \propto\cr\noalign{\kern2pt}\sim\cr\noalign{\kern-2pt}}}}}

\newcommand{\shifted}[1]{\tilde{#1}}

\newcommand{\comment}[1]{}
\renewcommand{\d}{\mathrm{d}}
\newcommand{\beq}{\begin{equation}}
\newcommand{\eeq}{\end{equation}}
\newcommand{\bea}{\begin{eqnarray}}
\newcommand{\eea}{\end{eqnarray}}
\newcommand{\bsp}{\begin{split}}
\newcommand{\esp}{\end{split}}

\newcommand{\hMpc}{\ h^{-1}\text{Mpc}}

\newcommand{\Perr}{P_\mathrm{err}}
\newcommand{\kmax}{k_\mathrm{max}}

\renewcommand{\vec}[1]{\bm{#1}}

\newcommand{\vx}{\vec x}

\newcommand{\vO}{\vec{\mathcal{O}}}

\newcommand{\vk} {{\boldsymbol k}}
\newcommand{\vp} {{\boldsymbol p}}
\newcommand{\vq} {{\boldsymbol q}}

\newcommand{\vpsi}{\vec \psi}

\newcommand{\tG}{\tilde{\mathcal{G}}}
\newcommand{\G}{\mathcal{G}}

\newcommand{\ihMpc}{\; h\text{Mpc}^{-1}}

\newcommand{\hMsun}{\; h^{-1}\text{M}_\odot}

\newcommand{\la}{\left\langle}
\newcommand{\ra}{\right\rangle}

\definecolor{darkgreen}{RGB}{0,120,0}

\newcommand{\secref}[1]{Section \ref{se:#1}}

\newcommand{\eq}[1]{(\ref{eq:#1})} 
\newcommand{\eqq}[1]{Eq.~(\ref{eq:#1})} 
\newcommand{\fig}[1]{Fig.~\ref{fig:#1}} 
 
\newcommand{\app}[1]{Appendix~\ref{app:#1}}

\newcommand{\mnras}{Mon.\ Not.\ R.\ Astron.\ Soc.}

\newcommand{\jcap}{JCAP}
\newcommand{\apjl}{APJL~}
\newcommand{\physrep}{Physics Reports}
\newcommand{\aap}{Astronomy \& Astrophysics}
\newcommand{\aj}{The Astronomical Journal}

\ifshowcomments
\newcommand{\MS}[1]{{\color{darkgreen}{MS: #1}}}
\newcommand{\MSNEW}[1]{#1}
\newcommand{\Marko}[1]{\textcolor{blue}{(Marko: #1)}}
\newcommand{\va}[1]{\textcolor{green}{(VA: #1)}}
\newcommand{\MZ}[1]{{\color{red}{MZ: #1}}}
\else
\newcommand{\MS}[1]{}
\newcommand{\Marko}[1]{}
\newcommand{\va}[1]{}
\newcommand{\MZ}[1]{}
\fi

\usepackage{colortbl}
\definecolor{rp}{cmyk}{0.2, 1, 0.6, 0}
\definecolor{green2}{cmyk}{0, 1, 0.5, 0}
\definecolor{lightgreen}{cmyk}{0.2, 0, 0.2, 0.2}
\definecolor{lightgray}{cmyk}{0.1,0.2,0,0.1}
\definecolor{lightgray2}{cmyk}{0.4,0.4,0,0.8}
\definecolor{black}{cmyk}{1.0,1.0,1.0,1.0}

\definecolor{lightgreen}{cmyk}{0.2, 0, 0.2, 0.2}
\definecolor{lightgray}{cmyk}{0.1,0.2,0,0.1}
\definecolor{lightgray2}{cmyk}{0.1,0.1,0,0.1}

\makeatletter
\newlength{\apb@width}
\newcommand{\autoparbox}[2][c]{\settowidth{\apb@width}{#2}\parbox[#1]{\apb@width}{#2}}

\makeatother

\def\beq{\begin{equation}}
\def\eeq{\end{equation}}

\def\bea{\begin{eqnarray}}
\def\eea{\end{eqnarray}}

\def\d{{\rm d}}

\def\beq{\begin{equation}}
\def\eeq{\end{equation}}
\def\bea{\begin{eqnarray}}
\def\eea{\end{eqnarray}}

\def\d{{\rm d}}

\def\d{{\rm d}}

\def\G{{\cal G}}

\def\0{{\boldsymbol 0}}
\def\k{{\boldsymbol{k}}}
\def\q{{\boldsymbol{q}}}
\def\p{{\boldsymbol{p}}}

\def\x{{\boldsymbol{x}}}

\DeclareRobustCommand{\SkipTocEntry}[4]{}

\newcommand{\vev}[1]{\langle #1 \rangle}

\begin{document}

\title{Modeling Biased Tracers at the Field Level}

\author{Marcel Schmittfull}
\affiliation{Institute for Advanced Study, Princeton, NJ 08540, United States}
\author{Marko Simonovi\'c}
\affiliation{Theoretical Physics Department, CERN, 1 Esplanade des Particules, Geneva 23, CH-1211, Switzerland}
\affiliation{Institute for Advanced Study, Princeton, NJ 08540, United States}
\author{Valentin Assassi}
\affiliation{Institute for Advanced Study, Princeton, NJ 08540, United States}
\author{Matias Zaldarriaga}
\affiliation{Institute for Advanced Study, Princeton, NJ 08540, United States}

\date{\today}
\begin{abstract}
In this paper we test the perturbative halo bias model at the field level. 
The advantage of this approach is that any analysis can be done without sample variance if the same initial conditions are used in simulations and perturbation theory calculations. 
We write the bias expansion in terms of modified bias operators in Eulerian space, designed such that the large bulk flows are automatically resummed and not treated perturbatively.
Using these operators, the bias model accurately matches the Eulerian density of halos in N-body simulations.
The mean-square model error is close to the Poisson shot noise for a wide range of halo masses and it is rather scale-independent, with scale-dependent corrections becoming relevant at the nonlinear scale.
In contrast, for linear bias the mean-square model error can be higher than the Poisson prediction by factors of up to a few on large scales, and it becomes scale dependent already in the linear regime.
We show that by weighting simulated halos by their mass, the mean-square error of the model can be further reduced by up to an order of magnitude, or by a factor of two when including $60\%$ mass scatter. 
We also test the Standard Eulerian bias model using the nonlinear matter field measured from simulations and show that it leads to a larger and more scale-dependent model error than the bias expansion based on perturbation theory. 
These results may be of particular relevance for cosmological inference methods that use a likelihood of the biased tracer  at the field level, or for initial condition and BAO reconstruction that requires a precise estimate of the large-scale potential from the biased tracer density.
\end{abstract}

\maketitle

\section{Introduction}
The bias expansion forms the basis for the analytical description of the clustering of biased tracers on large scales (for a recent review, see \cite{BiasReview1611}). There are many checks in the literature showing that it works well at the level of summary statistics such as the power spectrum of halos, cross-spectra of halos with the matter density, and different higher-point correlation functions (recent studies include, e.g., \cite{Chan1201,Tobias1201,Saito1405,Hoffmann1607,2016arXiv160900717F,Modi1612,AbidiBaldauf1802}).
In this paper, we explore how well the bias expansion can match simulations at the field level. 
This is closely related to previous studies on the stochasticity of biased tracers (e.g., \cite{1999ApJ...520...24D,1999ApJ...522...46T,1999ApJ...525..543M,1999ApJ...518L..69T,Blanton2000,BlantonEtAl2000,2001MNRAS.320..289S,Yoshikawa2001,2012ApJ...750...37J,SeljakWarren2004,BonoliPen0810,Seljak0904,Hamaus1004,CaiBernsteinSheth1007,RothPorciani2011,Baldauf1305,Modi1612,GinzburgDesjacques1706}), but requires modifications for nonlinearly biased tracers, as we shall see.
\vskip 4pt

The main motivation for testing the bias expansion at the field level is that it is more stringent than a comparison of summary statistics: A model that predicts the simulated halo or galaxy density correctly for all pixels also predicts all summary statistics correctly;  the reverse is generally not true. Also,  overfitting, which can be a potential issue when fitting bias predictions for summary statistics, is not a concern, because at the field level all pixels or Fourier mode phases must be fitted, and the fit is not dominated by the Fourier modes with highest signal-to-sample-variance-noise.
\vskip 4pt 

A second, closely related motivation is to avoid sample variance.
 It is difficult to use correlation functions to accurately test the bias expansion, because the correlation functions are subject to sample variance in the simulation volume. 
That sample variance is not present when comparing prediction and simulation at the field level for the same initial conditions.
This simplifies quantifying the accuracy and regime of validity of the bias expansion.
Specifically, it enables simulations of moderate volumes with accurate mass and spatial resolution to characterize the bias expansion and its error with a precision corresponding to the cosmic variance of surveys that cover a much larger volume. 
The model error determined in this way can then inform cosmological analyses of galaxy survey data, for example by predicting the fiducial stochastic model error or shot noise to be included in the likelihood.
\vskip 4pt 

A third motivation is that a bias model that works at the field level could turn out useful for other applications.
For example, it could serve as a forward model predicting the halo density given a linear initial density, which is one of the ingredients of cosmological inference methods that use a likelihood of the biased tracer at the field level \cite{2010MNRAS.406...60J,2010MNRAS.407...29J,JascheWandelt1203,Kitaura1203,2013ApJ...772...63W,Uros1706,JascheLavaux1706,Chirag1805,Schmidt1808}. 
Or it could help to optimize initial condition and BAO reconstruction, which requires a precise estimate of the large-scale potential field from the biased tracer density (see \cite{EisensteinRec} and references therein, and \cite{TassevRec,Marcel1508,KeselmanNusser2017,IsoRecZhu1611ThreeD,IsoRecYu1703Halos,IsoRecWang1703BAO,Marcel1704,2018PhRvD..97b3505S,HadaEisenstein1804,Chirag1805,2018arXiv180908135B,2018arXiv180910738S} for more recent developments).  
\vskip 4pt 

Two major goals of our analysis are to check how well the bias expansion describes the simulated overdensity of dark matter halos, which we will refer to as ``true'' halo overdensity $\delta_h^{\rm truth}$, and to measure the amplitude and the scale dependence of the residual noise. These two questions are tightly related to each other. To illustrate this, let us consider the simplest model with the linear bias~$b_1$
\beq
\label{eq:simple_bias}
\delta_h^{\rm truth} = b_1 \delta + \epsilon \;,
\eeq
where $\delta$ is the nonlinear dark matter field. The stochastic term $\epsilon$ in this formula must be present, since we do not expect that the relation between dark matter and halos is perfectly deterministic \cite{1999ApJ...520...24D,1999ApJ...522...46T,1999ApJ...525..543M,1999ApJ...518L..69T,Blanton2000,BlantonEtAl2000,2001MNRAS.320..289S,Yoshikawa2001,2012ApJ...750...37J}. The best possible $b_1$ that describes the halo density field can be found by minimizing the mean-square difference $\langle|\delta_h^{\rm truth} - b_1 \delta|^2\rangle$, leading to the usual formula 
\beq
\label{eq:b1optimalsimple}
b_1(k) = \frac{\langle\delta_h^{\rm truth}(\k)\delta^*(\k)\rangle}{\langle|\delta(\k)|^2\rangle} \;.
\eeq
If the fields $\delta_h^{\rm truth}$ and $\delta$ share the same initial conditions, the measurement of $b_1(k)$ can be done without sample variance. Notice that the bias measured in this way is a function of $k$. One way to argue how well the linear bias model works is to ask up to which scales $b_1(k)$ is a constant. A significant scale dependence is a sign that higher order corrections must be included. 
\vskip 4pt 

An equally relevant question is how big an error we make, using the best fit values for bias parameters (in our simple example, $b_1(k))$. The power spectrum of this model error, or noise (sometimes also referred to as stochasticity \cite{SeljakWarren2004,BonoliPen0810,Seljak0904,Hamaus1004,CaiBernsteinSheth1007,Baldauf1305,Modi1612,GinzburgDesjacques1706}), is for the linear bias model given by
\beq
\label{eq:Pnoise}
P_{\rm err}(k) \equiv \langle |\delta_h^{\rm truth}(\k) - b_1(k) \delta(\k)|^2 \rangle 
= \langle |\epsilon(\vk)|^2 \rangle 
= \langle|\delta_h^{\rm truth}(\k)|^2\rangle - \frac{\langle\delta_h^{\rm truth}(\k)\delta^*(\k)\rangle^2}{\langle|\delta(\k)|^2\rangle} \;,
\eeq
where in the last equality we have used \eqq{b1optimalsimple}. 
The naive expectation for the large-scale amplitude of $P_{\rm err}$ is that it is close to Poisson noise $1/\bar n\equiv V/N_{\rm particles}$, which is expected when randomly sampling the continuous density with pointlike particles. However, the amplitude of the noise measured in simulations is larger than $1/\bar{n}$ for low-mass halos, and smaller than $1/\bar {n}$ for high-mass halos \cite{CasasMiranda2002,2011MNRAS.415..383M,Baldauf1305,Modi1612,GinzburgDesjacques1706}. The noise can also have a significant scale dependence, even at relatively large scales. In some cases, the amplitude of the noise on mildly nonlinear scales can differ from the amplitude in the low-$k$ limit even by tens of percent. Large amplitude and large scale dependence, if real, are dangerous, because they can significantly impact the inference of cosmological parameters.
\vskip 4pt 

One possible interpretation of these results is that the scale dependence of the noise is due to the higher order terms in the bias expansion. Indeed, in definition \eqref{eq:simple_bias}, the noise field $\epsilon$ contains operators constructed from matter fields that are not included in the model. Even though one may naively think that the higher order terms are irrelevant at large scales, as we are going to see they can significantly change the behavior of the noise even in the low-$k$ limit. Therefore, a more appropriate relation between dark matter and halos on large scales is \cite{Perko1610,BiasReview1611,Schmidt1808}
\beq
\delta_h^{\rm truth} \,=\, \delta_h^{\rm model}[\delta] + \epsilon \;,
\eeq
where $\delta_h^{\rm model}[\delta]$ stands for the model based on perturbative bias expansion.\footnote{For simplicity, throughout the paper we will also use the notation $\delta_h \equiv \delta_h^{\rm model}$ when the confusion with the simulated halo density field is not possible.} The success of the perturbative description can then be rephrased as the question of whether or not including higher orders in perturbation theory leads to a $P_{\rm err}(k)$ that has an amplitude closer to the Poisson noise and no significant scale dependence up to the nonlinear scale. To test whether the noise of the perturbative bias models has these properties, we estimate $\epsilon$ as the field difference between the true halo density, obtained for example from an N-body simulation, and the perturbation theory prediction,
\begin{align}
  \label{eq:ModelError}
  \hat\epsilon \,\equiv \, \delta_h^\mathrm{truth} - \delta_h^{\rm model} \;.
\end{align}
This model error vanishes on average, $\langle\hat\epsilon\rangle=0$, and its power spectrum, 
\begin{align}
  \label{eq:PerrDef}
  \Perr(k)\, \equiv\, \langle|\hat\epsilon(\vk)|^2\rangle \;,
\end{align}
describes the mean-square deviation of a Fourier mode $\delta_h^\mathrm{truth}(\vk)$ from the bias model prediction $\delta_h^{\rm model}(\vk)$.
For linear bias this definition coincides with \eqq{Pnoise}.
If the higher order operators in the bias expansion are included in the model $\delta_h^\mathrm{model}[\delta]$, the model error $\hat\epsilon$ in Eqs.~\eq{ModelError} and \eq{PerrDef} is free from these higher order bias terms. 
It only contains other higher order bias terms, which are not included in the model, and stochastic noise terms.
We are going to show that, as a consequence, the model error power spectrum becomes more flat and has an amplitude closer to the Poisson prediction. 
This is because the higher order bias operators not included in the model make only small $k$-dependent contributions to the model error, and $k$-dependent corrections to the stochastic noise become only relevant on small scales.
We will discuss these points in more details throughout the paper. 
\vskip 4pt 

One technical challenge in predicting the halo overdensity at the field level in perturbation theory is the treatment of large infrared (IR) displacements (bulk flows). Comparisons with simulations are naturally done in Eulerian space in the final conditions, but the large displacements are treated perturbatively in Standard Eulerian perturbation theory. This causes significant decorrelation of the predicted fields and simulations even on perturbative scales. To solve this problem we introduce a bias expansion using a new basis of  Eulerian bias operators
that fully include the Zel'dovich displacement.
We call these operators {\em shifted} operators and define them as
\begin{equation}
\label{eq:ShiftedOps}
\shifted{\mathcal O}(\k) \equiv \int\d^3\q\ {\cal O}(\q)\,e^{-i\k\cdot(\q+\vpsi_1(\q))} \;,
\end{equation}
where $\q$ is a Lagrangian coordinate in the initial condition, $\vpsi_1$ is the Zel'dovich displacement field, and $\mathcal{O}(\q)$ is any of the standard bias operators written in Lagrangian coordinates. We will show that the bias expansion defined in this way successfully describes the number density of halos in simulations, without the aforementioned decorrelation. Furthermore, we will show that the correlation functions of the shifted operators are closely related to the standard IR-resummed Eulerian counterparts, making a clear connection to the usual one-loop power spectrum for biased tracers. 
\vskip 4pt

Another important technical point is that the bias parameters obtained by minimizing the difference between the true halo density field and the model do not generally correspond to the physical (renormalized) bias parameters measured from the low-$k$ limits of the $n$-point functions. The reason for this is that the shifted operators $\shifted O_i$ depend on small scales and they are not renormalized bias operators. We make the choice of not smoothing the density fields for two reasons. First, our goal is not to merely measure the bias parameters, but to push the bias expansion to its limits and see how high in $k$ we can in principle go, maintaining a good correlation with the halo density field. Second, we want to see how much of the halo density field on large scales can be explained by the Fourier modes that are not in the perturbative regime. The main advantage of our approach is that it leads to a lower model error than using the bias parameters defined and measured in the standard way. Given a fixed survey volume, this, in principle, leads to more powerful measurements of cosmological parameters. 
\vskip 4pt 

One may be tempted to argue that, instead of using an analytical bias model, one could directly use a full N-body simulation as the forward model for cosmological parameter inference \cite{Uros1706}.
This would capture small-scale modes and should therefore reduce the model error.
But a critical component of this simulation-based approach is to obtain the halo density from the simulated nonlinear dark matter density. 
Applying a halo finder to the simulated dark matter density has led to complications because it renders the forward-model non-differentiable \cite{Uros1706,Chirag1805}.
A possible solution is to use a neural network or other machine learning techniques to approximate the result of non-differentiable halo finding algorithms with a differentiable model \cite{Chirag1805}.
However, such an approximation typically makes some stochastic error.
If an analytical bias model can predict the halo density field from the dark matter field with a similar error, then it may be a potentially simpler and useful alternative to a neural net.
\vskip 4pt

\MSNEW{Nonlinear bias has been tested against simulations at the field level before \cite{RothPorciani2011}. Using a different method to fit bias parameters (fitting scatter plots of the  model and simulation halo density smoothed on a particular scale as opposed to minimizing the mean square model error for each wavenumber $k$), and using the local Standard Eulerian bias model, Ref.~\cite{RothPorciani2011} found the scatter of the predicted halo density around the simulated halo density to be much larger than expected by the Poisson shot noise prediction. We confirm this result for Standard Eulerian bias and provide possible explanations in \secref{EulerianFailure}. This motivates us to employ a bias model different from the Standard Eulerian one when working at the field level. }
\vskip 4pt

The paper is organized as follows.
We introduce the bias model in terms of shifted operators in \secref{BiasModel}, where we also describe our method to fit bias parameters at the field level and list other bias models that we will compare against simulations.
\secref{Numerics} presents the numerical implementation of the bias model and the N-body simulations.
We then compare the model against simulations in position space in \secref{PositionResults} and in Fourier space in \secref{FourierResults}, where we analyse the size and scale dependence of the model error, and the size of the bias terms contributing to the bias model.
\secref{PTTransferFunctions} presents a perturbative description of the transfer functions associated with the shifted bias operators that we use, and perturbative fits of these transfer functions.
The rest of the paper consists of two standalone sections. 
In the first of these two sections, \secref{EulerianRelation}, we discuss the relation to Standard Eulerian bias and perturbation theory. Specifically, we show in \secref{EulerianFailure} that the Standard Eulerian bias expansion fails to describe biased tracers at the field level. In \secref{IRConnection} we discuss the connection with the usual IR-resummed power spectrum.
In the second standalone section, \secref{MassWeighting}, we explore how an extension of our work using mass-weighted halos can reduce the stochastic noise and therefore the error of the bias model.
We end by summarizing the main results in \secref{Conclusions}.
All technical details are described in appendices.

\MSNEW{For readers interested mainly in our theoretical investigation and its relation to previous work and other models, see Sections~\ref{se:BiasModel},  \ref{se:PTTransferFunctions} and \ref{se:EulerianRelation}, as well as appendices \ref{app:DeltaZShiftedOps} and \ref{app:CubicOperators}. For readers interested in measures of success when comparing against simulations, see Sections~\ref{se:PositionResults} and \ref{se:FourierResults} and \app{PerfMeasures}.  For readers interested in the numerical implementation of the model at the field level, see Section~\ref{se:Numerics} and \app{orth}. }

\section{Bias Model at the Field Level}
\label{se:BiasModel}
The perturbative approach to halo biasing has a long history, and has been well studied both in Eulerian and Lagrangian space (see the review \cite{BiasReview1611} and references therein, including, e.g., \cite{Matsubara0807,McDonaldRoy0902,Matsubara1102,Sheth1207,Carlson1209,Matsubara1304,Biagetti1310,Valentin1402,Leonardo1406,MSZ1412,Matsubara1604,Vlah1609,Modi1612}). However, most of the focus has been on the prediction of summary statistics, such as the power spectrum or the bispectrum of biased tracers. In contrast, in this paper we are interested in perturbation theory predictions at the level of realizations.
In this section we describe a perturbative bias model that we are going to use to make comparisons with simulations. 
\vskip 4pt

\subsection{Bias Expansion in Terms of Shifted Operators}
Describing dark matter or biased tracers at the field level is a nontrivial challenge for perturbation theory. For instance, it is well known that the large IR displacements (bulk flows) induced by long modes {\em cannot} be treated perturbatively. If they were, the positions of particles computed in perturbation theory would be off by as much as $\mathcal O(10)\;h^{-1}{\rm Mpc}$ compared to their true values. This means that the density field obtained from N-body simulations and the one computed treating the large IR displacements perturbatively (using the same initial conditions) would be completely uncorrelated on scales smaller than $\mathcal O(10)\;h^{-1}{\rm Mpc}$.\footnote{It is important to stress that the effect of this decorrelation is much more dramatic at the field level than for the correlation functions. This is due to the general statement that the effects of bulk flows have to cancel in equal time $n$-point functions \cite{Peloso:2013zw,Kehagias:2013yd,Creminelli:2013mca,Creminelli:2013poa}. The only exception to this theorem are cases in which there are sharp features in the correlation function, such as the BAO peak. For example, the only effect of large displacements on the power spectrum is to smooth out the BAO wiggles (or spread the BAO peak in real space two-point function) \cite{Senatore:2014via,Baldauf:2015xfa,Vlah:2015zda,Blas:2016sfa,Senatore:2017pbn}, while the smooth part of the power spectrum at small scales remains unchanged.} This is precisely what happens in Standard Eulerian perturbation theory, making it deficient for the description of realizations of dark matter or halo density fields. We will come back to the details of this failure of Standard Eulerian perturbation theory in \secref{EulerianFailure}.
\vskip 4pt

On the other hand, in Lagrangian perturbation theory the large IR displacements are naturally taken into account. However, this framework has a different problem. It predicts only the nonlinear displacement field $\vpsi$ and not the density field $\delta$. Going from one to the other is a nontrivial step. Given that the relation between $\delta$ and $\vpsi$ is very nonlinear, even a very good knowledge of the displacement field up to some scale does not guarantee that the density field will be correct up to the same scale with the same precision~\cite{TobiasDMDisplacement1505,TobiasDMDensity1507}.
\vskip 4pt 

In this paper we present one possible perturbative description that circumvents these problems by  constructing a bias expansion tailored to describe biased tracers at the field level. We put forward the following requirements: 
\begin{enumerate}[label=(\alph*)]
\item The bias expansion must be perturbative;
\item The bias operators have to be written in Eulerian space, given that we are comparing theoretical predictions and simulations of the final Eulerian density field;
\item The large IR displacements have to be treated non-perturbatively.
\end{enumerate}
Our strategy to achieve all of these goals is to combine the virtues of Eulerian and Lagrangian descriptions into a hybrid scheme. We start with the description of biased tracers in Lagrangian space. The displacement field is then split into the dominant linear contribution and smaller higher order corrections. The nonlinear corrections to $\vpsi$ are treated perturbatively, while the linear piece is kept in the exponent. In this way, the dominant part of the large displacements can be treated exactly, and the resulting operators once written in Eulerian space are automatically IR-resummed. In the rest of this section we give the details of this construction. 
\vskip 4pt 

\MSNEW{We first motivate the construction by considering the proto-halo density at Lagrangian position $\vq$, which can be} modeled using a bias expansion in the linear Lagrangian-space density $\delta_1(\q)$:
\begin{align}
\label{eq:LagrBias}
\delta_h^{\mathsmaller{\rm L}}(\q) = b_1^{\mathsmaller{\rm L}}\,\delta_1(\q)&\,+\,b_2^{\mathsmaller{\rm L}}\,(\delta^2_1(\q)-\sigma_1^2)\, +\,  b_{\G_2}^{\mathsmaller{\rm L}}\G_2(\q) + \cdots \;,
\end{align}
where $b_1^{\mathsmaller{\rm L}}$, $b_2^{\mathsmaller{\rm L}}$, $b_{\G_2}^{\mathsmaller{\rm L}}, \ldots$ are Lagrangian bias parameters, $\sigma_1^2$ is the r.m.s.~fluctuation of the linear density field
\begin{align}
\label{eq:rmsLinearDelta}
\sigma_1^2 = \la\delta_1^2(\vq)\ra = \int_0^\infty \frac{\d k}{2\pi^2}\, k^2 P_{11} (k) \;, 
\end{align}
and the tidal operator $\G_2(\q)$ is defined as\footnote{The basis of operators at second order (and higher orders) in perturbation theory is not unique. One of the advantages of working with $\{\delta_1^2,\G_2\}$ is that the auto-power spectrum of $\G_2$ and its cross-spectrum with $\delta_1^2$ vanish in the low-$k$ limit. This simplifies our analysis and helps to disentangle relevant contributions to the shot noise in the low-$k$ limit. For other common choices of the basis operators and their relation to $\{\delta_1^2,\G_2\}$ see \cite{BiasReview1611}.}
\begin{equation}
\label{eq:G2realspace}
\G_2(\q) \equiv \left[ \frac{\partial_i\partial_j}{\partial^2} \delta_1(\q) \right]^2 - \delta^2_1(\q) \;.
\end{equation}
The representation of this operator in momentum space is given by
\begin{equation}
\label{eq:G2momentumspace}
\mathcal{G}_2(\vk) = 
\int_{\vp} 
\left[\frac{(\vp\cdot (\vk-\vp))^2}{p^2|\vk-\vp|^2}-1\right] \delta_1(\vp)\,\delta_1(\vk-\vp)\;.
\end{equation}
Notice that in our notation $\int_{\vp} \equiv \int \d^3\p/ (2\pi)^3$. For the rest of this section we will also use $\delta_h\equiv\delta_h^{\rm model}$. In the bias expansion~\eq{LagrBias} we kept only terms up to second order in perturbation theory. We will continue to work at this order throughout this section, because it is sufficient for introducing notation and motivating the bias model that we are going to use to make comparisons with simulations. The higher order or higher derivative operators needed for the consistent one-loop calculation can be straightforwardly included. We will come back to this in~\secref{PTTransferFunctions}. 
\vskip 4pt 

The bias expansion in Eq.~\eqref{eq:LagrBias} is in Lagrangian space. To go to Eulerian space, let us start from Eq.~\eqref{eq:LagrBias} and include the gravitational evolution. The gravitational evolution is encoded in the nonlinear displacement field\footnote{We are assuming that the halos are formed in the the initial conditions and displaced by $\vpsi$. In reality the evolution is more complicated and in general nonlocal in time. However, it can be shown that these complications can be rewritten such that they only change the values of bias coefficients in perturbative approach to halo clustering (for more details see~\cite{MSZ1412,Leonardo1406}). For this reason we proceed with the simplified picture of halo formation and evolution.}, such that the Eulerian coordinates $\vx$ of a halo at the initial position $\q$ are given by $\vx=\vq+\vpsi(\vq)$. The overdensity generated in this way is given by
\begin{align}
1+\delta_h(\vx) = \int\d^3 \vq\, (1+\delta_h(\vq)) \,\delta_D(\vx-\vq-\vpsi(\vq)) \;,
\end{align}
where $\delta_D$ is the Dirac delta. The Fourier transform of this field in Eulerian space is
\begin{align}
\label{eq:defFurierEulerianHalo}
\delta_h(\vk) \equiv \int\d^3\vx\,(1+\delta_h(\vx))\,e^{-i\vk\cdot\vx} = \int\d^3 \vq\, (1+\delta_h(\vq))\, e^{-i\vk\cdot(\vq+\vpsi(\vq))} \;.
\end{align}
For simplicity, in this equation and in the rest of the paper we restrict the range of momenta to $k\neq 0$, so that the zero modes or mean density do not enter our formulas. The nonlinear displacement from Lagrangian to Eulerian position can be expanded in a perturbative series $\vpsi = \vpsi_1 +\vpsi_2 + \cdots$. At first order, we have the well-known Zel'dovich approximation \cite{Zeldovich1970}
\begin{align}
\label{eq:psi1}
\vpsi_1(\vq) = \int_\vk e^{i\vk\cdot\vq} \, \frac{i\vk}{k^2} \, \delta_1(\vk) \;.
\end{align}
The second-order displacement can be written as
\begin{align}
\label{eq:psi2}
\vpsi_2(\vq) = -\frac{3}{14} \int_\vk e^{i\vk\cdot\vq} \, \frac{i\vk}{k^2} \, \mathcal{G}_2(\vk) \;.
\end{align}
Using the perturbative description of the nonlinear displacement field and expanding the exponent $e^{-i\k\cdot\vpsi(\q)}$ in Eq.~\eqref{eq:defFurierEulerianHalo} it is possible to recover the usual Standard Eulerian bias expansion. This procedure also fixes the relation between Lagrangian bias parameters and their Standard Eulerian counterparts. Of course, this is not a surprise, as we expect the two descriptions to agree order by order in perturbation theory. 
\vskip 4pt 

On the other hand, we do not want to expand the full nonlinear displacement. We are going to keep the largest part $\vpsi_1(\q)$ exponentiated and expand only the higher-order terms.\footnote{Let us define $W(k)$ to be a low-pass filter, compared to the wavelength of a Fourier mode $\delta_1(\k)$. For a given wavenumber $k$, the linear displacement can be split into the long-wavelength and short-wavelength part: $\vpsi_1 = \vpsi_1^L + \vpsi_1^S$, where $\vpsi_1^L = W(k) \vpsi_1$ and $\vpsi_1^S = (1 - W(k)) \vpsi_1$. The effect of $\vpsi_1^L$ on the short modes is fixed by the Equivalence Principle. Therefore, strictly speaking, only $\vpsi_1^L$ should be kept exponentiated and in any perturbative calculation $\vpsi_1^S$ has to be expanded order by order in perturbation theory. The error in our formulas introduced by keeping the full $\vpsi_1$ in the exponent is always higher order in $\vpsi_1^S$ than terms we calculate. Also, this error is mainly relevant on small scales. To keep the formulas simple, we decide not to do the long-short splitting in our calculation.} In this way, the largest part of the problematic IR displacements is not expanded in perturbation theory. With this in mind, we can rewrite Eq.~\eqref{eq:defFurierEulerianHalo} as
\begin{align}
\label{eq:FurierEulerianHaloWithPsi1}
\delta_h(\vk) = \int\d^3 \vq\, \Big( 1+ b_1^{\mathsmaller{\rm L}}\,\delta_1(\q) \,+\,b_2^{\mathsmaller{\rm L}}\,(\delta^2_1(\q)-\sigma_1^2)\, & +\, b_{\G_2}^{\mathsmaller{\rm L}}\G_2(\q) + \cdots \nonumber \\
& - i\k\cdot\vpsi_2(\q) + \cdots \Big) \, e^{-i\vk\cdot(\vq+\vpsi_1(\vq))} \;,
\end{align}
where the new contributions come from expanding the second (and higher) order displacement field in the exponent. It is important to stress that at leading order this new term can be expressed through the second order operator $\G_2$ (see Eq.~\eqref{eq:psi2}). Therefore, at second order in perturbation theory, expanding the nonlinear terms in the displacement field $\vpsi(\q)$ only shifts some of the standard Lagrangian bias parameters by a calculable constant. We will give more details about higher order terms in~\secref{PTTransferFunctions}.  
\vskip 4pt

The previous expression motivates us to write down the bias expansion in Eulerian space in terms of {\em shifted operators} defined as
\begin{equation}
\label{eq:Oshifted}
\shifted{\mathcal O}(\k) \equiv \int\d^3\q\ {\cal O}(\q)\,e^{-i\k\cdot(\q+\vpsi_1(\q))} \;,
\end{equation}
where $\mathcal{O}\in \{1, \ \delta_1, \ \delta_2\equiv(\delta_1^2 -\sigma_1^2), \ \G_2 , \ \ldots\}$.\footnote{Notice that these shifted fields are not just given by a translation of the position argument because they  implicitly include the inverse of the determinant of the Jacobian $\partial x_i/\partial q_j$ due to the coordinate transformation. This is similar to the Zel'dovich density, which is given by a uniform field in Lagrangian space shifted by $\vpsi_1(\vq)$.} We stress again some of the advantages of using an expansion in this basis:
(a) The shifted operators are written in Eulerian space and therefore allow for easy comparisons with simulations and quantification of their importance. 
(b) The large displacement terms $\vpsi_1(\q)$ are kept resummed, which is crucial for comparisons with simulations at the field level. Notice that this also implies that in this description the BAO wiggles are properly suppressed (the BAO peak is spread). However, the model is still perturbative in small quantities, such as derivatives of the linear displacement $\delta_1 = - \nabla \cdot \vpsi_1$. The power spectrum calculated using the shifted operators is identical on large scales to the standard 1-loop result with IR-resummation. (c) The shifted operators are easy to generate on a 3-d grid for a given initial condition realization on a 3-d grid, by shifting properly weighted particles from Lagrangian to Eulerian coordinates using the Zel'dovich displacement (see \secref{Numerics} below). 
\vskip 4pt 

One term in the previous equations that has a somewhat special role is the shift of a uniform density, $\mathcal{O}=1$. This contribution to $\delta_h(\k)$ is equal to the Zel'dovich density field
\begin{align}
\label{eq:Zel'dovichDensity}
\delta_Z(\vk) \equiv \int\d^3 \vq \, e^{-i\vk\cdot(\vq+\vpsi_1(\vq))} \;.
\end{align}
It is fixed by dynamics, and it is not a part of the bias expansion in the usual sense (it has no free parameters). However, the Zel'dovich density $\delta_Z(\vk)$ can also be expanded in the basis of shifted operators (see \app{DeltaZShiftedOps}),
\begin{align}
\label{eq:deltaZshifted}
\delta_Z(\vk) = \shifted\delta_1(\vk) + \frac{1}{2} \shifted{\mathcal{G}}_2(\vk) - \frac 13 \shifted{\mathcal{G}}_3(\vk) +\cdots \;,
\end{align}
where $\shifted{\mathcal{G}}_3$ is a cubic operator analogous to $\shifted{\mathcal{G}}_2$ (see \app{CubicOperators}).
In other words, $\delta_Z(\vk)$ can be absorbed in the bias expansion by simply changing the bias parameters. Of course, this is just a choice, and there is nothing wrong in keeping $\delta_Z$ explicitly in the formulas. As we are going to see later, different choices may be more appropriate for different applications. Let us point out that in the formula~\eqref{eq:deltaZshifted} the displacements $\vpsi_1(\q)$ are treated exactly. In other words, the exponential $e^{-i\vk\cdot \vpsi_1(\vq)}$ is never expanded in $\vpsi_1(\q)$. The only expansion parameter is the {\em derivative} of the displacement, $\nabla\cdot\vpsi_1(\q) = -\delta_1(\q)$, which is a small quantity.\footnote{This may seem counterintuitive at the first sight, because there are no derivatives of the displacement field in Eq.~\eqref{eq:Zel'dovichDensity}. However, they do appear once the momentum $\k$ in $e^{-i\k\cdot\vpsi_1(\q)}$ is written as a derivative with respect to $\q$. A much easier derivation of Eq.~\eqref{eq:deltaZshifted} is in real space, as presented in \app{DeltaZShiftedOps}.} This is consistent with the way the shifted operators are defined. 
\vskip 4pt

Using the basis of shifted operators \eq{Oshifted} we can therefore write the bias expansion of the halo density field in Eulerian coordinates, up to second order in perturbation theory, as
\begin{equation}
\label{eq:deltah_shifted_operators}
\delta_h(\k) = b_1 \, \shifted \delta_1(\vk) +  b_2 \, \shifted \delta_2 (\vk) +  b_{\G_2} \, \shifted \G_2(\vk) + \cdots \;.
\end{equation}
This is the main result of this section. Notice that the new bias parameters $b_i$ differ from the original Lagrangian biases $b_i^{\mathsmaller{\rm L}}$ by a constant. This difference comes from expanding the nonlinear part of the displacement (Eq.~\eqref{eq:FurierEulerianHaloWithPsi1}) and writing the Zel'dovich density field in terms of shifted operators (Eq.~\eqref{eq:deltaZshifted}). We give the explicit relation of $b_i$ and $b_i^{\mathsmaller{\rm L}}$ in~\secref{PTTransferFunctions}. 
\vskip 4pt

Equation~\eqref{eq:deltah_shifted_operators} has a structure similar to the usual Standard Eulerian bias expansion
\begin{equation}
\label{eq:deltah_Eulerian}
\delta_h(\k) = b_1^{\mathsmaller{\rm E}} \, \delta(\vk) + b_2^{\mathsmaller{\rm E}} \, \delta_2 (\vk) + b_{\G_2}^{\mathsmaller{\rm E}} \, \G_2(\vk) + \cdots \;,
\end{equation}
where $\delta_2(\k)$ is the Fourier transform of the squared Eulerian density $\delta^2(\vx)$ (as opposed to $\tilde\delta_2(\vk)$, which is obtained by squaring in Lagrangian coordinates and then transforming to Eulerian coordinates using \eqq{Oshifted}). Notice that all fields in \eqq{deltah_Eulerian} are nonlinear.
In contrast, in the expansion \eq{deltah_shifted_operators} all operators are expressed in terms of the {\rm linear} field $\delta_1$, which, as we are going to see, is more suitable for describing biased tracers at the field level.

\vskip 4pt
Another virtue of the expansion \eq{deltah_shifted_operators} is that the theoretical calculation of the  power spectrum is quite straightforward (see \secref{PkShifted}). 
It involves the calculation of the power spectra of shifted operators, which have a familiar form, for instance
\begin{equation}
\langle \shifted{\mathcal O}_i(\k) \shifted{\mathcal O}^*_j(\k) \rangle = \int d^3\q \,\langle {\cal O}_i(\q) {\cal O}_j(0) \, e^{-i\k\cdot(\vpsi_1(\q) - \vpsi_1(0))} \rangle e^{-i\k\cdot\q} \;.
\end{equation}
The expression on the r.h.s.~is common in Lagrangian perturbation theory. This connection is not surprising, given that we started our derivation in Lagrangian space. Even though we have come to the definition of the shifted operators using a different motivation, a lot of literature already exists on the power spectrum of biased tracers in Lagrangian perturbation theory (e.g., \cite{Matsubara0807,Matsubara1304}). In this paper we are going to use some results presented there. For some recent developments, such as Convolution Lagrangian Effective Field Theory, see for example~\cite{Vlah:2015sea,Vlah1609,Modi1706,2018arXiv180505304A} and references therein.

\subsection{Promoting Bias Parameters to Transfer Functions}
So far we wrote the bias expansion in terms of shifted operators keeping only terms up to second order in perturbation theory. If we want to describe the density field of biased tracers deeper in the nonlinear regime, we have to include higher order terms. For instance, even for the evaluation of the one-loop power spectrum one has to keep all cubic operators. Let us take a closer look at this example
\begin{equation}
\label{eq:deltah_shifted_operators_O3}
\delta_h(\k) =  b_1 \, \shifted \delta_1(\vk) + b_2 \, \shifted \delta_2 (\vk) + b_{\G_2} \, \shifted \G_2(\vk) + \sum_i b_3^i \, \shifted{\mathcal O}_3^i(\k) \;,
\end{equation}
where $\shifted{\mathcal O}_3^i$ is a set of cubic operators and $b_3^i$ are the corresponding bias parameters. At lowest order in perturbation theory the cubic operators correlate only with $\shifted \delta_1$. We can split the cubic operators into parts parallel and orthogonal to $\shifted \delta_1$,
\begin{equation}
\shifted{\mathcal O}_3^i(\k) = \frac{\vev{\shifted\delta_1^*(\k) \shifted{\mathcal O}_3^i(\k)}}{\vev{|\shifted\delta_1 (\k)|^2}} \; \shifted \delta_1(\k) + \left(\shifted{\mathcal O}_3^i(\k) - \frac{\vev{\shifted\delta_1^*(\k) \shifted{\mathcal O}_3^i(\k)}}{\vev{|\shifted\delta_1 (\k)|^2}} \; \shifted \delta_1(\k) \right) \equiv \frac{\vev{\shifted\delta_1^*(\k) \shifted{\mathcal O}_3^i(\k)}}{\vev{|\shifted\delta_1 (\k)|^2}} \; \shifted \delta_1(\k) + \shifted{\mathcal O}_3^{i\perp}(\k)\;.
\end{equation}
In this way, allowing for a scale-dependent bias parameter $b_1(k)$, we can write
\begin{equation}
\delta_h(\k) = b_1(k) \, \shifted \delta_1(\vk) + b_2 \, \shifted \delta_2 (\vk) + b_{\G_2} \, \shifted \G_2(\vk) + \sum_i b_3^i \, \shifted{\mathcal O}_3^{i\perp}(\k) \;.
\end{equation} 
At one-loop order, the new cubic operators are orthogonal to all other fields.
This implies that even the bias expansion up to second order in the fields, with the appropriate $b_1(k)$, is sufficient to describe the density field with the correct one-loop power spectrum. Allowing for scale-dependent bias parameters effectively allows us to reduce the order in perturbation theory that we need to describe the density field of biased tracers at a given order in perturbation theory. 
\vskip 4pt

This example provides motivation to promote all bias parameters to $k$-dependent functions
\begin{equation}
\delta_h(\k) = b_1(k) \, \shifted \delta_1(\vk) + b_2(k) \, \shifted \delta_2 (\vk) + b_{\G_2}(k) \, \shifted \G_2(\vk) + \cdots \;,
\end{equation} 
in order to take into account as much nonlinearity as possible. This expression can be compared to realizations of N-body simulations. Calculating the operators with the same initial conditions, the sample variance can be canceled~\cite{TobiasDMDisplacement1505}. The bias functions can be measured from the condition that the difference between realizations in simulations and theory is minimal. This procedure allows us to ask a very general question: How much of the real halo density field can be described with a few leading-order operators, even beyond the perturbative regime? In a setup this general, a perturbation theory-inspired model can be considered successful if it leads to small (close to Poisson) and scale-independent mean-square model error.
\vskip 4pt

When fitting the above model to a halo density at the field level, the bias coefficients $b_i$ are correlated with each other because the shifted fields $\shifted\delta_1$, $\shifted\delta_2$, and $\shifted{\mathcal{G}}_2$ are correlated among themselves (they are defined using the same initial conditions and the same displacement field $\vpsi_1$). When interpreting the bias parameters, it is useful to change the basis to avoid this correlation. We therefore rotate the shifted operators to mutually orthogonal fields using the Gram-Schmidt algorithm:
\begin{align}
  \label{eq:GramSchmidt}
\shifted\delta_1^\perp(\vk) &= \shifted\delta_1(\vk) \;, \\
  \shifted\delta_2^\perp(\vk) &=
\shifted\delta_2(\vk) + M_{10}(k)\shifted\delta_1(\vk) \;, \\
\shifted{\mathcal{G}}_2^\perp(\vk) &= 
\shifted{\mathcal{G}_2}(\vk) + M_{20}(k)\shifted\delta_1(\vk)+M_{21}(k)\shifted\delta_2(\vk) \;.
\end{align}
The Gram-Schmidt rotation matrix $M_{ij}(k)$ is $M_{10}(k)=-P_{\shifted\delta_2\shifted\delta_1}(k)/P_{\shifted\delta_1\shifted\delta_1}(k)$ etc., and can be computed using a Cholesky decomposition of the $3\times 3$ correlation matrix between the three shifted fields $\{\shifted\delta_1, \shifted\delta_2, \shifted{\mathcal{G}_2}\}$ in every $k$-bin as described in \app{orth}.
The bias expansion in this orthogonal basis is then
\begin{align}
\label{eq:bias_expansion_orthogonal}
\delta_h(\vk) &= \beta_1(k)\ \shifted\delta_1(\vk) +\beta_2(k) \ \shifted\delta_2^\perp(\vk) + \beta_{\mathcal{G}_2}(k)\ \shifted{\mathcal{G}}_2^\perp(\vk) + \cdots \; .
\end{align}
These new bias parameters, or transfer functions, $\beta_i(k)$ are independent from each other. We can therefore add higher-order operators using the same procedure without changing any of the lower-order bias parameters, which is a useful property.
In our framework, where transfer functions are determined by minimizing the mean-square model error at the field level, the change of basis, i.e., going from $b_i$ to $\beta_i$, does not change the predicted halo density; it merely provides a more convenient way to interpret the numerical values of bias parameters. 
Also notice that the first parameter remains unchanged, $\beta_1(k) = b_1(k)$. In \secref{PTTransferFunctions} we will present one-loop perturbation theory predictions  for $\beta_i(k)$ and compare against measurements of $\beta_i(k)$ from N-body simulations.

\subsection{Relation to Renormalized Bias Parameters}
\label{se:RenormalizationIssues}
Before we close this section listing all bias models that we use in the paper, we get back to an important point that we have only briefly mentioned in the introduction: The low-$k$ limit of the transfer functions $\beta_i(k)$ does not necessarily approach the values of physical (renormalized) bias parameters. This means that the bias parameters we measure at the field level are not generally expected to be the same as the bias parameters measured from correlation functions of the halo density field. In the terminology of renormalization, what we measure at the field level is closer to ``bare'' bias parameters. These biases depend on the cutoff scale, or the way the small scales are regulated. For example, as we are going to see, using the linear or the nonlinear matter density field to construct bias operators leads to very different transfer functions in the low-$k$ limit. One easy way to see why this happens is to take a look at the expression for a transfer function obtained using the minimization described above. If we assume that the basis of operators is orthogonal, we can write
\begin{equation}
\label{eq:betaikexample}
\beta_i(k) = \frac{\vev{\delta_h^{\rm truth}(\k) \shifted{\mathcal O}_i^{\perp*}(\k)}}{\vev{|\shifted{\mathcal O}_i^\perp(\k)|^2}} \;.
\end{equation}
The power spectrum in the denominator in general involves loops, and therefore it is obviously dependent on how the high-$k$ modes are treated. The usual way to deal with this issue is to renormalize the bias operators, subtracting the cutoff-dependent counterterms~\cite{Valentin1402}. Away from the perturbative regime and at the field level this becomes challenging. Take, for example, the operator $\delta_2$. The power spectrum of this operator is constant in the low-$k$ limit. This constant comes from integrating very short scales and can be always absorbed by the free amplitude of the shot noise in the power spectrum. However, this is not possible at the field level. If we add an independent field with constant power spectrum to the model with the hope to fix the problem, it can only give a positive definite contribution to the model error power spectrum, making the model worse. 
\vskip 4pt 

At this point it is important to clarify the relation to other works (see for example \cite{LazeyrasSchmidt1712, AbidiBaldauf1802}) in which similar techniques were exploited to measure the physical bias parameters. The idea is that the bias parameters can be measured by projecting the halo density field on the basis of bias operators, leading to equations very similar to \eqq{betaikexample}. One major difference is that the bias operators in \cite{LazeyrasSchmidt1712, AbidiBaldauf1802} are constructed from the {\em smoothed} density field. The smoothing scale $R$ is chosen to ensure that only the Fourier modes in the perturbative regime contribute and it is typically $R\sim \mathcal O(10)\; {\rm Mpc}$ at $z=0$.\footnote{In principle, the bigger the smoothing scale $R$, the less sensitive the results are to the nonlinear corrections. In practice, the choice of the smoothing scale is dictated by the volume of N-body simulations and convergence tests.} In this way it is indeed possible to measure the low-$k$ limit of the transfer functions and rigorously prove that they can be identified with the renormalized bias parameters. 
\vskip 4pt 

However, this program is somewhat orthogonal to our goals in this paper. We do not necessarily restrict to the perturbative regime $k\ll k_{\rm NL}$, but we want to test how well we can reproduce the halo density field even around the nonlinear scale. Using the smoothed density field to construct the basis operators would imprint the smoothing scale in all our calculations and lead to significant decorrelation with the halo density field already around $k\sim 0.1 \ihMpc$. In this context, keeping the short scales in the bias operators seems to lead to better results.
We therefore do not apply any smoothing to the fields.\footnote{The only exception is $\delta_3(\vq)$, which, as we discuss below, is smoothed with a sharp $k$ filter at $\kmax=0.5\ihMpc$. There is also an implicit smoothing of all fields due to the cell size $\Delta x\simeq 1\hMpc$ of the Eulerian grid, but this is only relevant on very small scales.}
The price that we have to pay for this choice is that the low-$k$ limit of the transfer functions does not correspond to bias parameters defined in the usual way. 
\vskip 4pt 

Let us finish by saying that one important exception in this discussion is the linear bias. In this case
\begin{equation}
\beta_1(k) = \frac{\vev{\delta_h^{\rm truth}(\k) \shifted{\delta}_1^*(\k)}}{\vev{|\shifted{\delta}_1(\k)|^2} } \;.
\end{equation}
The low-$k$ limit of this expression coincides with the usual definition of the renormalized linear bias, since the power spectrum in the denominator approaches $P_{11}(k)$. Therefore, we do expect to find that $b_1 = \beta_1(k\to 0)$ is indeed the same as inferred from the power spectrum or separate universe simulations.

\subsection{List of Bias Models}
\label{se:BiasList}
When comparing against simulations we will mostly use the bias expansion in terms of shifted operators described above, but sometimes we will also show comparisons with other bias expansions.   
The following list provides an overview over all bias models that we will use for the analysis.

\begin{itemize}

\item Quadratic bias model:
\begin{align}
\label{eq:quadratic_model}
\delta_h(\vk) &= \beta_1(k)\ \shifted\delta_1(\vk) +\beta_2(k) \ \shifted\delta_2^\perp(\vk) + \beta_{\mathcal{G}_2}(k)\ \shifted{\mathcal{G}}_2^\perp(\vk) \; .
\end{align}
This is our perturbation theory prediction described above for the density field of biased tracers in a realization. We are going to use this, or the cubic extension described below, as the reference model for comparisons with simulations and with other biasing schemes.

\item Linear bias model:
\begin{align}
\label{eq:linear_model}
\delta_h(\vk) = \delta_Z(\vk) + b_1^{\mathsmaller{\rm L}}(k) \ \shifted \delta_1(\vk) \; .
\end{align}
We include this model in the analysis to study how the second order terms in \eqq{quadratic_model} affect results, particularly the amplitude and scale dependence of the model error.
The transfer functions in this model approach the usual linear Lagrangian bias parameters on large scales. 
This is because we have kept the Zel'dovich density $\delta_Z$ explicitly in the formula. At leading order in perturbation theory there is no reason not to replace $\delta_Z$ with $\shifted \delta_1$ (see \eqq{deltaZshifted}). However, the second order contributions in $\delta_Z$, which are fixed by the gravitational evolution and come with fixed coefficients, can be significantly larger than the second order bias contributions (depending on halo mass). Dropping them would then affect the model error (shot noise) of the linear bias model, making it larger and more scale dependent. For this reason we choose to keep $\delta_Z$ in the formula. In other words, the linear bias model as we choose to write it here is the best possible one-parameter model that we can use in comparisons with realizations. This is a conservative choice because the impact of the additional second order terms in Eq.~\eqref{eq:quadratic_model} compared to Eq.~\eqref{eq:linear_model} is minimized. Even then, as we will see, the second order bias terms will be quite significant. 

\item Cubic bias model:
\begin{align}
\label{eq:partially_cubic_model}
\delta_h(\vk) &= \beta_1(k)\ \shifted\delta_1(\vk) +\beta_2(k) \ \shifted\delta_2^\perp(\vk) + \beta_{\mathcal{G}_2}(k)\ \shifted{\mathcal{G}}_2^\perp(\vk) + \beta_3(k) \ \shifted\delta_3^\perp (\k) \; .
\end{align}
Another possible modification is to include additional operators in the bias model. Here we include the shifted cubic term $\delta_1^3(\vq)$, ignoring all other contributions at the same order. Strictly speaking this choice is not consistent with perturbation theory and we should not trust this model on small scales where the two-loop terms become important. However, our motivation to keep $\delta_1^3$ is due to the fact that we want to study the impact of this operator on the amplitude of the shot noise in $k\to 0$ limit. As it turns out, in the basis of cubic operators $\{\delta_1^3, \delta_1\G_2, \G_3, \Gamma_3 \}$, the only operator that has a constant contribution to its auto power spectrum in the large-scale limit is $\delta_1^3$. Therefore, unlike in the case of correlation functions, at the level of realizations it {\em does} make sense to add a subset of bias operators at the given order in perturbation theory, as long as they can have a large contribution on very large scales. We find that adding $\delta_1^3$ is most effective when we remove small-scale modes from $\delta_1$ before cubing the field; we therefore apply a smoothing to $\delta_1$ with sharp cutoff at $k_\mathrm{max}=0.5\ihMpc$ when computing $\delta_1^3$ (none of the other fields are smoothed because their auto-power spectra are less UV sensitive).\footnote{With Gaussian smoothing the model can be improved further for high-mass halos, but this typically increases the  scale-dependence of the transfer function associated with $\delta_1^3$. The sensitivity of $\delta_1^3$ on smoothing suggests that a more systematic investigation of the optimal smoothing of this term could improve the bias model. Including the full set of allowed cubic operators can lead to further improvements, but also requires more bias parameters.}

\item Standard Eulerian bias model:
\begin{align}
\label{eq:eulerian_model}
\delta_h(\k) = \beta_1^{\mathsmaller{\rm E}}(k) \, \delta(\vk) + \beta_2^{\mathsmaller{\rm E}}(k) \, \delta_2^\perp (\vk) + \beta_{\G_2}^{\mathsmaller{\rm E}}(k) \, \G_2^\perp(\vk) \; .
\end{align}
This is the standard expression for the density field of biased tracers using Standard Eulerian bias. This model assumes that we can perfectly model the fully nonlinear dark matter density field $\delta$. In practice, we measure this from N-body simulations, i.e.~we use the best Standard Eulerian bias model we could ever hope for. Notice that the second order operators are also evaluated using the nonlinear field and they are orthogonal to each other and $\delta$. We are going to compare both first and second order terms with simulations. We have already discussed some shortcomings of modeling $\delta$ with the Standard Eulerian perturbation theory. As we are going to see, using the full nonlinear density field from N-body simulations also has its own problems. We will get back to these issues in \secref{EulerianFailure}, in which we will also consider possible modifications of this model by smoothing $\delta$ or replacing $\delta$ by the perturbative dark matter density.

\end{itemize}
\vskip 4pt

For each of the bias models listed above, we allow the bias parameters or transfer functions $\beta_i^{\mathsmaller{\rm X}}(k)$ to be free functions of wavenumber $k$. We will measure them from simulations as described in the next section and show that they are smooth functions. On large scales the $k$-dependence of these functions can be predicted using perturbation theory with a few free parameters. The number of these free parameters is the same as the number of usual bias parameters.

\section{Numerical Implementation}
\label{se:Numerics}

To test these bias expansions against simulations we proceed as follows.
We first draw a Gaussian linear density from a fiducial linear power spectrum, computed with CAMB \cite{camb} for a flat $\Lambda$CDM cosmology with $\Omega_m=0.3075, \Omega_bh^2=0.0223, \Omega_ch^2=0.1188, h=0.6774$, $\sigma_8=0.8159$, and $n_s=0.9667$ based on Planck 2015 \cite{Planck15Params}.
Using this linear density, we evaluate each halo bias model on a 3-d grid in Eulerian coordinates, and compare this against the halo density obtained from N-body simulation initialized with the same linear density.
We then compute the difference between the model and simulation density, which is free of sample variance and directly measures the error of the bias model in Eulerian coordinates.\footnote{Alternatively, the comparison between the model and simulation can be performed in Lagrangian space by evaluating the model in Lagrangian space and tracing simulated halos back to their Lagrangian positions (e.g., \cite{Modi1612,AbidiBaldauf1802}); converting this Lagrangian-space modeling error to Eulerian space is nontrivial though, which is why we evaluate model and simulations directly in Eulerian space.}
Before showing the results of this, let us briefly discuss in more detail how the model density and simulations are generated.

\subsection{Halo Bias Model on 3-D Grid in Eulerian Space}
\label{se:BiasOnGrid}

To evaluate the linear, quadratic and cubic bias models in Eqs.~\eq{quadratic_model}, \eq{linear_model}, and \eq{partially_cubic_model}, we must evaluate the shifted operators \eq{ShiftedOps} on a 3-d grid in Eulerian coordinates.
To do this, we generate a uniform catalog with $1536^3$ particles located at the vertices $\vq$ of a regular $1536^3$ grid in a 3-d box with side length $L=500\hMpc$, corresponding to a particle separation of $\Delta q=0.33\hMpc$.
We then displace each particle, $\vq\rightarrow \vq+\vpsi_1(\vq)$, where $\vpsi_1(\vq)$ is the linear displacement in Lagrangian coordinates $\vq$ from \eqq{psi1}, rescaled linearly to redshift $z=0.6$ using the linear growth function $D(z)$.
To compute the shifted operator corresponding to $\mathcal{O}(\vq)=1$, we paint the displaced particles to a $512^3$ grid using the standard cloud-in-cell (CIC) algorithm, so that each grid cell stores the number of nearby particles, weighted by the distance of each particle from the cell center.
The corresponding overdensity is the Zel'dovich density $\delta_Z(\vx)$ in Eulerian coordinates. 
Notice that this procedure is the same as when initializing N-body simulations from a regular grid, except that the displacement is evaluated at late time, $z=0.6$.
\vskip 4pt 

To generate the shifted linear density $\tilde\delta_1$, we proceed in a similar way.
We again start with the uniform catalog of $1536^3$ particles, but now assign each particle an artificial mass given by $\delta_1(\vq)$, rescaled linearly to $z=0.6$.
(Notice that the density of this catalog is $\delta_1(\vq)$.)
We displace these particles using $\vq\rightarrow \vq+\vpsi_1(\vq)$ as before. 
To paint the resulting catalog to a grid, we modify the CIC painting scheme such that now each particle contributes to nearby grid cells with the usual CIC distance weight multiplied by the mass of each particle.
We sum these masses, without dividing by the number of particles that contribute to each cell, so that nearby particles with equal mass (i.e.,~particles that originate from a region in Lagrangian space where $\delta_1(\vq)$ is constant)
can cluster and create a density that is larger than the mass of these particles.
This ensures that the volume factor given by the determinant of the Jacobian $\partial x_i/\partial q_j$ between Eulerian and Lagrangian coordinate systems is included in $\tilde \delta_1$, and that the mean density remains unchanged.
The shifted squared density $\tilde\delta_2$ and shifted tidal field $\tG_2$ are computed similarly, using $\delta_1^2(\vq)$ or $\G_2(\vq)$ for the particle mass.
\vskip 4pt 

Next, the fields entering the model are orthogonalized using the Gram-Schmidt procedure in \eqq{GramSchmidt}. Details specific to this orthogonalization procedure are described in \app{orth}.
Finally, we compute all power spectra between these orthogonalized model contributions and the true halo density obtained from an N-body simulation started from the same linear density, get the optimal model transfer functions $\beta_i(k)$ using linear regression \eq{LinRegr}, and sum up the model contributions weighted by the transfer functions.

\subsection{Phase-Matched N-body Simulations}

The phase-matched N-body simulations are generated as follows.
Using the same initial linear Gaussian density as above, initial particle positions and velocities at $z=99$ are set up using the Zel'dovich approximation for $1536^3$ dark matter particles in a $L=500\hMpc$ box.
These particles are evolved to redshift $z=0.6$ using the TreePM N-body code \texttt{MP-Gadget} \cite{MPGadgetWebsite,MPGadgetDOI}, with $N_\mathrm{mesh}=3072$ for the particle-mesh (PM) grid. The code makes about $4200$ time steps to reach $z=0.6$.
The mass of each dark matter particle is $2.94\times 10^9\,h^{-1}\mathrm{M}_\odot$.
\vskip 4pt 

In the resulting dark matter snapshot we identify halos using the standard friends-of-friends (FOF) algorithm with linking length of $0.2$ using \texttt{nbodykit} \cite{nbodykitPaper,nbodykitWebsite}.
We require halos to have at least 25 dark matter particles, corresponding to a minimum halo mass of $7.4\times 10^{10}\,h^{-1}\mathrm{M}_\odot$; the heaviest halo weighs about $1.3\times 10^{15}\,h^{-1}\mathrm{M}_\odot$.
We define four halo mass bins with  number densities roughly corresponding to different future experiments as indicated in Table~\ref{tab:MassBins}.
For each mass bin we compute the halo density on a $512^3$ grid using standard CIC painting.
\vskip 4pt 

\begin{table}[tbp]
\centering
\renewcommand{\arraystretch}{1.1}
\begin{tabular}{@{}lllll@{}}
\toprule
$\log M[h^{-1}\mathrm{M}_\odot]$ & \phantom{} & $\bar n\,[(h^{-1}\mathrm{Mpc})^{-3}]$ & \phantom{} & $\bar n$ is comparable to  \\ 
\colrule 
$10.8-11.8$  && $4.3\times 10^{-2}$ && LSST \cite{LSSTScienceBook,LSSTwebsite}, Billion Object Apparatus \cite{1604.07626} \\
$11.8-12.8$  && $5.7\times 10^{-3}$ && SPHEREx \cite{Spherex1412,spherexWebsite} \\
$12.8-13.8$  && $5.6\times 10^{-4}$ && BOSS CMASS \cite{SDSSwebsite}, DESI \cite{DESIFDRDoc,DESIwebsite}, Euclid \cite{EuclidWhitePaper,euclidECWebsite,euclidESAWebsite} \\
$13.8-15.2$  && $2.6\times 10^{-5}$ && Cluster catalogs 
 \\
\botrule
\end{tabular}
\caption{Simulated halo populations at $z=0.6$.
}
\label{tab:MassBins}
\end{table}

To estimate uncertainties, we generate five independent realizations of the linear density using different random seeds, and generate the bias expansion density and simulations for each of these five realizations.
Whenever we compare model and simulations we first compute their difference for each random seed and then average the result over the five realizations, to avoid sample variance.
\vskip 4pt 

We will refer to these simulations as the ground truth, and we will ask how well the analytic halo bias expansion can describe them. 
Of course, the simulations could be made more realistic by populating the halos with galaxies and including redshift space distortions, but we will restrict ourselves to halos in real space in this work.

\subsection{Determining Bias Transfer Functions}
\label{se:MeasureBias}

To compute the bias transfer functions $\beta_i(k)$  we minimize the mean-square model error defined in \eqq{PerrDef},
\begin{align}
  \label{eq:29}
P_\mathrm{err}(k)
&=
\frac{1}{N_\mathrm{modes}(k)} 
\sum_{\vk,|\vk|\approx k}|\delta_h^\mathrm{truth}(\vk)-\delta_h^{\rm model}(\vk)|^2,
\end{align}
in every $k$ bin.
This minimization is meaningful because $\Perr$ is non-negative and vanishes if and only if the amplitude and phases of all Fourier modes match perfectly,
\begin{align}
  \label{eq:23}
\Perr(k)=0
 \quad\Leftrightarrow \quad 
\delta_h^\mathrm{truth}(\vk)=\delta_h^{\rm model}(\vk)\,\textrm{ for all }\vk\textrm{ with }|\vk|\approx k.
\end{align}
Since all bias expansions that we consider are of the form
\begin{align}
\label{eq:deltaModel}
\delta_h^{\rm model}(\vk) = c(\vk)+\sum_i \beta_i(k) \mathcal{O}_i(\vk),
\end{align}
i.e.~linear in the bias transfer functions $\beta_i$, 
the minimization of $\Perr(k)$ in each $k$ bin is equivalent to linear regression or ordinary least squares in each $k$ bin, which gives
\begin{align}
\label{eq:LinRegr}
  \vec{\beta}(k) = \mathbb{O}^{-1}(k)\big\langle 
\bm{\mathcal{O}}(\vk)
\left[\delta_h^\mathrm{truth}(\vk)-c(\vk)\right]^*\big \rangle.
\end{align}
Here, $\mathbb{O}_{ij}(k)\equiv \langle \mathcal{O}_i(\vk) \mathcal{O}_j^*(\vk)\rangle$ is the covariance matrix between the model operators $\mathcal{O}_i$ in a $k$ bin, and $\mathbb{O}^{-1}(k)$ is the inverse of this matrix in that $k$ bin.\footnote{Different $k$ bins are uncorrelated because all model operators $\mathcal{O}_i$ are statistically isotropic and homogeneous.
}
As described above we orthogonalize these model operators using Gram-Schmidt orthogonalization \eq{GramSchmidt} so that the covariance matrix is diagonal for every $k$.  
\MSNEW{These scale-dependent transfer functions yield the model with the lowest possible noise when compared against the simulated halo density. }
We then fit these orthogonalized transfer functions using perturbation theory as described in \secref{PTTransferFunctions} below\MSNEW{, and test if the noise is close to the minimal one and can be described by a constant.}
\vskip 4pt 

Similarly to the measured model error, the transfer functions determined in this way avoid sample variance.
Related methods have also been used to model the displacement field \cite{TobiasDMDisplacement1505},  the nonlinear dark matter density \cite{TobiasDMDensity1507,Taruya1807}, or the 21cm radiation from reionization \cite{1806.08372}.
While one could include regularization or prior terms like $\sum_i\beta_i^2$ in the minimization, we find no need for this if fields are orthogonalized.

\MyFloatBarrier
\section{Simulation Results in Position Space}
\label{se:PositionResults}

We start the comparison of the bias models against simulations in position space in this section, turning to Fourier space in the subsequent section.

\subsection{Two-Dimensional Slices}

\begin{figure}[tp]
\centering
\includegraphics[width=0.6\textwidth]{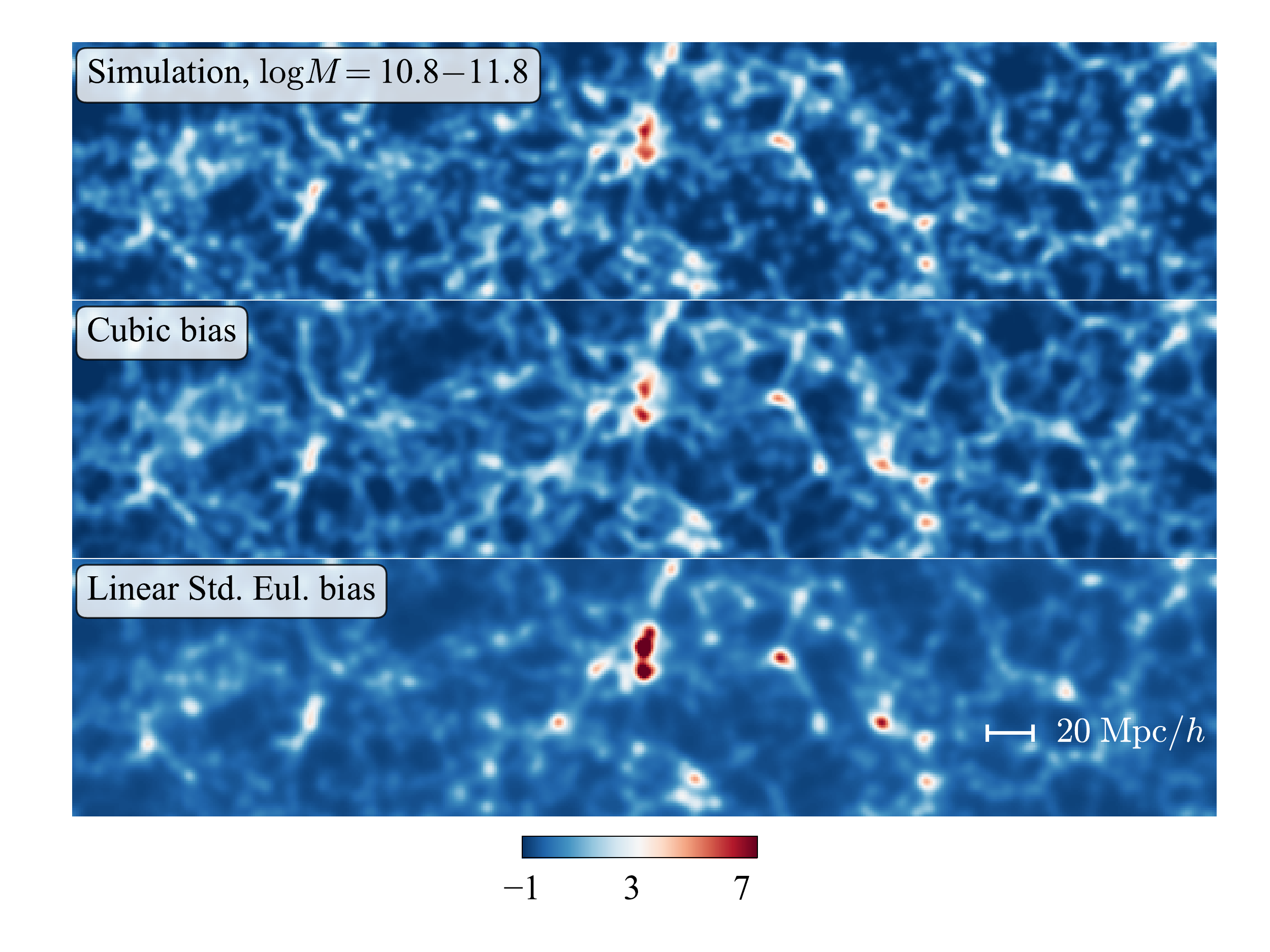}
\caption{2-d slices of the overdensity $\delta_h(\vx)$ of  simulated $10^{10.8}-10^{11.8}\hMsun$ halos (top), compared with the cubic bias model (center), and the linear Standard Eulerian bias model (bottom).
Each panel is $500\hMpc$ wide and $110\hMpc$ high, and each density is smoothed with a $R=2\hMpc$ Gaussian, $W_R(k)=\exp[-(kR)^2/2]$.
The colorbar indicates the values of this smoothed overdensity $\delta_h(\vx)$.
}
\label{fig:slices1}
\end{figure}

\fig{slices1} shows two-dimensional slices of the 3-d overdensity of halos $\delta_h(\vx)$ in one of the simulations, compared with two of the bias models. 
This shows that the cubic bias model provides an accurate description of the density contrast of these halos, with minor differences only visible on rather small scales.
The linear Standard Eulerian bias provides a less accurate description, but still gets most of the structure on large scales right.
\vskip 4pt 

\begin{figure}[tp]
\centering
\includegraphics[width=0.6\textwidth]{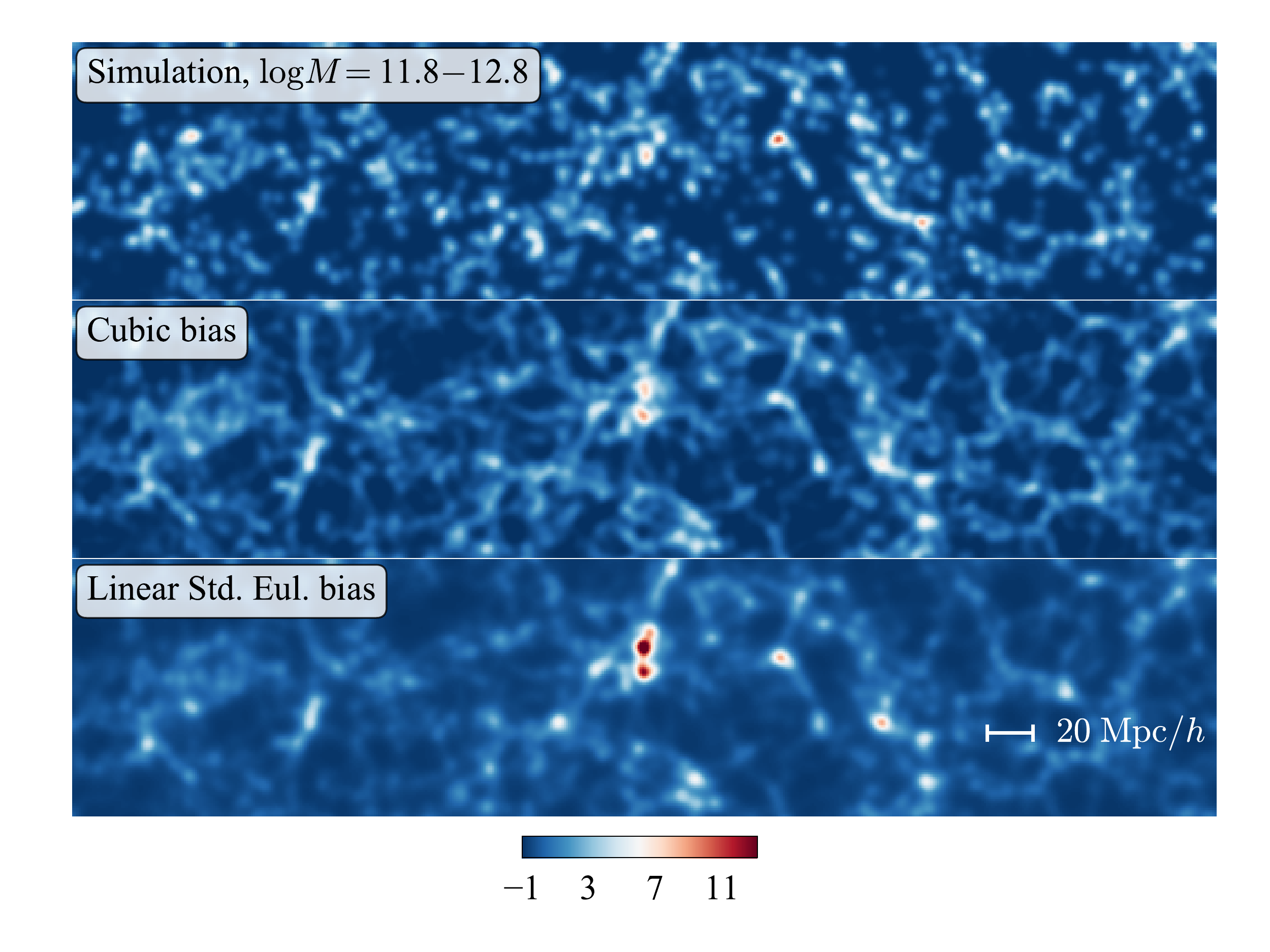}
\caption{Same as \fig{slices1} but for more massive and less abundant $10^{11.8}-10^{12.8}\hMsun$ halos.}
\label{fig:slices2}
\end{figure}

For more massive and less abundant halos, we obtain \fig{slices2}.  
The cubic model is less successful for these halos, especially on small scales.
For example, the model predicts a large spherical overdensity up from the center of the slice, but this does not exist for these halos in the simulation; in many other regions the model tends to underpredict the peaks of the true halo overdensity.  This is even more severe for the linear Standard Eulerian bias model, and for more massive halo populations.
On large scales, however, the models still work well, as we will see more clearly when we turn to Fourier space later.

\subsection{One-Point Probability Distribution}
\label{se:1ptPDF}

\begin{figure}[tp]
\centering
\includegraphics[width=0.8\textwidth]{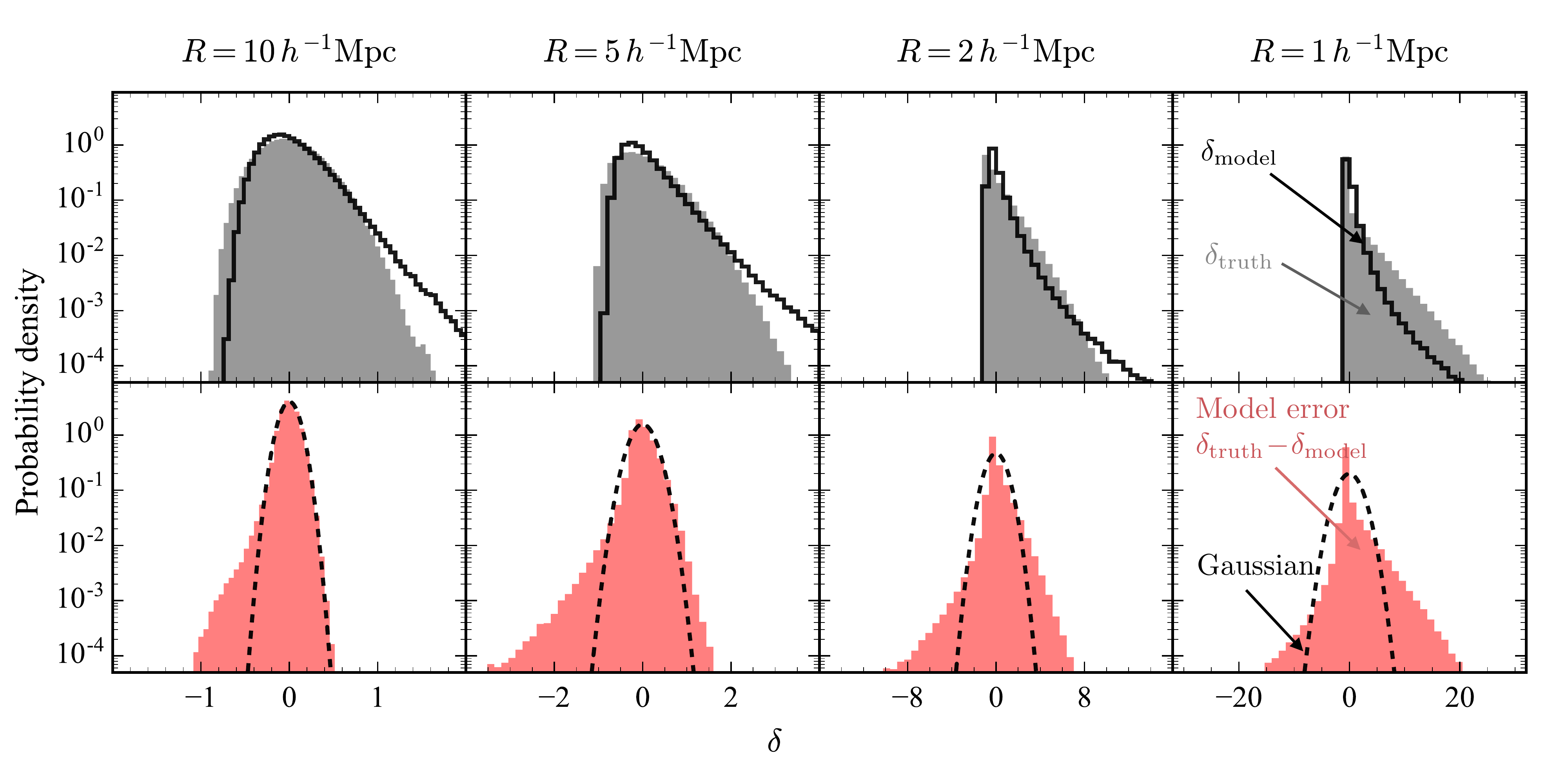}
\caption{Histogram of  simulated $10^{11.8}-10^{12.8}\hMsun$ halo overdensity (grey shaded area in upper panel) and the linear Standard Eulerian bias model generated on a 3-d grid (solid black curve in upper panel).
Different  columns represent different Gaussian smoothing scales $R$ applied to these densities.
Lower panels show the model error, $\hat\epsilon=\delta_h^\mathrm{truth}-\delta_h^\mathrm{model}$ (red shaded area), and a Gaussian curve (black dashed) using the sample variance of the model error.
All curves are normalized to integrate to unity.
The variance, skewness, and kurtosis of the true halo overdensity and the model error are reported in Tables~\ref{tab:FullVarSkewKurt} and \ref{tab:ErrorVarSkewKurt}.
The linear model tends to overpredict the peaks and underpredict the troughs of the halo density. 
}
\label{fig:histLinearModel}
\end{figure}

\begin{figure}[tp]
\centering
\includegraphics[width=0.8\textwidth]{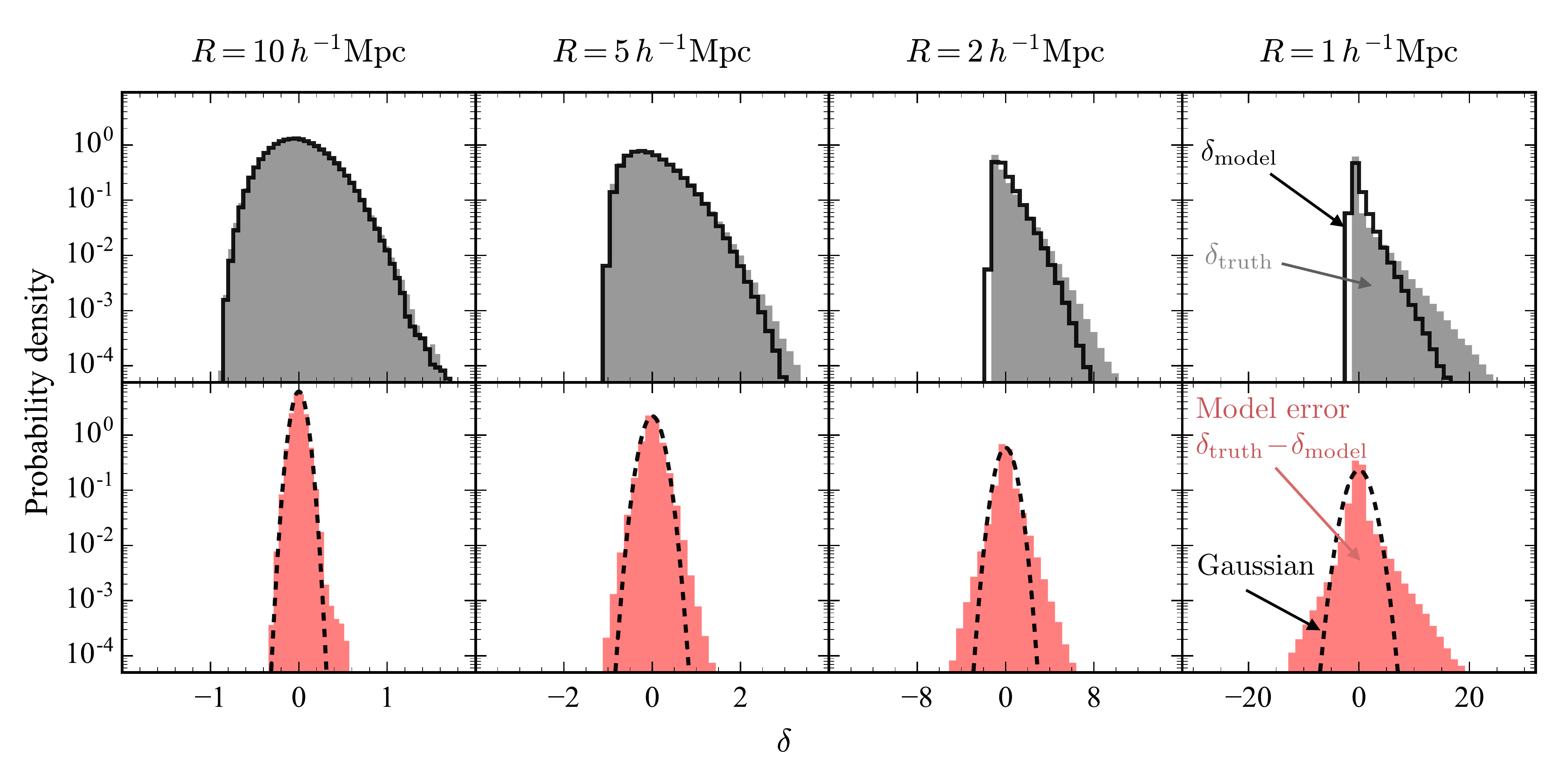}
\caption{Same as \fig{histLinearModel} but using the cubic  bias model. 
This provides a better description of the one-point pdf of the simulated halos than the linear model in \fig{histLinearModel}, showing that nonlinear bias terms improve the description of the one-point pdf as expected.
}
\label{fig:histCubicModel}
\end{figure}

\begin{table}[tbp]
\centering
\renewcommand{\arraystretch}{1.1}
\begin{tabular}{@{}ccccccccccccccccccc@{}}
\toprule
 & \phantom{} & \multicolumn{5}{c}{Variance[$\delta_h$]} & \phantom{$\quad$} &  \multicolumn{5}{c}{Skewness[$\delta_h$]} & \phantom{$\quad$} & \multicolumn{5}{c}{Kurtosis[$\delta_h$]} \\ 
$R$
& \phantom{$\quad$} & Linear & \phantom{} &  Cubic & \phantom{} & Truth 
& \phantom{} & Linear & \phantom{} &  Cubic & \phantom{} & Truth 
& \phantom{} & Linear & \phantom{} &  Cubic & \phantom{} & Truth 
\\ 
\colrule 
10  && 0.29 && 0.3 && 0.31 &&     0.97 && 0.33 && 0.35  &&    1.9 && -0.051 && 0.0012   \\
5  && 0.5 && 0.53 && 0.56 &&      2.0 && 0.78 && 0.83 &&      8.3 && 0.53 &&  0.7 \\
2  && 0.91 && 1.1 && 1.2 &&      5.2 && 1.6 &&  2.0 &&        70 &&  3.6  && 5.4 \\
1  && 1.3  && 1.7 &&  2.4 &&      11 && 2.7 &&   3.9 &&       320 && 12  && 20 \\
\botrule
\end{tabular}
\caption{Variance, skewness, and kurtosis of the halo density from the linear Standard Eulerian bias model, the cubic model, and the simulations (``Truth''), for $10^{11.8}-10^{12.8}\hMsun$ halos, after smoothing the density with different smoothing scales $R$ (given in units of $\hMpc$).
The skewness and kurtosis are computed as $\langle\delta^3\rangle/\langle\delta^2\rangle^{3/2}$ and $\langle\delta^4\rangle/\langle\delta^2\rangle^2-3$, respectively;  both vanish for a Gaussian distribution.
}
\label{tab:FullVarSkewKurt}
\end{table}

\begin{table}[tbp]
\centering
\renewcommand{\arraystretch}{1.1}
\begin{tabular}{@{}ccccccccccccc@{}}
\toprule
 & \phantom{} & \multicolumn{3}{c}{Variance[$\hat\epsilon$]} & \phantom{$\quad$} &  \multicolumn{3}{c}{Skewness[$\hat\epsilon$]} & \phantom{$\quad$} & \multicolumn{3}{c}{Kurtosis[$\hat\epsilon$]} \\ 
$R$
& \phantom{$\quad$} &   Linear & \phantom{} & Cubic 
& \phantom{} &   Linear & \phantom{} & Cubic 
& \phantom{} &   Linear & \phantom{} & Cubic 
\\ 
\colrule 
10  && 0.099 && 0.065  && -0.42&& 0.13 &&  3.2&& 0.68  \\ 
5  && 0.25 && 0.18&& -0.51 && 0.31  && 11&& 2.7  \\
2  && 0.85&& 0.67&&    0.0081 && 0.92 &&  34&& 8.8 \\
1  && 2 && 1.7 &&  1.7&& 2.3 &&  53&& 22 \\
\botrule
\end{tabular}
\caption{Variance, skewness, and kurtosis of the model error $\hat\epsilon=\delta_h^\mathrm{truth}-\delta_h^\mathrm{model}$ for the linear Standard Eulerian bias model and for the cubic model, for different smoothing scales $R$.  The skewness and kurtosis of the model error tend to be smallest for large smoothing scale $R$, as expected.
}
\label{tab:ErrorVarSkewKurt}
\end{table}

To get a more global view of the position-space halo density we estimate its one-point probability distribution by computing the histogram of the halo density for different smoothing scales.  \fig{histLinearModel} compares the simulations against the linear Standard Eulerian bias model evaluated on the 3-d grid, while \fig{histCubicModel} compares against the  cubic bias model. We focus on the  halos corresponding to \fig{slices2} where we found clearly visible differences between models and simulations.  
The variance, skewness and kurtosis of the densities shown in the histograms are listed in Table~\ref{tab:FullVarSkewKurt} for the full simulated and modeled densities, and in Table~\ref{tab:ErrorVarSkewKurt} for the model error.
\vskip 4pt 

The linear Standard Eulerian bias model tends to  underpredict troughs and overpredict peaks of the halo density, as shown in \fig{histLinearModel}.  The model error is not Gaussian for any of the shown smoothing scales; in particular its kurtosis is larger than 1 for all smoothing scales.
\vskip 4pt 

The cubic model provides a more accurate description of the halo density pdf, as shown in \fig{histCubicModel}.
This emphasizes the importance of using nonlinear bias terms even on rather large scales.
Still, the cubic model tends to underpredict the peaks of the true halo density, especially on small scales. 
This agrees with \fig{slices2} where the model also underpredicts the simulated density in more regions than it overpredicts it (considering only overdense regions that are easiest to pick up by eye).
Related to this, the variance, skewness and kurtosis of the cubic model halo density are similar to that of the true simulated density, especially for large smoothing scale (see Table~\ref{tab:FullVarSkewKurt}).
The model error of the cubic model looks most Gaussian for large smoothing scales, but it is never completely Gaussian, with a skewness of $0.13$ and a kurtosis of $0.68$ even for $R=10\hMpc$ smoothing.
Most of this is caused by the tails of the distribution, i.e.~by outliers of $\epsilon$.
Quantifying the non-Gaussianity of the error in more detail, for example by measuring bispectra, would be interesting. 
In what follows we will only consider the power spectrum of the error however.

\FloatBarrier
\section{Simulation Results in Fourier Space}
\label{se:FourierResults}

The one-point pdf and histograms shown above quantify the number of pixels where model and simulation density have the same value. \MSNEW{Even if there is a good match between model and simulations, the densities might not be spatially coherent and differ at the pixel by pixel level \cite{RothPorciani2011}.}
To test this, we turn to Fourier space and compute two performance measures quantifying the size of the model error mode by mode: 
First, in \secref{ResultsPerr}, we compute the model error power spectrum $\Perr(k)=\langle|\delta_h^{\rm truth}(\vk)-\delta_h^{\rm model}(\vk)|^2\rangle$ for the simulated halos as introduced in the introduction.
Second, in \secref{rcc}, we discuss the cross-correlation coefficient 
\begin{align}
  \label{eq:rccDef}
  r_{cc}(k) \equiv \frac{\langle\delta_h^{\rm model}(\vk)[\delta_h^{\rm truth}(\vk)]^*\rangle}{\left(
\langle|\delta_h^{\rm model}(\vk)|^2\rangle
\langle|\delta_h^{\rm truth}(\vk)|^2\rangle
\right)^{1/2}}
\end{align}
between Fourier modes of the model and simulated (truth) halo density.
As we are going to see in \secref{CosmoReln}, the size of the model error $\Perr$ and the cross-correlation coefficient $r_{cc}$ are directly related to the amount of cosmological information that can be extracted when using the model to describe a measurement of the halo density. (Also, $\Perr$ and $r_{cc}$ are closely related to each other by relations given in \app{PerfMeasures}.)
\vskip 4pt

Following these results on the size of the model error and the cosmological constraining power, we proceed in \secref{ScalDepnPerr} to investigate the scale dependence of the model error, which, if ignored, can lead to biases of cosmological parameter measurements.
In particular, we determine the maximum wavenumber $k_{\rm max}$ up to which it is safe to assume a scale-independent model error power spectrum or shot noise.

\vskip 4pt
We end the section by showing how large the contribution from the different bias terms is to the total model as a function of wavenumber, demonstrating the importance of including nonlinear bias terms.

\vskip 4pt
Throughout the section, $P_{\rm model}$ and $P_{\rm truth}$ refer to the halo power spectrum of the model and simulations, respectively.
As described in the introduction, our measurements differ quantitatively from previous measurements of stochasticity because we work at the field level and include nonlinear bias terms in the perturbative model.

\MyFloatBarrier
\subsection{Size of the Model Error}

\subsubsection{Model Error Power Spectrum}
\label{se:ResultsPerr}

\begin{figure}[htp]
\centering
\includegraphics[width=0.7\textwidth]{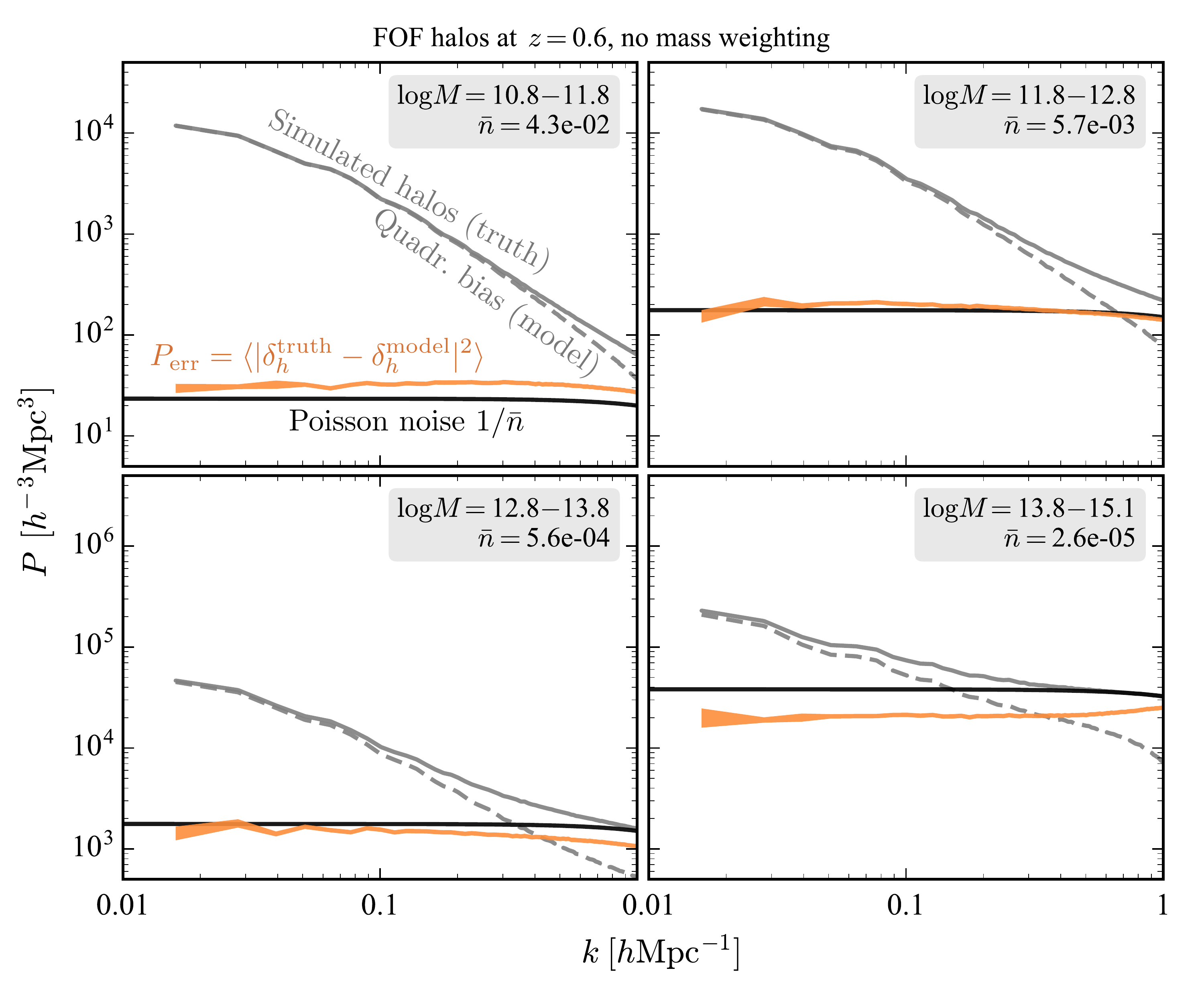}
\caption{Broadband power spectra of the simulated halo overdensity (solid grey), the best-fit quadratic bias model of that overdensity (dashed grey), and the field-level difference between simulation and model (orange), which represents the mean-square model error or error power spectrum $\Perr(k)=\langle|\delta_h^{\rm truth}(\vk)-\delta_h^{\rm model}(\vk)|^2\rangle$. Different panels show different halo mass bins.
The amplitude of the model error is larger than the Poisson prediction for low mass halos and smaller for high mass halos, and it is rather scale-independent.
The results are averaged over five independent simulations.  The uncertainty of $\Perr$ estimated from the scatter between these simulations is indicated by the width of the shaded orange region at low $k$, and it is smaller than that at high $k$.
}
\label{fig:M0PkBB}
\end{figure}

\fig{M0PkBB} shows the broadband power spectra of the four halo mass bins of simulated halos, and the best-fit model for one of the bias models introduced above (the quadratic bias model). 
The mean-square difference between the simulation and model density, given by the error power spectrum $\Perr(k)$, is shown in orange.
It is rather flat as a function of $k$, and it deviates from the Poisson prediction by up to a factor of 2, depending on halo mass.
\vskip 4pt 

\begin{figure}[htp]
\centering
\includegraphics[width=0.7\textwidth]{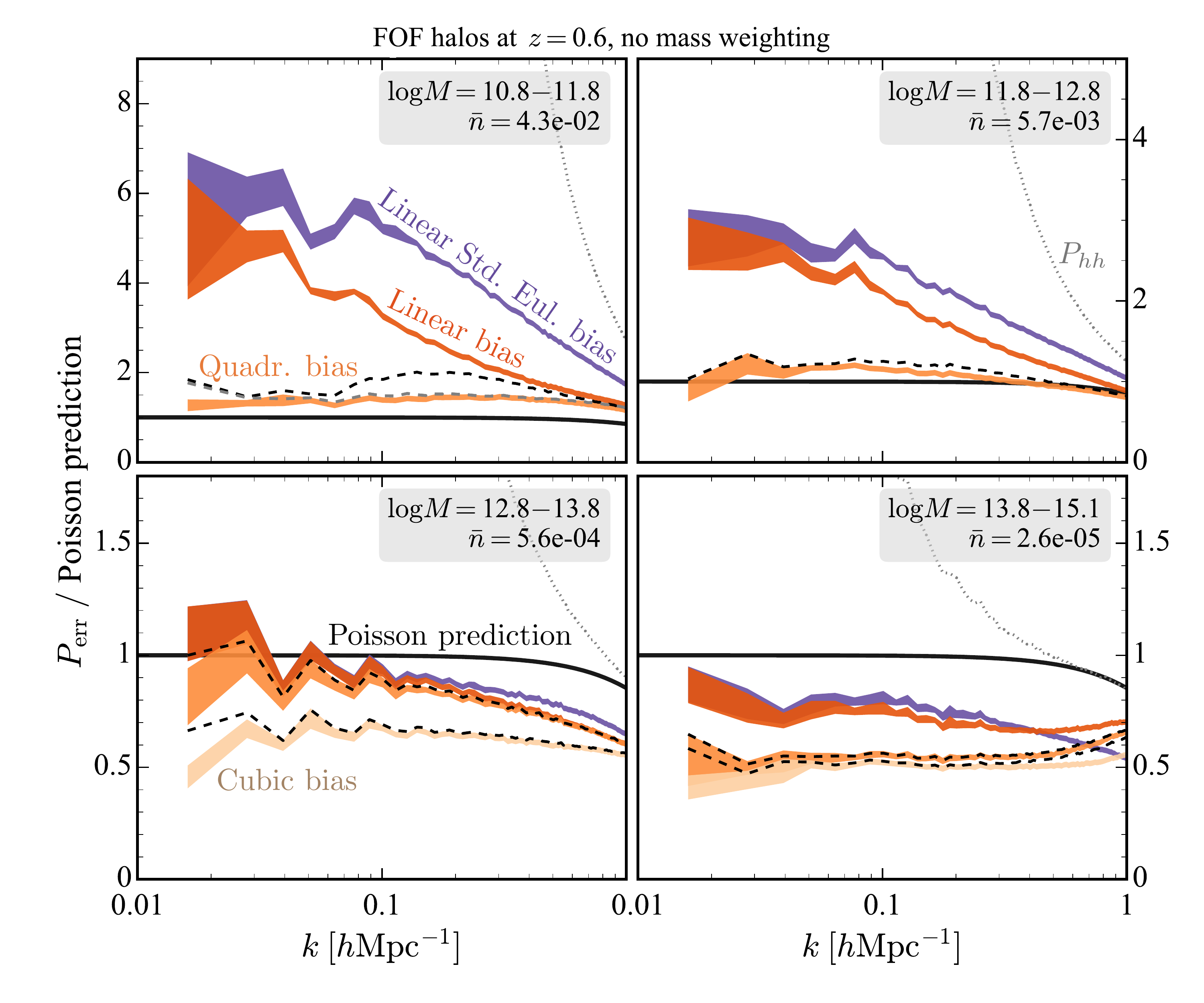} 
\caption{Mean-square model error $\Perr(k)=\langle|\delta_h^{\rm truth}(\vk)-\delta_h^{\rm model}(\vk)|^2\rangle$ of different halo bias models, divided by the Poisson prediction $1/\bar n$.
Different panels show different halo mass bins; different colors represent different bias models.
For all mass bins the quadratic and cubic bias models have the smallest large-scale model error and the smallest scale dependence.
Shaded areas represent the 1$\sigma$ credibility interval if bias transfer functions are allowed to be free functions of $k$ (the uncertainty is computed as the standard error of the mean estimated from the scatter between the five independent simulations).
If we instead fit these transfer functions using five $k$-independent parameters $b_1$, $c_s$, $b_{\Gamma_3}$, $b_2$, and $b_{\mathcal{G}_2}$, we obtain the dashed curves for the quadratic and cubic bias models.
For the densest halo sample (top left panel), keeping $\delta_Z$ as an extra field in the quadratic model without a transfer function yields the grey dashed curve when fitting the other transfer functions with the theory model.
For the two most massive halo samples (lower panels), we include the cubic bias model with $\shifted \delta_3(\k)$, which helps to describe these halos.
The small suppression of all curves at high $k$ is due to the CIC window used to paint particles to the grid.
}
\label{fig:noiseM0}
\end{figure}

Our goal is to study the amplitude and the scale dependence of the model error in more detail and also for the other halo bias models introduced previously in \secref{BiasList}. 
For this purpose we show $\Perr$ divided by the Poisson prediction $1/\bar n$ in \fig{noiseM0}.
\vskip 4pt

Let us first discuss the low-mass halos, $M\leq 10^{12.8}\,h^{-1}\mathrm{M}_\odot$.
We find that for the linear bias models,  the mean-square model error is larger than the Poisson prediction by a factor of a few, and it is rather scale-dependent, even on large scales.
In contrast,  the mean-square model error of the  quadratic bias model deviates only by a few tens of percent from the Poisson prediction, and is rather scale-independent, with some scale dependence only visible at $k\gtrsim 0.2\ihMpc$.
This shows that including the quadratic bias terms $\tilde\delta_2$ and $\tilde\G_2$ reduces the mean-square model error on large scales by a factor of 4 to 6 and reduces its scale dependence.
\vskip 4pt 

For more massive halos and clusters, $M>10^{12.8}\,h^{-1}\mathrm{M}_\odot$, we find that  the mean-square model error of the quadratic and cubic bias models is smaller than the Poisson prediction $1/\bar n$ by up to a factor of 2, which is about $30\%$ smaller than the mean-square model error of the linear bias models for these halos.
Qualitatively similar sub-Poissonian errors for heavy halos have been reported in the literature before \cite{Baldauf1305,Modi1612,GinzburgDesjacques1706}.
This is theoretically expected because of the self-exclusion and clustering of halos \cite{CasasMiranda2002,Baldauf1305,Baldauf1510}, which violate the assumption of placing point particles randomly in space (sampling the continuous density uniformly).
Although less clearly visible than for the low-mass halos, the model error of the nonlinear bias models is again less scale-dependent than the model error of linear bias, which deviates  by tens of percent from a $k$-independent shot noise at $k\gtrsim 0.1\ihMpc$,
\vskip 4pt

The model errors shown in color in \fig{noiseM0} represent the minimum mean-square model error if  the transfer functions $\beta_i(k)$ of the bias models are allowed to be free functions of $k$, obtained using linear regression in each $k$-bin as described in \secref{MeasureBias} above.
If we instead restrict the functional form of these transfer functions to a theory prediction by fitting the linear regression transfer functions $\beta_i(k)$ using five $k$-independent parameters $b_1$, $c_s$, $b_{\Gamma_3}$, $b_2$, and $b_{\mathcal{G}_2}$ (see \secref{TkFits} below for details), we obtain the black dashed curves in \fig{noiseM0} in the case of the quadratic and  cubic bias model (for the latter we fit $\beta_3(k)$ with a constant sixth parameter).
This more conservative model error is only minimally larger than before, which shows that the  transfer functions can be well described with a 5- or 6-parameter fit as we are going to see in more detail in \secref{TkFits} below.
\vskip 4pt

For the lowest halo mass bin, shown in the top left panel of \fig{noiseM0}, we show two dashed lines corresponding to the quadratic bias model. The difference between them is whether or not the Zel'dovich density $\delta_Z$ is absorbed in the bias expansion using \eqq{deltaZshifted}. The grey dashed curve is obtained keeping $\delta_Z$ explicitly in the bias expansion as an extra field with the fixed transfer function. In this case the noise is somewhat different with respect to the standard second order bias model, which implies that $\shifted\G_3$ and higher-order terms in the expansion of the Zel'dovich field become important. This is not surprising, since the amplitude of the noise for the lowest halo mass bin is very small and comparable to $\langle |\shifted \G_3(\k)|^2\rangle$ around $k\sim 0.1\ihMpc$. Our results suggest that in the limit of very low shot noise it is better to keep $\delta_Z$ explicitly in the bias expansion because this leads to an error with smaller amplitude and scale dependence. One may wonder whether this is consistent, given that we are anyway neglecting higher order bias operators. One way to justify keeping the Zel'dovich density field explicitly is to note that the coefficients in the expansion of $\delta_Z$ in terms of shifted operators are possibly significantly larger than typical Lagrangian bias parameters. It would be interesting to further explore this question. However, in cases with realistic halo masses the difference between the two approaches is very small compared to the amplitude of the shot noise. 
\vskip 4pt

Of course, it is a well known result from the literature that nonlinear bias is required to describe summary statistics such as the galaxy power spectrum or bispectrum on mildly nonlinear scales.
For example, analyses of data from the recent SDSS BOSS galaxy survey found that nonlinear bias terms are required to model their measurements \cite{2014MNRAS.443.1065B,2015MNRAS.451..539G,Beutler1607,2017MNRAS.464.1640S,2017MNRAS.470.2617A}.
It is therefore not surprising that we also find nonlinear bias to be important when comparing at the field level.
What is more surprising is that $\tilde\delta_2$ has a nearly constant auto-power spectrum  on large scales (see \fig{M0ModelContrisBB} below), but nevertheless it describes part of the true halo density on large scales, substantially lowering the large-scale model error.
As we will find in \secref{EulerianFailure}, this is a consequence of working with the shifted operator $\tilde\delta_2$; when instead working with the squared nonlinear Eulerian dark matter density, as is done in the Standard Eulerian bias expansion, the resulting field is dominated by UV modes and does consequently not correlate well with the true halo density on large scales.
\vskip 4pt

Overall we have shown in this section that the quadratic bias model performs substantially better than the linear models, because its model error is smaller and less scale-dependent.
As we are going to discuss in \secref{EulerianFailure} later, the quadratic bias model also performs better than nonlinear Standard Eulerian bias models, because it avoids squaring the nonlinear dark matter density and expanding large bulk flows.

\subsubsection{Correlation Coefficient}
\label{se:rcc}

\begin{figure}[tp]
\centering
\includegraphics[width=0.7\textwidth]{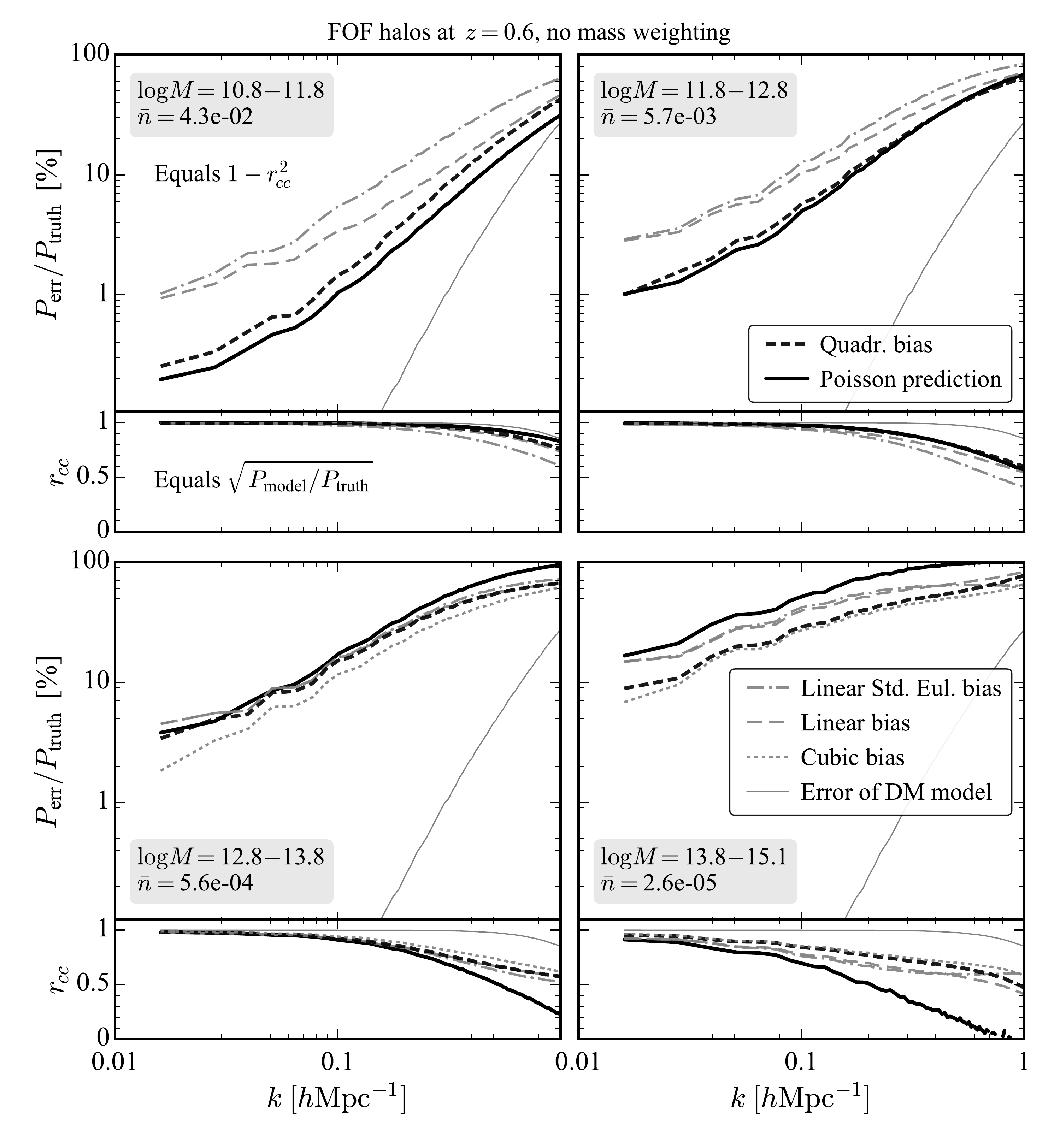}
\caption{Mean-square model error $P_\mathrm{err}(k)=\langle|\delta_h^\mathrm{truth}(\vk)-\delta_h^\mathrm{model}(\vk)|^2\rangle$ divided by the true halo power spectrum from simulations (upper subpanels), and  cross-correlation coefficient $r_{cc}=r_{cc}(\delta_h^\mathrm{truth},\delta_h^\mathrm{model})$ between simulations and model (lower subpanels).
This demonstrates that the correlation between model density and  halo density in simulations is similar to that expected from the Poisson prediction on all scales.
In detail, it is somewhat larger than that for low-mass halos and somewhat smaller for high-mass halos, because $\Perr$ deviates from $1/\bar n$ as shown previously.
The curves in the upper subpanels coincide with $1-r_{cc}^2$, and the curves in the lower panel are equal to $\sqrt{P_\mathrm{model}/P_\mathrm{truth}}$, because the model transfer functions minimize $\Perr$ (see \app{PerfMeasures}).
The  cubic bias model is only shown for the two heaviest halo samples because they are identical to the quadratic  bias model for the other halo samples.
}
\label{fig:M0_1mr2}
\end{figure}

A related question is how well the model density is correlated with the simulated halo density. 
This is shown in \fig{M0_1mr2}.
For the lightest halos, we find that the model and simulated halo density are more than $75\%$ correlated at $k\leq 1\ihMpc$, and more than $99.5\%$ correlated at $k\leq 0.08\ihMpc$ for the quadratic bias model, which is similar to the  level of correlation expected from Poisson shot noise for the number density of these halos.
For the heavier and less abundant halo samples, the cross-correlation coefficient is lower, as expected, because the shot noise is larger.
The linear bias models are less correlated with the simulated halo density than the quadratic and cubic bias model are, reflecting their larger model error (this is best seen in the upper panels of \fig{M0_1mr2} which show $1-r_{cc}^2$).
\vskip 4pt

Perhaps surprisingly, the quadratic bias model is more than $50\%$ correlated with the simulated halo density for all halo mass bins up to $k=1\ihMpc$.
This implies that, even on such small scales, which are expected to be well inside the one-halo term regime, there is significant information about the phases of the linear initial conditions that are used to generate the bias model density.
One might think that this is impossible, at least for the most massive halos, because the radius of these halos is larger than $1\hMpc$ and modes smaller than the halo radius are virialized, which should destroy any memory of the initial conditions.
A possible explanation could be that we know the center of mass positions of these massive halos very accurately (to a fraction of the halo radius), which can probe modes on scales smaller than the radius of these halos.\footnote{We can tell if the halos are located at positions $x$ and $y$ or $x$ and $y+\Delta$, where $\Delta$ is only limited by the resolution with which we can measure center of mass halo positions and not directly by the radius of the halos, as long as the halos are separated by a few halo radii, which is usually the case given the low number density of very massive halos.}
\vskip 4pt 

\fig{M0_1mr2} also shows the mean-square model error divided by the power spectrum of the simulated halos, which represents the fractional mean-square error of the model and coincides with $1-r_{cc}^2$ (see \app{PerfMeasures}).
For the quadratic bias model, this fractional mean-square error is less than $1\%$ for the lowest halo mass bin on large scales $k\leq 0.1\ihMpc$.
This means that the rms fluctuations of the model Fourier modes around the truth are less than $10\%$ at $k\leq 0.1\ihMpc$ for these halos.
The error increases on smaller scales and for the heavier, less abundant halos, as expected. 
\vskip 4pt 

In addition to the stochastic model error, the bias model is expected to fail on small scales because of missing 2-loop terms.
To get a rough estimate of their size,  \fig{M0_1mr2} also shows the error that the cubic bias model makes when predicting the fully nonlinear dark matter field measured from the N-body simulations, which is essentially free from shot noise (thin solid grey curve).
For the densest halo sample this error becomes comparable to the error of the bias model in describing the halo density on very small scales, but at all other scales and for all other halo samples the error of the dark matter model is much smaller.
This suggests that the model error is dominated by stochastic noise rather than missing higher order terms in the bias expansion, at least for the heavier halo samples.

\subsubsection{Relation to Cosmological Information Content}
\label{se:CosmoReln}

In the last two subsections, we have characterized the size of the model error $\hat\epsilon=\delta_h^{\rm truth}-\delta_h^{\rm model}$ and the cross-correlation coefficient $r_{cc}$ between truth and model. 
But how is this related to the cosmological information one would hope to extract when applying this model to a measurement of the halo density?
As we are going to show in this section, the size of the model error discussed above determines the amount of cosmological information one can extract from the halo density relative to the total information one would get with a perfect model.
\vskip 4pt

To see this, let us first write the true halo density as the sum of the model density and the model error,
\begin{align}
  \label{eq:24}
  \delta_h^{\rm truth} = \delta_h^{\rm model}+\hat\epsilon,
\end{align}
and assume that the model is evaluated for the optimal transfer functions that minimize $\Perr=\langle|\hat\epsilon|^2\rangle$ and enforce $\langle\delta_h^{\rm model}\hat\epsilon\rangle=0$.
Since we know how this model density depends on the linear density and therefore on cosmology, we can use it to measure cosmological parameters.  
In contrast, we do not attempt to use any potential cosmology information of the model error $\hat\epsilon$ --- otherwise we would include it in the model density.
The model error therefore acts as an uncorrelated noise contribution to the field.\footnote{In our approach the model error $\hat\epsilon$ has two contributions:
First, stochastic noise terms, which cannot be predicted given the initial condition Fourier modes on large scales.
Second, higher-order bias terms not included in the model, or more specifically, the components of these higher-order bias terms that are orthogonal to any bias term in the model (so they cannot be absorbed by transfer functions; for example $\shifted{\mathcal{G}}_3^\perp$ is part of $\hat\epsilon$ for our models).
These orthogonal higher-order bias terms do depend on cosmology, but to make use of this we would have to include them in the model.
All cosmological information that we extract from an observation of $\delta_h^{\rm truth}$ is therefore contained in $\delta_h^{\rm model}$, and the model error $\hat\epsilon$ acts as a noise contribution.
Notice that the model error is uncorrelated with the model density, $\langle\delta_h^{\rm model}\hat\epsilon\rangle=0$, because both the stochastic and orthogonal higher-order terms are orthogonal to all terms in $\delta_h^{\rm model}$. As a consequence, $P_{\rm truth}=P_{\rm model}+\Perr$.
}
The size of the model error relative to the size of the true halo density therefore determines how noisy the field is and how much cosmological information we can extract from it.
In the last two subsections we have quantified this by comparing the noise power, $\Perr=\langle|\hat\epsilon|^2\rangle$, against the power of the measurable true density, $P_{\rm truth}=\langle|\delta_h^{\rm truth}|^2\rangle$.
\vskip 4pt

To  illustrate this more clearly, consider the amplitude $A$ of the model power spectrum, $P_{\rm model}(k)\rightarrow A P_{\rm model}(k)$, as a proxy for the cosmological information content. 
How well can we determine $A$ given a measurement of $\delta_h^{\rm truth}$ modeled with $\delta_h^{\rm model}$?
This is given by the Fisher information
\begin{align}
  \label{eq:FAA1}
  F_{AA}\,=\, \sum_k 
\frac{\partial P_{\rm model}(k)}{\partial A}
\left[\frac{2\,P^2_{\rm truth}(k)}{N_{\rm modes}(k)} \right]^{-1}
\frac{\partial P_{\rm model}(k)}{\partial A}
\,=\,\sum_k \frac{N_{\rm modes}(k)}{2}\left(\frac{P_{\rm model}(k)}{P_{\rm truth}(k)}\right)^2\;,
\end{align}
where we assumed a diagonal Gaussian covariance given the observed power spectrum $P_{\rm truth}$.
We have also assumed that there are no other parameters in the analysis (in practice, one would usually need at least one parameter to describe $\Perr$, and marginalizing over this can degrade $F_{AA}$ in \eqq{FAA1} \cite{TobiasTheoError1602}).
\vskip 4pt

If the model were perfect, i.e.,~$\Perr=0$ and $P_{\rm model}=P_{\rm truth}$, \eqq{FAA1} would give the optimal amount of information, which is determined by the total number of Fourier modes.
For an imperfect model, $P_{\rm model}<P_{\rm truth}$, the ratio $P_{\rm model}/P_{\rm truth}$ in \eqq{FAA1} determines how much less information one gets per Fourier mode.
The square root of this is shown in the lower subpanels of \fig{M0_1mr2} above.
This therefore shows how close the bias expansion gets to keeping the information on the model amplitude $A$, which contains cosmological information; on large scales, it typically keeps $99\%$ or more of the cosmological information, but it retains less on smaller scales.

\vskip 4pt
Using the results from \app{PerfMeasures}, \eqq{FAA1} can be rewritten in several ways if transfer functions are chosen such that $\Perr$ is minimized.
For example,
 \begin{align}
\label{eq:FAA2}
  F_{AA}
\,=\,\sum_k \frac{N_{\rm modes}(k)}{2}
\,r_{cc}^4(k).
\end{align}
This shows that the amount of information on the amplitude $A$ is given by the correlation coefficient between the perturbative bias model (whose cosmology dependence or dependence on $A$ we know) and the true density.
We can also rewrite this as
\begin{align}
\label{eq:FAA3}
  F_{AA}
\,=\,\sum_k \frac{N_{\rm modes}(k)}{2}
\left[1-(1-r_{cc}^2(k))\right]^2\;.
\end{align}
The first term in square brackets corresponds to the optimal amount of information; the second term, $1-r_{cc}^2$, represents the fractional amount of information we lose if the perturbative model and truth are not perfectly correlated (if they are perfectly correlated, we lose no information because $1-r_{cc}^2=0$; if they are completely uncorrelated we lose $100\%$ of the cosmological information because $1-r_{cc}^2=1$).  
This is indicated by the upper subpanels in \fig{M0_1mr2} above.
This shows that the bias expansion with optimal transfer functions loses $0.2\%$ of the cosmological information at $k\simeq 0.02\ihMpc$ for the lightest and most abundant halos, while it loses more of that information on smaller scales and for the heavier and less abundant halos.
\vskip 4pt

A third way to rewrite the above formula follows from $1-r_{cc}^2=\Perr/P_{\rm truth}$ (assuming optimal transfer functions):
\begin{align}
  \label{eq:15}
   F_{AA}\,=\, \sum_k \frac{N_{\rm modes}(k)}{2}\,\left(
1 - \frac{P_{\rm err}}{P_{\rm truth}}
\right)^2\;.
\end{align}
Similarly to before, the first term in brackets gives the optimal amount of information, and the second term represents the penalty we get if the field level model error $\Perr$ is large, which is the case if stochastic noise terms or higher order bias terms not included in the model are large.
\vskip 4pt

In practice, when analyzing data from a galaxy survey, the noise power spectrum $\Perr$ is not known \emph{a priori} --- we only know this for the particular set of halos that we selected from our simulations and compared against the model density, and it is difficult to determine which halos exactly host the galaxies observed by a survey and what noise power spectrum they have.
This reflects an important difference between large-scale structure and CMB data analysis: 
For the CMB, the noise power spectrum is known if the detector noise of the experiment is known, and the noise bias it imprints on  CMB auto-power spectra can be subtracted, or it can be avoided by using cross-correlations.
In contrast, for large-scale structure, the theoretical model itself has a noise, which imprints a noise bias ($\Perr$) on the measured galaxy power spectrum. Its amplitude -- and potential scale dependence -- depend on the sample of galaxies under consideration; since they are unknown in general, the amplitude and scale dependence of $\Perr$ need to be marginalized over and $\Perr$ cannot simply be subtracted from the measured galaxy power.
Our goal in this paper is to characterize the model error and the induced power spectrum noise bias for simulated halos. 
This can serve as a guide for the expected amplitude and scale dependence of the noise bias of the galaxy power spectrum in a real survey, and, as explained above, it quantifies the amount of cosmological information retained by the bias expansion.

\subsection{Scale Dependence of the Model Error}
\label{se:ScalDepnPerr}

The above Fisher information represents the inverse \emph{variance} with which parameters like the model amplitude $A$ can be measured when modeling a measurement of the halo density with the bias expansion.
A different question is whether the resulting parameter measurements are also \emph{unbiased}.
This is not determined by the fractional size of the model error or noise relative to the size of the true halo density, but by our ability to describe the expectation value of the true halo power spectrum (or any other observable).
For that, we need to parametrize the noise power spectrum $\Perr(k)$, and ask how accurate that parametrization is, i.e.~how well the sum of $P_{\rm model}(k)$ and the parametrized $\Perr(k)$ matches the observable $P_{\rm truth}(k)$.

\vskip 4pt
A common and simple choice for data analyses is to parametrize the model error with a scale-independent constant, $\Perr(k)={\rm const}$. 
This approach is correct if $\Perr(k)$ is really independent of scale, which is
expected theoretically on scales much larger than the typical size of halos (e.g., \cite{Perko1610,Schmidt1808}),  and this is indeed what we found for the nonlinear bias models  on large scales.
But on small scales, the measured model error does depend on scale.
If that scale dependence is sufficiently strong, ignoring it  can potentially bias cosmological parameter measurements, because the scale dependence of the error could be misinterpreted as a cosmological signal.
To account for this, we either need a more general parametrization of $\Perr(k)$, or we need to exclude from data analyses all small scales $k>k_{\rm max}$ where the scale dependence of the model error is significant.
In the next subsections we will investigate the latter approach in more detail, quantifying the scale dependence of the model error and determining the $k_{\rm max}$ up to which it is safe to assume a constant $\Perr(k)$.

\subsubsection{Simulation Results}

\begin{figure}[tp]
\centering
\includegraphics[width=0.7\textwidth]{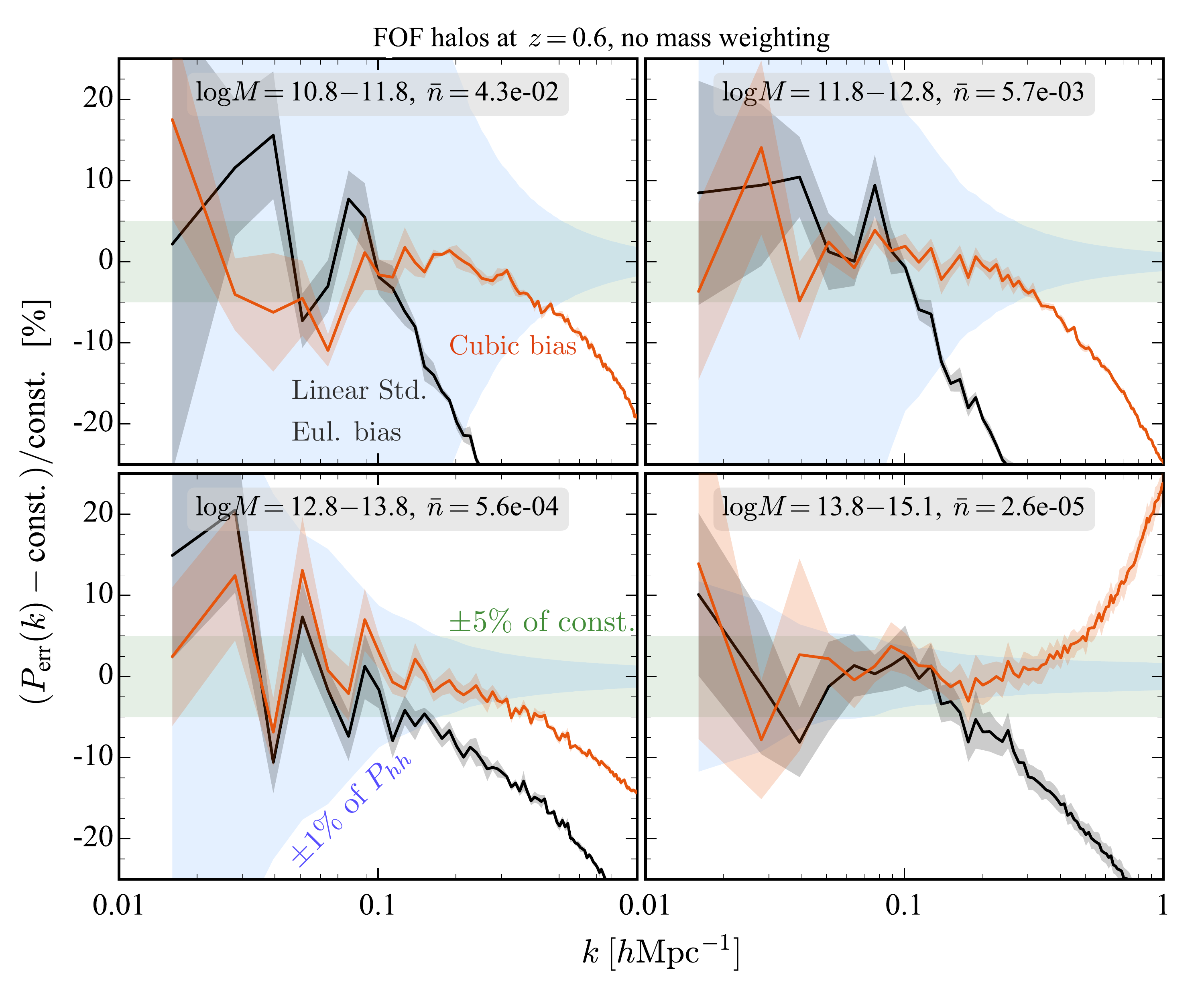}
\caption{Fractional deviation of the mean-squared model error $\Perr(k)$ from a constant in $k$, for the linear Standard Eulerian bias (black) and cubic bias (orange).
Including the nonlinear bias terms makes $\Perr(k)$ flatter at $k\gtrsim 0.1\ihMpc$, thereby 
pushing the wavenumber where the deviation of $\Perr$ from a constant exceeds $\pm 1\%$ of the halo power spectrum $P_{hh}$ (blue region), or $\pm 5\%$ of the low-$k$ constant (green region), to higher $k$.
As a consequence, including nonlinear bias terms can extend the $k$-range usable in a data analysis that assumes a $k$-independent model error or shot noise, although at the price of introducing  more free bias parameters.
In this figure and the next ones, the linear Standard Eulerian bias model uses a free transfer function $\beta_1(k)$; the cubic model instead uses the 6-parameter theory fit described in the main text, and includes the full $\delta_Z$ field.
The low-$k$ constant against which we compare is computed by tripling the width of $k$ bins to $\Delta k\simeq 0.038\ihMpc$ to reduce the noise of the measured $\Perr$, averaging over realizations, and computing the mean of this rebinned $\Perr(k)$ at $k<0.15\ihMpc$.
Shaded regions around solid curves represent the 1$\sigma$ uncertainty estimated from the scatter between the five simulations.
}
\label{fig:M0_PerrLowKSimple}
\end{figure}

\begin{figure}[tp]
\centering
\includegraphics[width=0.7\textwidth]{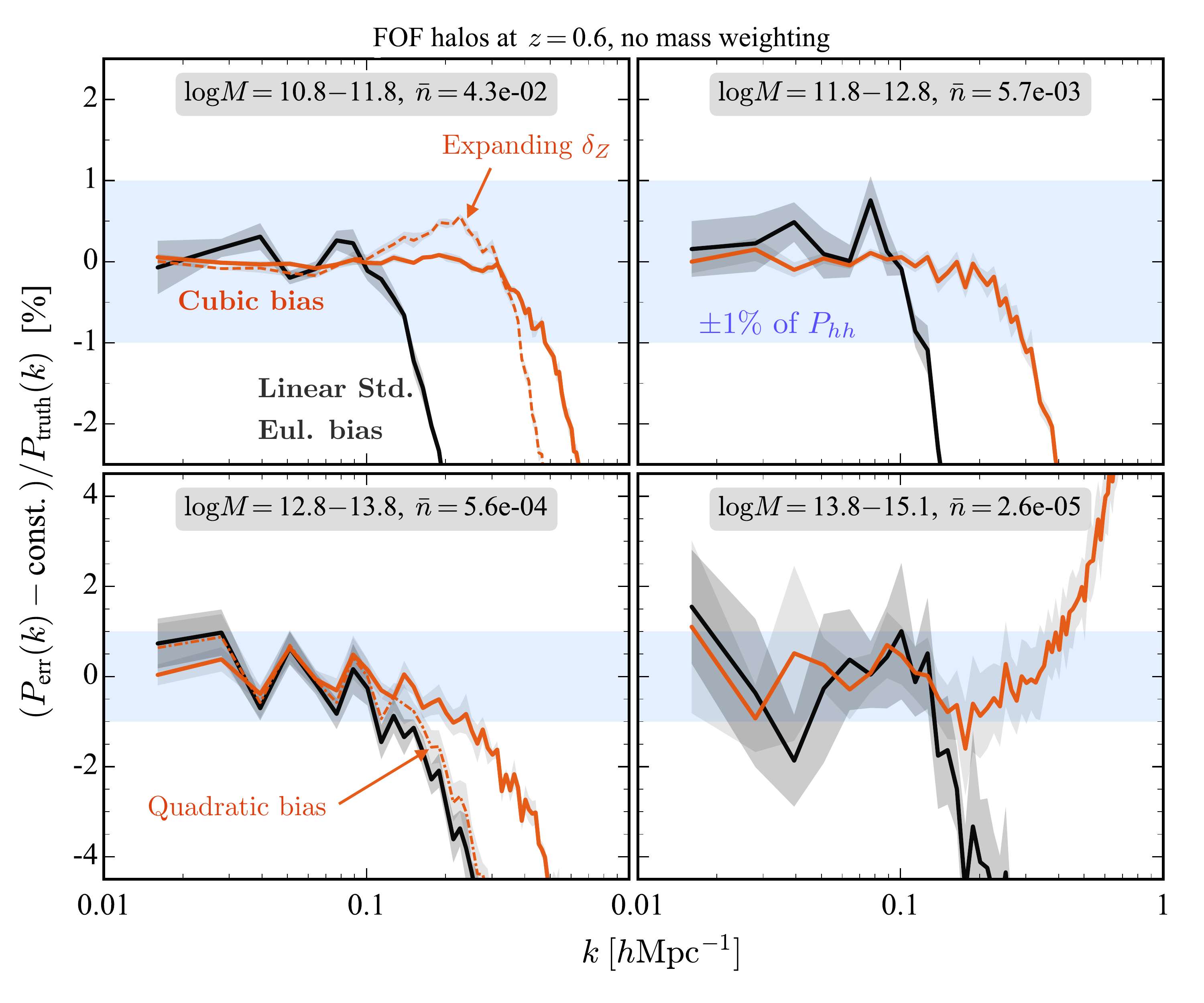}
\caption{Deviation of the model error power spectrum $P_\mathrm{err}(k)$ from a constant in $k$, relative to the halo power spectrum $P_{\rm truth}$, for the linear Standard Eulerian bias (black) and cubic  bias model (orange).
The blue band shows $\pm 1\%$ of $P_{\rm truth}$.
This shows that by including nonlinear bias terms the deviation of the model error from a constant in $k$ becomes relevant compared to the halo power spectrum at higher $k$ than for linear bias, allowing smaller scales to be included in an analysis that assumes a constant model error power spectrum.
The cubic model shown in solid orange includes the full $\delta_Z$.  
Expanding $\delta_Z$ gives a slightly larger error at $k\simeq 0.2-0.3\ihMpc$ for the halos in the top left panel (as expected theoretically for such low levels of noise; see \secref{ResultsPerr}), but does not visibly affect the curves for the halos in the other panels.
Dropping the cubic $\shifted\delta_3$ term from the model increases the scale dependence of $\Perr$ at $k\gtrsim 0.1\ihMpc$ for the halos in the lower left panel, but does not visibly affect $\Perr$ for the halos in the other panels.
As before, shaded regions around solid curves represent the 1$\sigma$ uncertainty estimated from the scatter between the five simulations.
}
\label{fig:M0_Perr-const_over_Phh}
\end{figure}

Let us start by quantifying the scale dependence of the model error.
\fig{M0_PerrLowKSimple} shows the fractional deviation of the measured model error power spectrum $P_\mathrm{err}(k)$ from the constant low-$k$ component of $\Perr$, for the linear Standard Eulerian (black) and  cubic (orange) bias model.
For the linear Standard Eulerian model, $P_\mathrm{err}(k)$ starts to deviate from a constant by more than $5\%$ at $k\simeq 0.1-0.15\ihMpc$; for the cubic model, that only happens at $k\simeq 0.3-0.4\ihMpc$ for the three low halo mass bins, and at $k\simeq 0.5\ihMpc$ for the most massive bin.
Including the nonlinear bias terms therefore increases the $k$-range where $\Perr$ is approximately constant by a factor of more than 2. 
\vskip 4pt 

We can also ask how large this scale dependence of $\Perr$ is compared to the amplitude of the measured halo power spectrum, $[\Perr(k)-\mathrm{const}]/P_{\rm truth}(k)$.
This is shown in  \fig{M0_Perr-const_over_Phh}.
For the linear Standard Eulerian bias model, the scale dependence of $\Perr$ exceeds $1\%$ of $P_{\rm truth}$ (shown in blue in \fig{M0_Perr-const_over_Phh}) again around $k\simeq 0.1-0.15\ihMpc$.
In contrast, the flatter $\Perr$ of the cubic model exceeds $1\%$ of $P_{\rm truth}$ only at $k\simeq 0.25-0.5\ihMpc$, depending on halo mass.
\vskip 4pt 

These results depend only mildly on details of the quadratic or cubic bias model.
Expanding the Zel'dovich density $\delta_Z$ in the bias model using \eqq{deltaZshifted} only has a visible effect for the lowest halo mass bin where the number density is so large that $\Perr\simeq 30\;h^{-3}{\rm Mpc}^3$ at low $k$, which is sufficiently small that corrections from expanding $\delta_Z$ become relevant.
The cubic term $\shifted \delta_3$ only affects the model error of the $10^{12.8}-10^{13.8}\hMsun$ halos; for these halos, the quadratic bias parameter $\beta_2$ is close to crossing zero and is smaller than the linear and cubic bias parameters, so that the cubic bias is relatively more important (see also \secref{SizeOfBiasTerms} and Figures \ref{fig:M0ModelContris} and \ref{fig:M0Tk} below). 
Other than that, the scale dependence of the model error with full or expanded $\delta_Z$ and with or without the cubic bias term is rather similar.

\subsubsection{Detectability of the Scale Dependence of the Model Error}
\label{se:PractRelevanceOfPerrKDepn}

The scale dependence of the model error is only relevant if it is strong enough to be statistically detectable by galaxy surveys, because in that case it may bias cosmological parameters if unaccounted for.
We therefore compute the significance with which any scale dependence of $\Perr$ could be detected in the halo power spectrum for a survey covering a volume $V_\mathrm{survey}$ and using Fourier modes up to $k_{\rm max}$. This is given by
\begin{align}
  \label{eq:SNRPerr}
\mathrm{SNR}^2(P_\mathrm{err}\neq\mathrm{const.})
\;=\;\frac{V_\mathrm{survey}}{2}
\sum_{k=k_\mathrm{min}}^{k_\mathrm{max}}
\frac{\Delta k\, k^2}{2\pi^2}\,
\left(\frac{P_\mathrm{err}(k)-\mathrm{const.}}{P_{\rm truth}(k)}\right)^2.
\end{align}
Here, a Gaussian covariance is assumed for the measured halo power spectrum $P_{\rm truth}$, and the sum is over $k$ bins with width $\Delta k$ (the result does not depend on $\Delta k$ if the binning is sufficiently fine).
As expected, the significance of the scale dependence of the model error is determined by the size of the scale dependence relative to the amplitude of the measured halo power spectrum (shown in \fig{M0_Perr-const_over_Phh}), and it increases with the survey volume and with the highest included wavenumber $k_\mathrm{max}$, because these determine the number of 3-d Fourier modes.
Importantly, this is the best-case scenario for the model error because we assume all bias parameters to be perfectly known (by matching the field level prediction against the simulations).
\vskip 4pt

\begin{figure}[tp]
\centering
\includegraphics[width=0.7\textwidth]{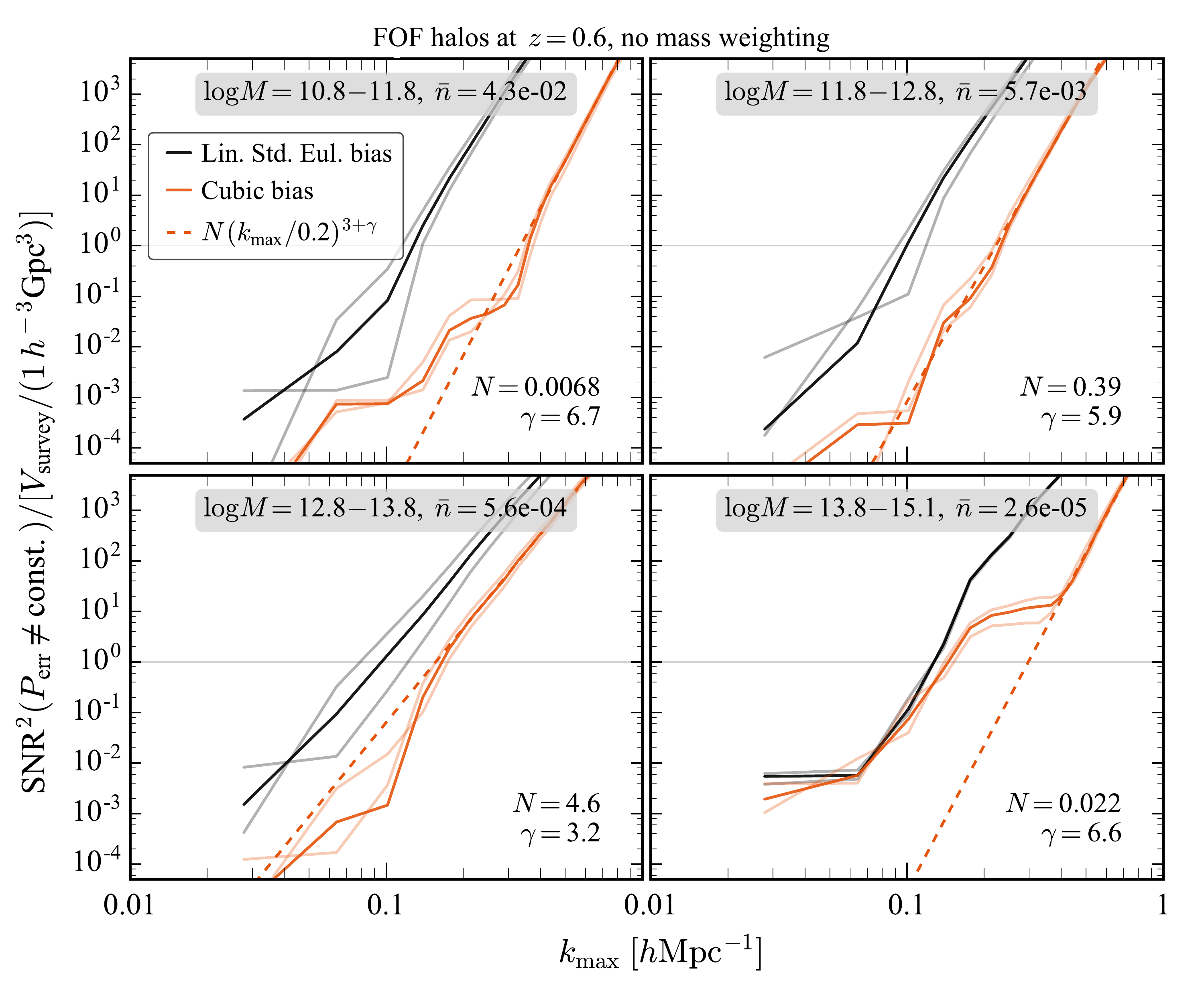}
\caption{Cumulative squared signal-to-noise-ratio for detecting a deviation of $\Perr$ from a constant in $k$, $\mathrm{SNR}^2(\Perr\ne \mathrm{const})$, in a survey volume of $1\,h^{-3}\mathrm{Gpc}^3$, as a function of $k_\mathrm{max}$, for the linear Standard Eulerian (solid black) and the cubic (solid orange) bias models.
Including nonlinear bias terms reduces the scale dependence of the model error so that it becomes only detectable at higher $k_\mathrm{max}$; this allows one to use more Fourier modes in an analysis that assumes a $k$-independent model error, at the price of having more model parameters.
For high $k_\mathrm{max}$ the $\mathrm{SNR}^2$ roughly follows a power law (dashed orange), as discussed in \secref{SNRFit}.
To suppress noise in our measurement of $\Perr$ the constant part of $P_\mathrm{err}(k)$ is determined by re-binning $P_\mathrm{err}$ and $P_{\rm truth}$ to three times wider $k$ bins than before,  averaging the result over the five simulations, and then taking the average of $\Perr$ at $k< 0.1\ihMpc$ for the linear model and $k< 0.15\ihMpc$ for the cubic model. For comparison, the semi-transparant lines show results when averaging only at $k< 0.06\ihMpc$ or $k< 0.15\ihMpc$ for the linear model, or at $k< 0.1\ihMpc$ or $k< 0.2\ihMpc$ for the cubic model.
The linear Standard Eulerian model uses the fully nonlinear dark matter density as an input, and allows the linear bias transfer function to be an arbitrary function of $k$, which provides an optimistic estimate for how well the linear Standard Eulerian bias can do.
The cubic model instead uses the 6-parameter theory fit of the transfer functions $\beta_i(k)$ described in the main text, and only requires the linear density as an input. 
We use the cubic model that includes the full $\delta_Z$ field.
}
\label{fig:SNRkmax}
\end{figure}

\begin{figure}[p]
\centering
\includegraphics[width=0.8\textwidth]{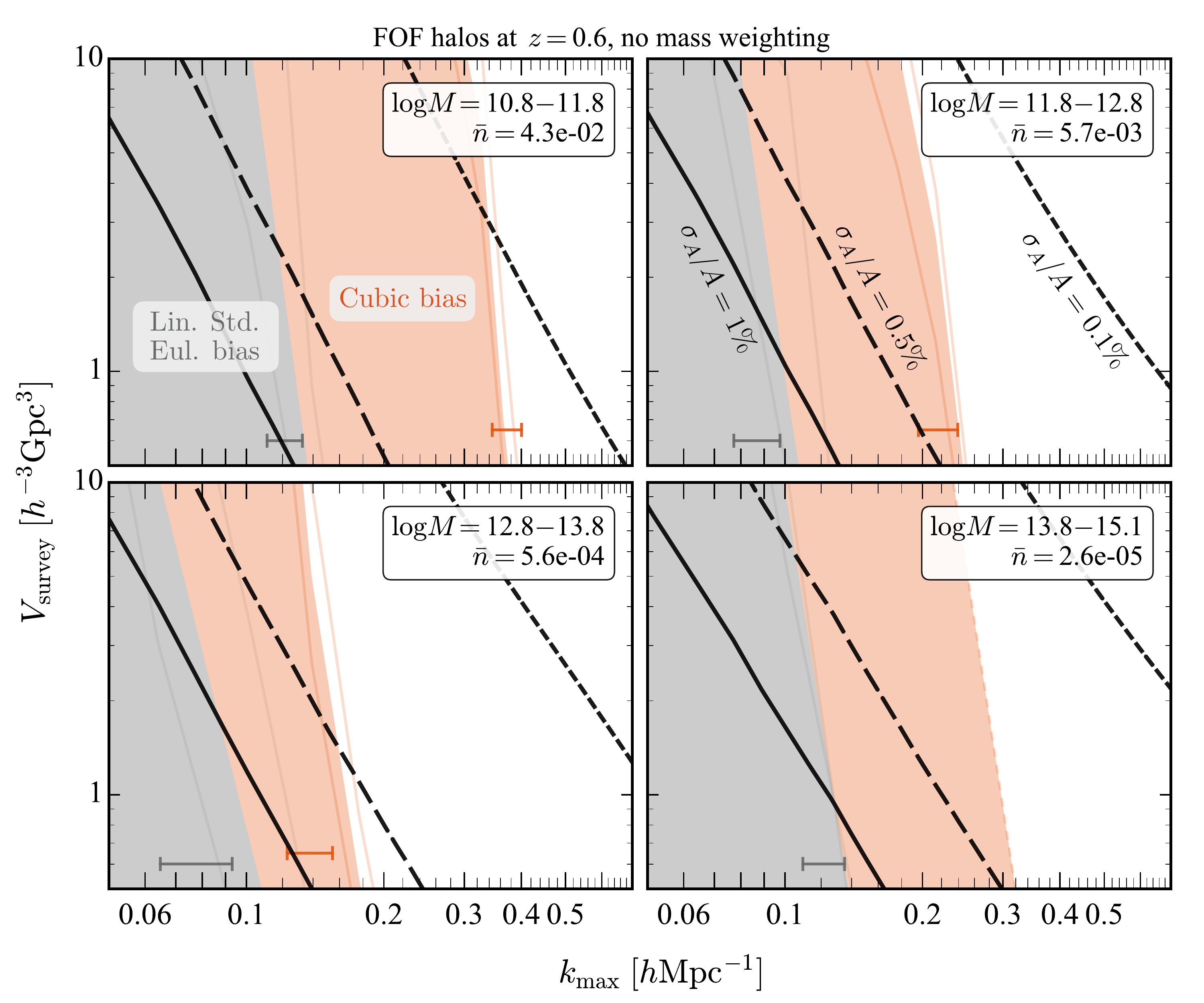}
\caption[]{Relevance of the scale dependence of the model error as in Table~\ref{tab:Kmax}, but for more survey volumes. 
For $k_{\rm max}$ and $V_{\rm survey}$ within shaded regions, it is safe to assume a scale-independent model error, $\Perr={\rm const}$.  Outside shaded regions, the scale dependence of the model error is detectable with more than 1$\sigma$, so that assuming $\Perr={\rm const}$ can bias cosmological parameters by 1$\sigma$ or more.
For the linear Standard Eulerian bias the threshold is typically at $\kmax\simeq 0.1-0.15\ihMpc$ (grey region).
In contrast, for the cubic bias model the maximum wavenumber $\kmax$ is higher by a factor of 2 to 3 (orange region).
For comparison, the bold black lines show what $k_\mathrm{max}$ is required for a given volume to measure the amplitude $A$ of the model power spectrum to $1\%$ (solid), $0.5\%$ (dashed), or $0.1\%$ (short-dashed), assuming Gaussian cosmic variance and that all other parameters (six parameters for transfer functions and one for $P_\mathrm{err}$) are perfectly known.
In the case of a $10\,h^{-3}\mathrm{Gpc}^3$ volume, the total number of halos in the four panels is $\bar n V_\mathrm{survey}=$ 430 M, 57 M, 5.6 M, and 0.26 M.
All results apply to a single redshift at $z=0.6$.
\\
Because of noise in our determination of $\Perr$, the shaded regions are somewhat uncertain. This is indicated by the semi-transparent lines, which show results when the $k$ range for determining the constant part of $\Perr$ is changed in the same way as in \fig{SNRkmax}.
Alternative estimates for the breakdown of the bias models are shown as solid horizontal lines at the bottom of each panel. These are computed as the value of $k$ where $|P_\mathrm{err}(k)-\mathrm{const}|/P_{\rm truth}(k)$ first exceeds $0.5\%$ or $1\%$ for the linear Standard Eulerian bias model, or $0.2\%$ and $0.5\%$ for the cubic model (to reduce the impact of noise we use a spline smoothed $P_\mathrm{err}$).
\\
For the linear Standard Eulerian bias we assume that the transfer function $\beta_1(k)$ is a free function $k$, whereas for the cubic bias we use the 6-parameter fit of the transfer functions $\beta_i(k)$ described in the main text.
For the cubic model in the lower right panel we use the fitting formula from \fig{SNRkmax} because it looks more robust in that case; otherwise we use the full $k$ dependence shown in \fig{SNRkmax}.
}
\label{fig:kmaxVsurvey}
\end{figure}

\begin{table}[tbp]
\centering
\renewcommand{\arraystretch}{1.1}
\begin{tabular}{@{}ccccccc@{}}
\toprule
&&&& \multicolumn{3}{c}{$\kmax\,[h\mathrm{Mpc}^{-1}]$}  \\
$\log M[h^{-1}\mathrm{M}_\odot]$ & \phantom{} & $\bar n\,[(h^{-1}\mathrm{Mpc})^{-3}]$ & \phantom{$\quad$} &  Lin.~Std.~Eul. & \phantom{} & Cubic \\ 
\colrule 
$10.8-11.8$  && $4.3\times 10^{-2}$ && 0.1 (0.14) && 0.3 (0.37) \\
$11.8-12.8$  && $5.7\times 10^{-3}$ && 0.08 (0.1) && 0.18 (0.24) \\
$12.8-13.8$  && $5.6\times 10^{-4}$ && 0.07 (0.1) && 0.13 (0.18) \\
$13.8-15.2$  && $2.6\times 10^{-5}$ && 0.1 (0.14) && 0.24 (0.32) \\
\botrule
\end{tabular}
\caption{Maximum wavenumber $\kmax$ when a scale dependence of the model error can be detected with 1$\sigma$ in a $10\,h^{-3}\mathrm{Gpc}^3$  volume (or in a $0.5\,h^{-3}\mathrm{Gpc}^3$ volume, shown in brackets), for the linear Standard Eulerian and the cubic bias models.
The $k_{\rm max}$ of the cubic model is typically 2 to 3 higher than that of linear Standard Eulerian bias.  This can improve measurements of cosmological parameters that affect the galaxy power spectrum at these scales, for example the sum of neutrino masses.
The values carry a significant uncertainty due to the noise in our measurement of the model error, as shown in \fig{kmaxVsurvey}.
}
\label{tab:Kmax}
\end{table}

We  evaluate \eqq{SNRPerr} for $V_{\rm survey}=1\,h^{-3}\mathrm{Gpc}^3$ as a function of $k_\mathrm{max}$ in \fig{SNRkmax}.
This shows that the scale dependence of $\Perr(k)$ cannot be detected for $\kmax<0.1\ihMpc$, but it becomes a 1$\sigma$ effect around $\kmax\simeq 0.1\ihMpc$ for the linear Standard Eulerian bias model and at higher $\kmax$ for the cubic model. 
\vskip 4pt 

\fig{kmaxVsurvey} shows this critical value of $\kmax$ when a deviation from a scale-independent model error becomes a 1$\sigma$ effect as a function of the survey volume.
We also summarize this in Table~\ref{tab:Kmax} for two survey volumes.
\vskip 4pt

For the linear Standard Eulerian bias we find the following. For the lowest halo mass bin, the scale dependence becomes significant for $\kmax\simeq 0.1\ihMpc$ in a $10\,h^{-3}\mathrm{Gpc}^3$ volume or for $\kmax\simeq 0.14\ihMpc$ in a $0.5\,h^{-3}\mathrm{Gpc}^3$ volume.
For the highest halo mass bin, the critical $\kmax$ is similar, while for the intermediate mass halos  it is somewhat lower, $\kmax\simeq 0.07-0.1\ihMpc$.
For the cubic bias model, the model error is much less scale dependent, and its scale dependence becomes a 1$\sigma$ effect only at higher $\kmax$, typically around $\kmax\simeq 0.15-0.3\ihMpc$ (see Table~\ref{tab:Kmax}).
\vskip 4pt 

The cubic model therefore extends the $k$ range where an analysis with scale-independent model error may in principle be safe by a factor of 2 to 3 compared to the linear Standard Eulerian bias model.
In principle, this could reduce cosmic variance error bars of parameters by a factor of $\sqrt{N_\mathrm{modes}}\propto\kmax^{3/2}\simeq 3-5$.
To illustrate the resulting increase in constraining power, we consider an idealized example where all parameters are fixed except for the overall amplitude $A$ of the clustering part of the halo power spectrum, and ask how well this amplitude can be measured.
For this purpose, bold black lines in \fig{kmaxVsurvey} show for what volume and $\kmax$ the amplitude $A$ can be constrained to $1\%$, $0.5\%$, or $0.1\%$ (the lines are obtained similarly to \eqq{SNRPerr} but summing over $(P_\mathrm{model}/P_\mathrm{truth})^2$).
\vskip 4pt

For the lowest halo mass bin, using the linear Standard Eulerian bias with a scale-independent model error can safely constrain the amplitude $A$ to $0.5\%$ if the volume is larger than $\sim 3\,h^{-3}\mathrm{Gpc}^3$ (corresponding to volumes where the black long-dashed curve in \fig{kmaxVsurvey} is in the grey shaded area).
In contrast, using cubic bias for these halos, the amplitude can be constrained to $0.1\%$ if the volume is larger than $\sim 3\,h^{-3}\mathrm{Gpc}^3$, because the model error is less scale-dependent.
\vskip 4pt

For the more massive halos with a more realistic number density, the scale dependence of the error becomes relevant on larger scales, so that the amplitude cannot be constrained as well, typically $\sigma_A/A\simeq 0.7-1\%$ for linear Standard Eulerian bias and $\sigma_A/A\simeq 0.3-0.5\%$ for cubic bias in a $3\,h^{-3}\mathrm{Gpc}^3$ volume.
Therefore, parameter constraints that rely on measuring the halo power spectrum at redshift $z=0.6$ with subpercent level precision while assuming a scale-independent model error or shot noise will require  nonlinear bias terms, or a rather large volume, $V_\mathrm{survey}>10\,h^{-3}\mathrm{Gpc}^3$.\footnote{As a caveat, the scale dependence of the model error could have a functional form that gets absorbed by nuisance parameters rather than cosmological parameters. 
In that case cosmological parameters would remain unbiased even when including scales where the scale dependence of $\Perr$ is significant.
Relying on such a coincidence may be challenging though.
}
\vskip 4pt 

We can also ask what volume and $k_{\rm max}$ are needed to constrain the model amplitude $A$ with a certain precision.
Let us pick the halo sample in the upper right panel in \fig{kmaxVsurvey}.
If this is modeled with linear Standard Eulerian bias, a $1\%$ amplitude measurement is possible by using all modes up to $k_\mathrm{max}=0.1\ihMpc$ in a $1\,h^{-3}\mathrm{Gpc}^3$ volume.
Or one could observe a larger volume, so that there are more 3-d modes per $k$, and use a lower $k_\mathrm{max}$. 
Using $k_{\rm max}>0.1\ihMpc$ for these halos with the linear Standard Eulerian bias model would require accounting for scale-dependent corrections of $\Perr$.
Measuring the amplitude to $0.5\%$ in a volume smaller than $10\,h^{-3}\mathrm{Gpc}^3$ requires a higher $k_\mathrm{max}$. When assuming $\Perr={\rm const}$, this is only possible with the cubic bias model.
A $0.1\%$ amplitude measurement  is not possible with these halos when assuming $\Perr={\rm const}$, except maybe for very large volumes.
\vskip 4pt

In reality, the signal one is after may only come from some range of scales, and there may be degeneracies between parameters, so that cosmological parameters will be less well constrained than the simple model amplitude $A$ that we used above.
We also emphasize that the $\kmax$ values reported here are only approximate estimates because of uncertainty in our measurement of $\Perr(k)$ from only five simulations. 
This is indicated by thin grey lines in \fig{kmaxVsurvey}, which show how the critical $\kmax$ changes when computing the low-$k$ constant part of $\Perr(k)$ from a different $k$ range.
\vskip 4pt

We emphasize that the primary goal of this section was to investigate whether there is any physical scale dependence of the model error when describing simulated halos with the best possible bias expansion based on the shifted operators we include, and how relevant this may be. 
If such a scale dependence of $\Perr$ is too strong, it is wrong to account for it with scale-independent rescaling of the shot noise in data analyses, and we have quantified for what scales and volume this is the case.
In practice there can be other reasons and systematics that may require a lower $\kmax$ cutoff in data analyses.

\subsubsection{Interpretation of the Scale Dependence of the Model Error}
\label{se:SNRFit}

The critical $\kmax$ values shown in \fig{kmaxVsurvey} and Table~\ref{tab:Kmax} above depend only weakly on the survey volume. The reason for this is that the assumption of a scale-independent model error breaks down abruptly once a critical $\kmax$ value is exceeded, which is the case because the detectability of a scale dependence of the model error scales strongly with $\kmax$, as shown in \fig{SNRkmax}.
This strong scaling with $\kmax$ can be understood as follows. 
On scales larger than the typical size $R_M$ of a halo, the true stochastic part of the modeling error 
$\epsilon_\mathrm{stoch}$
can be written as \cite{MSZ1412,Perko1610}
\begin{align}
  \label{eq:9}
  \epsilon_{\rm stoch}(\vx) = d_1\epsilon_0(\vx) + d_2 \epsilon_0(\vx)\delta_0(\vx) + \cdots
+ \bar d_1 R_M^2 \nabla^2 \epsilon_0(\vx) + \bar d_2 R_M^2 \nabla^2\left[\epsilon_0(\vx)\delta_0(\vx)\right] + \cdots,
\end{align}
where $d_i$ and $\bar d_i$ are $k$-independent parameters, and  $\epsilon_0(\vx)$ is a stochastic noise field with $P_{\epsilon_0\epsilon_0}(k)=\mathrm{const}\sim 1/\bar n$ and $\langle\epsilon_0 \mathcal O_i\rangle=0$ for all model operators $\mathcal O_i$.
The power spectrum of $\epsilon_0(\vx)\delta_0(\vx)$ is constant in $k$,
\begin{align}
  \label{eq:10}
  P_{\epsilon_0\delta_0,\epsilon_0\delta_0}(k) = \int_\vp P_{\epsilon_0\epsilon_0}(|\vk-\vp|) P_{11}(p)
=P_{\epsilon_0\epsilon_0}\sigma_{1}^2.
\end{align}
On scales larger than the typical size of a halo, $k\ll k_M\sim 1/R_M$, the stochastic contribution to the error power spectrum therefore takes the form  \cite{Perko1610,Schmidt1808}
\begin{align}
P_{\rm stoch}(k) =  c_1+c_2k^2+\mathcal{O}\left((k/k_M)^4\right).
\end{align}
Even if $P_\mathrm{\epsilon_0\epsilon_0}$ had some $k$ dependence, expanding $P_{\epsilon_0\epsilon_0}(|\vk-\vp|)$  in $k/p\ll 1$ on large scales would again lead to a constant term and a correction scaling like $k^2$. In addition to the stochastic term, the error power spectrum also has a part related to the higher order contributions in the bias model
\begin{align}
\label{eq:PerrScaling}
P_{\rm err}(k)= P_{\rm stoch}(k) + P_{\rm 2-loop} (k) + \ldots \;.
\end{align}
Let us briefly discuss the relative size of the different contributions to the noise power spectrum. The $k^2$ correction to $P_{\rm err}$ is roughly given by
\begin{equation}
c_2k^2 \sim \frac1{\bar n} \left( \frac{k}{k_M} \right)^2 \sim  \frac1{\bar n} \left(k R_M \right)^2 \sim \frac1{\bar n} \left(k R_{M_{\rm min}} \right)^2 \left( \frac{M}{M_{\rm min}} \right)^{\frac 23} \sim \frac1{\bar n_{\rm min}} \left( \frac{k}{k_{\rm NL}} \right)^2 \left( \frac{M}{M_{\rm min}} \right)^{\frac 53} \;.
\end{equation}
In this formula, $M_{\rm min}$ is a typical mass of a halo in the lowest mass bin, and $1/\bar n_{\rm min}$ is the corresponding shot noise. In the last step, we have for simplicity assumed that $R_{M_{\rm min}}\sim 1/k_{\rm NL}$ and that the number density of halos scales like $n\sim M^{-1}$, which is consistent with the halo mass function in the relevant range of masses. On the other hand, the two-loop contribution at $z=0.6$ is roughly given by~\cite{TobiasTheoError1602}
\begin{equation}
P_{\rm 2-loop} (k) \,\sim\, 0.3\; b_1^2(M) \left( \frac{k}{k_{\rm NL}} \right)^{3.3} P(k) \;,
\end{equation}
where $P(k)$ is the nonlinear matter power spectrum. Comparing the two contributions we find that the two-loop power spectrum is sub-dominant when 
\begin{equation}
0.3 \; b_1^2(M) \left( \frac{k}{k_{\rm NL}} \right)^{1.3} \left( \frac{M}{M_{\rm min}} \right)^{-\frac 53} (\bar n_{\rm min}P(k)) \;<\; 1 \;.
\end{equation}
For the lowest mass bin the number on the l.h.s.~of this inequality is of order $\mathcal O({\rm few})$. This is expected, as we have already seen that the two-loop contribution has a size similar to the shot noise for the least massive halos. However, for higher halo masses the relevant number quickly becomes much less than 1. Even though the linear bias increases with mass, this is not nearly as fast as needed to compensate the strong mass dependence $M^{-5/3}$. For this reason it is safe to assume that for the halo masses relevant for future LSS surveys the stochastic $k^2$ contribution dominates over the two-loop contribution to $\Perr$, so that the following relation holds on scales in the perturbative regime: $P_\mathrm{err}(k)-c_1\propto k^2$. Then, the $\mathrm{SNR}^2$ for the scale dependence of $\Perr$, given by \eqq{SNRPerr}, scales as
\begin{align}
  \label{eq:11}
  \mathrm{SNR}^2(P_\mathrm{err}\neq\mathrm{const.})
\;\propto\;
\sum_{k=k_\mathrm{min}}^{\kmax} k^2 \left(\frac{k^2}{k^n}\right)^2
\;\propto\; \kmax^{3}\times\kmax^{4-2n},
\end{align}
where the $\kmax^3$ factor represents the number of modes, and  we assumed $P_{\rm model}(k)\propto k^{n}$ in the range of wavenumbers that dominates the sum in \eqq{SNRPerr}. Typically, $n\simeq -1.5$ at $k\simeq 0.2\ihMpc$, so that we expect $\mathrm{SNR}^2\propto \kmax^{3+\gamma}$ with $\gamma\simeq 7$.
Indeed, fitting the $\mathrm{SNR}^2$ shown in \fig{SNRkmax} with a power law in $\kmax$,
\begin{align}
  \label{eq:6}
  \mathrm{SNR}^2(P_\mathrm{err}\neq\mathrm{const.})
\;\approx \; N\,\left(\frac{V_\mathrm{survey}}{1\,h^{-3}\mathrm{Gpc}^3}\right)\,
\left(\frac{k_\mathrm{max}}{0.2\ihMpc}\right)^{3+\gamma},
\end{align}
with dimensionless fitting parameters  $N$ and $\gamma$,
we find that $\gamma\simeq 6-7$ provides an acceptable fit for  the cubic model, in agreement with the expected value of $\gamma\simeq 7$.
An exception to this are the halos in the lower left panel of \fig{SNRkmax},  for which we find $\gamma\simeq 3.2$;
this might be an indication of missing bias terms for these halos, which represent a somewhat special case as we already saw above because the cubic bias term is so important for these halos. Because of the steep scaling with $\kmax$, deviations from a scale-independent model error become almost immediately relevant once a critical value of $\kmax$ is exceeded, without much dependence on the survey volume, as noted above.
\vskip 4pt

One may wonder whether fitting the measured $\Perr(k)$ with $c_1+c_2k^2$ provides a better fit than fitting $\Perr(k)$  with just a constant $c_1$. 
As expected from the scaling arguments above, we find indeed that adding the $c_2k^2$ term allows to describe the measured $\Perr(k)$ up to higher $k$.
However, depending on the range of scales and on the mass bin, it is not entirely clear whether this is due to the expected $k^2$ term or higher-order bias terms that we have ignored in the model and that contribute to $\Perr(k)$ in a way that might mimic $k^2$ over a small $k$ range. 
After all, in the previous analysis we have only kept the leading two-loop term and our estimates break down close to $k_{\rm NL}$. 
Additional confusion comes from the fact that including $c_2k^2$ to describe $\Perr$ requires adding another free parameter $c_2$, so that the $k$ range where the fit holds always gets extended just because there is more flexibility to fit the measured $\Perr(k)$. 
Another problematic aspect of this measurement is that in some cases $\Perr$ is orders of magnitude smaller than the halo power spectrum so that our determination of $\Perr(k)-c_1$ has significant noise because of the small number of simulations used.
With more simulations it may be possible to establish the presence of the $k^2$ term in $\Perr(k)$ or higher order bias terms more conclusively. 
\vskip 4pt

At this stage it is therefore not clear whether adding a $k^2$ term to the noise or including higher order bias parameters leads to a larger improvement of cosmological constraining power for different ranges of scales and halo mass.
We leave this question for future work.
At a practical level, it is of course always possible to include a $k^2$ term in the noise, marginalize over its amplitude, and see if it improves cosmological constraining power.
Even if the $k^2$ term only happens to describe missing two-loop terms (without exploiting their cosmology dependence), it may extend the $k$ range where the model fits the data, potentially improving the cosmological constraining power from the other terms in the model whose cosmology dependence is included.

\subsection{Size of the Bias Terms}
\label{se:SizeOfBiasTerms}

\subsubsection{Relative Size}

\begin{figure}[tp]
\centering
\includegraphics[width=0.7\textwidth]{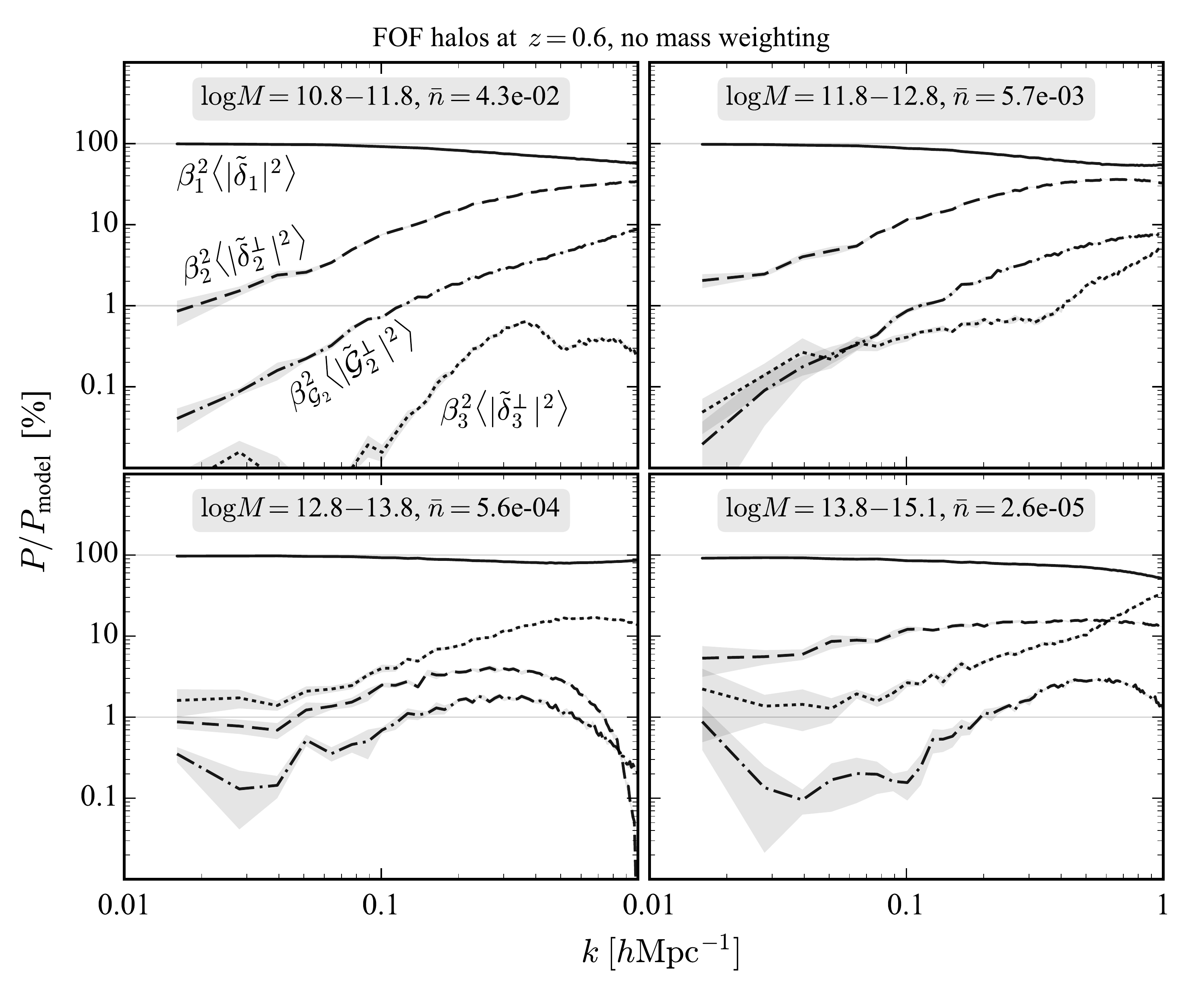}
\caption{Fractional size of bias terms contributing to the best-fit cubic bias model.
The linear bias term is clearly the dominant contribution for all halo samples on all scales. 
The next most important term is typically the quadratic bias term $\shifted\delta_2^\perp$.
It typically contributes $10\%$ in power at $k\simeq 0.1\ihMpc$.
As expected, it contributes less ($1-10\%$ in power), on larger scales, $0.02\ihMpc\le k< 0.1\ihMpc$, while it contributes more (up to $50\%$ in power) on smaller scales,  $k>0.1\ihMpc$.
An exception are the halos in the lower left panel, where this term is smaller (as expected for this range of halo mass), and the cubic term $\shifted\delta_3^\perp$ is more important, contributing between $2\%$ and $20\%$ in power.
The shifted quadratic tidal bias term $\shifted{\mathcal{G}}_2^\perp$ is always smaller than $\shifted\delta_2^\perp$, but it still contributes more than $1\%$ in power at $k\gtrsim 0.1\ihMpc$.
Grey shaded areas represent the standard error of the mean estimated from five independent realizations.
We use the model where $\delta_Z$ is expanded and absorbed by bias operators using \eqq{deltaZshifted}.
Results are averaged over five simulations and the shaded regions represent the 1$\sigma$ uncertainty estimated from the scatter between them.
}
\label{fig:M0ModelContris}
\end{figure}

We have demonstrated that nonlinear bias terms can lead to a substantial reduction of the model error at the field level. 
But which of the nonlinear bias terms are most important? 
To answer this, we compute the contributions of the individual bias terms to the complete model power spectrum, for the case of the cubic bias model, which is the most successful model that we have tested.
We work with the orthogonalized bias operators so that there are no contributions from cross-spectra between different bias terms.
\vskip 4pt 

\fig{M0ModelContris} shows the  contribution of each bias term relative to the total halo power spectrum model. 
As expected, the linear bias term is clearly the most important one, contributing more than $50\%$ in power to the total model at all wavenumbers and for all halo masses, and more than $98\%$ in power on large scales for all halos except the most massive ones. 
\vskip 4pt 

For the two low-mass halo bins, the second largest contribution is the quadratic bias $\beta_2\tilde\delta_2^\perp$.
It contributes about $1-2\%$ to the model power spectrum on large scales $k= 0.02\ihMpc$, 
about $8-10\%$ at $k=0.1\ihMpc$, and about $20\%$ at $k=0.2\ihMpc$.
The quadratic tidal term and the cubic term contribute about an order of magnitude less in power, but still contribute with more than $1\%$ to the total model power spectrum at $k>0.1\ihMpc$, especially the quadratic tidal term.
\vskip 4pt 

For the more massive $10^{12.8}-10^{13.8}\hMsun$ halos the cubic term is the second largest contribution for all wavenumbers, contributing $2\%$ to the power spectrum on large scales and up to $20\%$ on smaller scales.
The quadratic bias contributes less,  but it is still between $1\%$ and $4\%$ of the total power spectrum on most scales.
The quadratic tidal term is somewhat smaller than that, reaching a maximum contribution of about $2\%$ at $k=0.2-0.4\ihMpc$.
This is roughly the halo mass range where we expect the quadratic bias to cross zero while the linear and cubic bias are large, so it is not surprising that the quadratic contribution is smaller than that of the cubic term for these halos.
\vskip 4pt 

For the most massive halos, with $10^{13.8}-10^{15.1}\hMsun$, the quadratic bias term is the second largest contribution at $k<0.6\ihMpc$, representing $5\%$ to $15\%$ of the total model power spectrum, and the cubic term takes over at higher $k$.  The cubic term is again larger than $2\%$ of the total power spectrum on all scales.
\vskip 4pt 

This shows that different bias terms dominate on different scales for different halo masses, which is not surprising because the bias parameters change with halo mass. Also, their contribution to the total model power spectrum is often larger than $1\%$, which supports the finding above that they should be included when modeling the halo power spectrum to $1\%$ or better.
\vskip 4pt 

Given the size of the individual bias terms we can also estimate how well bias parameters need to be known to be able to predict the halo power spectrum with a given precision \cite{TobiasTheoError1602}.  
For example, for the two low-mass halo bins, modeling the halo power spectrum to $1\%$ at $k=0.1\ihMpc$ ($k=0.2\ihMpc$) requires $\beta_2$ to be known to  $10\%$ ($5\%$), because $\beta_2\shifted\delta_2^\perp$ contributes $10\%$ ($20\%$) in power.  The shifted quadratic tidal bias contributes less, so knowing $\beta_{\mathcal{G}_2}$ to $50\%$ is sufficient to model the halo power spectrum to $1\%$ at $k\le 0.2\ihMpc$.

\subsubsection{Absolute Size}

\begin{figure}[tp]
\centering
\includegraphics[width=0.7\textwidth]{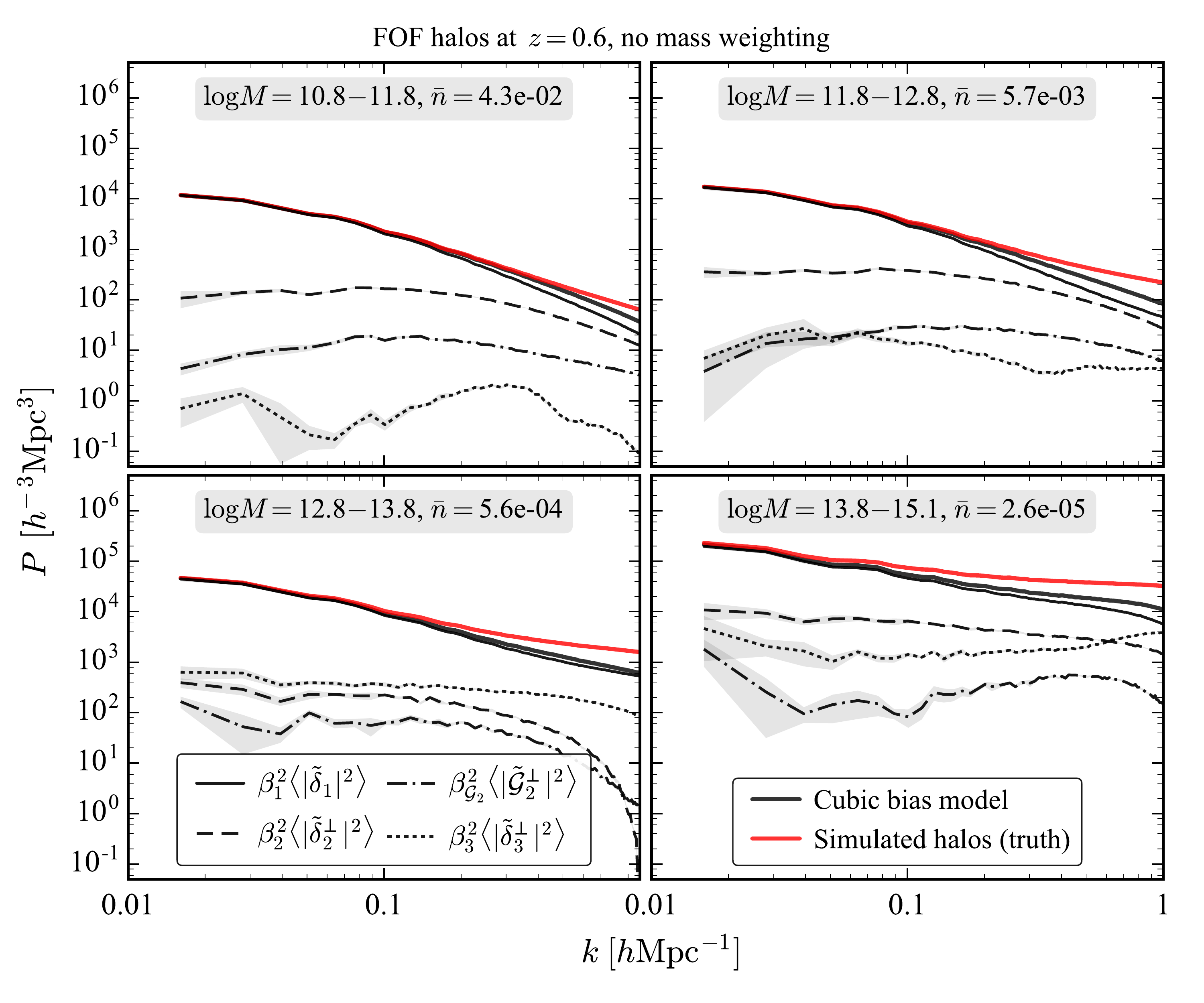}
\caption{Similar to \fig{M0ModelContris} but showing the broadband shape of the bias terms contributing to the cubic model.
}
\label{fig:M0ModelContrisBB}
\end{figure}

To see the broadband scale dependence of the different bias terms contributing to the power spectrum, 
\fig{M0ModelContrisBB} shows the absolute power spectra of the bias terms.
The power spectra of the quadratic and cubic bias terms, $\tilde\delta_2^\perp$ and $\tilde\delta_3^\perp$, are mostly constant in $k$ at low $k$, and become only visibly $k$-dependent at $k\gtrsim 0.1\ihMpc$.
If we were to perform a data analysis using only the halo power spectrum, we would not be able to separate such constant terms from the constant part of the model error or shot noise, i.e.~it seems challenging to constrain $\beta_2$ and $\beta_3$ using measurements of the halo power spectrum at $k\lesssim 0.1\ihMpc$. 
This would be a potential concern, because such a data analysis might not be able to identify a reduced model error for the correct bias parameter values.
The $k$ dependence of quadratic and cubic bias terms at higher $k$ might come as a rescue, as might a different type of data analysis that uses a  likelihood at the field level or higher-order statistics like the bispectrum.
\vskip 4pt 

Figures \ref{fig:M0ModelContris} and \ref{fig:M0ModelContrisBB} also show that the tidal quadratic bias term $\shifted{\mathcal{G}}_2^\perp$ contributes more than $1\%$ in power only at $k\gtrsim 0.1\ihMpc$, and it contributes typically a few times less in power than the quadratic bias term $\shifted\delta_2^\perp$, although both terms are at the same order in perturbation theory.
This might just be a consequence of the ordering we use to orthogonalize bias operators.
An alternative explanation would be that Lagrangian space proto-halos may be less sensitive to the quadratic tidal field than to the local quadratic bias term, which may also be the reason why the quadratic tidal bias of Lagrangian proto-halos has only recently been identified with simulations \cite{Modi1612,AbidiBaldauf1802}.

\section{Transfer Functions in Perturbation Theory}
\label{se:PTTransferFunctions}
The main purpose of this section is to derive the $k$-dependence of the transfer functions $\beta_i(k)$ at leading order in perturbation theory. At the level of a single realization this means that we have to keep all relevant cubic operators in the bias expansion that contribute to the one-loop power spectrum. The steps in the derivation are the same as in~\secref{BiasModel}. In this section we use the simplified notation $\delta_h\equiv\delta_h^{\rm model}$.

\subsection{Bias Expansion Including Cubic Operators}
Let us begin with the expression for the halo density field in Lagrangian space, keeping all cubic operators in the bias expansion
\begin{align}
\delta_h^{\mathsmaller{\rm L}}(\q)\ =\ b_1^{\mathsmaller{\rm L}}\,\delta_1(\q)&\,+\,b_2^{\mathsmaller{\rm L}}\,[\delta_2(\q)-\sigma_1^2]\, +\,  b_{\G_2}^{\mathsmaller{\rm L}}\G_2(\q)\nonumber\\[4pt]
&\, +\,b_3^{\mathsmaller{\rm L}}\,\delta_3(\q)\,+\,  b_{\G_2\delta}^{\mathsmaller{\rm L}}\,[{\G_2\delta}](\q)\, +\, b_{\G_3}^{\mathsmaller{\rm L}}\, \G_3(\q)\,+\, b_{\Gamma_3}^{\mathsmaller{\rm L}}\,\Gamma_3(\q)\ .
\end{align}
The explicit formulas for the cubic operators can be found in \app{CubicOperators}. As before, to go to Eulerian coordinates we have to displace halos using the nonlinear displacement field $\vpsi$
\beq
\label{eq:delta_h_full}
\delta_h(\k) = \int \d^3\q \; (1+ \delta^{\mathsmaller{\rm L}}_h(\q) )e^{-i\k \cdot (\q + \boldsymbol\psi(\q))} \;.
\eeq
Our goal is to rewrite this expression in terms of shifted operators
\beq
\tilde{\cal O}(\k) \equiv \int\d^3\q\ {\cal O}(\q)\,e^{-i\k\cdot(\q+\vpsi_1(\q))}\ .
\eeq
To achieve this, we split the displacement field in the halo density \eqq{delta_h_full} into the linear piece~$\vpsi_1$, which is kept in the exponent, and the nonlinear correction $\bar\vpsi\equiv \vpsi_2+\vpsi_3+\cdots$, which is expanded. 
Keeping all terms up to third order, as required for the one-loop power spectrum, we get\hskip 1pt\footnote{For simplicity, we will suppress the argument $\q$ in all fields.}
\beq
\delta_h(\k) = \int \d^3\q \; e^{-i\k \cdot \boldsymbol\psi_1}  (1+ \delta_h^{\mathsmaller{\rm L}})(1+\bar\vpsi \cdot\nabla)e^{-i\k \cdot\q} \;.
\eeq
Notice that we have rewritten $\k$ which multiplies $\bar\vpsi$ as a derivative with respect to $\q$. After integration by parts, the derivative can act on $\vpsi_1$, $\delta_h^{\mathsmaller{\rm L}}$ and $\bar\vpsi$, so we get
\beq
\delta_h(\k) = \int d^3\q \; e^{-i\k \cdot \boldsymbol\psi_1}  \Big[ 1+ \delta_h^{\mathsmaller{\rm L}} + i k_a (1+\delta_h^{\mathsmaller{\rm L}}) \partial_b \psi_1^a \, \bar\psi^b - (1+\delta_h^{\mathsmaller{\rm L}}) \nabla \cdot \bar\vpsi -\bar\vpsi\cdot\nabla\delta_h^{\mathsmaller{\rm L}} \Big] e^{-i\k \cdot\q} \;.
\eeq
The third term in the square brackets can be further simplified. The product $\partial_b\psi_1^a\bar\psi^b$ starts at third order in perturbation theory and we can neglect $\delta_h$. One $k_a$ that appears in this term can be written as a derivative with respect to $q_a$. After another integration by parts and keeping terms of order 3 or less, we find
\beq
\delta_h(\k) = \int \d^3\q \; \Big[ 1+ \delta_h^{\mathsmaller{\rm L}} + \partial^b \psi_1^a \, \partial_a \psi_{2b} - (1+\delta_h^{\mathsmaller{\rm L}}) \nabla \cdot\vpsi_2 - \nabla\cdot\vpsi_3 -(1+b_1^{\mathsmaller{\rm L}})\vpsi_2 \cdot\nabla\delta_1 \Big] e^{-i\k \cdot (\q + \boldsymbol\psi_1)} \;.
\eeq
Each term in this formula has the form of a shifted operator \eq{Oshifted}. We have already mentioned that the shift of the uniform density, i.e.~the Zel'dovich density, can be written in terms of bias operators (see \app{DeltaZShiftedOps}). Similarly, the other terms $\{\partial^b \psi_1^a \, \partial_a \psi_{2b},\; \nabla\cdot\vpsi_2,\; \nabla\cdot\vpsi_3\}$ can each be expressed as linear combinations of bias operators. We leave the details of this calculation for \app{CubicOperators}. As a result, the effect of these terms is to effectively change the values of the bias parameters in the original Lagrangian-space bias expansion in $\delta_h^{\mathsmaller{\rm L}}$. On the other hand, the last term ${\cal S}_3\equiv \vpsi_2\cdot\nabla\delta_1$ is a shift term of the linear field by $\vpsi_2$, which {\em cannot} be expressed in terms of bias operators. It is a new contribution that has to be taken into account separately. However, the coefficient of this term is completely fixed by the equivalence principle---i.e.~this term does not come with an extra free parameter. 
\vskip 4pt
Putting everything together, the bias model becomes
\begin{align}
\delta_h\ =\  (1+b_1^{\mathsmaller{\rm L}})\,\shifted\delta_1 &\,+\,b_2^{\mathsmaller{\rm L}}\,\shifted\delta_2\, +\, b_{\G_2}\shifted\G_2\nonumber\\[4pt]
&\, +\,b_3^{\mathsmaller{\rm L}}\,\shifted\delta_3\,+\, b_{\G_2\delta}\,[\shifted{\G_2\delta}]\, +\, b_{\G_3}\,\shifted\G_3\,+\, b_{\Gamma_3}\,\shifted\Gamma_3\,-\,(1+b_1^{\mathsmaller{\rm L}})\shifted{\cal S}_3\ .\label{eq:biasshift}
\end{align}
Notice that the local bias parameters remain unchanged in this process (apart from the linear bias $b_1^{\mathsmaller{\rm L}}$, which, as expected, increases by 1 when going to Eulerian space). Indeed, this is a general statement that is true at all orders in perturbation theory. The simplest way to see this is to consider the limit of the dark matter density field, when all Lagrangian bias parameters are equal to zero. In this limit, all local operators $\shifted \delta_n$ (i.e., shifted $\delta^n(\vq)$) with $n\ge 2$ have to drop from the expression for the nonlinear dark matter density field because otherwise the power spectrum would have nonphysical shot noise contributions. The other, nonlocal bias parameters are related to the original Lagrangian ones in the following way:
\begin{align}
\label{eq:shifted_b_lagrangian_b_1}
&b_{{\cal G}_2}\ =\ \frac{2}7 + b^{\mathsmaller{\rm L}}_{\G_2}\\
\label{eq:shifted_b_lagrangian_b_2}
&b_{{\cal G}_2\delta}\ =\ -\frac{3}{14}(1+b^{\mathsmaller{\rm L}}_1) + b^{\mathsmaller{\rm L}}_{\G_2\delta} \\
\label{eq:shifted_b_lagrangian_b_3}
&b_{{\cal G}_3}\ =\ -\frac{2}9 + b^{\mathsmaller{\rm L}}_{\G_3}\\
\label{eq:shifted_b_lagrangian_b_4}
&b_{{ \Gamma}_3}\ =\ \frac{1}6 + b^{\mathsmaller{\rm L}}_{\Gamma_3} \ .
\end{align}

\subsection{Transfer Functions}
In Sections~\ref{se:PositionResults} and \ref{se:FourierResults}, we compared the halo density field with the bias model
\beq
\label{eq:bias_model_transfer}
\delta_h(\k)\ =\ \beta_1(k)\shifted\delta_1(\k)\, +\, \beta_2(k)\shifted\delta_2^{\perp}(\k)\, +\, \beta_{\G_2}(k)\shifted\G_{2}^\perp(\k)\ .
\eeq
We can use the perturbative bias expansion derived in the previous section, \eqq{biasshift}, to predict the shapes of the transfer functions $\beta_1(k)$, $\beta_2(k)$ and $\beta_{\G_2}(k)$ at one-loop order in perturbation theory as follows. 
\vskip 4pt

Let us first consider the cubic operators $\shifted{\cal O}_3$ in \eqq{biasshift}. These operators can easily be decomposed into the part parallel to $\shifted\delta_1$ and the part orthogonal to $\shifted\delta_1$
\beq
\shifted{\cal O}_3^\parallel(\k)\, \equiv\, \frac{\vev{\shifted\delta_1\shifted{\cal O}_3}}{\vev{\shifted\delta_1 \shifted\delta_1}} \shifted\delta_1(\k)\ ,  \qquad \shifted{\cal O}_3^\perp(\k)\, \equiv\, \shifted{\cal O}_{3}(\k)-\frac{\vev{\shifted\delta_1\shifted{\cal O}_3}}{\vev{\shifted\delta_1 \shifted\delta_1}} \shifted\delta_1(\k)\ .
\eeq
At one-loop level, the orthogonal cubic fields are also effectively orthogonal to second-order operators, because their correlation is higher order in perturbation theory. This means that the fields $\shifted{\cal O}_3^\perp(\k)$ do not contribute to the one-loop power spectrum and we can drop them from our expressions. In a similar fashion, we can project $\shifted\delta_2$ and $\shifted\G_2$ to $\shifted\delta_1$ and make the remaining fields orthogonal to each other. It is then straightforward to compute the transfer functions in terms of the bias parameters:
\begin{align}
\beta_1(k)&\ =\ b_1 + b_2 \frac{\vev{\shifted\delta_1 \shifted \delta_2}}{\vev{\shifted \delta_1 \shifted \delta_1}}+ b_{\G_2} \frac{\vev{\shifted\delta_1 \shifted \G_2}}{\vev{\shifted\delta_1 \shifted\delta_1 }} + \sum_i b_i \frac{\vev{\shifted\delta_1 \shifted {\cal  O}^i_3}}{\vev{\shifted\delta_1 \shifted\delta_1 }}\ ,\\
\beta_2(k)&\ =\ b_2 + b_{\G_2} \frac{\vev{\shifted\delta_2^\perp \shifted\G_{2}^\perp}}{\vev{\shifted\delta_2^\perp \shifted\delta_2^\perp}}\ ,\label{eq:beta2}\\[8pt]
\beta_{\G_2}(k)&\ =\ b_{\G_2}\ ,
\end{align}
where $b_1=b_1^{\mathsmaller{\rm L}}+1$, $b_{2,3}=b_{2,3}^{\mathsmaller{\rm L}}$ and other bias parameters are given by Eqs.~\eqref{eq:shifted_b_lagrangian_b_1}-\eqref{eq:shifted_b_lagrangian_b_4}. Indeed, it is easy to check that with these transfer functions the power spectrum of the halo density field in Eq.~\eqref{eq:bias_model_transfer} exactly agrees with the usual one-loop halo power spectrum (up to the fact that the IR resummation is automatically included if one uses shifted operators). 
\vskip 4pt

These transfer functions can be  further simplified in practice. For example, not all cubic operators contribute to $\beta_1(k)$ with independent shapes. At leading order in perturbation theory the only two nontrivial $k$-dependent terms come from $\shifted{\Gamma}_3$ and $\shifted{\mathcal S}_3$. The other operators give constant contributions, which are degenerate with $b_1$.\footnote{Note that this degeneracy is exact with standard fields. With shifted fields there can be some $k$-dependence due to the nontrivial effects of the displacement field. However, any such $k$-dependent contribution must be at higher order in perturbation theory and thus we neglect it.} Once this is taken into account, $\beta_1(k)$ becomes
\begin{equation}
\beta_1(k)\ =\ b_1 + b_2 \frac{\vev{\shifted\delta_1 \shifted \delta_2}}{\vev{\shifted \delta_1 \shifted \delta_1}}+ b_{\G_2} \frac{\vev{\shifted\delta_1 \shifted \G_2}}{\vev{\shifted\delta_1 \shifted\delta_1 }} + b_{\Gamma_3} \frac{\vev{\shifted\delta_1 \shifted {\Gamma}_3}}{\vev{\shifted\delta_1 \shifted\delta_1 }} - b_1 \frac{\vev{\shifted\delta_1 \shifted {\mathcal S}_3}}{\vev{\shifted\delta_1 \shifted\delta_1 }} \ .
\end{equation}
Notice that $b_1$ in this formula is different from its starting value due to the degenerate contributions from cubic operators. The new value corresponds to the so called renormalized bias $b_1$. The important point is that we kept the same parameter multiplying the operator $\shifted{\mathcal S}_3$. Even though this may not be obvious from just a few leading orders in perturbation theory, this choice is imposed by the fact that $\shifted{\mathcal S}_3$ comes from the shift of the halo density field by the second order displacement. This term is fixed and has no extra free parameters, even when renormalization is taken into account. Finally, we have to add a $k^2$ term to the transfer function $\beta_1(k)$ with a free coefficient. In analogy with the EFT counterterm for the one-loop matter power spectrum we label this parameter $c_s^2$ even though this counterterm is there to absorb all UV contributions from correlation functions of the form $\vev{\shifted\delta_1 \shifted{\mathcal{O}_3} }$ and the bias coefficients from the higher-derivative bias operators such as $\nabla^2 \delta$. 
\vskip 4pt

Let us now turn to the second transfer function. This expression can be further simplified. The first step is to write
\begin{equation}
\vev{\shifted\delta_2^\perp \shifted\delta_2^\perp} = \vev{\shifted\delta_2 \shifted\delta_2} - \frac{\vev{\shifted\delta_2 \shifted\delta_1}^2}{\vev{\shifted\delta_1 \shifted\delta_1}} \;,
\end{equation}
which implies that at one loop $\vev{\shifted\delta_2^\perp \shifted\delta_2^\perp} = \vev{\shifted\delta_2 \shifted\delta_2}$ because the second term is higher order in perturbation theory. For the same reason, at large scales we can replace $\vev{\shifted\delta_2^\perp \shifted\G_{2}^\perp}$ with $\vev{\shifted\delta_2 \shifted\G_{2}}$.
As a result, we can write the transfer function $\beta_2(k)$ as follows
\beq
\beta_2(k)\, =\, b_2 + b_{\G_2}\frac{\vev{\shifted\delta_2 \shifted\G_{2}}}{\vev{\shifted\delta_2 \shifted \delta_2}}\ .
\eeq
In the limit $k\to 0$ the numerator of the second term scales like $\mathcal O(k^2)$ while the denominator approaches a constant. Therefore, the second term vanishes on very large scales. Notice that this contribution is not suppressed by loop factors because both the numerator and denominator are of the same order in perturbation theory. For this reason, when the transfer functions are measured at not-so-large scales where the scaling $\mathcal O(k^2)$ is not valid, the second term is not necessarily negligible. However, because of the large constant contribution to $\vev{\shifted\delta_2 \shifted \delta_2}$, the second term turns out always to be small enough, given the size of the higher loop corrections that we are neglecting and final error bars with which we determine the bias parameters. 
\vskip 4pt 

To summarize, we use the following minimal model to fit the $k$-dependent transfer functions
\begin{align}
\label{eq:beta1Pred}
\beta_1(k)&\ =\ b_1 + c_s^2 k^2 + b_2 \frac{\vev{\shifted\delta_1 \shifted \delta_2}}{\vev{\shifted \delta_1 \shifted \delta_1}}+ b_{\G_2} \frac{\vev{\shifted\delta_1 \shifted \G_2}}{\vev{\shifted\delta_1 \shifted\delta_1 }} + b_{\Gamma_3} \frac{\vev{\shifted\delta_1 \shifted {\Gamma}_3}}{\vev{\shifted\delta_1 \shifted\delta_1 }} - b_1 \frac{\vev{\shifted\delta_1 \shifted {\mathcal S}_3}}{\vev{\shifted\delta_1 \shifted\delta_1 }} \ ,\\
\beta_2(k)&\ =\ b_2 \ , \qquad {\rm and} \qquad \beta_{\G_2}(k)\ =\ b_{\G_2} \ .
\end{align}
This model has 5 free parameters, the same as the one-loop power spectrum. When we use the cubic bias model, we add one extra parameter, $b_3$, which is fitted from the low-$k$ limit of $\beta_3(k)$.

\subsection{Power Spectra of Shifted Fields from Theory and on a Grid}
\label{se:PkShifted}

To fit the transfer functions with \eqq{beta1Pred} we need to calculate the power spectra $\langle \shifted{\mathcal O}_a \shifted{\mathcal O}_b \rangle$ of shifted operators that enter \eqq{beta1Pred}. 
As we already mentioned, this calculation is the same as in~\cite{Vlah:2015sea,Vlah1609}, and more details can be found there. Here we summarize only the main steps. Let us start from the definition
\begin{equation}
\langle \shifted{\mathcal O}_a \shifted{\mathcal O}_b \rangle (\k) \ =\ \int d^3\q \; e^{-i\k\cdot \q} \big\langle \mathcal O_a(\q) \; \mathcal O_b (0) \; e^{-i \k \cdot ( \boldsymbol\psi_1(\q)- \boldsymbol\psi_1(0)) } \big\rangle \;.
\end{equation}
The strategy to evaluate the expectation value is to bring the operators $\mathcal O_a$ and $\mathcal O_b$ to the exponent and use the cumulant theorem. This can be achieved using the following trick
\begin{equation}
\langle \shifted{\mathcal O}_a \shifted{\mathcal O}_b \rangle (\k) \ =\ i\frac{d}{d\lambda} \int d^3\q \; e^{-i\k\cdot \q} \big\langle e^{-i \left[ \k \cdot ( \boldsymbol\psi_1(\q)- \boldsymbol\psi_1(0)) + \lambda \mathcal O_a(\q) \mathcal O_b (0) \right] } \big\rangle \Big|_{\lambda=0} \;.
\end{equation}
The cumulant theorem reads
\beq
\langle e^{-i X} \rangle = {\rm Exp} \left( \sum_{n=0}^\infty \frac{(-i)^n}{n!} \langle X^n \rangle_c \right)  \;,
\eeq
and since we are interested only in terms at leading order in $\lambda$, the final expression for the expectation value of the exponential is given by
\begin{align}
\langle e^{-i \left[ \k \cdot \Delta \boldsymbol\psi + \lambda \mathcal O_{ab} \right]} \rangle \ = \ & {\rm Exp}  \left( -\frac12 k_ik_j \langle \Delta \boldsymbol\psi^i \Delta \boldsymbol\psi^j \rangle_c \right) \left( 1 - i\lambda\langle \mathcal O_{ab} \rangle_c -\lambda k_i \langle \Delta \boldsymbol\psi^i \mathcal O_{ab} \rangle_c + \frac{i\lambda}{2} k_i k_j \langle \Delta \boldsymbol\psi^i \Delta \boldsymbol\psi^j \mathcal O_{ab} \rangle_c \right. \nonumber \\
& \left. + \frac{\lambda}{6} k_i k_j k_m \langle \Delta \boldsymbol\psi^i \Delta \boldsymbol\psi^j  \Delta \boldsymbol\psi^m \mathcal O_{ab} \rangle_c - \frac{i\lambda}{24} k_i k_j k_m k_n \langle \Delta \boldsymbol\psi^i \Delta \boldsymbol\psi^j  \Delta \boldsymbol\psi^m  \Delta \boldsymbol\psi^n \mathcal \mathcal O_{ab} \rangle_c + O(\lambda^2) \right) \;,
\end{align}
where $\Delta \boldsymbol\psi \equiv \boldsymbol\psi_1(\q)- \boldsymbol\psi_1(0)$ and $\mathcal O_{ab} \equiv \mathcal O_a(\q) \mathcal O_b (0)$. Notice that we have truncated the sum. The reason is that at one-loop level the operator $\mathcal O_{ab}$ can be at most fourth order in perturbation theory and therefore it can be contracted with at most four displacement fields $\Delta\vpsi$ in a connected $n$-point function. Furthermore, for different terms in transfer functions the operator $\mathcal O_{ab}$ can be proportional to even or odd powers of initial density fields and some of the correlation functions vanish for Gaussian initial conditions. We will refer to the first non-vanishing term proportional to $\lambda$ as the leading-order term (LO) and higher order terms in perturbation theory as next-to-leading (NLO) and next-to-next-to-leading orders (NNLO). 
\vskip 4pt
\begin{figure}[htp]
\centering
\includegraphics[width=0.5\textwidth]{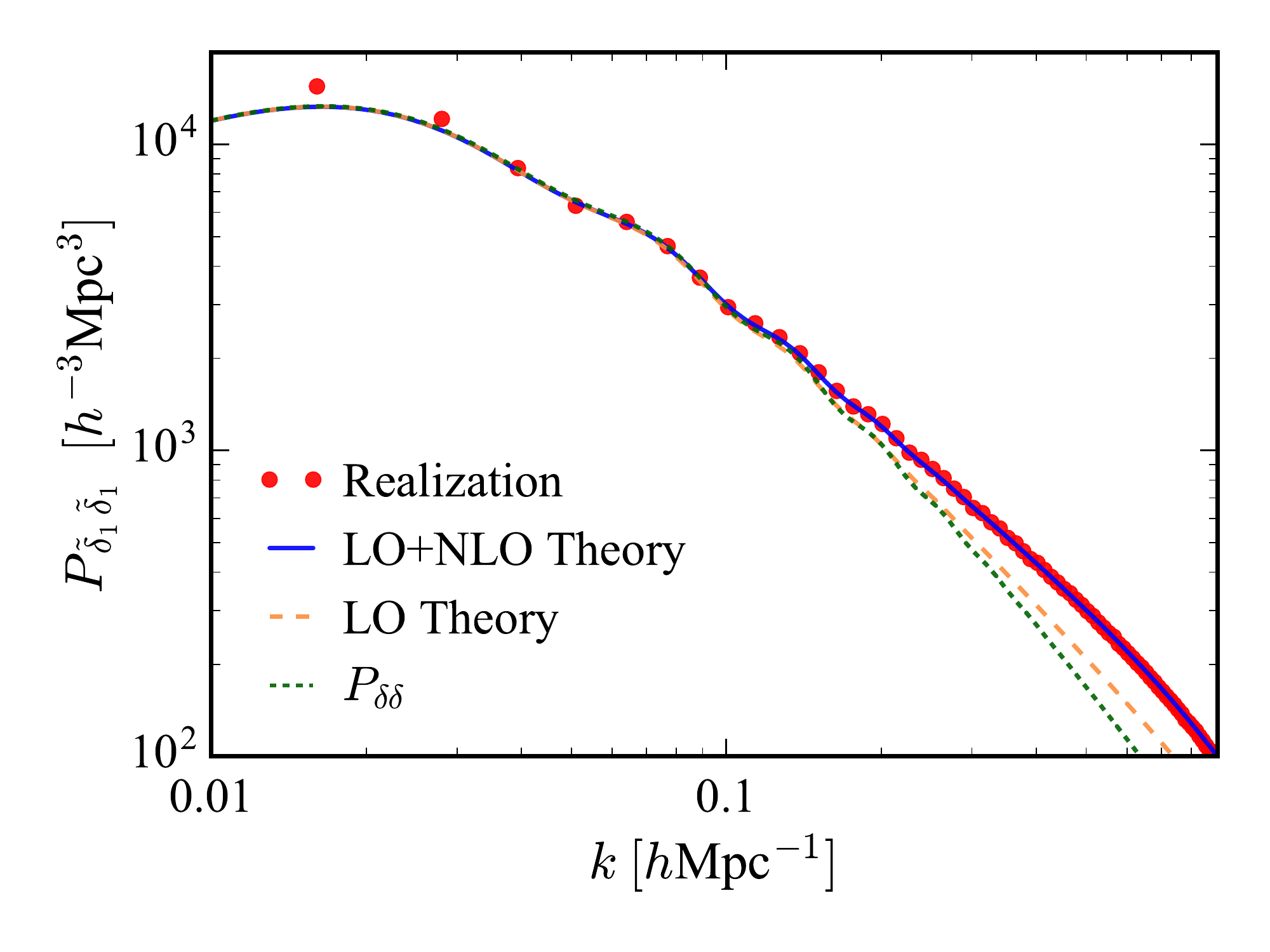}
\caption{Comparison of the power spectrum of a realization of the linear density field shifted on a grid (red points) and theoretical predictions for the mean (continuous curves). The green dotted line is the linear power spectrum. The orange dashed line is the power spectrum of the shifted density field evaluated keeping only the LO term. The blue solid line represents the exact result when the NLO term is taken into account. The NLO correction is important for getting the correct prediction.
In contrast to all previous figures, the theory curves are not computed for the same realization as the simulation but as integrals over the mean linear theory power spectrum from CAMB.  
The theory and realization on the grid therefore differ by cosmic variance, especially on large scales.
}
\label{fig:ModelP_dd}
\end{figure}

Let us take a look at the simplest example of $\langle \shifted \delta_1 \shifted \delta_1 \rangle$. In this case only two terms are non-vanishing
\begin{equation}
i\frac{d}{d\lambda}\langle e^{-i \left[ \k \cdot \Delta \boldsymbol\psi + \lambda \delta_1(\q) \delta_1(0) \right]} \rangle \Big|_{\lambda=0} = \left( \langle \delta_1(\q) \delta_1(0) \rangle - \frac12 k_i k_j \langle \Delta \boldsymbol\psi^i \Delta \boldsymbol\psi^j \delta_1(\q) \delta_1(0) \rangle_c \right) {\rm Exp}  \left( -\frac12 k_ik_j \langle \Delta \boldsymbol\psi^i \Delta \boldsymbol\psi^j \rangle_c \right) \;.
\end{equation}
All connected $n$-point functions on the r.h.s~of this equation involve only linear fields and can be calculated by going to momentum space. The result is a function of $\q$ and one can do the Fourier transform to calculate the power spectrum. Fig.~\ref{fig:ModelP_dd} shows the power spectrum prediction with and without the NLO corrections. The points represent measurements from a realization of the linear density field shifted by $\vpsi_1$. The agreement when NLO corrections are included is quite good. Notice that in this example the NLO term is of  one-loop order and has to be kept for consistency. 
\vskip 4pt
\begin{figure}[htp]
\centering
\includegraphics[width=0.45\textwidth]{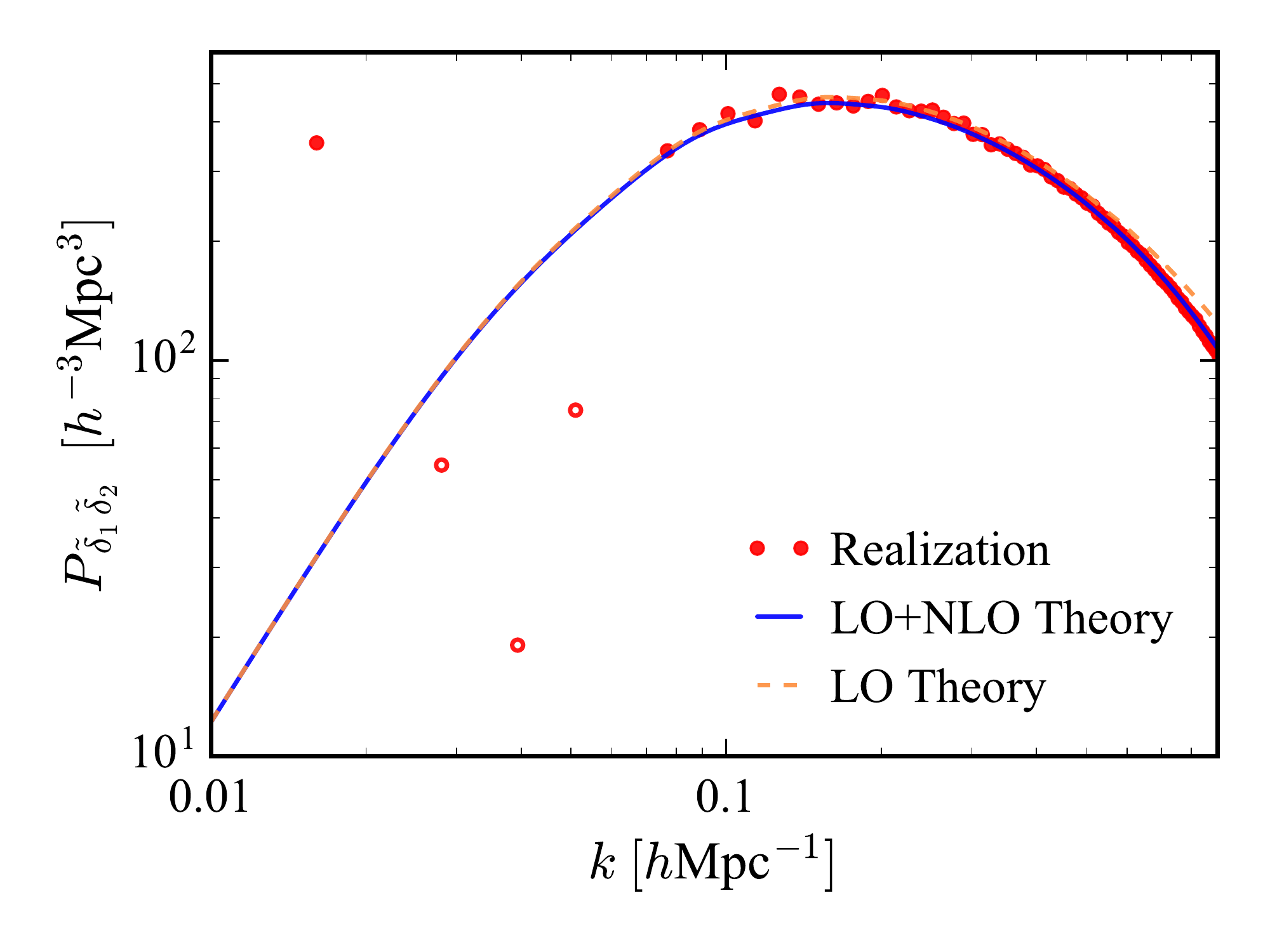}
\hspace{5pt}
\includegraphics[width=0.45\textwidth]{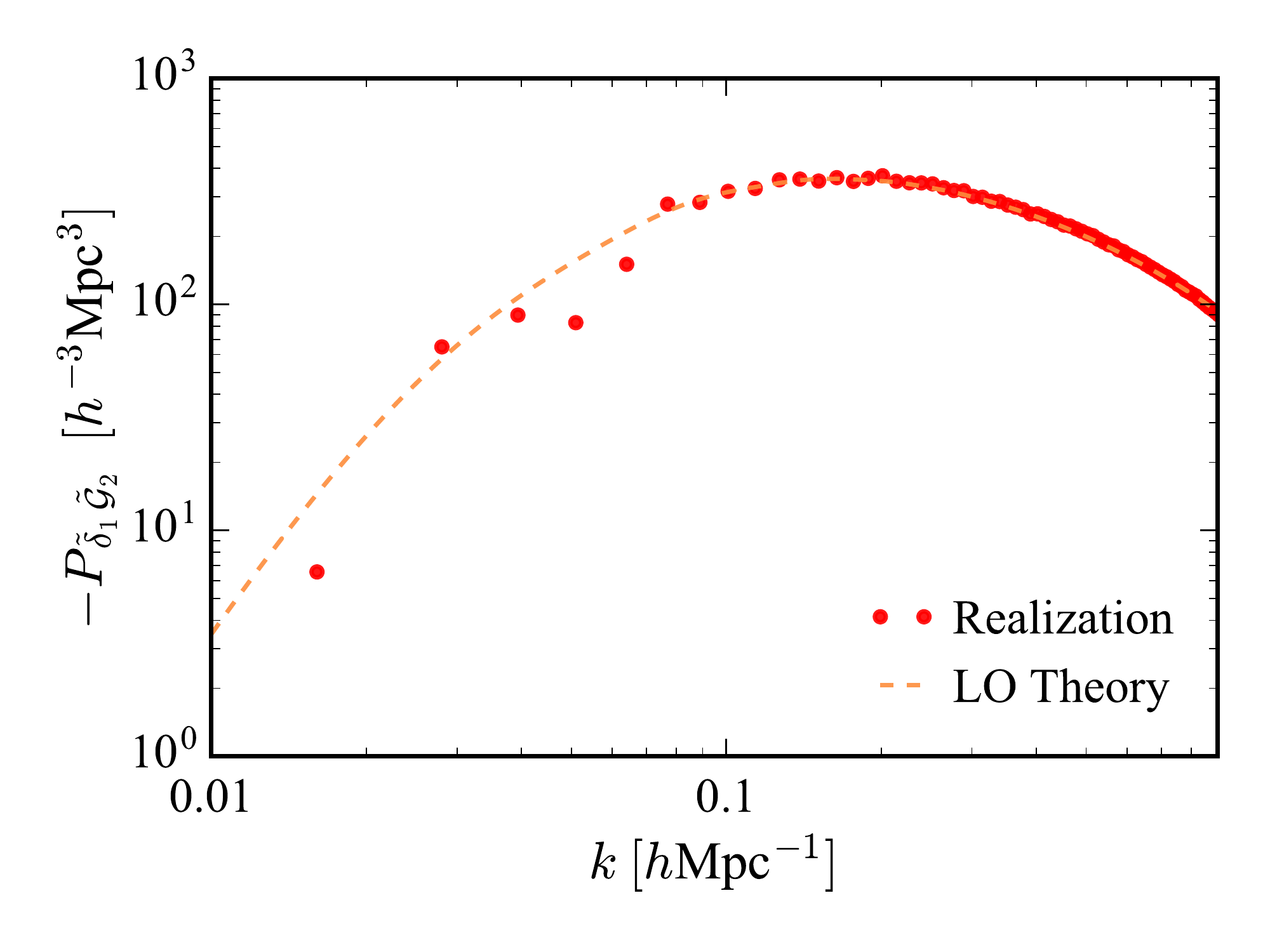}
\caption{The cross spectra of the shifted linear density field with shifted second order operators. Red points are measurements from a realization, with unfilled symbols representing negative values. The orange dashed curve are predictions keeping only the LO terms in the formulas. The LO predictions agree quite well with the measurements. The blue solid line shows the result that includes the NLO corrections. While these are important to get the exact result, such as in the example with $\langle \shifted\delta_1 \shifted{\delta}_2 \rangle$, they are quite small. Notice that the size of these cross spectra is smaller than $\langle \shifted\delta_1 \shifted\delta_1 \rangle$ on all scales.
As in the previous figure, the theory curves are evaluated as integrals over the linear theory power spectrum and not matched to the realization of the simulation.
}
\label{fig:ModelP_dd2_andP_dG2}
\end{figure}

Other correlators of interest for transfer functions are those where the shifted linear field is correlated with a shifted second order operator, such as $\langle \shifted\delta_1\shifted\delta_2 \rangle$ or $\langle \shifted \delta_1 \shifted{\G}_2 \rangle$. These correlation functions would vanish for non-shifted operators  in the case of Gaussian initial conditions. Because of the correlations induced by the linear displacement field $\vpsi_1$, they are not zero but the LO terms are roughly of the order of the one-loop corrections. For example,
\begin{align}
i\frac{d}{d\lambda} & \langle e^{-i \left[ \k \cdot \Delta \boldsymbol\psi + \lambda \delta_1(\q) \delta_2(0) \right]} \rangle \Big|_{\lambda=0} = \nonumber \\
& \qquad \left( - i k_i \langle \Delta \boldsymbol\psi^i \delta_1(\q) \delta_2(0) \rangle_c + \frac{\lambda}{6} k_i k_j k_m \langle \Delta \boldsymbol\psi^i \Delta \boldsymbol\psi^j  \Delta \boldsymbol\psi^m \delta_1(\q) \delta_2(0) \rangle_c \right) {\rm Exp}  \left( -\frac12 k_ik_j \langle \Delta \boldsymbol\psi^i \Delta \boldsymbol\psi^j \rangle_c \right) \;.
\end{align}
A similar expression can be written for the correlation with $\shifted{\G}_2$. The correlation functions in the brackets can be again straightforwardly calculated. Notice that the NLO term in this formula is higher order in perturbation theory because it is proportional to the cubic power of the linear power spectrum. Indeed, these NLO corrections are small. In Fig.~\ref{fig:ModelP_dd2_andP_dG2} we compare the theoretical prediction with LO and NLO with the measurement of the cross spectra of shifted fields in a given realization. As we can see, the NLO corrections are indeed small compared to one loop terms. Given that all our analysis is valid only at one-loop level, the NLO terms in these correlation functions can be neglected. 
\vskip 4pt

Finally, in a similar way one can calculate the other two correlation functions $\langle \shifted\delta_1 \shifted{\Gamma}_3 \rangle$ and $\langle \shifted\delta_1 \shifted{\mathcal{S}}_3\rangle$ in the transfer function $\beta_1(k)$. In this case there are three terms that survive in the cumulant expansion, but the LO term is already of one-loop order and the higher order contributions can be ignored.

\MyFloatBarrier
\subsection{Fitting the Transfer Functions from Simulations}
\label{se:TkFits}

In the previous sections we have derived the transfer functions in perturbation theory. Working at the one-loop level, for the cubic bias model we got
\begin{align}
\label{eq:beta1Fit}
\beta_1(k)&\ =\ b_1 + c_s^2 k^2 + b_2 \frac{\vev{\shifted\delta_1 \shifted \delta_2}}{\vev{\shifted \delta_1 \shifted \delta_1}}+ b_{\G_2} \frac{\vev{\shifted\delta_1 \shifted \G_2}}{\vev{\shifted\delta_1 \shifted\delta_1 }} + b_{\Gamma_3} \frac{\vev{\shifted\delta_1 \shifted {\Gamma}_3}}{\vev{\shifted\delta_1 \shifted\delta_1 }} - b_1 \frac{\vev{\shifted\delta_1 \shifted {\mathcal S}_3}}{\vev{\shifted\delta_1 \shifted\delta_1 }} \ ,
\end{align}
while the other three transfer functions are constant
\begin{align}
\label{eq:beta23Fit}
\beta_2(k)&\ =\ b_2 \ , \qquad \beta_{\G_2}(k)\ =\ b_{\G_2} \qquad {\rm and} \qquad \beta_3(k) = b_3  \ .
\end{align}
In this section we compare this theoretical prediction with the free transfer functions $\beta_i(k)$ measured from simulations by minimizing the mean-square model error.
\vskip 4pt 

\begin{figure}[tp]
\centering
\includegraphics[width=0.7\textwidth]{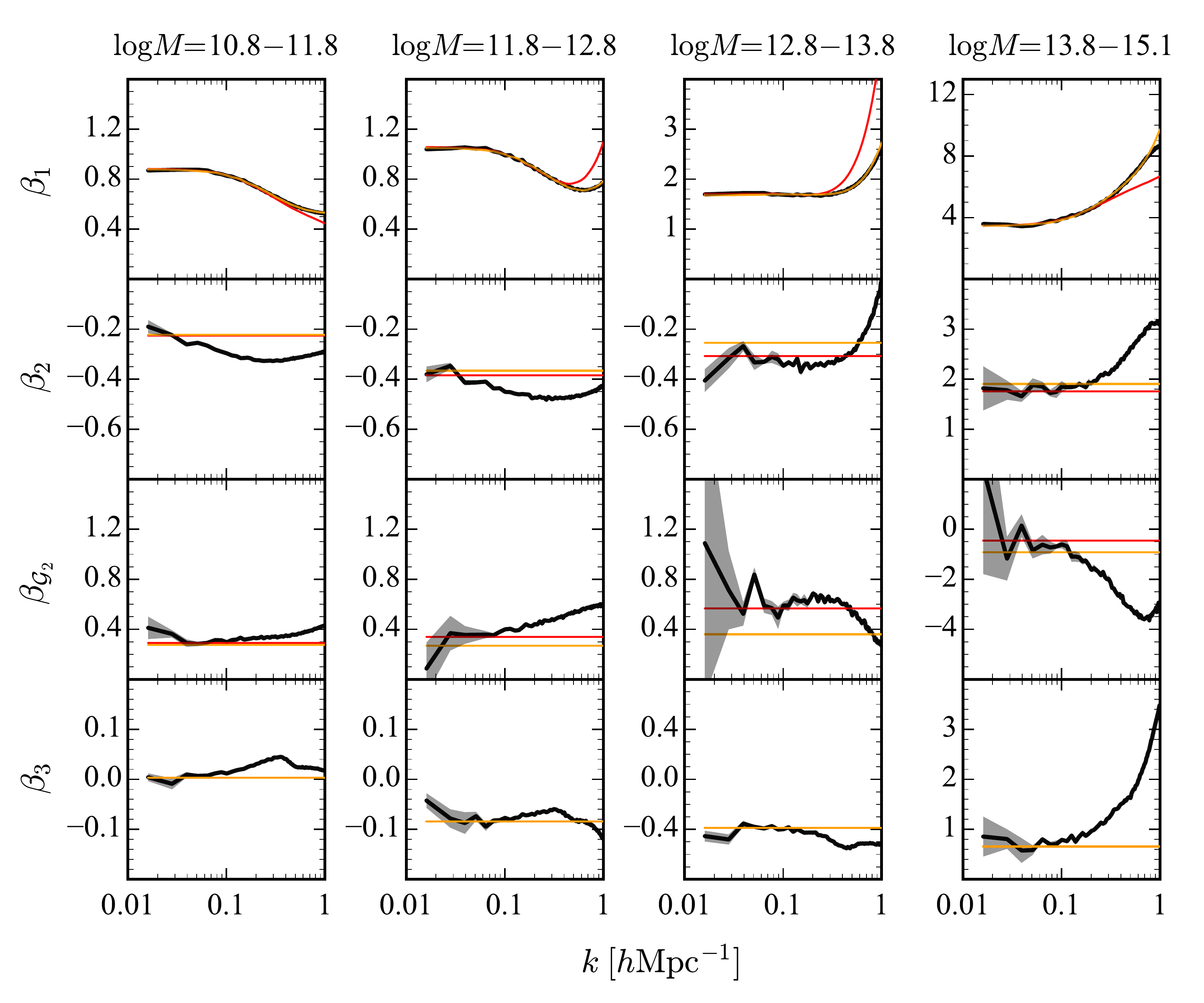}
\caption{Transfer functions $\beta_i(k)$ of the cubic bias model, $\delta_h= \beta_1 \shifted\delta_1+\beta_2\shifted\delta_2^\perp+\beta_{\mathcal{G}_2}\shifted{\G}_2^\perp+\beta_3 \shifted {\delta}_3^\perp$, for the four mass bins.
Treating all $k$ bins as independent and minimizing the power of the model error in each $k$ bin gives the black lines, with uncertainty shown in grey (estimated from the scatter between the five independent simulations). 
When fitting these transfer functions with the theoretical model described in the text, using six $k$-independent parameters, we obtain the red and orange lines. 
These fits either include theoretical errors, effectively restricting the fitting region to low $k$ (red), or they are fitted up to higher $k$ without theoretical errors (orange).
The model error is almost the same for the free transfer functions and the smooth theory fits (see dashed lines in \fig{noiseM0} above). }
\label{fig:M0TkNoZ}
\end{figure}

\begin{figure}[tp]
\centering
\includegraphics[width=0.7\textwidth]{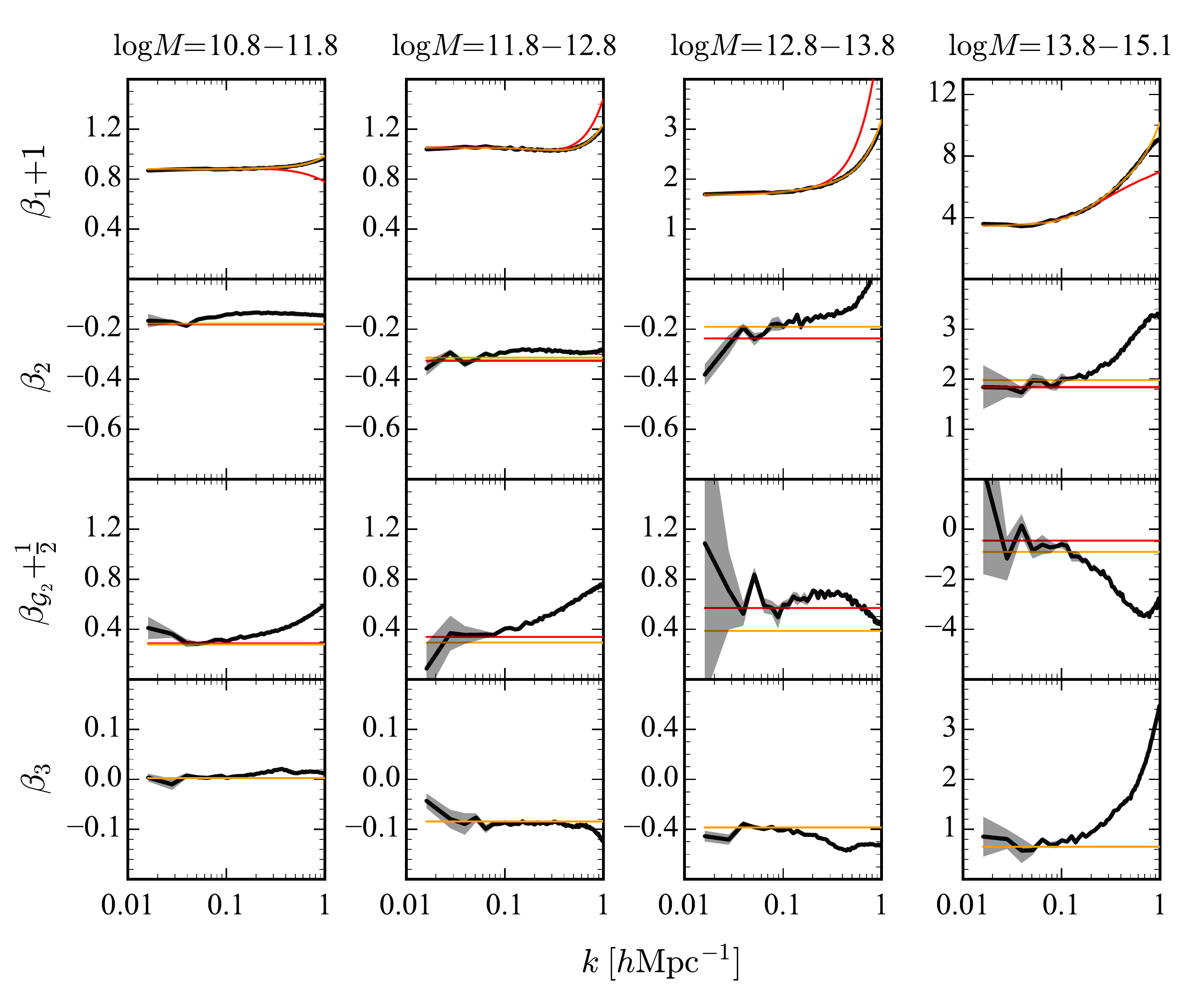}
\caption{Same as \fig{M0TkNoZ} but for bias model where $\delta_Z$ is kept explicitly in the expansion of the halo field.  This changes the transfer functions at high $k$, because contributions to $\delta_Z$ that are not absorbed by the bias terms through \eqq{deltaZshifted} become relevant.  The theory fit of $\beta_1(k)$ captures this well though. The model error, shown in \fig{noiseM0} above, is the same with or without $\delta_Z$ in the model, except for the lowest halo mass bin, where the model error is small enough that correction terms from expanding $\delta_Z$ become visible. We add $1$ and $1/2$ to $\beta_1$ and $\beta_{\mathcal{G}_2}$ for easier comparison against the bias model without $\delta_Z$, noting that $\delta_h=\delta_Z+\beta_1\tilde\delta_1+\cdots\approx (\beta_1+1)\tilde\delta_1+\beta_2\tilde\delta_2^\perp+(\beta_{\mathcal{G}_2}+1/2)\tilde\G_2^\perp+\beta_3\tilde\delta_3^\perp$.
}
\label{fig:M0Tk}
\end{figure}

One obvious problem with fitting of the transfer functions is that the perturbation theory expressions are valid only in the low-$k$ limit, while most of the constraining power in the fit comes from the small scales. To solve this problem and avoid overfitting, we use the prescription from \cite{TobiasTheoError1602} to consistently include theoretical errors in the estimate of the bias parameters. The covariance matrix used to calculate $\chi^2$ can then be written as a sum of two terms:
\begin{equation}
{\rm Cov}(k_i,k_j) = C_{noise}(k_i) \, \delta_{ij} + \delta\beta (k_i) \, \delta\beta (k_j) \, e^{-\frac12 \frac{(k_i-k_j)^2}{\Delta k^2}} \;,
\end{equation}
where $k_i$ is the wavenumber in bin $i$. The first term is related to the noise in the halo power spectrum. For example, in the case of $\beta_1(k)$ this contribution in the low-$k$ limit can be estimated as
\begin{equation}
C_{noise}(k) = \frac{1}{N_{sim}} \frac{k_F^3}{4\pi k^2 \delta k} \frac{P_{\rm err}}{P_{11}(k)} \;,
\end{equation}
where $N_{sim}$ is the number of simulations, $k_F$ is the fundamental mode of a simulation box, and $\delta k$ is the width of the bins. In practice our model for $C_{noise}(k)$ is based on a fit of the scatter between five simulations in the measurement of each of the transfer functions. In this way our estimate of this contribution is valid on all scales and for each $\beta_i(k)$. As we already explained, the second term in the covariance matrix comes from the theoretical uncertainties. Our estimate for the size of the one and two-loop corrections to the transfer functions is
\begin{equation}
\delta\beta\big|_{\rm 1-loop} = b_1 \left( \frac{D(z)}{D(0)} \right)^2 \left( \frac{k}{0.3\ihMpc} \right)^{1.8} \;, \qquad {\rm and} \qquad \delta\beta \big|_{\rm 2-loop} = b_1 \left( \frac{D(z)}{D(0)} \right)^4 \left( \frac{k}{0.45\ihMpc} \right)^{3.3} \;.
\end{equation}
Given that in our perturbative model $\beta_1(k)$ is calculated up to one-loop, we use the two-loop error in the fit. For fitting the other transfer functions, we use the one-loop contributions to the covariance. Finally, the parameter $\Delta k$ is the coherence length of the transfer functions. In other words, this is the typical scale at which the transfer functions vary with $k$. Given that they are quite smooth, we choose $\Delta k = 0.2\ihMpc$. Our fits are not sensitive to the choice of $\Delta k$ as long as it remains in a reasonable range of values. We choose $k_{\rm max} = 0.5\ihMpc$ for $\beta_1$ and $k_{\rm max} = 0.2\ihMpc$ for all other transfer functions. Given our theoretical errors, the values of the best fitted bias parameters saturate well before $k_{\rm max}$.
\vskip 4pt 

The result of fitting the transfer functions using the procedure described so far is shown in \fig{M0TkNoZ} (red lines). The $k$ dependence of $\beta_1(k)$ is well described at $k\le 0.3-0.4\ihMpc$ when using theoretical errors in the fitting procedure. The constant pieces of the nonlinear bias transfer functions $\beta_2(k)$, $\beta_{\mathcal{G}_2}(k)$, and $\beta_3(k)$ are in reasonable agreement with those measured from simulations at low $k$. The typical relative error of the fitted parameters is roughly $1\%$ for $b_1$ and roughly $10\%$ for all other parameters. As we have shown in \fig{noiseM0} above,  the model error changes only minimally when using the theory fits instead of the full transfer functions  from simulations for the cubic model, confirming that a perturbative description of the transfer functions is sufficiently accurate for our purposes.
\vskip 4pt

If the Zel'dovich density is kept explicitly and not absorbed by shifted bias operators using \eqq{deltaZshifted}, the  transfer functions remain unchanged at low $k$ (except for the expected offset of 1 for $\beta_1$ and $1/2$ for $\beta_{\mathcal{G}_2}$), but they change their shape at high $k$, as shown in \fig{M0Tk}. This is because the higher order corrections to \eqq{deltaZshifted} become important at high $k$. However, the theory prediction for the transfer functions with theoretical errors are flexible enough to capture this difference. 
\vskip 4pt 

\begin{figure}[tp]
\centering
\includegraphics[width=0.4\textwidth]{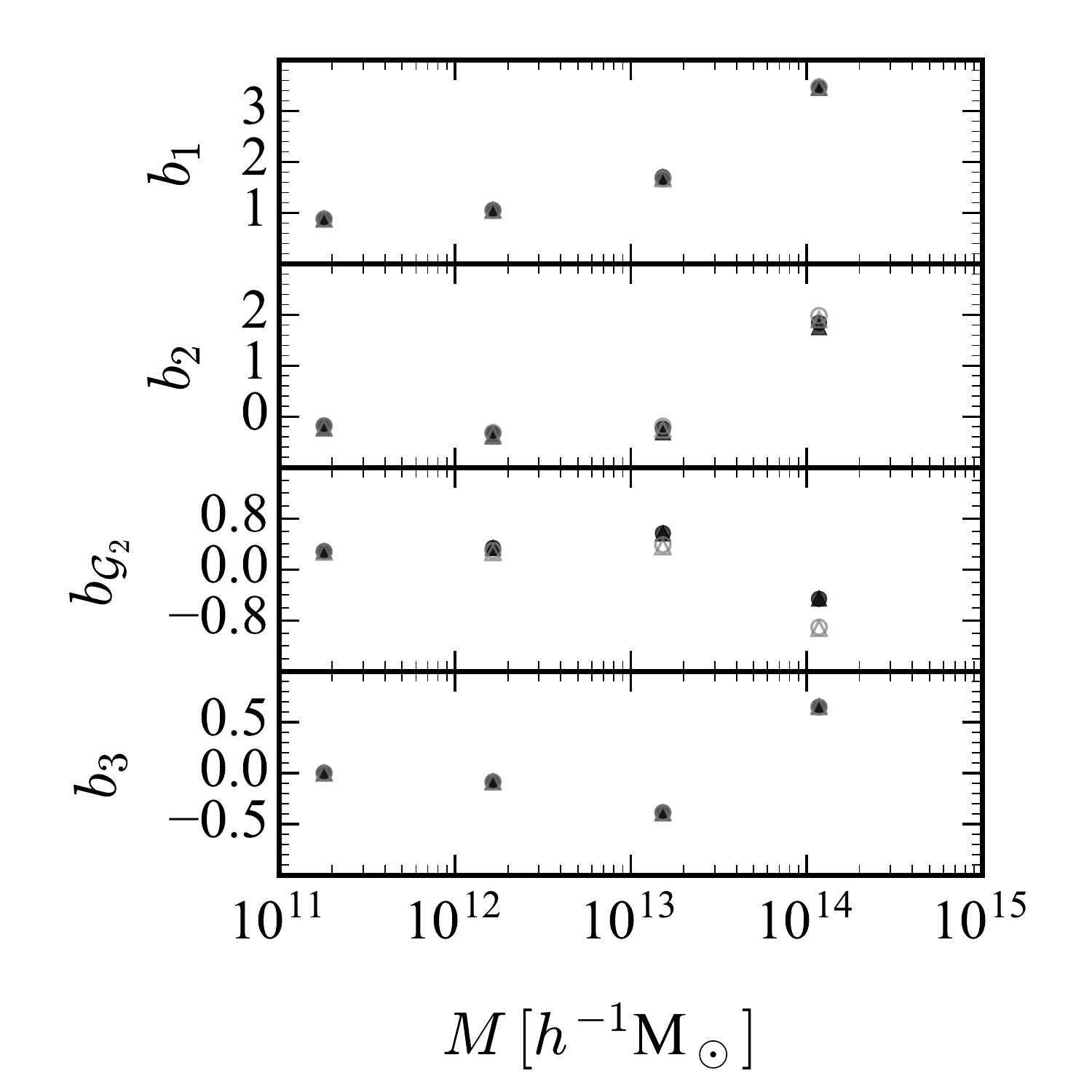}
\caption{Best-fit bias parameters as a function of halo mass, obtained by fitting the transfer functions $\beta_i(k)$  shown in  \fig{M0TkNoZ} and \fig{M0Tk} using Eqs.~\eq{beta1Fit} and \eq{beta23Fit}.
We show results when using the cubic model with the full Zel'dovich density (circles), and when absorbing it using bias operators (triangles). In the former case, we plot $b_1+1$ and $b_{\mathcal{G}_2}+1/2$ to simplify the comparison. 
Filled symbols use theoretical errors in the fitting process, while open symbols are without theoretical errors and go to higher $k$.
The marginalized $1\sigma$ uncertainty is typically $\sim 1\%$ for $b_1$ and $\sim 10\%$ for all other parameters.
}
\label{fig:biasOfM}
\end{figure}

So far we have used the perturbation theory predictions in a rigorous way, accompanied with the appropriate estimate of the neglected higher order corrections. This effectively restricts the range of applicability of perturbative results to $k\leq 0.4 \ihMpc$, which is a bit smaller than the rough estimate for the nonlinear scale $k_{\rm NL}$ at $z=0.6$. In what follows we use the perturbation theory prediction for $\beta_1(k)$ in a different way---just as an ansatz for the fitting function. We modify our fitting procedure to remove any theoretical error for $\beta_1(k)$ and extend the range of wavenumbers to $k_{\rm max}=0.8 \ihMpc$. The theoretical errors and range of scales remain the same for other transfer functions. In other words, the point of this exercise is to test whether the functional form of $\beta_1$ with 5 free parameters has enough freedom to fit the full shape of the measured transfer function. The results are shown in \fig{M0TkNoZ} and \fig{M0Tk} as orange lines. We can see that extending the fit to higher $k$ without accounting for theoretical error improves the fit of $\beta_1(k)$ almost up to $k\simeq 1\ihMpc$, although in some cases this gives a worse fit of the constant part of the higher order transfer functions, which have a smaller effect on the model error than corrections to $\beta_1(k)$.
\vskip 4pt

\begin{table}[tbp]
\centering
\renewcommand{\arraystretch}{1.1}
\begin{tabular}{@{}lrrrrrrrrrrrr@{}}
\toprule
$\log M[h^{-1}\mathrm{M}_\odot]$ & \phantom{ } &  
\multicolumn{1}{c}{$1+b_1$} & \phantom{ } & \multicolumn{1}{c}{$c_s$} & \phantom{ } & 
\multicolumn{1}{c}{$b_{\Gamma_3}$}
&\phantom{ } & \multicolumn{1}{c}{$b_2$} &\phantom{ } & \multicolumn{1}{c}{$1/2+b_{\mathcal{G}_2}$} &\phantom{ } & 
\multicolumn{1}{c}{$b_3$}  \\ 
\colrule 
$10.8-11.8$  && 0.88 (0.88)   && -0.17 (0.059)  && 0.26 (0.30)  && -0.18 (-0.18)   && 0.29 (0.28)   && 0.0023 (0.0023)  \\
$11.8-12.8$  && 1.05 (1.05)  && 0.48 (0.23)    && 0.26 (0.26)  && -0.33 (-0.31)  && 0.34 (0.29)  && -0.085 (-0.085)  \\
$12.8-13.8$  && 1.70 (1.67)  && 3.5 (1.2)     && 0.23 (0.061)  && -0.24 (-0.19)  && 0.57 (0.39)  && -0.39 (-0.39)  \\
$13.8-15.2$  && 3.46 (3.47)  && -0.34 (3.0)     && 0.055 (1.2)   && 1.8 (2.0)   && -0.46 (-0.90)  && 0.65 (0.65)  \\
\botrule
\end{tabular}
\caption{Best-fit bias parameters fitting the transfer functions in Eqs.~\eq{beta1Fit} and \eq{beta23Fit}, when using the cubic bias model with the full Zel'dovich density, $\delta_h=\delta_Z+\beta_1\shifted\delta_1+\cdots$.
Numbers in brackets show results when ignoring the theoretical error when fitting the transfer functions, effectively fitting to higher $k$.
For the relation to usual Lagrangian bias parameters see Eqs.~\eq{shifted_b_lagrangian_b_1}-\eq{shifted_b_lagrangian_b_4}.
The marginalized $1\sigma$ uncertainty is typically $\sim 1\%$ for $b_1$ and $\sim 10\%$ for all other parameters.
}
\label{tab:TkFitTableFullZ}
\end{table}

\begin{table}[tbp]
\centering
\renewcommand{\arraystretch}{1.1}
\begin{tabular}{@{}lrrrrrrrrrrrr@{}}
\toprule
$\log M[h^{-1}\mathrm{M}_\odot]$ & \phantom{ } &  
\multicolumn{1}{c}{$b_1$} & \phantom{ } & \multicolumn{1}{c}{$c_s$} & \phantom{ } & 
\multicolumn{1}{c}{$b_{\Gamma_3}$}
&\phantom{ } & \multicolumn{1}{c}{$b_2$} &\phantom{ } & \multicolumn{1}{c}{$b_{\mathcal{G}_2}$} &\phantom{ } & 
\multicolumn{1}{c}{$b_3$}  \\ 
\colrule 
$10.8-11.8$  && 0.88 (0.88)  && 0.012 (0.078) && 0.60 (0.59) && -0.22 (-0.22)  && 0.29 (0.27)  && 0.0029 (0.0029) \\
$11.8-12.8$  && 1.06 (1.05) && 0.65 (0.25)    && 0.57 (0.57) && -0.38 (-0.37) && 0.34 (0.27) && -0.084 (-0.084) \\
$12.8-13.8$  && 1.70 (1.67) && 3.7 (1.2)       && 0.53 (0.35) && -0.31 (-0.25) && 0.57 (0.36) && -0.39 (-0.39) \\
$13.8-15.2$  && 3.46 (3.47) && -0.14 (3.0)     && 0.32 (1.4) && 1.8 (1.9)  && -0.46 (-0.92) && 0.66 (0.66) \\
\botrule
\end{tabular}
\caption{Same as Table~\ref{tab:TkFitTableFullZ} but absorbing the Zel'dovich density with bias operators using \eqq{deltaZshifted}. 
Again, numbers in brackets show results when ignoring the theoretical error when fitting the transfer functions, effectively fitting to higher $k$, and the marginalized $1\sigma$ uncertainty is typically $\sim 1\%$ for $b_1$ and $\sim 10\%$ for all other parameters.
}
\label{tab:TkFitTableExpandZ}
\end{table}

The best-fit values for the bias parameters are shown in \fig{biasOfM} and in Tables~\ref{tab:TkFitTableFullZ} and \ref{tab:TkFitTableExpandZ}. The marginalized $1\sigma$ uncertainty is typically $\sim 1\%$ for $b_1$ and $\sim 10\%$ for all other parameters.
As expected, the linear bias increases with halo mass, from $b_1\simeq 0.9$ for the lowest halo mass bin to $b_1\simeq 3.5$ for the most massive halos; the local quadratic and cubic bias parameters $b_2$ and $b_3$ are negative for low and intermediate mass halos, and become large and positive for the more massive halos.
The quadratic tidal bias parameter is positive for low masses and negative for the most massive halos.
These trends broadly agree with theoretical expectations and previous measurements of bias parameters in the literature using  different measurement techniques \cite{Modi1612,LazeyrasSchmidt1712,AbidiBaldauf1802}. 
However, let us again stress that we expect $b_1$ to be the only bias parameter that is equal to its renormalized value, measured for example from the power spectrum. The other bias parameters can be different from the values inferred from the correlation functions. The fact that $b_2$ or $b_{\G_2}$ are close to their renormalized values indicates that our prescription for building shifted operators is not very sensitive to very high-$k$ modes. It would be interesting to see if this remains true at higher orders in perturbation theory and we leave this question for future work.
\vskip 4pt 

In most cases the best-fit parameters are similar with or without theoretical errors included in the fitting procedure, and with or without absorbing $\delta_Z$ in the bias operators. 
An exception are  $c_s$ and $b_{\Gamma_3}$, which are fitted only from the $k$-dependence of $\beta_1(k)$ and which vary significantly between fitting procedures. This is due to a strong degeneracy of these two parameters when fitting only $\beta_1(k)$. This degeneracy could be broken by including the shifted $\Gamma_3$ operator in the bias expansion on the grid and measuring its transfer function. We do not attempt to do this here, noting that the transfer functions are fitted sufficiently well for our purposes independently of the fitting method.

\subsection{Power Spectrum Model Using Approximate Transfer Functions}
\label{se:PkModel}

In \secref{FourierResults} above we focused on the model error power spectrum $\Perr$ assuming optimal transfer functions that minimize $\Perr(k)$ in every $k$ bin. 
This led to several simplifications, including a relation between $\Perr$ and the cross-correlation coefficient $r_{cc}$ between the model and truth (see \app{PerfMeasures}). 
How do these results change if we instead use the approximate transfer functions from this section? In particular, what are the implications for modeling of the power spectrum and inference of cosmological parameters?
We explore this by calculating the impact of using approximate transfer functions $\beta_i' = \beta_i + \Delta\beta_i$, where $\beta_i$ are the optimal transfer functions that minimize $\Perr$, and $\beta_i'$ are their approximation using the perturbation theory fits (or any other basis of smooth functions).
\vskip 4pt

Let us begin with the power spectrum of the model error. It is easy to show that the new $P_{\rm err}$ is given by
\begin{equation}
P_{\rm err}' = P_{\rm err}+ \sum_i \left( \frac{\Delta\beta_i}{\beta_i} \right)^2 \beta_i^2\, \vev{|\mathcal O^\perp_i|^2} \;,
\end{equation}
where for simplicity we have suppressed the explicit dependence on $k$. The relative change can be written as 
\begin{equation}
\frac{\Delta P_{\rm err}}{P_{\rm err}} = \sum_i f_i \left( \frac{\Delta\beta_i}{\beta_i} \right)^2 \frac{P_{\rm model}}{P_{\rm err}}\;,
\end{equation}
where $f_i$ is the fraction of the model halo power spectrum $P_{\rm model}$ that comes from the operator $\mathcal O_i^\perp$. The typical values of $f_i$ are $f_i\sim 1$ for the shifted linear field and $f_i\leq 0.1$ for higher order operators (see \fig{M0ModelContris}). Close to the nonlinear scale where the perturbation theory fits break down the error power spectrum is roughly a few times smaller than the halo power spectrum (depending on the halo mass). Therefore, if the transfer functions can be fitted better than $\mathcal O(10\%)$, the change of $P_{\rm err}$ compared to the optimal value is smaller than $\mathcal O(1\%)$. 
\vskip 4pt

The situation is different if we are interested in the model for the halo power spectrum. In this case 
\begin{equation}
P_{\rm model}' = P_{\rm model} + 2\sum_i \frac{\Delta\beta_i}{\beta_i} \,\beta_i^2\vev{|\mathcal O^\perp_i|^2}\, + \mathcal O \left(\left( \frac{\Delta\beta_i}{\beta_i} \right)^2 \right) \;.
\end{equation}
Notice that the leading correction is linear in $\Delta\beta_i/\beta_i$, and the model for the halo power spectrum is therefore more sensitive to the error in the transfer functions. In cosmological parameter inference, in order not to get biased results, we would have to marginalize over the uncertainty $\Delta P_{\rm model}$. In other words, this uncertainty acts like an extra noise. It is then interesting to compare $\Delta P_{\rm model}$ and $P_{\rm err}$ and see which one is expected to dominate. It is easy to see that 
\begin{equation}
\frac{\Delta P_{\rm model}}{P_{\rm err}} = 2\sum_i f_i \frac{\Delta\beta_i}{\beta_i} \frac{P_{\rm model}}{P_{\rm err}}\;.
\end{equation}
Plugging in typical numbers we can see that this ratio can be easily of order 1 (or even higher for the low mass halos) if the transfer functions are not approximated to better than $\mathcal O(10\%)$ at all scales of interest for $\beta_1$, or ${\rm few} \times \mathcal O(10\%)$ for other transfer functions. In our perturbation theory fits this is marginally achieved.  
\vskip 4pt 

\begin{figure}[tp]
\centering
\includegraphics[width=0.7\textwidth]{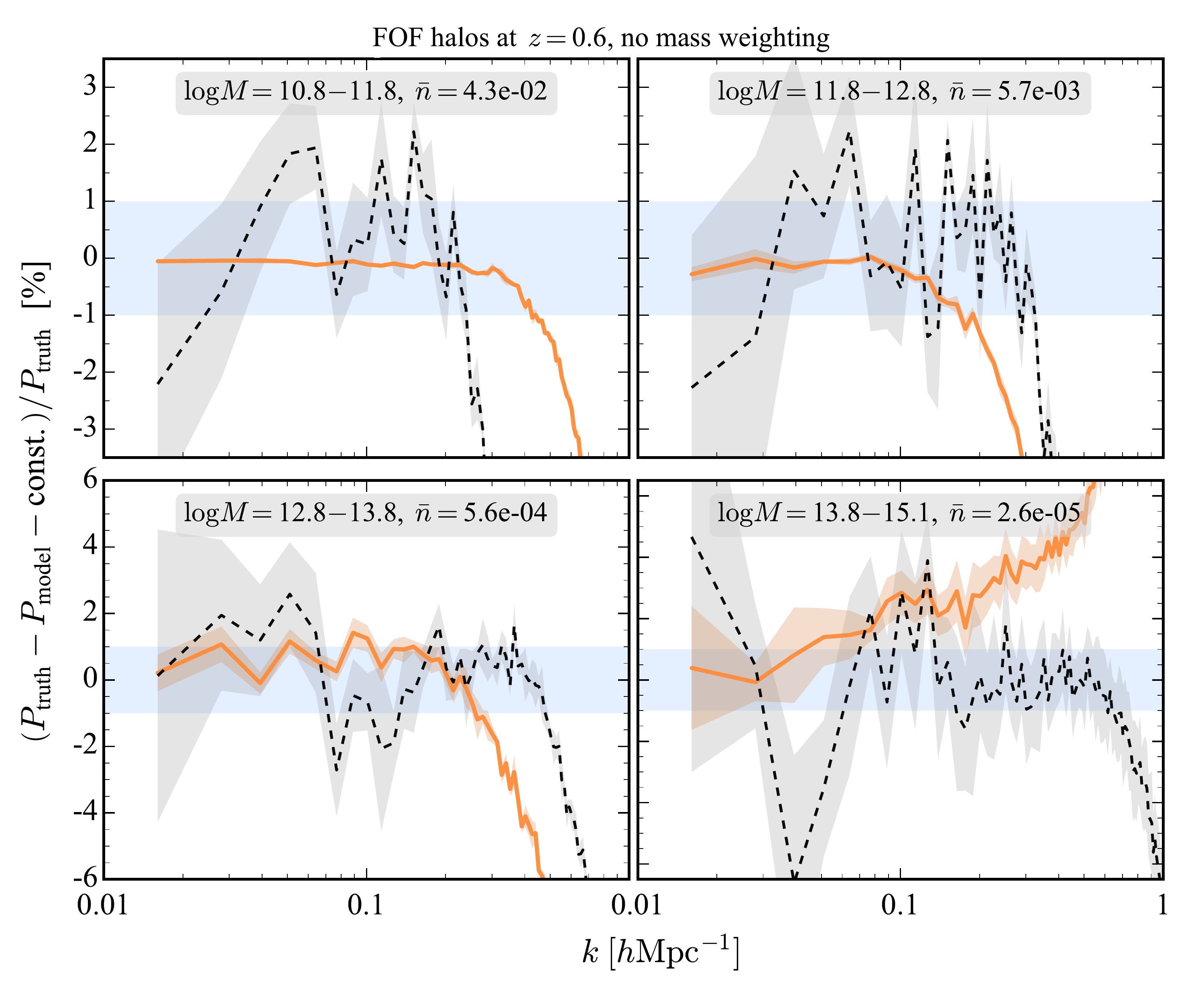}
\caption{Comparison of the power spectrum of the quadratic bias model against the true halo power spectrum measured in simulations.
The orange solid curves assume best-fit transfer functions that minimize $\Perr(k)$ in every $k$ bin for each realization. 
In contrast, the black dashed curves use the five parameter approximation of the ensemble-averaged transfer functions using perturbation theory as described in \secref{PkModel}, designed to fit both the best-fit transfer functions that minimize $\Perr$ and the true halo power spectrum.
The five parameter fit  describes the true halo power spectrum at the $1\%$ to $2\%$ level up to $k\simeq 0.2-0.6\ihMpc$, depending on halo mass, although the uncertainty, estimated from the scatter between the five independent simulations and shown by shaded regions, is considerable. 
}
\label{fig:PkModel}
\end{figure}

It is important to point out that the transfer function fits are dominated by the low-$k$ data, particularly when the theoretical errors are included. This leads to a relatively large variance on the best-fit parameters as explained in the previous section. It is then interesting to ask whether in this range it is possible to find a set of bias parameters that gives the correct model for the halo power spectrum. To answer this question we fit the measured power spectrum using the same bias model as used for the transfer functions. Importantly, we restrict the range of possible bias parameters to be compatible with the fits of the transfer functions. The results are shown in \fig{PkModel} where the relative error for the halo power spectrum is plotted as a function of $k$. The solid orange line corresponds to the model with the optimal transfer functions (minimizing $\Perr$). The dashed black line corresponds to the perturbation theory model with five bias parameters determined to give the best possible fit to the halo power spectrum while still being compatible with the fits of the transfer functions. Only for the lowest mass bin the optimal transfer functions perform better than the perturbation theory fits. In all other cases, the simple fits are sufficient to model the measured halo power spectrum accurately on perturbative scales.
For these halos (i.e., for all but the lightest halos), the approximate transfer functions describe the true halo power spectrum slightly better at high $k$ than the optimal transfer functions that minimize $\Perr$; this is expected because the approximate transfer functions are fitted such that  $\Perr$ is close to the minimal value and at the same time the model power spectrum is close to the true halo power spectrum, whereas the optimal transfer functions only minimize $\Perr$.\footnote{One could choose to ignore $\Perr$ and determine bias transfer functions such that $P_{\rm model}+{\rm const}$ matches the truth power spectrum as well as possible. However, this would necessarily increase $\Perr$ compared to what we obtain with transfer functions that minimize $\Perr$.  Such an increased model error would act as a larger noise which would in general degrade the cosmological information content (see \secref{CosmoReln} above).   }

\section{Relation to Standard Eulerian Perturbation Theory}
\label{se:EulerianRelation}
In this section we discuss how the above results relate to previous approaches and calculations in the literature based on Standard Eulerian bias and Standard Eulerian perturbation theory. 
Specifically, we will demonstrate in \secref{EulerianFailure} that the Eulerian bias expansion fails to describe halos at the field level, and we will discuss the connection with the usual IR-resummation in Standard Eulerian Perturbation Theory in \secref{IRConnection}. 

\subsection{Failure of the Standard Eulerian Bias Expansion at the Field Level}
\label{se:EulerianFailure}

The Standard Eulerian bias model is given by
\begin{equation}
\delta_h(\k) = b_1^{\mathsmaller{\rm E}} \, \delta(\vk) + b_2^{\mathsmaller{\rm E}} \, \delta_2 (\vk) + b_{\G_2}^{\mathsmaller{\rm E}} \, \G_2(\vk) + \cdots \;,
\end{equation}
where all operators are evaluated using the nonlinear matter density field $\delta$. In the perturbative approach, $\delta$ is calculated using Standard Eulerian perturbation theory. Alternatively, the nonlinear matter density field can be measured from simulations. In this section we are going to show that both these approaches face problems when theoretical predictions are compared to simulations at the level of realizations.

\subsubsection{Standard Eulerian Bias Using the Perturbative Matter Density Field}
Let us begin by using the perturbative nonlinear matter density field $\delta$ as the  input for the Standard Eulerian bias model. As in the rest of the paper, we restrict ourselves to operators up to second order and promote bias parameters to $k$-dependent transfer functions. The model for the halo density field then reads
\begin{equation}
\label{eq:explicit_quadratic_eulerian}
\delta_h(\k) = b_1^{\mathsmaller{\rm E}}(k) \big(\delta_1(\vk)+ \delta^{[2]}(\k) \big) + b_2^{\mathsmaller{\rm E}}(k) \, \delta_1^2 (\vk) + b_{\G_2}^{\mathsmaller{\rm E}}(k) \, \G_2[\delta_1](\vk) + \cdots \;,
\end{equation}
where we explicitly wrote the operators at second order in perturbation theory. For example, the second order density field can be calculated in a realization using a simple convolution
\begin{equation}
\delta^{[2]}(\k) = \int_{\p_1\p_2} (2\pi)^3\delta^D(\k-\p_1-\p_2) \ F_2(\p_1,\p_2) \ \delta_1(\p_1) \ \delta_1(\p_2) \;,
\end{equation}
where the $F_2$ kernel is given by
\begin{equation}
F_2(\p_1,\p_2) = \frac 57 +\frac 12 \frac{\p_1\cdot\p_2}{p_1 p_2} \left( \frac{p_1}{p_2} + \frac{p_2}{p_1} \right) + \frac27 \frac{(\p_1\cdot \p_2)^2}{p_1^2p_2^2} \;.  
\end{equation}
Notice that the displacements (the second term in $F_2$) are treated perturbatively. We have already emphasized that this is the reason why Standard Eulerian perturbation theory fails on small scales in describing a realization of the density field of biased tracers. To illustrate this point more quantitatively, let us consider a very simple Universe where the true halo density field is {\em exactly} given by a linear bias:
\beq
\delta_h(\k) = b_1^{\mathsmaller{\rm E}} \delta(\k) + \epsilon(\k)\ ,
\eeq
where $\epsilon$ accounts for stochasticity.
If we fit this halo density field to the model $\delta_h^{\rm (m)}$ based on the linear theory $\delta_1(\k)$ with a scale-dependent transfer function
\beq 
\delta_h^{\rm (m)}(\k) = \beta_1^{\mathsmaller{\rm E}}(k)\delta_1(\k)\ ,
\eeq 
linear regression gives
\begin{align}
\beta_1^{\mathsmaller{\rm E}}(k) \equiv \frac{\vev{\delta_h(\k)\delta_1^*(\k)}}{P_{11}(k)}	& = b_1^{\mathsmaller{\rm E}} \frac{\vev{\delta(\k)\delta_1^*(\k)}}{P_{11}(k)}\ .
\end{align}
Let us compute the r.h.s.~of this equation using Standard Eulerian perturbation theory. On large scales we expect $\beta_1^{\mathsmaller{\rm E}}(k)$ to be close to $b_1^{\mathsmaller{\rm E}}$ with corrections of order $P_{\rm loop}/P_{11}$. However, at next-to-leading order, we find that the transfer function is
\beq 
\beta_1^{\mathsmaller{\rm E}}(k)=b_1^{\mathsmaller{\rm E}} \left(1+\frac{P_{13}(k)}{P_{11}(k)}\right)\;,
\eeq 
where $P_{13}$ is one of the two contributions to the matter power spectrum at one loop $P_{\rm loop}\equiv 2P_{13} + P_{22}$~\cite{Bernardeau:2001qr}. Famously, due to a large contribution from the IR shift terms, $P_{13}$ is much larger than $P_{\rm loop}$~\cite{Senatore:2014via}, and being large and negative causes a significant decay of the transfer function even on scales larger than the nonlinear scale. This decorrelation means that even in the perturbative regime the model fails to predict the halo density field. As a result, the residual noise becomes large and strongly scale-dependent. We find
\begin{align}
\label{eq:P22Perr}
\Perr(k) = \langle{|\hat\epsilon(\k)|^2}\rangle &\ \equiv\ \langle|\delta_h(\k)-\beta_{1}^{\mathsmaller{\rm E}}(k)\delta_{1}(\k)|^2\rangle\ =\ \frac{\alpha}{\bar n}+(b_1^{\mathsmaller{\rm E}})^2 P_{22}(k)\ .
\end{align}
Of course, the residual noise gets corrections from higher-order loop contributions too. However, the $P_{22}$ term is already much larger than the naive expectation---the one-loop power spectrum. To conclude, if Standard Eulerian perturbation theory is used to predict the realization of the halo density field, we expect to find a model error which becomes large and strongly scale dependent around the nonlinear scale.
\vskip 4pt

\begin{figure}[tp]
\centering
\includegraphics[width=0.48\textwidth]{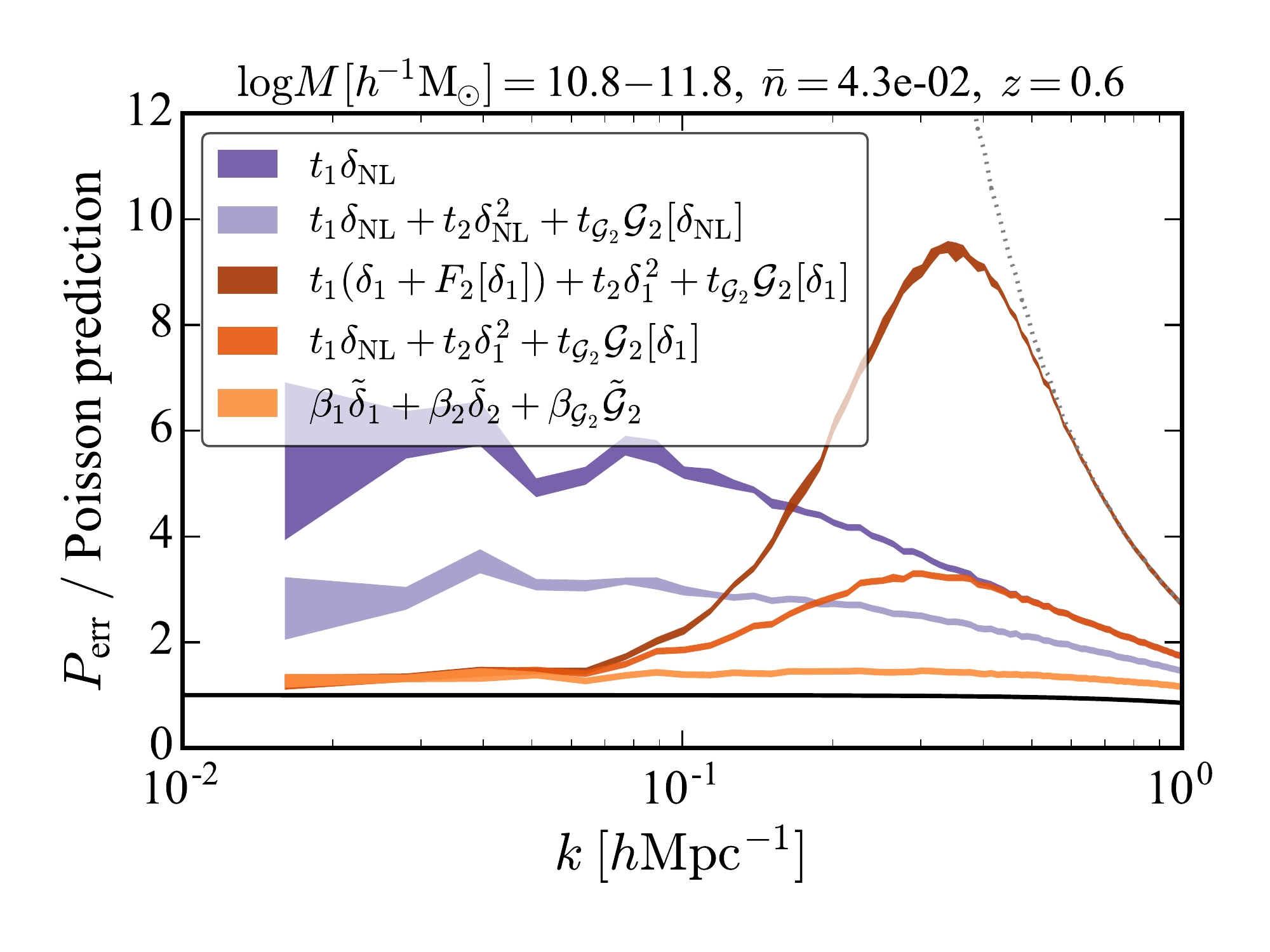}
\hskip 2pt
\includegraphics[width=0.48\textwidth]{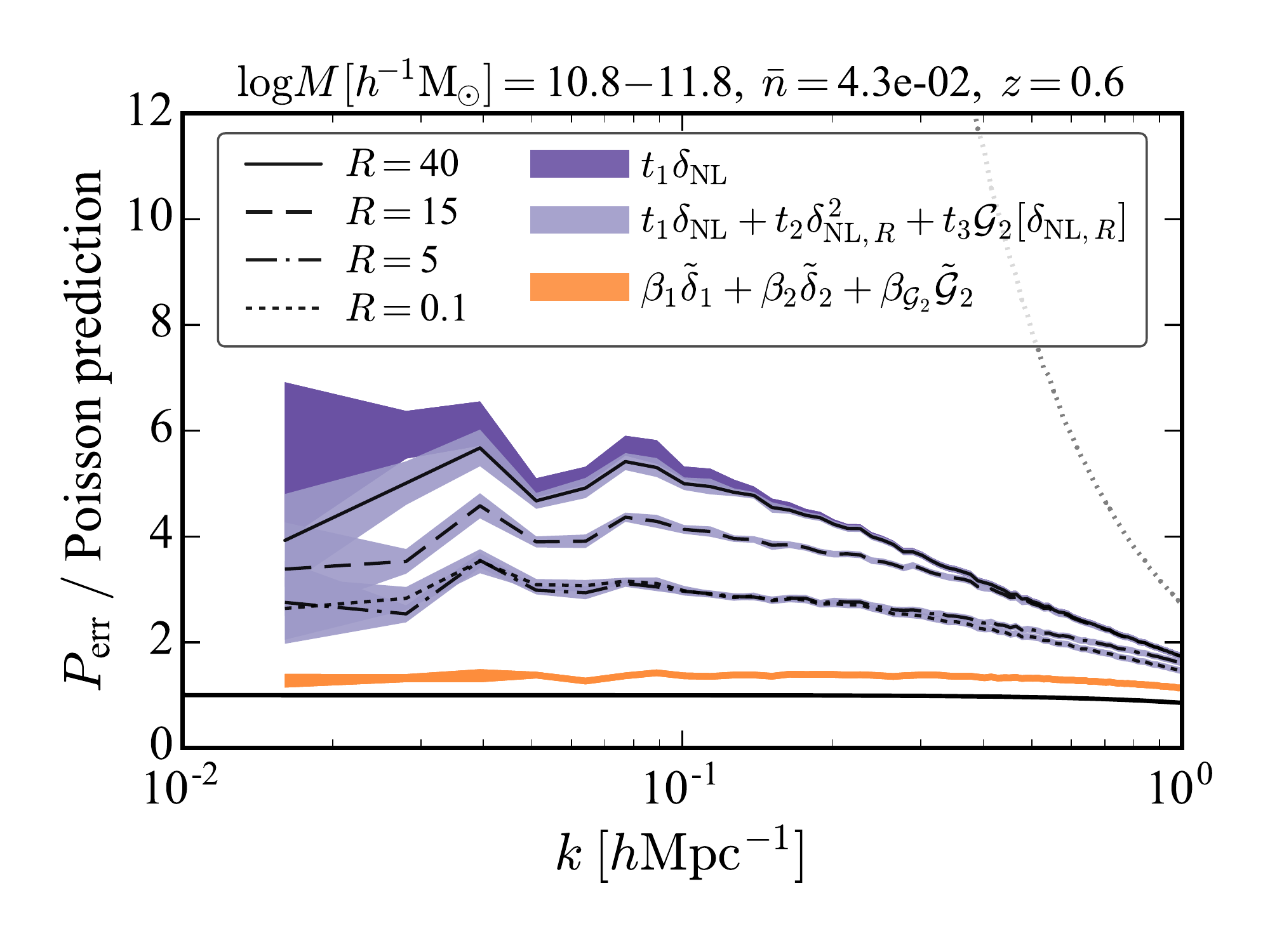}
\caption{\emph{Left panel:}
Model error power spectrum for Standard Eulerian bias models, for the lowest halo mass bin.
Using the nonlinear dark matter $\delta_\mathrm{NL}$ from simulations as the input for the  Standard Eulerian bias model (purple) creates a large error on large scales because it involves squaring $\delta_\mathrm{NL}$, which is rather UV-sensitive.
Alternatively, using the perturbative dark matter density  as the input to the bias model (dark orange) is treating large bulk flows perturbatively, which causes a decorrelation between the model and the true halo density that shows up as a bump in the model error at $k\gtrsim 0.1\ihMpc$.
The quadratic model with shifted bias operators (bright orange) avoids both of these issues by squaring the linear density in Lagrangian space, where this operation is less UV sensitive, and then shifting the resulting field to Eulerian space to achieve coherence  with the Eulerian-space halo density of the simulations.
\emph{Right panel:} Similar, but with Gaussian smoothing applied to $\delta_{\rm NL}$ before computing the quadratic bias operators. For larger smoothing scale $R$, the model error becomes larger because we keep less of the small-scale modes in $\delta_{\rm NL}^2$ that describe the large-scale halo density.
Gaussian smoothing does therefore not resolve the above issues of Standard Eulerian bias.
In both panels,  the width of the shaded regions at low $k$ represents the 1$\sigma$ uncertainty estimated as the standard error of the mean of the five independent simulations; at high $k$, the uncertainty is smaller than the width of the curves.
}
\label{fig:EulerianModelsNoise}
\end{figure}

To test this expectation we use the model in Eq.~\eqref{eq:explicit_quadratic_eulerian} and compare it to simulations. The plot of the power spectrum of the model error normalized to the Poisson prediction is shown in Fig.~\ref{fig:EulerianModelsNoise}. As we expect, this model works very well at large scales, and in the limit $k\to 0$ the noise is close to the Poisson expectation. However, already around $k\sim 0.1\; h{\rm Mpc}^{-1}$ the noise becomes scale-dependent and sharply rises. This is due to the decorrelation of the predicted and simulated halo density fields at these scales. In the high $k$ limit, when the transfer functions approach zero, the power spectrum of the model error by definition approaches the halo power spectrum (black dotted curve). This creates a characteristic bump in the noise curve. Notice that the same quadratic model written in terms of shifted operators performs much better and has the constant noise practically all the way to $k\sim 1\;h{\rm Mpc}^{-1}$.

\subsubsection{Standard Eulerian Bias Using the Matter Density Field from Simulations}
One may be tempted to think that a simple way to fix the problem from the previous section is to use the nonlinear matter density field measured in simulations rather than the perturbative prediction. After all, the N-body simulations provide us with the best possible dark matter field that we can hope for. How does the Standard Eulerian bias model work in this case?
\vskip 4pt 

Let us begin with the linear Standard Eulerian bias. Its model error is shown in Fig.~\ref{fig:EulerianModelsNoise} in dark purple. 
It is a few times larger than the Poisson prediction, especially on very large scales, which is not surprising because we only use a linear bias term.
The next step is to include the second order bias operators, that is to use the following model
\begin{equation}
\delta_h(\k) = b_1^{\mathsmaller{\rm E}}(k) \delta(\vk) + b_2^{\mathsmaller{\rm E}}(k) \, \delta^2 (\vk) + b_{\G_2}^{\mathsmaller{\rm E}}(k) \, \G_2[\delta](\vk) \;,
\end{equation}
where all operators are evaluated using the nonlinear dark matter field measured from simulations. The resulting model error is shown by the light purple line in Fig.~\ref{fig:EulerianModelsNoise}. While the error is quite flat, its amplitude is still a few times larger than the Poisson expectation. In particular, in the low-$k$ limit, the error is much larger than for the quadratic bias model based on perturbation theory. This implies that, even on very large scales, the bias model based on the nonlinear matter density field fails to predict the realization of halos.
\vskip 4pt

This observation brings us back to the discussion of bare vs renormalized bias parameters. As we already explained, using different prescriptions for small scale modes leads to different results for transfer functions. The short modes can have a significant impact on the long-wavelength fluctuations through the nonlinear effects. This effect is amplified when using the true nonlinear dark matter field, where the short modes have a large amplitude. In particular, the low-$k$ limit of the halo power spectrum is dominated by the term
\begin{equation}
\label{eq:low-k-limit_d2d2}
P_{\rm model}(k\to 0) \sim 2(b_1^{\mathsmaller{\rm E}})^2 \int_{\p} P^2(p)\;,
\end{equation}
where $P(k)$ is the nonlinear matter power spectrum. The integral, related to the variance of the square of the density field, is large and dominated by the short modes. In other words, the quadratic term $\delta^2$ in the bias expansion is producing a large shot noise at large scales. When fitted to simulations at the level of realizations, the minimization procedure will favor very small values for $b_2^{\mathsmaller{\rm E}}$ to compensate for this large noise. On the other hand, if the optimal $b_2^{\mathsmaller{\rm E}}$ is chosen to be very small, then the quadratic operator $\delta^2$ essentially does not contribute to the model for the halo density field and consequently the noise of this model is higher than expected. 
\vskip 4pt

To confirm that the second order terms are the real cause for the issue, we also use a hybrid model where the field multiplying $b_1^{\mathsmaller{\rm E}}$ is nonlinear, while the second order operators are calculated using $\delta_1$
\begin{equation}
\label{eq:HybridModel}
\delta_h(\k) = b_1^{\mathsmaller{\rm E}}(k) \delta(\vk) + b_2^{\mathsmaller{\rm E}}(k) \, \delta_1^2 (\vk) + b_{\G_2}^{\mathsmaller{\rm E}}(k) \, \G_2[\delta_1](\vk) \;.
\end{equation}
The error of this model is shown in the left panel in Fig.~\ref{fig:EulerianModelsNoise} in orange. As expected, on large scales this model performs as well as the other perturbative models. On small scales the error is not flat, even though the amplitude of the ``bump" is much smaller than before. The residual scale dependence is due to the improper treatment of the large IR displacements in the linear field which enters the second order operators. This is resolved when using the shifted operators \eq{ShiftedOps} for the bias expansion as we do in the other sections of the paper. 
\vskip 4pt

The final question that we can ask is how the results look if the problematic high-$k$ modes are removed from the model. This can be achieved by constructing the bias operators using the smoothed nonlinear density field. The limit of large smoothing scale is particularly important, because in this limit the low-$k$ values of the transfer functions have to match the renormalized bias parameters (this fact was used in~\cite{LazeyrasSchmidt1712, AbidiBaldauf1802} to infer the values of biases at the field level). What kind of model error do we get in this case? The left panel of Fig.~\ref{fig:EulerianModelsNoise} shows the answer. Larger smoothing scales lead to larger model errors in the low-$k$ limit. In other words, in order to explain the halo density field on large scales, it is better to keep the full nonlinear density field than to smooth it out. These results suggest that the description of the halo density field using the renormalized bias parameters and operators is less optimal than the basis $\shifted O_i$ that we use in this paper. 
\vskip 4pt 

In conclusion, both the Standard Eulerian perturbation theory and the Standard Eulerian bias model have problems when compared to realizations of the halo density field, \MSNEW{confirming the results of \cite{RothPorciani2011}.} The perturbative approach expands shift terms which leads to a decorrelation on short scales and a large scale-dependence of the model error around $k\simeq 0.2\ihMpc$, while using the nonlinear matter density field from simulations amplifies the effects of very short modes and leads to a large model error even in the low-$k$ limit. Crucially, in both cases, some information from the short scales has to be kept in the model. Smoothing the nonlinear matter field always leads to a larger error.

\subsection{Connection to the IR-resummation in Standard Eulerian Perturbation Theory}
\label{se:IRConnection}
So far we have argued that in order to make a perturbative prediction for the realization of the density field of dark matter or biased tracers one has to work with shifted operators. However, at the level of the transfer functions or predictions for the power spectra, only the correlation functions of shifted operators appear. It is then natural to ask how these correlation functions relate to the more familiar counterparts in IR-resummed Standard Eulerian perturbation theory where the large bulk flows are also treated nonperturbatively. This question has been explored previously (see for instance~\cite{Vlah:2015sea}) and in this section we review the main arguments and give some further details. We will begin with the simplest case of dark matter only and then move to biased tracers. 

\subsubsection{Dark Matter}
The nonlinear dark matter field is given by the same expression as $\delta_h$ where all Lagrangian bias parameters are set to zero
\begin{align}
\shifted \delta \ =\  \shifted\delta_1 &\,+\, \frac27 \, \shifted\G_2 \,-\, \frac3{14}\,[\shifted{\G_2\delta}]\, -\, \frac29 \,\shifted\G_3\,+\, \frac16 \,\shifted\Gamma_3\,-\,\shifted{\cal S}_3\ .
\end{align}
The power spectrum of this field up to one-loop order is given by
\begin{equation}
\shifted P(k) = \langle \shifted \delta_1 \shifted \delta_1 \rangle + \frac47 \langle \shifted \delta_1 \shifted{\G}_2 \rangle + \frac4{49} \langle \shifted{\G}_2\shifted{\G}_2 \rangle - \frac3{7} \langle \shifted \delta_1 [\shifted{\G_2\delta}] \rangle - \frac49 \langle \shifted \delta_1\shifted\G_3\rangle + \frac13 \langle \shifted \delta_1 \shifted\Gamma_3 \rangle - 2 \langle \shifted \delta_1 \shifted{\cal S}_3 \rangle \;. 
\end{equation}
Let us make a few comments about some of the terms in this expression. The kernel of the $\G_3$ operator is such that $\langle \delta_1 \G_3 \rangle$ vanishes. This implies that the cross spectrum of shifted operators $\langle \shifted \delta_1\shifted\G_3\rangle$ is non-vanishing only at the two-loop order and we can neglect this contribution. The cross spectrum $\langle \delta_1 [{\G_2\delta}] \rangle$ is proportional to $P_{11}(k)$
\begin{equation}
\langle \delta_1 [{\G_2\delta}] \rangle = -\frac 83 P_{11}(k) \int_0^\infty \frac{p^2dp}{4\pi^2}  P_{11}(p) \;.
\end{equation}
The corrections to this expression for the shifted fields are of the two-loop order and we will ignore them. In the standard calculation of the one-loop power spectrum for biased tracers this term renormalizes the linear bias $b_1$. However, given that in this case we are calculating the power spectrum of the dark matter field, this contribution has to cancel. Indeed, the cancellation is ensured by the contribution from $\shifted{\cal S}_3$. The symmetrized kernel of this operator is such that
\begin{equation}
F_{\shifted{\cal S}_3}^s (\k,\p,-\p) \Big|_{k\to 0} = \frac 4{21} + \mathcal O \left(\frac{k^2}{p^2} \right) \;. 
\end{equation}
This implies that the low-$k$ limit of the correlator $\langle \shifted \delta_1 \shifted{\cal S}_3 \rangle$ is given by
\begin{equation}
\langle \shifted \delta_1 \shifted{\cal S}_3 \rangle \Big|_{k\to 0} = \frac 47 P_{11}(k) \int_0^\infty \frac{p^2dp}{4\pi^2}  P_{11}(p) \;.
\end{equation}
This precisely cancels the contribution from $\langle \shifted \delta_1 [\shifted{\G_2\delta}] \rangle$ in the power spectrum. Therefore, the nontrivial terms that survive at one-loop order are
\begin{equation}
\label{eq:shiftedP1loopDM}
\shifted P(k) = \langle \shifted \delta_1 \shifted \delta_1 \rangle + \frac47 \langle \shifted \delta_1 \shifted{\G}_2 \rangle + \frac4{49} \langle \shifted{\G}_2\shifted{\G}_2 \rangle + \frac13 \langle \shifted \delta_1 \shifted\Gamma_3 \rangle - 2 \langle \shifted \delta_1 \shifted{\cal S}_3^{\rm new} \rangle \;. 
\end{equation}
where $\shifted{\cal S}_3^{\rm new}$ is derived from the $\shifted{\cal S}_3$ operator by subtracting the constant $4/21$ contribution from the kernel. This is the prediction for the one-loop IR-resummed power spectrum from a realization of the shifted fields. 
\vskip 4pt

\begin{figure}[htp]
\centering
\includegraphics[width=0.5\textwidth]{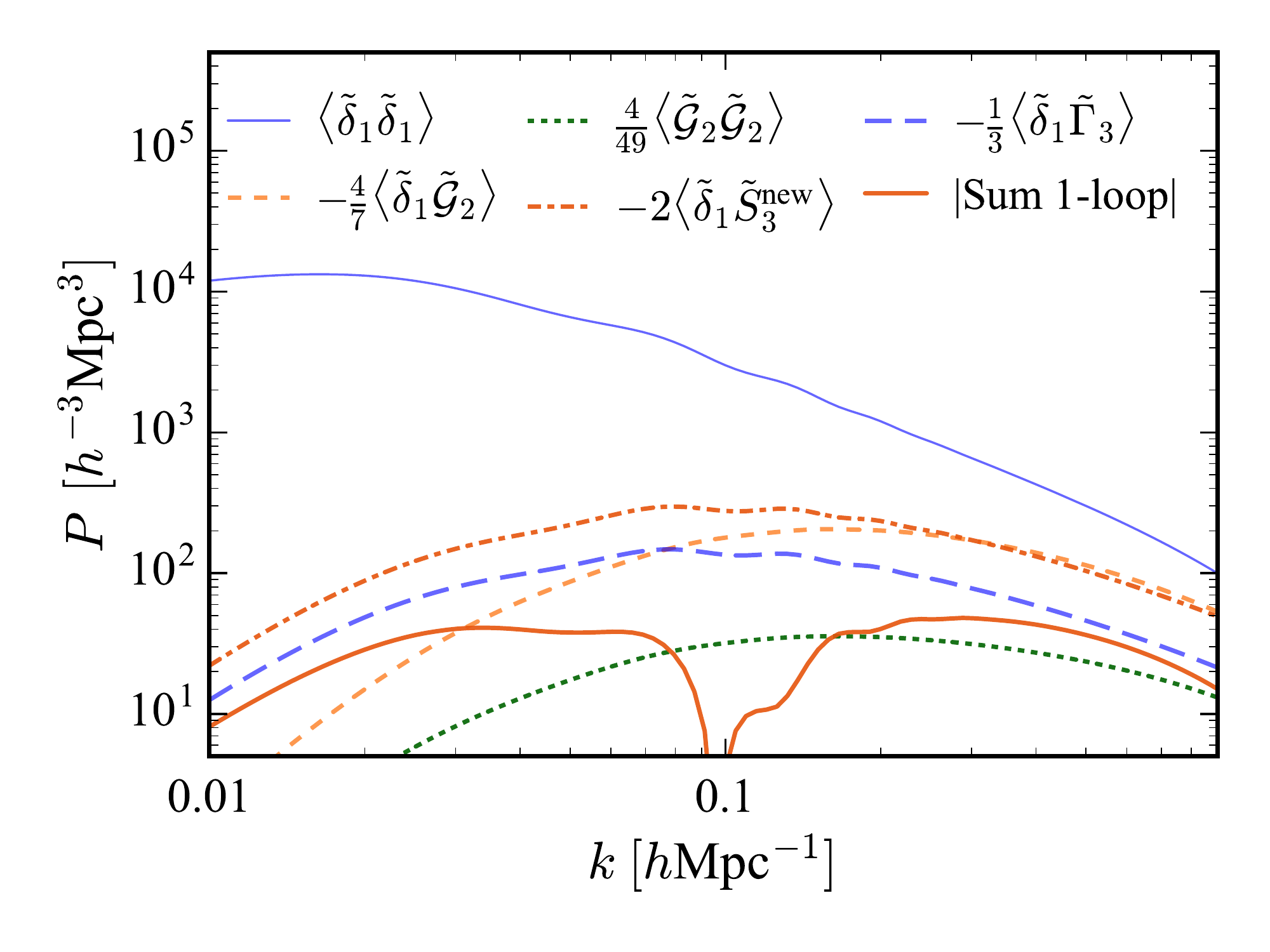}
\caption{Different contributions to the one-loop dark matter power spectrum evaluated using \eqq{shiftedP1loopDM}, using the mean linear theory power spectrum in integrals. The thin blue solid line is the power spectrum of the shifted linear density field. Different dotted and dashed lines are different one-loop contributions. The solid brown thick line is the sum of all one-loop terms.}
\label{fig:shiftedP_contributions}
\end{figure}

\fig{shiftedP_contributions} shows the different contributions to the power spectrum. The thin blue line is the power spectrum of the shifted linear field. The thick brown line is the sum of all four terms in the previous equation which represent the one-loop contributions.\footnote{Notice that there is a one-loop contribution in $\langle \shifted \delta_1 \shifted \delta_1 \rangle$ as well, which we do not write explicitly.} One interesting point to notice is that the total one-loop contribution is at least an order of magnitude smaller than the leading term in the power spectrum on all scales. This result is not surprising, since the expansion of the nonlinear density field in terms of shifted operators is closely related to the expansion of the nonlinear displacement field in Lagrangian perturbation theory, and it is well known that the one-loop power spectrum of the displacement field is smaller than the linear prediction on all scales. 
\vskip 4pt 

In what follows we are going to compare $\shifted P(k)$ to the usual one-loop IR-resummed power spectrum in Standard Eulerian perturbation theory. Before showing the details let us make some general comments. The shifted power spectrum $\tilde P(k)$ contains all terms of the Standard Eulerian perturbation theory up to one-loop. Therefore, the difference can be only two-loop and higher order contributions. Secondly, the large IR-displacements are resummed in $\shifted P(k)$ in the same way as in the usual IR-resummation, using the Zel'dovich displacement field $\vpsi_1$. This implies that the BAO wiggles must be suppressed in the same way. Indeed, we are going to show that both these expectations are correct.   
\vskip 4pt 

Let us begin with a brief summary of how the IR-resummed power spectrum is calculated. The starting point is to split the linear power spectrum in the smooth (non-wiggly) part $P_{11}^{\rm nw}(k)$ and the wiggly part that comes from the BAO oscillations $P_{11}^{\rm w}(k)$. Algorithms to do this splitting efficiently can be found in~\cite{Hamann:2010pw,Baumann:2017gkg}. The effects of the large displacements exactly cancel in the equal-time correlation functions if the power spectrum is smooth. Therefore, the non-wiggly part of the linear power spectrum can be used to evaluate the loop integrals in the usual way. On the other hand, the BAO wiggles are damped by the large displacements (the BAO peak is broadened in the real space correlation function). For this reason the wiggle part of the one-loop power spectrum evaluated using $P_{11}^{\rm w}(k)$ has to be suppressed by the appropriate exponential factor (for more details see \cite{Senatore:2014via,Baldauf:2015xfa,Vlah:2015zda,Blas:2016sfa,Senatore:2017pbn}). The final formula is given by
\begin{equation}
P^{\rm IR}(k) = P_{11}^{\rm nw}(k) + P_{\rm 1-loop}^{\rm nw}(k) + e^{-\Sigma_{\lambda k}^2 k^2}(1+\Sigma_{\lambda k}^2 k^2) P_{11}^{\rm w} + e^{-\Sigma_{\lambda k}^2 k^2} P_{\rm 1-loop}^{\rm w} \;,
\end{equation}
where 
\begin{equation}
\Sigma_{\Lambda}^2 = \frac1{6\pi^2} \int_0^\Lambda dp \ P_{11}(p)\left( 1 - j_0(p\ell_{\rm BAO}) +2 j_2(p\ell_{\rm BAO}) \right) \;.
\label{eq:sigma2definition}
\end{equation}
The parameter $\lambda$ in $\Sigma^2_{\lambda k}$ is usually chosen to be smaller than 1, in order to ensure that the displacements with a given wavenumber affect only the fluctuations on shorter scales. However, in our definition of shifted operators such a condition is not imposed, and for the purposes of the comparison we will use the $k$-independent $\Sigma_{\infty}^2$. In a $\Lambda$CDM-like cosmology the difference between the two definitions is small. 
\vskip 4pt

\begin{figure}[htp]
\centering
\includegraphics[width=0.45\textwidth]{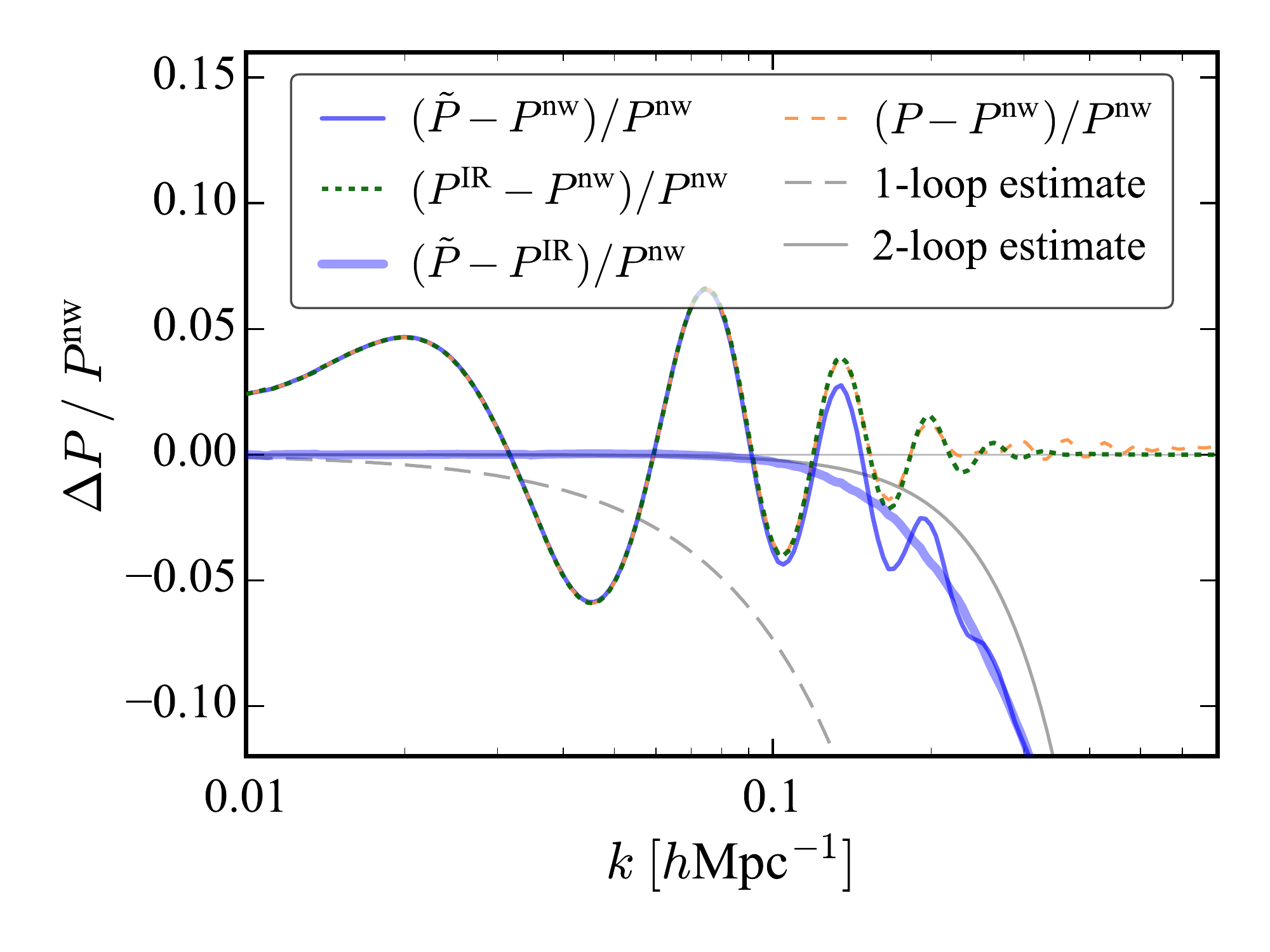}
\hspace{5pt}
\includegraphics[width=0.45\textwidth]{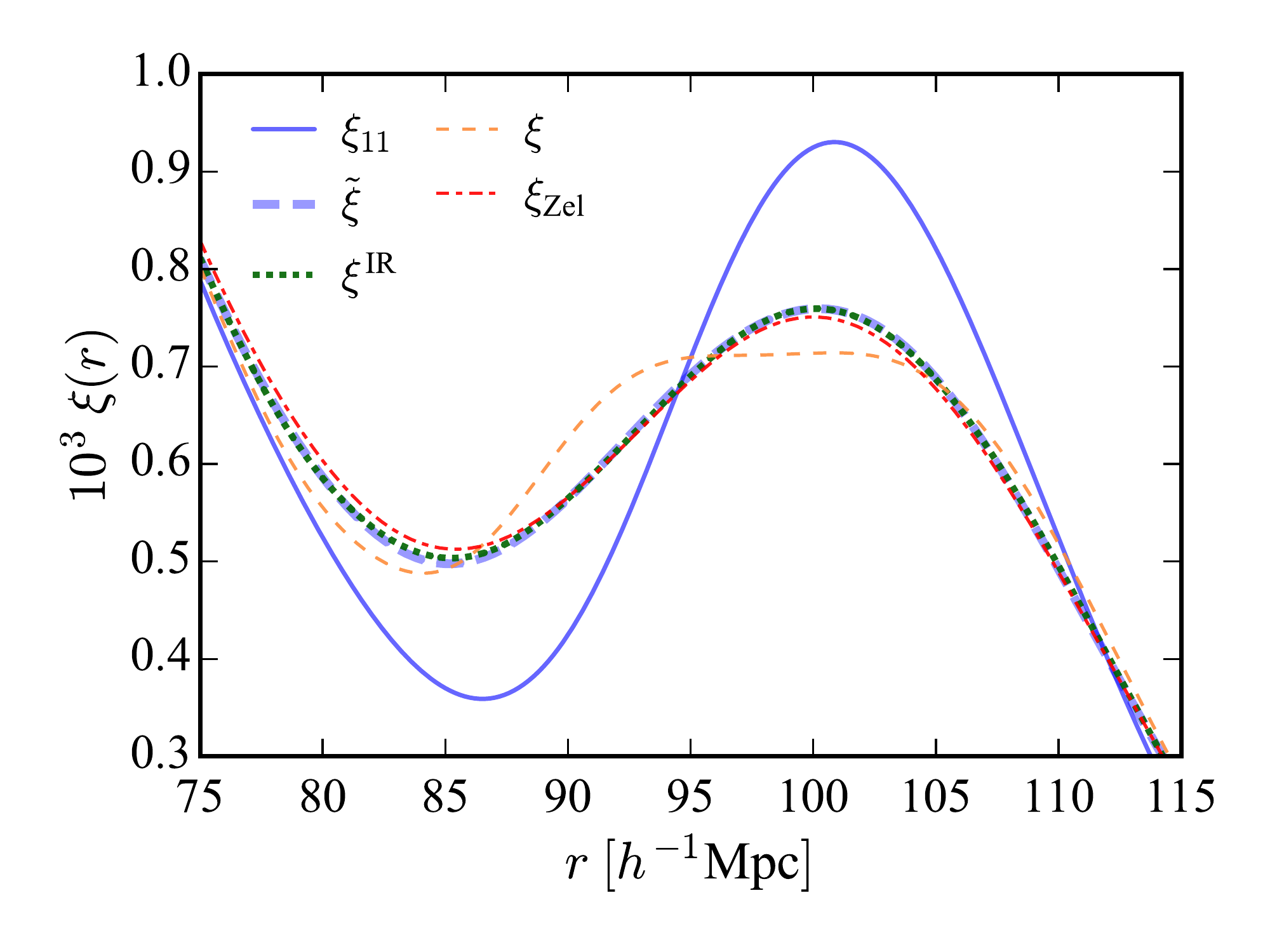}
\caption{Comparison of the IR resummation and shifted fields, for the power spectrum (left) and correlation function (right). All curves are evaluated using theory expressions involving the mean linear power spectrum without matching simulated realizations.}
\label{fig:comparison_IRresum}
\end{figure}

Figure~\ref{fig:comparison_IRresum} shows the comparison of the one-loop dark matter power spectrum calculated using the shifted operators and the standard formula for the IR-resummation. The agreement between the two is reasonably good. The left panel shows different power spectra normalized to the standard one-loop non-wiggle power spectrum. The thin dashed and solid gray lines are the estimate for the typical relative size of the one- and two-loop corrections respectively at $z=0.6$. We can see that the wiggles in the non-IR-resummed one-loop power spectrum are irregular, unlike the case with the IR-resummation. As expected, the difference between the broadband of $\shifted P(k)$ and the Standard Eulerian prediction is of the order of two-loop terms (within a factor of 2). Figure~\ref{fig:comparison_IRresum} also shows that the wiggles in $P^{\rm IR}(k)$ and $\shifted P(k)$ are identical since the relative difference $(\shifted P(k)-P^{\rm IR}(k))/P^{\rm nw}(k)$ is smooth (thick blue line). 
\vskip 4pt 

The other way to see that the wiggles in $P^{\rm IR}(k)$ and $\shifted P(k)$ are the same is to look at the correlation function in real space and focus on the BAO peak. This comparison is shown in the right panel of Fig.~\ref{fig:comparison_IRresum}. The correlation functions calculated using $\shifted P(k)$ and $P^{\rm IR}(k)$, labeled by $\shifted\xi(r)$ and $\xi^{\rm IR}(r)$ respectively, are almost identical. They correctly predict the broadening of the BAO peak, compared to the linear theory prediction $\xi_{11}(r)$. As expected, the correlation function that corresponds to the one-loop power spectrum without the IR-resummation $\xi(r)$ has a very irregular peak. For reference we also plot the prediction based on the Zel'dovich power spectrum, $\xi_{\rm Zel}(r)$, which is known to be in good agreement with simulations. 
\vskip 4pt 

In conclusion, the dark matter one-loop power spectrum calculated with shifted operators is indeed, up to two-loop corrections, identical to the IR-resummed one-loop Standard Eulerian prediction. This remains true for the halo one-loop power spectrum, as we discuss next. 

\subsubsection{Halos}
Let us now turn to the halo density field. Using results from the previous section we can rewrite it as
\begin{align}
\delta_h\ =\  b_1 \,\shifted\delta &\,+\,b_2\,\shifted\delta_2\, + \left( b_{\G_2} - \frac 27 b_1 \right) \shifted\G_2 \, +\,b_3 \,\shifted\delta_3\,+ \left( b_{\G_2\delta} +\frac 3{14} b_1 \right)[\shifted{\G_2\delta}]\, + \left( b_{\G_3} + \frac 29 b_1 \right) \shifted\G_3\,+ \left( b_{\Gamma_3} - \frac 16 b_1 \right) \shifted\Gamma_3\, .
\end{align}
Notice that $b_1$ multiplies the nonlinear shifted density field. For this reason the $\shifted{\mathcal{S}}_3$ operator is absent from the bias expansion and some bias parameters are modified. This expression is very similar to the Standard Eulerian bias expansion. We have already demonstrated that the power spectrum of $\shifted\delta$ is indeed close to the IR-resummed Standard Eulerian one-loop power spectrum. The same is true for the other correlation functions as well. 
\vskip 4pt

To see this more explicitly from the definition of shifted operators let us take a look at the auto and cross spectra of operators $\shifted{\mathcal O}\in\{\shifted{\delta}_2,\shifted{\G}_2\}$ as an example. We have argued that, neglecting the two-loop corrections, these spectra can be calculated at leading order in the following way
\begin{equation}
\langle \shifted{\mathcal O}_a \shifted{\mathcal O}_b \rangle (\k) \ =\ \int d^3\q \; e^{-i\k\cdot \q} \big\langle \mathcal O_a(\q) \; \mathcal O_b (0) \big\rangle \; {\rm Exp}  \left( -\frac12 k_ik_j \langle \Delta \boldsymbol\psi^i \Delta \boldsymbol\psi^j \rangle_c \right) \;.
\label{eq:example_d2G2}
\end{equation}
Following the arguments of~\cite{Baldauf:2015xfa} we are going to show that this expression is identical to the IR-resummed counterpart at the one-loop order. We can first write the correlation function under the integral as a sum of the smooth part and a feature at the BAO scale. Then the integral of the smooth part is dominated by $q\sim 1/k$. For this separation the typical size of the exponential factor can be approximated as 
\begin{equation}
{\rm Exp}  \left( -\frac12 k_ik_j \langle \Delta \boldsymbol\psi^i \Delta \boldsymbol\psi^j \rangle_c \right) \sim {\rm Exp}  \left( -\frac{k^3 P_{11}(k)}{6\pi^2} \right) \;.
\end{equation}
This approximation follows from the form of the correlation function of the relative displacement field which can be written as
\begin{equation}
\langle \Delta \boldsymbol\psi^i \Delta \boldsymbol\psi^j \rangle_c = \frac 1{\pi^2} \int_0^\infty dp \int_{-1}^1 d\mu\; \frac{p^i p^j}{p^2} P_{11}(p) \sin^2 \left(\frac{q p\mu}{2} \right) \;, 
\end{equation}
and noting that this integral peaks at $p\sim 1/q$ for a given $q$. The exponential factor therefore can be always neglected as long as we are in the pertutbative regime where $\frac{k^3 P_{11}(k)}{6\pi^2}\ll 1$. In conclusion, the featureless or smooth part of the power spectrum is identical for the shifted and ordinary operators at the one-loop level
\begin{equation}
\langle \shifted{\mathcal O}_a \shifted{\mathcal O}_b \rangle^{\rm nw} (\k) = \langle \mathcal O_a \mathcal O_b \rangle^{\rm nw} (\k) + {\rm ``two-loop \; corrections"}\;.
\end{equation}
\vskip 4pt

Let us now turn to the feature at the BAO scale. In this case the integral in \eqq{example_d2G2} has support only around $q\sim \ell_{\rm BAO}$ and the exponent, which is a smooth function of $q$, can be approximated at its value at $q\sim \ell_{\rm BAO}$. This leads to the following expression for the wiggle part of the power spectrum 
\begin{equation}
\langle \shifted{\mathcal O}_a \shifted{\mathcal O}_b \rangle^{\rm w} (\k) = e^{-\Sigma_{\infty}^2k^2}\langle \mathcal O_a \mathcal O_b \rangle^{\rm w} (\k) \;.
\end{equation}
where $\Sigma_\Lambda^2$ is exactly given by \eqq{sigma2definition}. We can therefore see that the IR-resummed power and cross spectra of the operators $\delta_2$ and $\G_2$ are indeed the same as the spectra of their shifted counterparts.\footnote{Notice that the constant low-$k$ contribution to $\langle \delta_2 \delta_2 \rangle$ is the same for the power spectrum of the shifted fields. Let us define $P_{\delta_2\delta_2}(0)\equiv \langle \delta_2(\k) \delta_2(-\k) \rangle_{k\to 0}$. The two-point function of this constant part is proportional to the Dirac delta function $\langle \delta_2(\q) \delta_2(0) \rangle = P_{\delta_2\delta_2}\delta^D(\q)$. Then we can write
\begin{equation}
\langle \shifted\delta_2 \shifted\delta_2 \rangle_{k\to 0} = \int d^3\q \; e^{-i\k\cdot \q} P_{\delta_2\delta_2}\delta^D(\q) \; {\rm Exp}  \left( -\frac12 k_ik_j \langle \Delta \boldsymbol\psi^i \Delta \boldsymbol\psi^j \rangle_c \right) = P_{\delta_2\delta_2} \;.
\end{equation}
} In this paper we are interested only in $\delta_2$ and $\G_2$ as these are the only bias operators that we keep in the expansion when we compare it to simulations. However, it is important to stress that this derivation holds more generally and the same conclusions apply to any one-loop contribution to the power spectrum of biased tracers. 
\vskip 4pt 

One important consequence of these results is that all correlators in formulas for the transfer functions can be replaced by the corresponding IR-resummed expressions, which are easier to calculate. In other words, the Standard Eulerian bias expansion 
\begin{align}
\delta_h(\k) = \beta_1^{\mathsmaller{\rm E}}(k) \, \delta(\vk) + \beta_2^{\mathsmaller{\rm E}}(k) \, \delta_2^\perp (\vk) + \beta_{\G_2}^{\mathsmaller{\rm E}}(k) \, \G_2^\perp(\vk) \; .
\end{align}
gives the same (up to two-loop error) power spectrum as our quadratic bias model based on the shifted operators using the following expressions
\begin{equation}
\beta_1^{\mathsmaller{\rm E}}(k) = \ b_1 + c_s^2 k^2 + b_2 \frac{\vev{\delta \delta_2}}{\vev{\delta_1 \delta_1}} + \left( b_{\G_2} -\frac 27 b_1 \right) \frac{\vev{\delta \G_2}}{\vev{\delta_1 \delta_1 }} + \left( b_{\Gamma_3} -\frac16 b_1 \right) \frac{\vev{\delta \Gamma_3}}{\vev{\delta_1 \delta_1 }} \;,
\end{equation}
\begin{equation}
\beta_2^{\mathsmaller{\rm E}}(k) = b_2 \;,
\end{equation}
\begin{equation}
\beta_{\G_2}^{\mathsmaller{\rm E}}(k) = b_{\G_2} - \frac 27 b_1 \;.
\end{equation}
In these formulas all power spectra are calculated in Standard Eulerian perturbation theory with IR-resummation and all bias parameters are as measured from the transfer functions using the shifted fields. 
\vskip 4pt 

Let us finish this section by making a comment about measuring the bias parameters from the power spectrum. We have just argued that Standard Eulerian perturbation theory with IR-resummation predicts the correct shape for the nonlinear power spectrum of biased tracers. The measurement of bias parameters then proceeds in the usual way leading to the usual results. On the other hand, we have also argued that the bias parameters with the lowest model error are different from those inferred from the correlation functions. How do we see this difference using the power spectrum? The answer is that measuring the bias parameters minimizing the model error and fitting the power spectrum are two different fitting procedures with different number of fitting parameters. For instance, when fitting the power spectrum it is common to combine all constant $k\to 0$ contributions from the bias operators such as $\delta_2$ with the noise power spectrum. In this way, the bias parameters are measured only from the $k$-dependence of different contributions. This is possible because all the constant terms are exactly degenerate with the Poisson noise, whose amplitude is fitted simultaneously with other parameters. On the other hand, in the minimization procedure we only fit for the bias parameters and the noise is entirely fixed by their best fit values. In other words, we cannot trade the contributions of different bias operators for the noise. Given that there is one less parameter to fit, the values of biases in the two fitting procedures must be different. Notice that the noise fitted from the power spectrum is always higher than that inferred from minimization. This can be of particular relevance when trying to measure cosmological parameters.

\FloatBarrier
\section{Extension: Halo Mass Weighting}
\label{se:MassWeighting}

So far we have only studied the halo \emph{number density}, and the error or stochastic noise when modeling this with a bias expansion.
But it is well known that the stochastic noise can be smaller for the halo \emph{mass density}, where each halo is weighted by its mass \cite{Seljak0904,Hamaus1004,CaiBernsteinSheth1007,Hamaus1104,Hamaus1207,Baldauf1305,Schmidt1511,2009MNRAS.394..398W}. 
More generally, any weighting of the halos that makes their overdensity more similar to the dark matter overdensity should reduce the stochastic noise relative to the dark matter-based bias expansion.
In particular, if we could weight each halo by the exact value of the dark matter density in the surrounding region, there would be no difference at all between the weighted halo density and the dark matter density, i.e.~the model error would vanish.
A related motivation is that the dark matter density satisfies momentum conservation, so the power spectrum of stochastic effects scales as $k^4$  on large scales \cite{ZELDOVICH1965241,Peebles1980,Goroff86,2008PhRvD..77b3533C,Schmidt1511} and cannot generate a white noise contribution to the large-scale linear dark matter power spectrum; weighting halos by their mass also imposes approximate momentum conservation, therefore suppressing the $k^0$ white noise of the halo number density on large scales \cite{Seljak0904}.
\vskip 4pt 

In practice, the efficiency of this mass weighting method is of course limited by how well halo mass, or the local value of the dark matter density, can be estimated from observables such as galaxy luminosities,  which is limited for example by scatter in the observable-to-halo-mass relation and the fact that halo mass is only a proxy for the local dark matter density.
Moreover, on small scales the error of any analytical model for the weighted halo density will never vanish because
terms that are not included in the model (e.g., two-loop contributions to the nonlinear dark matter density) would ultimately appear  in the measured model error. 
We defer a more complete analysis of halo mass weighting and nonlinear bias expansions to future work, and discuss only some simple simulation results to get a sense for how it can impact the stochastic model error.
\vskip 4pt 

It has been shown that weighting halos with $w(M)=\alpha_h+\alpha_MM$, where $M$ is the halo mass and $\alpha_h$ and $\alpha_M$ are constants, is a good approximation to the optimal halo mass weighting and can significantly reduce the stochasticity of a linear bias expansion \cite{Hamaus1004,CaiBernsteinSheth1007}.
This motivates us to work with a similar mass weighting scheme, but we generalize it by promoting the constants $\alpha_h$ and $\alpha_M$ to $k$-dependent transfer functions.
In the rest of the section we will describe this mass weighting method in more detail, and then present results from simulations.
\vskip 4pt

We will use the following notation in this section. 
$\delta_h^{\rm truth}$ is the true halo \emph{number density} of a simulation or galaxy survey data; $\delta_M$ is the true halo \emph{mass density} of a simulation or galaxy survey data, obtained by weighting each halo by its mass; $\delta_M^\perp$ is the  component of $\delta_M$ that is orthogonal to $\delta_h^{\rm truth}$; 
the \emph{weighted} or \emph{mass-weighted density} $\delta_h^{\rm obs}$ is a linear combination of simulated (observable) halo number and mass density---this combined field is what we regard as the observable; finally, $\delta_h^{\rm model}$ is the bias model that describes $\delta_h^{\rm obs}$.
The power spectra of $\delta_h^{\rm truth}$, $\delta_M^\perp$, and $\delta_h^{\rm obs}$ are called
$P_{\rm truth}$, $P_{MM}^\perp$, and $P_{\rm obs}$, respectively.

\subsection{Mass Weighting Method}
\label{se:massWeightingOptimization}

We will choose the mass weights such that the mean-square error between the weighted halo density (the observable) and the bias expansion (the model) is minimized in every $k$ bin.
To achieve this, we first rewrite the weighted halo density $\delta_h^{\rm obs}$, where each halo is weighted by $w(M)=\alpha_h+\alpha_MM$, as a linear combination of the measured halo number density $\delta_h^{\rm truth}$ and the measured halo mass density $\delta_M$.
Orthogonalizing the latter with respect to the former, so that $\langle\delta_h^{\rm truth}\delta_M^\perp\rangle=0$, and allowing for $k$-dependent weights, we have
\begin{align}
  \delta_h^{\rm obs}(\vk) = \alpha_h(k) \delta_h^{\rm truth}(\vk) + \alpha_M(k)\delta_M^\perp(\vk) \equiv \sum_\mu\alpha_\mu\delta_\mu.
\end{align}
Then, we minimize the mean-square model error
\begin{align}
\Perr(k) 
= \Big\langle\big|\delta_h^{\rm obs}(\vk)-\delta_h^{\rm model}(\vk)\big|^2\Big\rangle
=\Big\langle \big|\sum_\mu\alpha_\mu\delta_\mu-\sum_i \beta_i\, \shifted{\mathcal O}_i\big|^2 \Big\rangle
\end{align}
simultaneously with respect to the mass weights $\alpha_\mu(k)$ and the bias parameters $\beta_i(k)$ in every $k$ bin.\footnote{We will only consider models of the form $\delta_h^{\rm model}=\sum_i \beta_i \shifted{\mathcal O}_i$ whenever we apply mass weighting.}
A trivial but pathological solution is to set all $\alpha_\mu=0$ and $\beta_i=0$, which would give $\Perr=0$, but at the same time it would set the observable density $\delta_h^{\rm obs}$ to zero.
To avoid this, we minimize under the constraint that at least one of the mass weighting parameters has to be nonzero, say $\alpha_h\neq 0$.
Then,
\begin{align}
\Perr(k) = \alpha_h^2\,\Big\langle \big|
\delta_h^{\rm truth}+
\frac{\alpha_M}{\alpha_h}\delta_M^\perp
-\sum_{i} \frac{\beta_i}{\alpha_h} \shifted{\mathcal O}_i\big|^2 \Big\rangle.
\end{align}
For any $\alpha_h\neq 0$, the optimal $\alpha_M$ and $\beta_i$ are then found using linear regression analogously to \eqq{LinRegr}.
Afterward, $\alpha_h$ can be changed to any nonzero value, which changes the overall normalization of all coefficients.
We choose $\alpha_h$ such that the power spectrum of the weighted halo density, $P_{\rm obs}(k)$,  is equal to the power spectrum of the halo number density in the absence of mass weighting, 
i.e.~we impose
\begin{align}
  \label{eq:alphaNorm}
 P_\mathrm{obs}(k)\equiv\alpha_h^2(k)P_{\rm truth}(k)+\alpha_M^2(k)P_{MM}^\perp(k)=P_{\rm truth}(k). 
\end{align}
We therefore add information about the halo mass at the field level in such a way that the observable power spectrum remains unchanged.
We are going to apply this  mass weighting procedure to the simulated halos in the next section.
\vskip 4pt 

Note that changing the normalization condition would change the power spectra, but it would not affect the ratio $\Perr/P_\mathrm{obs}$ or the cross-correlation coefficient between $\delta_h^{\rm obs}$ and the model.
Also note that by adding the orthogonalized mass density $\delta_M^\perp$ as opposed to the mass density $\delta_M$ we ensure that adding this field cannot cancel the number density to set $\delta_h^{\rm obs}$ to zero (the coefficient $\alpha_M$ is only turned on when a part of the mass density that is uncorrelated with the number density helps to reduce the model error).
\vskip 4pt 

To get a better understanding for how the mass weighting operates, consider a toy model where the true halo number and mass density are
\begin{align}
  \label{eq:8}
  \delta_h^{\rm truth} &= \sum_i\beta_i \shifted{\mathcal{O}}_i + \epsilon_h,\\
  \delta_M &= \sum_i\gamma_i \shifted{\mathcal{O}}_i + \epsilon_M.
\end{align}
Then,
\begin{align}
  \label{eq:14}
  \delta_{\rm obs} = \alpha_h\left[\sum_i(\beta_i+s\gamma_i)\shifted{\mathcal{O}}_i + \epsilon_h+s\,\epsilon_M\right]
\end{align}
where $s\equiv\alpha_M^{\rm n.o.}/\alpha_h$ and $\alpha^{\rm n.o.}_M$ is the weight that would appear when not orthogonalizing, i.e.~writing $\delta_h^{\rm obs}=\alpha_h\delta_h^{\rm truth}+\alpha_M^{\rm n.o.}\delta_M=\alpha_h(\delta_h^{\rm truth}+s\delta_M)$.  The mass weighting procedure then amounts to choosing $\alpha_h$ and $s$ such that the power spectrum of  $\epsilon_h+s\,\epsilon_M$ is small while keeping the total power spectrum of $\delta_h^{\rm obs}$ unchanged.
If the fractional size of the stochastic error is different between the two fields\footnote{I.e., if $\langle|\epsilon_h|^2\rangle/\langle|\sum_i\beta_i\shifted{\mathcal{O}}_i|^2\rangle\,\neq\,\langle|\epsilon_M|^2\rangle/\langle|\sum_i\gamma_i\shifted{\mathcal{O}}_i|^2\rangle$, which makes sure that $\delta_h^{\rm truth}$ and $\delta_M$ do not just differ by a normalization factor.}, 
this is most efficient when $\epsilon_h$ and $\epsilon_M$ are correlated; in that case we can generally pick $s(k)$ such that the stochastic fields $\epsilon_h$ and $\epsilon_M$ cancel each other mode by mode, without entirely canceling the signal part involving the $\shifted{\mathcal{O}}_i$ operators.
The effectiveness of this is controlled by the correlation coefficient between $\epsilon_h$ and $\epsilon_M$, and by the fractional size of the stochastic noise relative to the signal.
We will get back to this in more detail in \secref{MassToy} below, but first we describe simulation results with mass weighting.

\subsection{Simulation Results}

\begin{figure}[tp]
\centering
\includegraphics[width=0.6\textwidth]{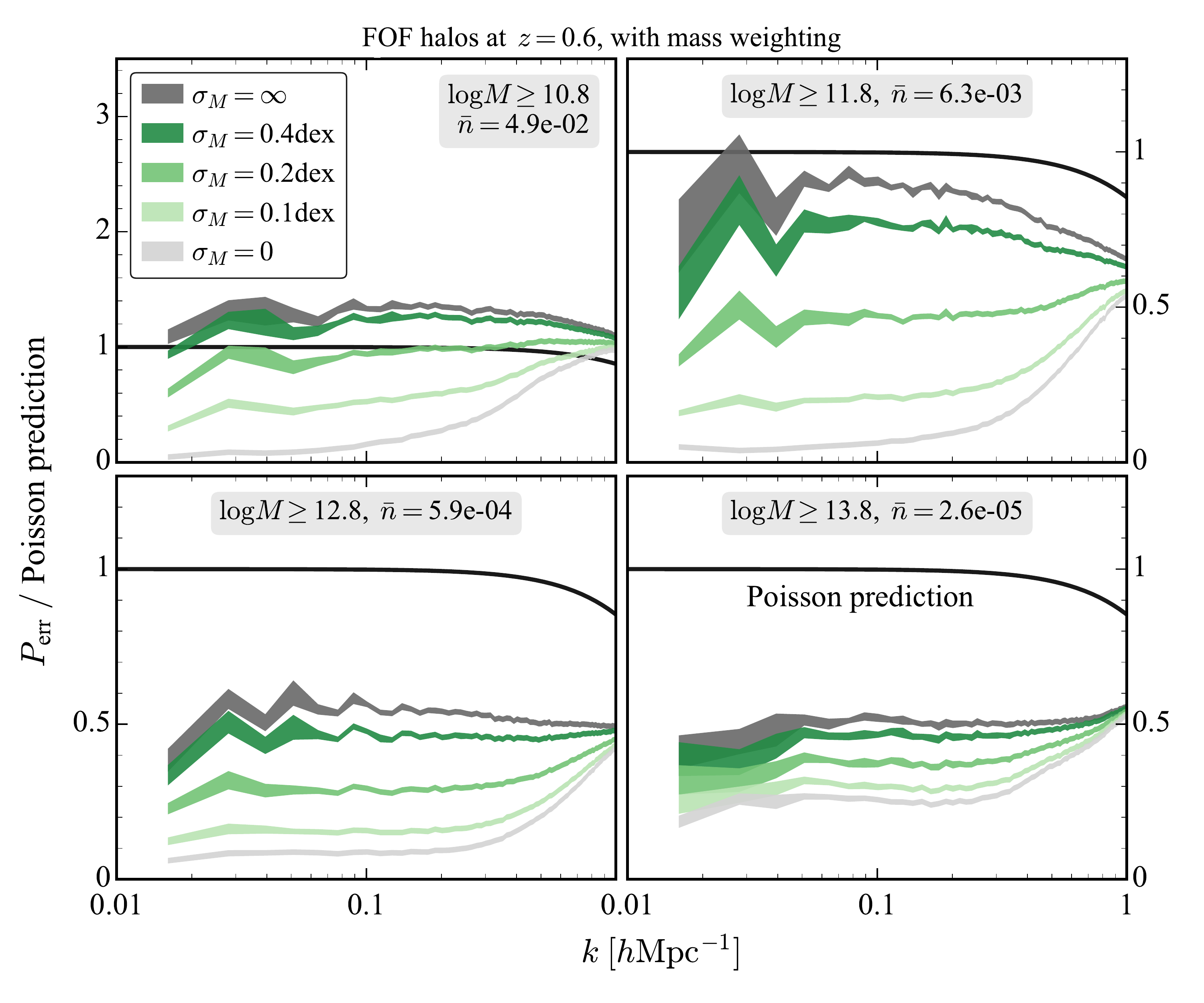}
\caption{Impact of mass weighting on the mean-square model error divided by the Poisson expectation, $\langle|\delta_h^{\rm obs}(\vk)-\delta_h^{\rm model}(\vk)|^2\rangle/(1/\bar n)$.
Here $\delta_h^{\rm obs}$ is a weighted sum of halo number and mass density, $\delta_h^{\rm obs}(\vk)=\alpha_h(k)\delta_h^{\rm truth}(\vk)+\alpha_M(k)\delta_M^\perp(\vk)$, extracted from simulations.
The model is the cubic bias model as before but without $\delta_Z$, i.e.~$\delta_h^{\rm model}=\sum_{i\in\{1,2,\mathcal{G}_2,3\}}\beta_i(k)\tilde\delta_i^\perp(\vk)$. 
The light grey curve assumes that the halo mass is known perfectly (as measured by the FOF halo finder), the green curves include a random scatter $\sigma_M$ in the halo masses, and the dark grey curve assumes no mass weighting, corresponding to the no-mass-weighting result presented previously in the paper.
Transfer functions  $\alpha_\mu(k)$ and $\beta_i(k)$ are optimized as free functions of $k$, with $\alpha_\mu$ satisfying the normalization condition \eq{alphaNorm}.
At low $k$, the width of the curves represents the uncertainty of $\Perr$ estimated from the scatter between the five independent simulations; at high $k$, the estimated uncertainty is smaller.
}
\label{fig:noiseM0M1}
\end{figure}

\begin{figure}[tp]
\centering
\includegraphics[width=0.6\textwidth]{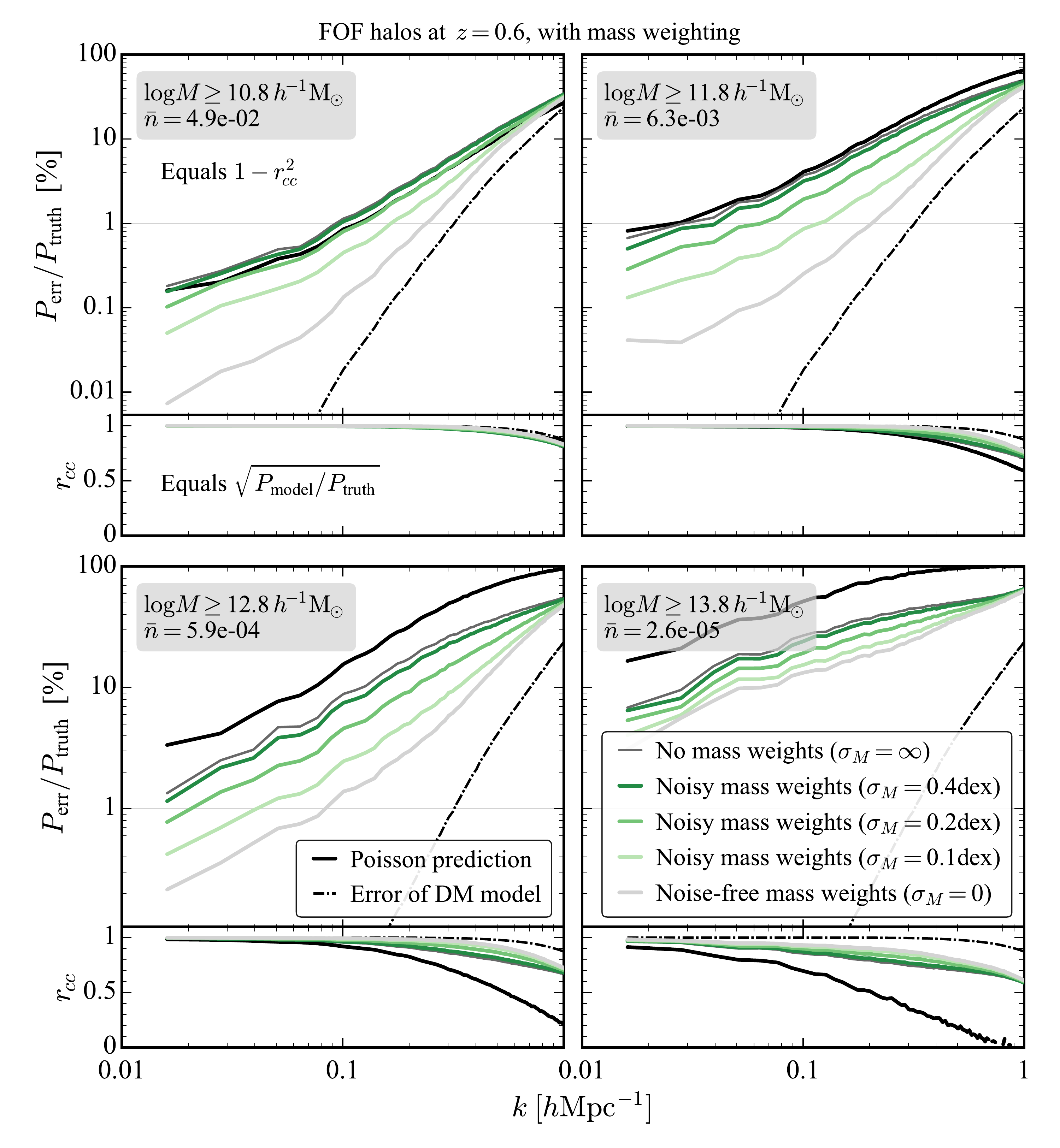}
\caption{Top panels: Impact of mass weighting on the mean square model error $\langle|\delta_h^{\rm obs}(\vk)-\delta_h^{\rm model}(\vk)|^2\rangle$ divided by the true mass-weighted halo density constructed from simulations, $\delta_h^{\rm obs}=\alpha_h\delta_h^{\rm truth}+\alpha_M\delta_M^\perp$ as described in the text, with normalization such that $\langle|\delta_h^{\rm obs}(\vk)|^2\rangle=P_{\rm truth}(k)$, where $P_{\rm truth}$ is the measured halo number density power spectrum.
Lower panels: Cross-correlation coefficient between best-fit cubic bias model and mass-weighted halo density from simulations.
This shows  that halo mass weighting can reduce the model error by more than an order of magnitude on large scales.
}
\label{fig:FractNoiseM0M1}
\end{figure}

Light grey curves in \fig{noiseM0M1} show the mean-square error $\Perr$ of the best cubic bias model to describe the mass-weighted halo density $\delta_h^{\rm obs}$ computed using exact FOF halo masses.
For all but the most massive halo bin, this mean-square model error is less than $10\%$ of the Poisson prediction $1/\bar n$ on large scales; for the most massive halo bin, $M\ge 10^{13.8}\hMsun$, it is about $25\%$ of the Poisson prediction.\footnote{The mass bins are similar to the last sections, using the same minimum halo masses for the four bins, but we do not impose any maximum halo mass cut for any of the bins (very massive halos are easy to observe and halo mass weighting should appropriately up- or down-weight those halos). }
Relative to no mass weighting (dark grey in \fig{noiseM0M1}), mass weighting therefore reduces the large-scale mean-square model error by a factor of 17 for the two densest halo populations, $M\ge 10^{10.8}\hMsun$ and $M\ge 10^{11.8}\hMsun$, by a factor of 7 for the heavier and rarer $M\ge 10^{12.8}\hMsun$ halos, and by a factor of 2 for the very massive $M\ge 10^{13.8}\hMsun$ halos.
\vskip 4pt 
 
To be more realistic, green curves in \fig{noiseM0M1}  include a log-normal scatter added to the FOF halo masses.\footnote{I.e., for each halo we replace $\ln M\rightarrow \ln M+\varepsilon - \sigma_\varepsilon^2/2$, where $\varepsilon$ is drawn from a normal distribution with zero mean and variance $\sigma_\varepsilon^2$; we subtract $\sigma_\varepsilon^2/2$ to ensure that the scatter does not change the average mass $\langle M\rangle$ of the halo population (note $\langle e^\varepsilon \rangle=e^{\sigma_\varepsilon^2/2}$).}  
We find that for a $0.4\,\mathrm{dex}$ (i.e., factor 2.5)  mass scatter, mass weighting is not effective and the model error is only marginally reduced compared to using just the halo number density.
For $0.2\,\mathrm{dex}$ (i.e., $60\%$) mass scatter, however, the large-scale $\Perr$ is reduced relative to no mass weighting by a factor of $1.5-2$ for the three low- and intermediate halo mass bins, and by a factor of 1.3 for the most massive halos.
For $0.1\,\mathrm{dex}$ (i.e., $26\%$) mass scatter, the large-scale $\Perr$ is reduced by a factor of $3-5$ for the low- and intermediate mass halos, and by a factor of 1.6 for the most massive halos.
So if we can determine halo masses with a scatter of $\sim 60\%$ or less,  this could reduce $\Perr$ by factors of a few for halo samples like ours.
\vskip 4pt 

What is the scale dependence of the model error after mass weighting? 
\fig{noiseM0M1} shows essentially no scale dependence  for $k\lesssim 0.1\ihMpc$, but there is a clear scale dependence at higher $k$, and this tends to be stronger than the scale dependence of $\Perr(k)$ without mass weighting.
This could be caused by two-loop terms that are missing in the model and therefore contribute to the measured $\Perr(k)$; after mass weighting, the stochastic noise contribution $P_{\epsilon_0\epsilon_0}$ to  $\Perr$ might be so small that the missing two-loop terms could be the dominant contribution to $\Perr$ at high $k$, especially when using a high number density of halos and assuming perfectly known halo mass.
Alternatively, the $k^2$ corrections to $\Perr$ might be larger after mass weighting.  Resolving this question is beyond the scope of this paper.
Note that in order to make use of the reduced model error on small scales, one would have to model this increased scale dependence of the model error or modify the bias model or mass weighting scheme to obtain a flatter $\Perr$.
\vskip 4pt

\fig{FractNoiseM0M1} shows the cross-correlation coefficient $r_{cc}(k)$ between the mass-weighted halo field $\delta_h^{\rm obs}$ and the best-fit cubic bias model, and shows the fractional mean-square model error $1-r^2_{cc}$.
Using exact FOF halo masses with no scatter, the correlation coefficient at $k\simeq 0.02\ihMpc$ is between $99.995\%$ and $99.9\%$ ($1-r_{cc}^2$ between $0.01\%$ and $0.2\%$)  for all but the most massive halo bin.
This is a substantial improvement over no mass weighting where the correlation coefficient at $k\simeq 0.02\ihMpc$ is between $99.9\%$ to $99.2\%$ ($1-r_{cc}^2$ between $0.2\%$ and $1.5\%$).
The correlation decreases on smaller scales and when adding scatter to the halo mass. Similarly as before, a scatter of $0.4\,\mathrm{dex}$ is too large for mass weighting to be effective, but with a scatter of $0.2\,\mathrm{dex}$ or $0.1\,\mathrm{dex}$ the mass weighting can substantially improve the cross-correlation coefficient, exceeding $99.5\%$ ($1-r_{cc}^2=1\%$) on large scales for all but the most massive halo mass bin.

\subsection{Contributions and Mass Weights}

\begin{figure}[tp]
\centering
\includegraphics[width=0.6\textwidth]{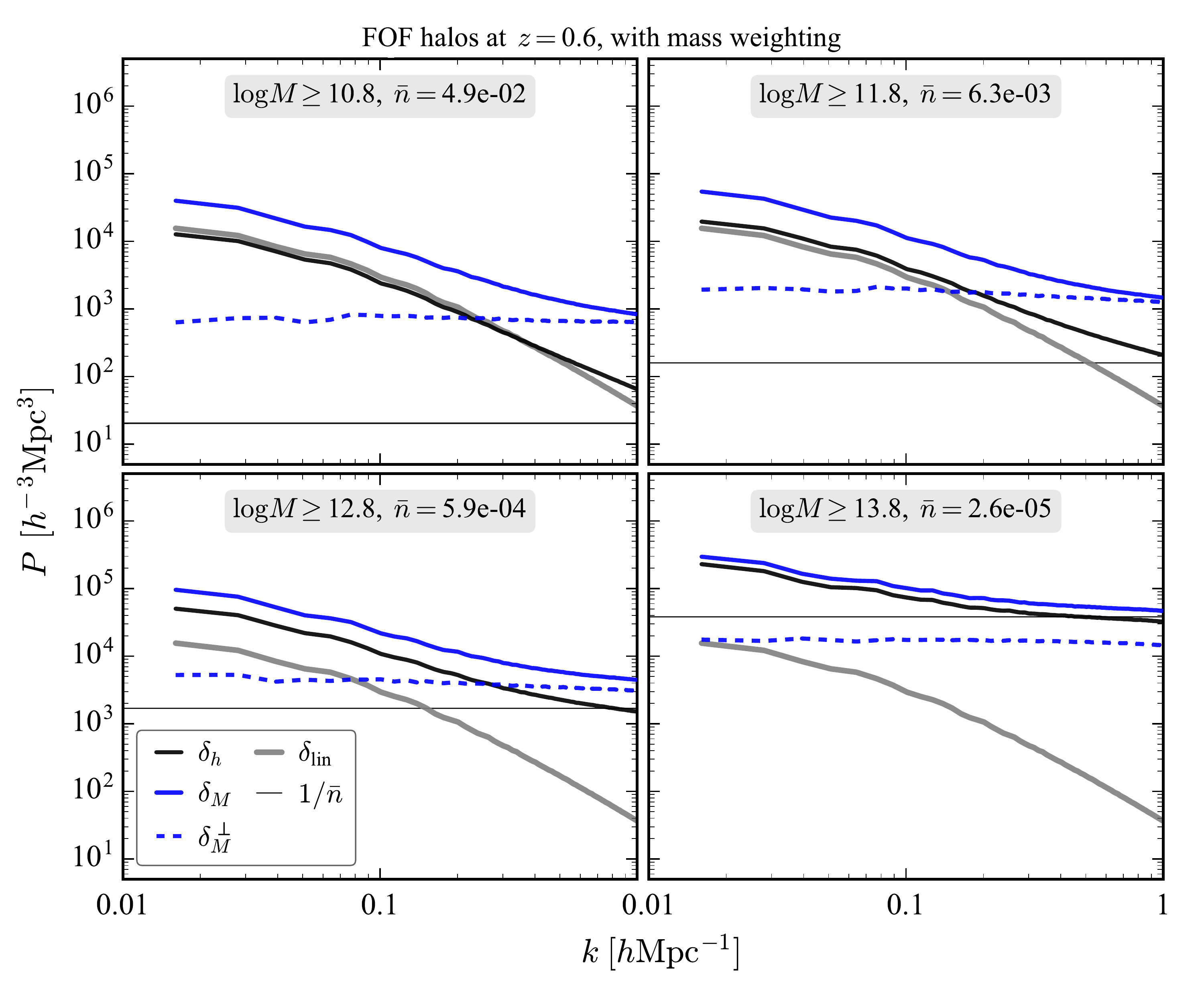}
\caption{Power spectra relevant for halo mass weighting.
The power spectrum of the halo mass density (solid blue)  has a shape similar to that of the halo number density (solid black), but it is more biased relative to the linear density (solid grey), because heavy halos are up-weighted. Orthogonalizing the halo mass density with respect to the halo number density leads to a flat power spectrum (dashed blue); this corresponds approximately to a combination of the stochastic noise terms of the halo number density and mass density, so adding or subtracting it from $\delta_h$ can cancel part of its stochastic noise.  No scatter is added to halo masses (doing so increases the constant, stochastic part).
}
\label{fig:MassWeightOrth}
\end{figure}

\begin{figure}[tp]
\centering
\includegraphics[width=0.6\textwidth]{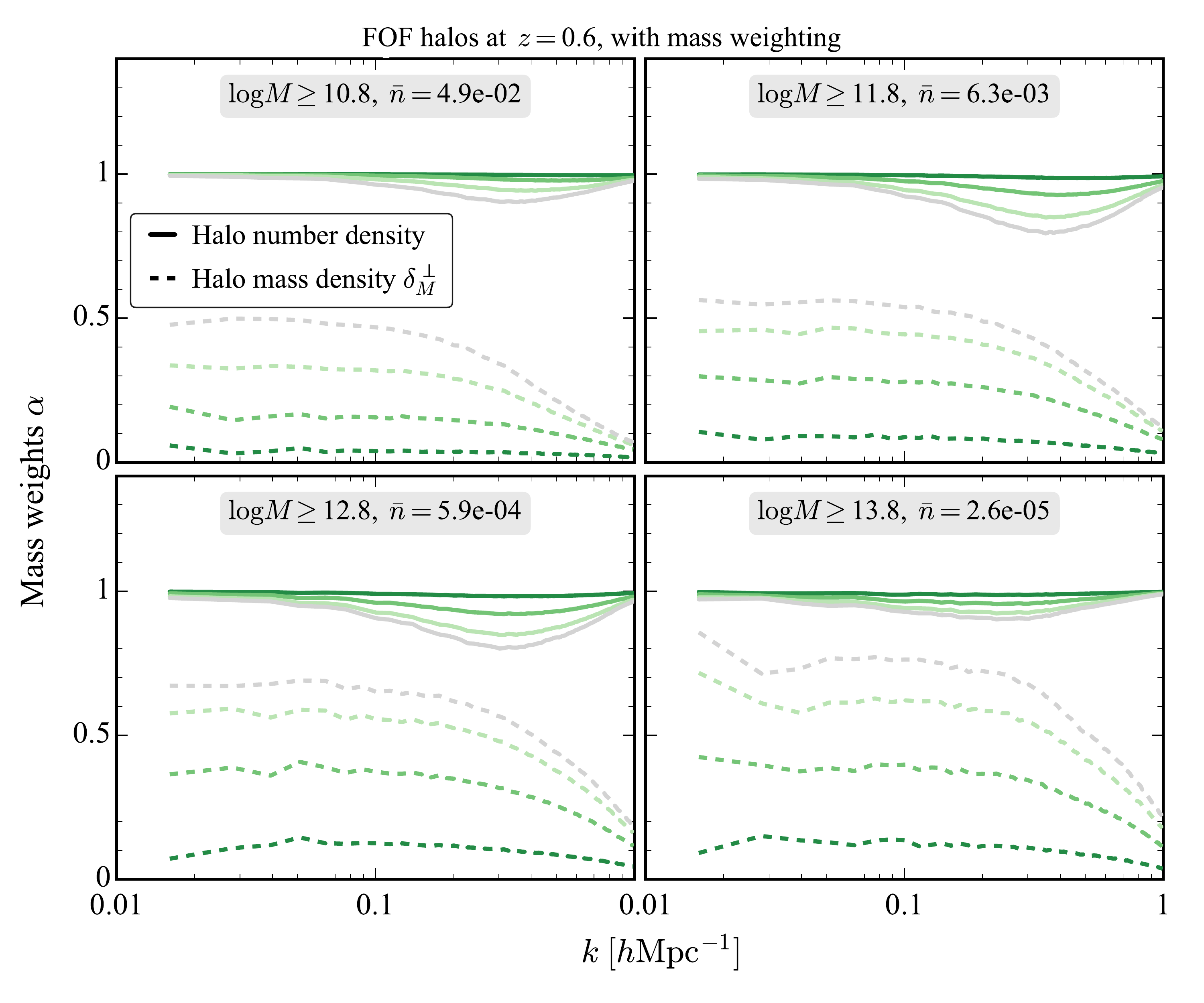}
\caption{Weights used for halo mass weighting.
The weight $\alpha_M$ of the orthogonalized halo mass density $\delta_M^\perp$ (dashed) is  constant on large scales and drops on small scales $k\gtrsim 0.2\ihMpc$. It decreases with increasing halo mass scatter (colors are the same as in Figures~\ref{fig:noiseM0M1} and \ref{fig:FractNoiseM0M1}, with no scatter in bright gray and 0.4 dex scatter in dark green).
 The weight $\alpha_h$ of the halo number density $\delta_h^{\rm truth}$ (solid) is fixed by $\alpha_M$ through the normalization constraint \eq{alphaNorm} of the weighted power spectrum; it is close to one on large and small scales, and around 0.8 to 0.9 on intermediate scales, to compensate for the significant contribution of $\alpha_M\delta_M^\perp$ to the total power spectrum on these scales (see the next figure).
}
\label{fig:Alphas}
\end{figure}

\begin{figure}[tp]
\centering
\includegraphics[width=0.6\textwidth]{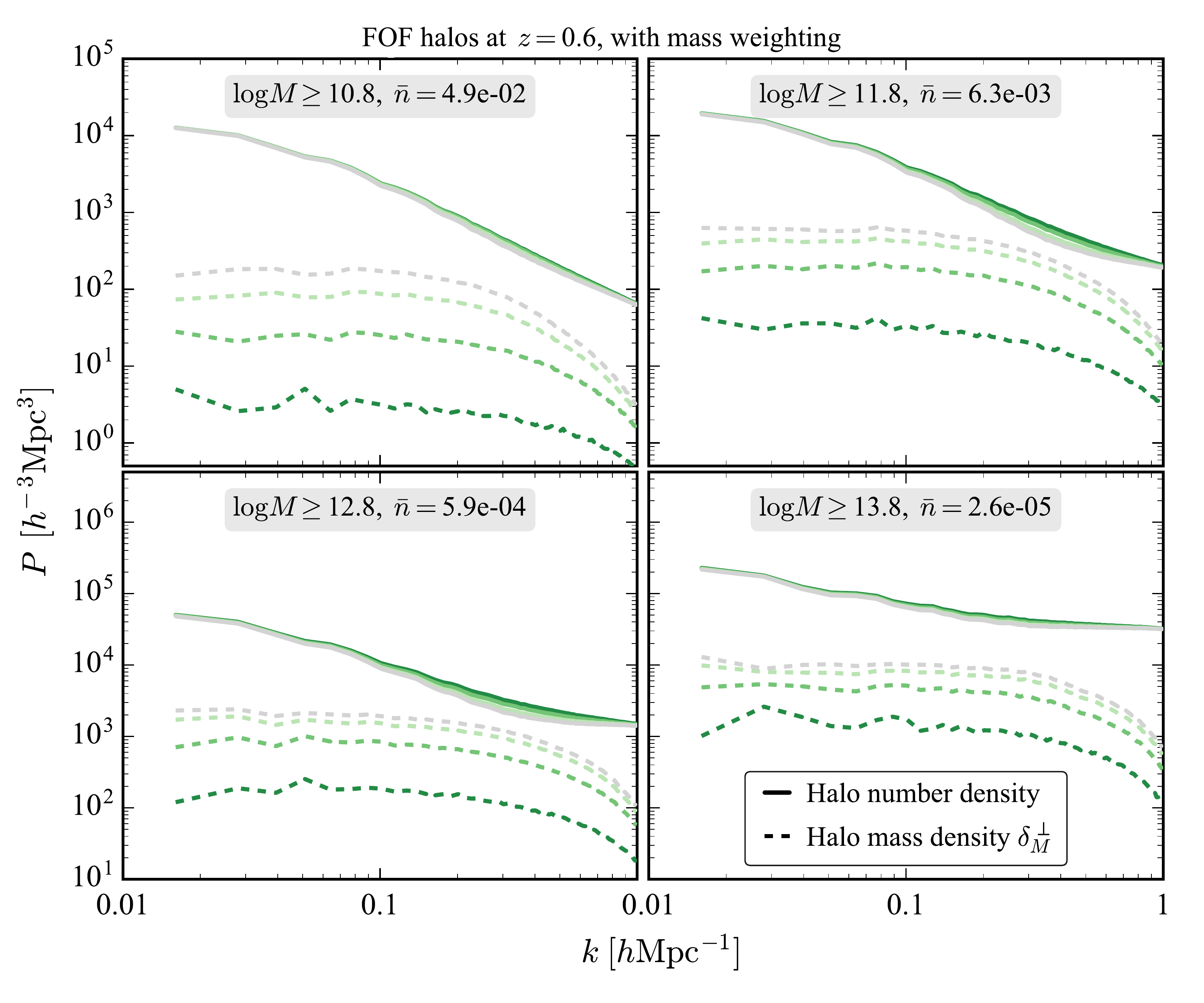}
\caption{Contribution $\alpha_h^2\langle|\delta_h^{\rm truth}|^2\rangle$ from the halo number density (solid), and $\alpha_M^2\langle|\delta_M^\perp|^2\rangle$ from the orthogonalized halo mass density (dashed), to the power spectrum of the weighted density $\delta_h^{\rm obs}=\alpha_h\delta_h^{\rm truth}+\alpha_M\delta_M^\perp$. 
Colors are the same as in the previous figures.  
}
\label{fig:AlphaContris}
\end{figure}

To see in more detail how the mass weighting operates, \fig{MassWeightOrth} shows power spectra of the halo number density $\delta_h^{\rm truth}$ and the halo mass density $\delta_M$. They have a similar $k$ dependence, but the mass density has a larger linear bias because heavy halos are up-weighted relative to the number density (and we use the same minimum halo mass cutoff for both densities).
In contrast, the power spectrum of the \emph{orthogonalized} mass density, $\delta_M^\perp=\delta_M-\langle\delta_h^{\rm truth}\delta_M\rangle/\langle|\delta_h^{\rm truth}|^2\rangle\times\delta_h^{\rm truth}$, is rather independent of scale.
To understand this, note that if $\delta_M$ were equal to the halo number density plus an additional white noise field, $\delta_M=\delta_h^{\rm truth}+\epsilon$, then $\delta_M^\perp=\epsilon$ and its power spectrum would consequently be flat.  
We will argue in the next section that we can indeed expect $\delta_M^\perp$ to be a combination of the stochastic noise fields of the halo number and mass density, whose power spectrum is again expected to be flat on large scales.
\vskip 4pt

The optimal mass weights $\alpha_h$ and $\alpha_M$ are shown in \fig{Alphas} as a function of $k$ for the four halo mass bins. 
Both weights are rather smooth functions of $k$.
The halo number density weight $\alpha_h$ is one on large and small scales, but decreases by 10$\%$ to 20$\%$ on intermediate scales, $0.1\ihMpc\lesssim k\lesssim 0.6\ihMpc$.
The weight $\alpha_M$ of the halo mass density is typically a few times smaller, and is constant at low $k$ but suppressed at  intermediate and small scales, $k\gtrsim 0.2\ihMpc$.  When increasing the mass scatter, $\alpha_M$ becomes smaller as expected because the information in the halo mass density is less useful.

\fig{AlphaContris} shows how much power the halo number density $\delta_h^{\rm truth}$ and the mass density $\delta_M^\perp$ contribute to the combined field.
The halo number density $\delta_h^{\rm truth}$ always dominates.
The contribution from the orthogonalized mass density $\delta_M^\perp$ typically reaches several tens of percent of the number density power around $k\simeq 0.3-0.4\ihMpc$, and is  approximately constant at $k\lesssim 0.2\ihMpc$ and drops at higher $k$.
This again suggests that $\delta_M^\perp$  contains mostly stochastic noise, which partially cancels the stochastic noise of $\delta_h^{\rm truth}$, reducing the stochastic noise of the combined field.

\subsection{Toy Model}
\label{se:MassToy}

To better understand the partial cancellation of the stochastic noise in the halo number and mass density, we
consider a simple toy model (similarly to \cite{Hamaus1004,Hamaus1104}), where the true halo number and mass density are given by 
\begin{align}
  \label{eq:toy1}
\delta_h^{\rm truth}&=b_h\delta_1+\epsilon_h,\\
  \label{eq:toy2}
\delta_M&=b_M\delta_1+\epsilon_M.
\end{align}
$\epsilon_X$ are stochastic noise terms that are uncorrelated with the model, $\langle\epsilon_X\delta_1\rangle=0$.
In that case,
\begin{align}
  \label{eq:deltaMPerp}
  \delta_M^\perp \;\approx\;\epsilon_M-\frac{b_M}{b_h}\epsilon_h.
\end{align}
Here and in what follows we ignore terms whose power spectrum is suppressed by a factor of $P_{\epsilon\epsilon}/(b^2P_{11})$, which is small at low $k$.
\eqq{deltaMPerp} shows  that the orthogonal component of the halo mass density is indeed a combination of the stochastic noise fields $\epsilon_h$ and $\epsilon_M$ of the halo number and mass densities.
We therefore expect its power spectrum to be flat, which is what we saw in \fig{MassWeightOrth} above. 
\vskip 4pt 

Forming a linear combination of $\delta_h^{\rm truth}$ and $\delta_M^\perp$ can then cancel part of the stochastic noise $\epsilon_h$ and reduce the model error of the combined field to the low levels found in \fig{noiseM0M1} above.
For the toy model in Eqs.~\eq{toy1} and \eq{toy2} we can see this explicitly as follows.
The weighted combined field is 
\begin{align}
  \label{eq:4}
  \delta_h^{\rm obs}\equiv\alpha_h\delta_h^{\rm truth} + \alpha_M\delta_M^\perp &\approx
\alpha_h\left[b_h\delta_1+\left(1-t\,\frac{b_M}{b_h}\right)\epsilon_h + t\,\epsilon_M\right],
\end{align}
where we defined $t\equiv \alpha_M^\perp/\alpha_h$.
The model error of this field is
\begin{align}
  \label{eq:ToyModelError}
\epsilon_\mathrm{obs}\equiv
\delta_h^{\rm obs}-\alpha_hb_h\delta_1 
&\approx
\alpha_h\left[\left(1-t\,\frac{b_M}{b_h}\right)\epsilon_h + t\,\epsilon_M\right].
\end{align}
The mass weighting then varies the parameter $t$ to obtain the linear combination of $\epsilon_h$ and $\epsilon_M$ in this equation such that its power spectrum is minimized, noting that $\alpha_h$ is fixed by the normalization.
Completing the square, we find for the error power spectrum at the optimal weight $t$:
\begin{align}
  \label{eq:PobsCalc}
  \frac{P_{\epsilon_{\rm obs}}(k)}{P_{\rm obs}(k)} & \approx
\;\sqrt{\frac{E_{hh}}{P_{11}}\,\frac{E_{MM}}{P_{11}}}\;
\frac{(1-r_{hM}^2)}{2(I-r_{hM})}.
\end{align}
Here,  we again ignored terms suppressed by a factor of $P_{\epsilon\epsilon}/(b^2P_{11})$;
we  defined the stochastic error power spectra
\begin{align}
  \label{eq:13}
E_{XY}(k)\equiv\left\langle\frac{\epsilon_X(\vk)}{b_X}\frac{\epsilon_Y^*(\vk)}{b_Y}\right\rangle
\end{align}
 for $X,Y\in\{h,M\}$, and the cross-correlation coefficient
$r_{hM}(k)=E_{hM}/(E_{hh}E_{MM})^{1/2}$ between $\epsilon_h$ and $\epsilon_M$;
and we also defined\footnote{Since $I\ge 1$ the denominator in \eqq{PobsCalc} is always non-negative. 
}
\begin{align}
  \label{eq:12}
  I(k) \equiv \frac{1}{2}\left(\sqrt{\frac{E_{hh}}{E_{MM}}} +\sqrt{\frac{E_{MM}}{E_{hh}}}\right) \,\ge 1.
\end{align}
\vskip 4pt 

\eqq{PobsCalc} shows that the noise power of the weighted combination of halo number and mass density is determined by the geometric mean of the noise power spectra $E_{hh}/P_{11}$ and $E_{MM}/P_{11}$  of the individual fields, and by the correlation coefficient $r_{hM}$ between the noise fields $\epsilon_h$ and $\epsilon_M$ of the halo number and mass density.
\vskip 4pt 

If $E_{MM}\ne E_{hh}$, \eqq{PobsCalc}  scales  as $1-r_{hM}^2$ in the limit $r_{hM}\rightarrow 1$. 
Therefore, the stochastic noise of the combined field $\delta_h^{\rm obs}$ becomes small if the stochastic noise of halo number and mass density are very correlated while their fractional contributions to the total halo number and mass density differ in amplitude  (so that $E_{MM}\ne E_{hh}$ and $I>1$). 
This makes sense intuitively: If two fields have a very correlated signal and very correlated noise, while their signal-to-noise ratios are different, we can combine the two fields such that the noise cancels identically while the signal remains nonzero. For example, if $\delta_M=2\delta_1+\epsilon$ and $\delta_h^{\rm truth}=\delta_1+\epsilon$, then $\delta_M-\delta_h^{\rm truth}=\delta_1$.
\vskip 4pt

The result \eq{PobsCalc} simplifies further if one error is much larger than the other, say $E_{MM}\ll E_{hh}$ (i.e., $\langle|\epsilon_M|^2\rangle/b_M^2\ll \langle|\epsilon_h|^2\rangle/b_h^2$). Then, 
\begin{align}
  \label{eq:2}
\frac{P_{\epsilon_{\rm obs}}(k)}{P_{\rm obs}(k)} & \;\approx\;
  \frac{E_{MM}}{P_{11}}\,\left(1-r_{hM}^2\right).
\end{align}
The fractional noise of the combined field is therefore given by the fractional noise of the low-noise field, times $(1-r_{hM}^2)$, which is small when the two noise fields $\epsilon_h$ and $\epsilon_M$  are very correlated.\footnote{Another interesting limit is $I=1$. This happens if and only if $E_{MM}=E_{hh}$, i.e.~when the fractional size of the stochastic noise fields is the same, $\langle|\epsilon_M|^2\rangle/b_M^2=\langle|\epsilon_h|^2\rangle/b_h^2$. In that case, \eqq{PobsCalc} becomes $P_{\epsilon_{\rm obs}}/P_{\rm obs}=E_{hh}/P_{11}\times (1+r_{hM})/2$.  
If the noise fields are perfectly correlated, $r_{hM}=1$, this gives $E_{hh}/P_{11}$. 
We therefore do not gain anything by combining the fields, which is not surprising because this case corresponds to $\delta_h^{\rm truth}$ and $\delta_M$ being identical up to an overall normalization factor, so combining $\delta_h$ and $\delta_M$ cannot reduce the fractional stochastic noise.
In contrast, if $\epsilon_h$ and $\epsilon_M$ were perfectly anti-correlated, $r_{hM}=-1$, the combined noise would vanish, which again makes sense because the noise cancels identically when adding the two fields while the signal does not.
In practice, we expect the halo number and mass density and their fractional stochastic noise to be different, i.e.~$E_{MM}\ne E_{hh}$, so that $I>1$.
}
\vskip 4pt

These arguments are related to the idea of canceling cosmic variance using two correlated tracers with different biases \cite{Uros0807}, which was one of the original motivations to study halo mass weighting \cite{Seljak0904}.  In that case, the argument is usually formulated as improving the Fisher information when regarding the relative bias $b_h/b_M$ between the two tracers as the signal of interest, whereas we  computed the reduction of the  stochastic noise term when combining the two tracer fields and modeling that combination with a bias expansion.  The final result, especially the $1-r^2_{hM}$ scaling as $r_{hM}\rightarrow 1$, is similar.
\vskip 4pt 

Of course the analytical arguments above assumed a simple toy model without nonlinear bias, and ignored corrections to \eqq{deltaMPerp} that can be important at intermediate and high $k$.
The simulation results, however, include the nonlinear bias terms of the cubic model and show that mass weighting works very efficiently in that case up to rather high $k$.
\vskip 4pt 

In conclusion, we confirm that halo mass weighting is a promising method to suppress the stochastic model error or shot noise, which could yield more powerful cosmological constraints if achievable in practice, but more work is required to characterize the most suitable bias model and its error when mass weighting is applied.\footnote{In addition to improving the bias model or modeling the scale dependence of $\Perr$, one might also want to employ a more realistic halo mass scatter, noting that mass estimates from the luminosity of massive halos ($M\gtrsim 10^{12}\hMsun$) can suffer from the flattening of the stellar-to-halo mass relation at high mass, while determining the halo mass of very faint (possibly satellite) galaxies is also challenging. 
A potential systematic offset in the assumed mean stellar-to-halo mass relation can also have an impact, likely leading to suboptimal weights.
Still, it would be interesting to study clustering for different cuts in luminosity or other galaxy and cluster properties (e.g., \cite{2001MNRAS.328...64N,2004ApJ...606..702T,2005ApJ...630....1Z,Guo1212}) to suppress stochasticity.
}

\FloatBarrier
\section{Summary}
\label{se:Conclusions}

Using a modified basis of operators in Eulerian space, we have constructed a model of  biased tracers at the density field level that accounts for bulk flows (large IR displacements) without a perturbative expansion in the displacement. 
We find that this model is able to describe the halo density obtained from N-body simulations accurately over a wide range in scale and halo mass.
Our main findings are as follows.
\begin{itemize}
\item To obtain coherent positions of particles in the model and simulations, it is important to use the bias expansion in terms of shifted operators  which keep large IR displacements resummed.
In contrast, Standard Eulerian bias  applied to the dark matter density from Standard Eulerian perturbation theory treats displacements perturbatively, leading to a decorrelation between model and simulations at wavenumbers $k\gtrsim 0.2\ihMpc$.
The Standard Eulerian bias expansion based on the fully nonlinear dark matter density (including full displacements) instead suffers from a large constant contribution to the power spectrum of $\delta_m^2$, which generally leads to a large model error even on very large scales.  
For these reasons, the bias expansion in terms of shifted operators is more suitable for modeling biased tracers at the field level.
Its power spectrum  agrees with that of the usual IR-resummed Standard Eulerian power spectrum.
\item Nonlinear bias operators are important to obtain a model error power spectrum $\Perr(k)\equiv\langle|\delta_h^{\rm truth}(\vk)-\delta_h^{\rm model}(\vk)|^2\rangle$ that is comparable to the Poisson prediction, $\Perr(k)=1/\bar n$.
For the quadratic and cubic bias models, the amplitude of $\Perr$ is a few tens of percent higher than the Poisson prediction for $M\lesssim 10^{13}\hMsun$ halos, and about a factor of 2 smaller than the Poisson prediction for heavier halos.
Without the nonlinear bias terms, the model error power spectrum is approximately five times larger for $M\lesssim 10^{13}\hMsun$ halos and $30\%$ for heavier halos, even on very large scales, $k<0.05\ihMpc$ (see \fig{noiseM0}).
\item The cross-correlation coefficient $r_{cc}$ between the modeled and simulated halo density is roughly consistent with the Poisson shot noise prediction for low and intermediate mass halos, reaching up to $r_{cc}=99.9\%$ on large scales. For heavy, $M\gtrsim 10^{13}\hMsun$  halos the correlation is better than expected from the Poisson prediction, remaining above $50\%$ up to $k\simeq 1\ihMpc$ (see \fig{M0_1mr2}).
\item For the simulated halo samples  at $z=0.6$ that we analyzed it would be safe to assume a scale-independent model error or shot noise up to $\kmax\simeq 0.13-0.3\ihMpc$, depending on halo mass, when analyzing a $10\;h^{-3}\mathrm{Gpc}^3$ volume, or up to $\kmax\simeq 0.18-0.37\ihMpc$ when analyzing a $0.5\;h^{-3}\mathrm{Gpc}^3$ volume.  
Without nonlinear bias terms, these $\kmax$ values are smaller by a factor of 2 to 3 because the measured model error is much more scale dependent (see Table~\ref{tab:Kmax} and \fig{kmaxVsurvey}).
\item On small scales, $k\gtrsim 0.3\ihMpc$, the model error depends strongly on scale. This could be due to expected $(k/k_M)^2$ corrections to the noise on scales comparable to the typical size of a halo, although the uncertainty of our measurements of the model error is too large to test this conclusively. Alternatively this could be caused by additional higher order bias terms that we did not include in the bias expansion.
\item The local quadratic bias term contributes typically $10\%$ in power at $k\simeq 0.1\ihMpc$ (see \fig{M0ModelContris}). Modeling the halo power spectrum to $1\%$ at $k=0.1\ihMpc$ therefore typically requires the quadratic bias parameter $\beta_2$ to be known or constrained to better than $10\%$.  For $M=10^{12.8}-10^{13.8}\hMsun$ halos, this contribution is smaller because $\beta_2$ is much smaller than the linear bias, but for these halos the contribution from the cubic local bias term is $5\%$ in power at $k=0.1\ihMpc$, so $\beta_3$ should be known within $20\%$.
\item The bias transfer functions $\beta_i(k)$ that we use in the bias  model can be  easily described with a 5- or 6-parameter fit based on theoretical predictions of the expected shape of these transfer functions (see \secref{PTTransferFunctions}). The number of free parameters of our model is therefore the same as in usual nonlinear bias expansions (i.e., one parameter $b_i$ for each included bias term, and $c_s$).
\item We confirm that the mean-square model error can be suppressed by an order of magnitude when weighting halos by their mass. This is similar to previous findings \cite{Seljak0904,Hamaus1004,CaiBernsteinSheth1007}, although our definitions differ somewhat because we include nonlinear bias terms as part of the model and not the stochasticity.
With a mass scatter of $0.2$ dex (i.e., $60\%$) or less, the stochastic mean-square model error can typically be suppressed by a factor of two or more, even for relatively low number density of $\bar n\simeq 5\times 10^{-4}\;h^3\mathrm{Mpc}^{-3}$ (see Figures~\ref{fig:noiseM0M1} and \ref{fig:FractNoiseM0M1}).
\end{itemize}
\vskip 4pt 

By demonstrating that dark matter halos in N-body simulations can be modeled accurately at the level of the density field realization we have provided a stringent test of the validity of the bias expansion.
This can be useful for several practical applications, including  forward-model inferences that rely on a density-level forward model in the likelihood,  cosmological analyses of power spectra measured from galaxy survey data where the model error enters as a noise contribution,
and the design of initial condition and BAO reconstruction algorithms from biased tracers.
We leave these and  other possible applications for future work.

\section*{Acknowledgements}
We would like to thank Mikhail Ivanov and Zvonimir Vlah for useful discussions, and Vincent Desjacques, Fabian Schmidt, and Uros Seljak for comments on an earlier version of the manuscript. 
The simulations and analyses used the public \texttt{MP-Gadget} TreePM code \cite{MPGadgetWebsite,MPGadgetDOI} developed by Yu Feng, and the public \texttt{nbodykit} toolkit \cite{nbodykitPaper,nbodykitWebsite} developed by Nick Hand and Yu Feng. They were generated with the Hyperion cluster at the Institute for Advanced Study. and we thank Lee Colbert for support.
M.~S.~gratefully acknowledges support from the Jeff Bezos Fellowship, the Corning Glass Works Foundation Fellowship, and the National Science Foundation.

\appendix

\section{Zel'dovich Density \texorpdfstring{$\delta_Z$}{} in Terms of Shifted Operators}
\label{app:DeltaZShiftedOps}
In this appendix we derive the expression for the Zel'dovich density field in terms of shifted operators. We start from the real space expression
\beq
\delta_Z (\x) = J^{-1}(\q) - 1 \;,
\eeq
where 
\beq
J^{-1}(\q) \equiv \left| \frac{\partial\x}{\partial\q} \right|^{-1} = \left| \delta_i^j + \partial_i\psi_1^j(\q) \right|^{-1} \;.
\eeq
The determinant can be expanded up to third order in PT
\begin{align}
J^{-1}(\q) &= 1 - \boldsymbol\nabla\cdot\boldsymbol\psi_1(\q) + \frac12 (\boldsymbol\nabla\cdot\boldsymbol\psi_1(\q))^2 - \frac16 (\boldsymbol\nabla\cdot\boldsymbol\psi_1(\q))^3 + \frac12 \partial_a\psi_1^b(\q)\;\partial_b\psi_1^a(\q) \nonumber \\
& \qquad - \frac12 \boldsymbol\nabla\cdot\boldsymbol\psi_1(\q)\; \partial_a\psi_1^b(\q)\; \partial_b\psi_1^a(\q) - \frac 13 \partial_a\psi_1^b(\q)\; \partial_b\psi_1^c(\q)\; \partial_c\psi_1^a(\q) \;.
\end{align}
Notice that all the fields are evaluated in $\q$, which solves the equation $\x= \q +\boldsymbol \psi_1 (\q)$. This means that we never expand fields in linear displacement, but only in derivatives of the linear displacement. This is consistent with the way the shifted operators are defined. The inverse Jacobian can be rewritten as 
\begin{align}
J^{-1}(\q) & = 1 +\delta_1(\q) + \delta_2(\q) + \frac 12 \G_2(\q) + \delta_3(\q) + \delta(\q) \G_2(\q) - \frac13 \G_3(\q) \;.
\end{align}
Therefore
\beq
\delta_Z (\x) =  \delta_1(\q) + \delta_2(\q) + \frac 12 \G_2(\q) + \delta_3(\q) + \delta(\q) \G_2(\q) - \frac13 \G_3(\q) \;.
\eeq

On the other hand the shifted operators can be expressed in the same way. Keeping only the terms up to third order in PT we get
\beq
\tilde\delta_1 (\x) = J^{-1}(\q) \delta_1(\q) = \delta_1(\q) + \delta_2(\q) + \delta_3(\q) + \frac12 \delta(\q) \G_2(\q) \;,
\eeq
\beq
\tilde\delta_2 (\x) = J^{-1}(\q) \delta_2(\q) = \delta_2(\q) + \delta_3(\q) \;,
\eeq
\beq
\tilde\G_2 (\x) = J^{-1}(\q) \G_2(\q) = \G_2(\q) + \delta(\q)\G_2(\q) \;.
\eeq
Combining these expressions we finally get 
\beq
\delta_Z(\k) = \tilde\delta_1(\k) + \frac12 \tilde\G_2(\k) - \frac13 \tilde\G_3(\k) \;. 
\eeq
Notice that the quadratic and cubic operators $\tilde\delta_2$ and $\tilde\delta_3$ cancel, and they do not contribute to the final expression. This is expected since the Zel'dovich density field cannot have shot noise.

\section{Relation Between Model Error Power Spectrum and Cross-Correlation Coefficient}
\label{app:PerfMeasures}

One statistic that we used in the main text to quantify the agreement between bias expansion and simulations  is the power spectrum of the model error $\Perr(k)$ as defined in \eqq{PerrDef}.
This measures, in a mean-square sense, what part of the simulated halo number density cannot be captured by the bias expansion.
Another statistic that we used is the cross-correlation coefficient $r_{cc}(k)$ between Fourier modes of the model and true simulated halo density as defined in \eqq{rccDef}.
This also describes how well the model describes phases of individual Fourier modes.
It is therefore not surprising that these two performance measures are closely related to each other as follows.
\vskip 4pt

Since all the halo bias models that we use are linear in bias parameters, the  minimization of $\Perr$ with respect to these bias parameters implies
\begin{align}
  \label{eq:MinimizeModelError}
\frac{\partial \langle\left|\delta_h^{\rm truth}(\k) -\delta_h^{\rm model}(\k) \right|^2\rangle}{\partial\beta_i(k)}=0
\quad\Rightarrow\quad
\langle (\delta_h^{\rm truth}(\k)-\delta_h^{\rm model}(\k) ) [\delta_h^{\rm model}(\k)]^*\rangle=0.
\end{align}
The model error of the best-fit model is therefore uncorrelated with the model,
which is optimal in the sense that the modeling error is entirely due to stochastic or higher order contributions to the true simulated density that cannot be captured by the model.
\eqq{MinimizeModelError} implies that \emph{for the best-fit model} we have $P_\mathcal{MM}=P_{\mathcal{MT}}$, where $\mathcal{M}\equiv\delta_{h}^{\rm model, best-fit}$ and $\mathcal{T}\equiv\delta_h^{\rm truth}$, and thus
\begin{align}
  \label{eq:30}
r_{cc}&=\frac{P_{\mathcal{MT}}}{\sqrt{P_\mathcal{MM}P_\mathcal{TT}}} 
=\sqrt{\frac{P_\mathcal{MM}}{P_\mathcal{TT}}} \;,
\end{align}
and
\begin{align}
  \label{eq:16}
  \Perr &\;=\; \langle|\mathcal{T}-\mathcal{M}|^2\rangle
\;=\; P_\mathcal{TT}-P_\mathcal{MM}
\;=\; P_\mathcal{TT}\,(1-r_{cc}^2) \;,
\end{align}
where $k$-arguments are suppressed.
This shows that the cross-correlation coefficient $r_{cc}$ between the best-fit model and the truth is identical to the square-root of the model power spectrum divided by the truth power spectrum, and the minimum mean-square model error $\Perr$ is proportional to  one minus the squared correlation coefficient between the best-fit model and the truth. 
This shows that $\Perr$ and $r_{cc}$ are closely related to each other.
\vskip 4pt 

It is important to stress that this is valid only for the best-fit model with the best fit-transfer functions. If we use 5- or 6-parameter fits to the transfer function the relation between the model error and the cross-correlation coefficient is not so simple anymore. Let us call $\mathcal M'$ the model that uses $\beta_i' = \beta_i + \delta\beta_i$, where $\beta_i$ are the best-fit transfer functions and $\beta_i'$ are their approximation using the perturbation theory fits. In this case a part of the error is correlated with the model, and therefore $P_\mathcal{M'M'}\neq P_{\mathcal{M'T}}$ (also see \secref{PkModel}).

\section{Gram-Schmidt Orthogonalization of Bias Operators}
\label{app:orth}

In this appendix we describe the procedure to orthogonalize the operators that enter the model for the halo density; for example, $\mathcal{O}_i(\vk)=(\tilde \delta_1(\vk),\tilde \delta_2(\vk),\tG_2(\vk))$ for the quadratic bias model.
We will assume that all model operators $\mathcal{O}_i$ are statistically isotropic and homogeneous fields so that different $k$ bins are independent from each other and we can apply the orthogonalization for each $k$ bin separately.
\vskip 4pt 

We start by  computing the covariance matrix $\mathbb{O}$ between the original non-orthogonal fields $\mathcal{O}_i(\vk)$, $i=1,\dots,n$, in every $k$ bin, 
\begin{align}
  \label{eq:32}
\mathbb{O}_{ij}(k)= \langle \mathcal{O}_i(\vk) \mathcal{O}_j^*(\vk)\rangle.   
\end{align}
In shorter notation, suppressing $k$ and $\vk$ arguments, we write $\mathbb{O}=\langle \vO\vO^\dagger\rangle$, where $\vO$ is a column vector with $n$ rows, and $\vO^\dagger=(\vO^*)^T$ is the conjugate transpose.
The corresponding correlation matrix is $C\equiv D\mathbb{O}D$, where $D_{ij}\equiv\delta_{ij}^K (\mathbb{O}_{ii})^{-1/2}$.
Performing its Cholesky decomposition gives a lower-triangular real $n\times n$ matrix $L$ such that
\begin{align}
  \label{eq:17}
  C = LL^T.
\end{align}
If we define rotated fields
\begin{align}
  \label{eq:18}
   \vO^\perp = W L^{-1}D\vO
\end{align}
with a real diagonal $n\times n$ normalization matrix $W$ to be determined later, we find
\begin{align}
  \label{eq:19}
  \langle \vO^\perp (\vO^{\perp})^\dagger\rangle 
\;=\;W L^{-1}D  \mathbb{O} D (L^{-1})^T W^T
\;=\;WW^T
\end{align}
which is a diagonal matrix, i.e.~$\langle \mathcal{O}_i^\perp (\mathcal{O}_j^\perp)^*\rangle=0$ for $i\neq j$ as desired.
Fixing the normalization $W$ such that $(WL^{-1}D)_{ii}=1$ (no sum) gives 
$W_{ij} = \delta_{ij}^K/(L^{-1}D)_{ii}$ (no sum).
The rotation from original fields $\vO$ to orthogonal fields $\vO^\perp$ is therefore
\begin{align}
  \label{eq:21}
  \vO^\perp = M\vO,
\end{align}
where the rotation matrix $M=WL^{-1}D$. Explicitly,
\begin{align}
  \label{eq:22}
  M_{ij} = \frac{(L^{-1})_{ij}(\mathbb{O}_{jj})^{-1/2}}{(L^{-1})_{ii}(\mathbb{O}_{ii})^{-1/2}}\qquad \text{(no sum)}.
\end{align}
\vskip 4pt

To compute $\mathcal{O}^\perp_i(\vk)=\sum_j M_{ij}(k)\mathcal{O}_j(\vk)$
for a given Fourier vector $\vk$ on a 3-d grid, we evaluate the rotation matrix $M_{ij}(k)$ at the same $k$-bin as that to which $\vk$ contributes when we compute power spectra, without any interpolation.
For $\mathbb{O}_{ij}$ we use the measured cross spectra of the shifted fields generated on the 3-d grid as described in \secref{BiasOnGrid}.
Validating our implementation, we find that the cross-correlation  $|\langle \mathcal{O}^\perp_i(\vk) \mathcal{O}^\perp_j(-\vk)\rangle|\,/\,[\langle |\mathcal{O}^\perp_i(\vk)|^2\rangle\langle |\mathcal{O}^\perp_j(\vk)|^2\rangle]^{1/2}\,\leq\, 10^{-5}$ for $i\neq j$ for all $k$ and all fields used in this paper.

\section{Cubic Operators}
\label{app:CubicOperators}
In this appendix we provide the explicit expressions for all cubic operators in the following model for the halo density field
\beq
\delta_h(\k) = \int \d^3\q \; \Big[ 1+ \delta_h^{\mathsmaller{\rm L}} + \partial^b \psi_1^a \, \partial_a \psi_{2b} - (1+\delta_h^{\mathsmaller{\rm L}}) \nabla \cdot\vpsi_2 - \nabla\cdot\vpsi_3 -(1+b_1^{\mathsmaller{\rm L}})\vpsi_2 \cdot\nabla\delta_1 \Big] e^{-i\k \cdot (\q + \boldsymbol\psi_1)} \;.
\eeq
To simplify formulas, throughout this section we will use the definition
\beq
\mathcal O (\k) = \int_{\p_1,\p_2,\p_3} (2\pi)^3 \delta^D(\k-\p_1-\p_2-\p_3) \, F_{\mathcal{O}}^s(\p_1,\p_2,\p_3) \, \delta_1(\p_1) \delta_1(\p_2) \delta_1(\p_3) \;,
\eeq
where $F_{\mathcal O}^s$ is a symmetrized kernel. The usual bias operators at third order have the following kernels
\beq
F_{\mathcal \delta_3}(\p_1,\p_2,\p_3) = 1\;,
\eeq
\beq
F_{\mathcal \G_2\delta }(\p_1,\p_2,\p_3) = \left( \frac{(\p_1\cdot\p_2)^2}{p_1^2p_2^2} -1 \right) \;,
\eeq
\beq
F_{\mathcal G_3}(\p_1,\p_2,\p_3) = - \frac1{2} \left( 1- 3 \frac{(\p_1\cdot\p_2)^2}{p_1^2p_2^2} + 2 \frac{(\p_1\cdot\p_2)(\p_1\cdot\p_3)(\p_2\cdot\p_3)}{p_1^2p_2^2p_3^2} \right) \; ,
\eeq
\beq
F_{\Gamma_3}(\p_1,\p_2,\p_3) = - \frac4{7} \left( \frac{(\p_1\cdot\p_2)^2}{p_1^2p_2^2} -1 \right) \left( \frac{((\p_1+\p_2)\cdot\p_3)^2}{|\p_1+\p_2|^2p_3^2} -1 \right) \; .
\eeq
\vskip 4pt

Let us begin with the third order displacement field. Neglecting the transverse part of the displacement which does not contribute to the one-loop power spectrum, we have
\beq
\boldsymbol \psi_3(\k) = \frac{i \k}{k^2} \int_{\p_1,\p_2,\p_3} (2\pi)^3 \delta^D(\k-\p_1-\p_2-\p_3) L_3^s(\p_1,\p_2,\p_3) \delta_1(\p_1) \delta_1(\p_2) \delta_1(\p_3) \;,
\eeq
where
\begin{align}
L_3(\p_1,\p_2,\p_3) & = \frac 5{42} \left[ 1 - \frac{(\p_1\cdot\p_2)^2}{p_1^2p_2^2} \right] \left[ 1 - \frac{((\p_1+\p_2)\cdot\p_3)^2}{|\p_1+\p_2|^2p_3^2} \right]
 - \frac1{18} \left[ 1- 3 \frac{(\p_1\cdot\p_2)^2}{p_1^2p_2^2} + 2 \frac{(\p_1\cdot\p_2)(\p_1\cdot\p_3)(\p_2\cdot\p_3)}{p_1^2p_2^2p_3^2} \right] \;.
\end{align}
It immediately follows that
\beq
F_{\nabla\cdot\vpsi_3}(\p_1,\p_2,\p_3) = - \frac 19 F_{\mathcal G_3}(\p_1,\p_2,\p_3) + \frac5{24}F_{\Gamma_3}(\p_1,\p_2,\p_3)  \;.
\eeq
We consider the operator $\partial_a \psi_2^b \partial_b \psi_1^a$ next. From the definition of $\vpsi_2$ it follows that 
\beq
F_{\partial_a \psi_2^b \partial_b \psi_1^a}(\p_1,\p_2,\p_3) = L_2(\p_1,\p_2) \frac{((\p_1+\p_2)\cdot\p_3)^2}{|\p_1+\p_2|^2p_3^2} \;,
\eeq
where
\beq
L_2(\p_1,\p_2) = \frac{3}{14} \left( 1 - \frac{(\p_1\cdot\p_2)^2}{p_1^2p_2^2} \right) \;.
\eeq
We can rewrite this kernel in the following way
\begin{equation}
F_{\partial_a \psi_2^b \partial_b \psi_1^a}(\p_1,\p_2,\p_3) = -\frac3{14} F_{\mathcal G_2\delta}(\p_1,\p_2,\p_3) + \frac 38 F_{\Gamma_3}(\p_1,\p_2,\p_3)\;.
\end{equation}
The last term in the bias model has the following kernel  
\beq
F_{\psi_2^b \partial_b \delta_1} = - \frac{(\p_1+\p_2)\cdot\p_3}{|\p_1+\p_2|^2}  L_2(\p_1,\p_2) \;.
\eeq
This term is a shift of the linear density field by $\boldsymbol\psi_2$. As expected, this shift contribution cannot be written in terms of cubic bias operators.

\bibliography{main}

\begin{thebibliography}{119}%
\makeatletter
\providecommand \@ifxundefined [1]{%
 \@ifx{#1\undefined}
}%
\providecommand \@ifnum [1]{%
 \ifnum #1\expandafter \@firstoftwo
 \else \expandafter \@secondoftwo
 \fi
}%
\providecommand \@ifx [1]{%
 \ifx #1\expandafter \@firstoftwo
 \else \expandafter \@secondoftwo
 \fi
}%
\providecommand \natexlab [1]{#1}%
\providecommand \enquote  [1]{``#1''}%
\providecommand \bibnamefont  [1]{#1}%
\providecommand \bibfnamefont [1]{#1}%
\providecommand \citenamefont [1]{#1}%
\providecommand \href@noop [0]{\@secondoftwo}%
\providecommand \href [0]{\begingroup \@sanitize@url \@href}%
\providecommand \@href[1]{\@@startlink{#1}\@@href}%
\providecommand \@@href[1]{\endgroup#1\@@endlink}%
\providecommand \@sanitize@url [0]{\catcode `\\12\catcode `\$12\catcode
  `\&12\catcode `\#12\catcode `\^12\catcode `\_12\catcode `\%12\relax}%
\providecommand \@@startlink[1]{}%
\providecommand \@@endlink[0]{}%
\providecommand \url  [0]{\begingroup\@sanitize@url \@url }%
\providecommand \@url [1]{\endgroup\@href {#1}{\urlprefix }}%
\providecommand \urlprefix  [0]{URL }%
\providecommand \Eprint [0]{\href }%
\providecommand \doibase [0]{http://dx.doi.org/}%
\providecommand \selectlanguage [0]{\@gobble}%
\providecommand \bibinfo  [0]{\@secondoftwo}%
\providecommand \bibfield  [0]{\@secondoftwo}%
\providecommand \translation [1]{[#1]}%
\providecommand \BibitemOpen [0]{}%
\providecommand \bibitemStop [0]{}%
\providecommand \bibitemNoStop [0]{.\EOS\space}%
\providecommand \EOS [0]{\spacefactor3000\relax}%
\providecommand \BibitemShut  [1]{\csname bibitem#1\endcsname}%
\let\auto@bib@innerbib\@empty
\bibitem [{\citenamefont {{Desjacques}}\ \emph {et~al.}(2018)\citenamefont
  {{Desjacques}}, \citenamefont {{Jeong}},\ and\ \citenamefont
  {{Schmidt}}}]{BiasReview1611}%
  \BibitemOpen
  \bibfield  {author} {\bibinfo {author} {\bibfnamefont {V.}~\bibnamefont
  {{Desjacques}}}, \bibinfo {author} {\bibfnamefont {D.}~\bibnamefont
  {{Jeong}}}, \ and\ \bibinfo {author} {\bibfnamefont {F.}~\bibnamefont
  {{Schmidt}}},\ }\href {\doibase 10.1016/j.physrep.2017.12.002} {\bibfield
  {journal} {\bibinfo  {journal} {\physrep}\ }\textbf {\bibinfo {volume}
  {733}},\ \bibinfo {pages} {1} (\bibinfo {year} {2018})},\ \Eprint
  {http://arxiv.org/abs/1611.09787} {arXiv:1611.09787} \BibitemShut {NoStop}%
\bibitem [{\citenamefont {{Chan}}\ \emph {et~al.}(2012)\citenamefont {{Chan}},
  \citenamefont {{Scoccimarro}},\ and\ \citenamefont {{Sheth}}}]{Chan1201}%
  \BibitemOpen
  \bibfield  {author} {\bibinfo {author} {\bibfnamefont {K.~C.}\ \bibnamefont
  {{Chan}}}, \bibinfo {author} {\bibfnamefont {R.}~\bibnamefont
  {{Scoccimarro}}}, \ and\ \bibinfo {author} {\bibfnamefont {R.~K.}\
  \bibnamefont {{Sheth}}},\ }\href {\doibase 10.1103/PhysRevD.85.083509}
  {\bibfield  {journal} {\bibinfo  {journal} {\prd}\ }\textbf {\bibinfo
  {volume} {85}},\ \bibinfo {eid} {083509} (\bibinfo {year} {2012})},\ \Eprint
  {http://arxiv.org/abs/1201.3614} {arXiv:1201.3614 [astro-ph.CO]} \BibitemShut
  {NoStop}%
\bibitem [{\citenamefont {{Baldauf}}\ \emph {et~al.}(2012)\citenamefont
  {{Baldauf}}, \citenamefont {{Seljak}}, \citenamefont {{Desjacques}},\ and\
  \citenamefont {{McDonald}}}]{Tobias1201}%
  \BibitemOpen
  \bibfield  {author} {\bibinfo {author} {\bibfnamefont {T.}~\bibnamefont
  {{Baldauf}}}, \bibinfo {author} {\bibfnamefont {U.}~\bibnamefont {{Seljak}}},
  \bibinfo {author} {\bibfnamefont {V.}~\bibnamefont {{Desjacques}}}, \ and\
  \bibinfo {author} {\bibfnamefont {P.}~\bibnamefont {{McDonald}}},\ }\href
  {\doibase 10.1103/PhysRevD.86.083540} {\bibfield  {journal} {\bibinfo
  {journal} {\prd}\ }\textbf {\bibinfo {volume} {86}},\ \bibinfo {eid} {083540}
  (\bibinfo {year} {2012})},\ \Eprint {http://arxiv.org/abs/1201.4827}
  {arXiv:1201.4827 [astro-ph.CO]} \BibitemShut {NoStop}%
\bibitem [{\citenamefont {{Saito}}\ \emph {et~al.}(2014)\citenamefont
  {{Saito}}, \citenamefont {{Baldauf}}, \citenamefont {{Vlah}}, \citenamefont
  {{Seljak}}, \citenamefont {{Okumura}},\ and\ \citenamefont
  {{McDonald}}}]{Saito1405}%
  \BibitemOpen
  \bibfield  {author} {\bibinfo {author} {\bibfnamefont {S.}~\bibnamefont
  {{Saito}}}, \bibinfo {author} {\bibfnamefont {T.}~\bibnamefont {{Baldauf}}},
  \bibinfo {author} {\bibfnamefont {Z.}~\bibnamefont {{Vlah}}}, \bibinfo
  {author} {\bibfnamefont {U.}~\bibnamefont {{Seljak}}}, \bibinfo {author}
  {\bibfnamefont {T.}~\bibnamefont {{Okumura}}}, \ and\ \bibinfo {author}
  {\bibfnamefont {P.}~\bibnamefont {{McDonald}}},\ }\href {\doibase
  10.1103/PhysRevD.90.123522} {\bibfield  {journal} {\bibinfo  {journal}
  {\prd}\ }\textbf {\bibinfo {volume} {90}},\ \bibinfo {eid} {123522} (\bibinfo
  {year} {2014})},\ \Eprint {http://arxiv.org/abs/1405.1447} {arXiv:1405.1447}
  \BibitemShut {NoStop}%
\bibitem [{\citenamefont {{Hoffmann}}\ \emph {et~al.}(2017)\citenamefont
  {{Hoffmann}}, \citenamefont {{Bel}},\ and\ \citenamefont
  {{Gazta{\~n}aga}}}]{Hoffmann1607}%
  \BibitemOpen
  \bibfield  {author} {\bibinfo {author} {\bibfnamefont {K.}~\bibnamefont
  {{Hoffmann}}}, \bibinfo {author} {\bibfnamefont {J.}~\bibnamefont {{Bel}}}, \
  and\ \bibinfo {author} {\bibfnamefont {E.}~\bibnamefont {{Gazta{\~n}aga}}},\
  }\href {\doibase 10.1093/mnras/stw2876} {\bibfield  {journal} {\bibinfo
  {journal} {\mnras}\ }\textbf {\bibinfo {volume} {465}},\ \bibinfo {pages}
  {2225} (\bibinfo {year} {2017})},\ \Eprint {http://arxiv.org/abs/1607.01024}
  {arXiv:1607.01024} \BibitemShut {NoStop}%
\bibitem [{\citenamefont {{Fujita}}\ \emph {et~al.}(2016)\citenamefont
  {{Fujita}}, \citenamefont {{Mauerhofer}}, \citenamefont {{Senatore}},
  \citenamefont {{Vlah}},\ and\ \citenamefont
  {{Angulo}}}]{2016arXiv160900717F}%
  \BibitemOpen
  \bibfield  {author} {\bibinfo {author} {\bibfnamefont {T.}~\bibnamefont
  {{Fujita}}}, \bibinfo {author} {\bibfnamefont {V.}~\bibnamefont
  {{Mauerhofer}}}, \bibinfo {author} {\bibfnamefont {L.}~\bibnamefont
  {{Senatore}}}, \bibinfo {author} {\bibfnamefont {Z.}~\bibnamefont {{Vlah}}},
  \ and\ \bibinfo {author} {\bibfnamefont {R.}~\bibnamefont {{Angulo}}},\
  }\href@noop {} {\bibfield  {journal} {\bibinfo  {journal} {ArXiv e-prints}\ }
  (\bibinfo {year} {2016})},\ \Eprint {http://arxiv.org/abs/1609.00717}
  {arXiv:1609.00717} \BibitemShut {NoStop}%
\bibitem [{\citenamefont {{Modi}}\ \emph
  {et~al.}(2017{\natexlab{a}})\citenamefont {{Modi}}, \citenamefont
  {{Castorina}},\ and\ \citenamefont {{Seljak}}}]{Modi1612}%
  \BibitemOpen
  \bibfield  {author} {\bibinfo {author} {\bibfnamefont {C.}~\bibnamefont
  {{Modi}}}, \bibinfo {author} {\bibfnamefont {E.}~\bibnamefont {{Castorina}}},
  \ and\ \bibinfo {author} {\bibfnamefont {U.}~\bibnamefont {{Seljak}}},\
  }\href {\doibase 10.1093/mnras/stx2148} {\bibfield  {journal} {\bibinfo
  {journal} {\mnras}\ }\textbf {\bibinfo {volume} {472}},\ \bibinfo {pages}
  {3959} (\bibinfo {year} {2017}{\natexlab{a}})},\ \Eprint
  {http://arxiv.org/abs/1612.01621} {arXiv:1612.01621} \BibitemShut {NoStop}%
\bibitem [{\citenamefont {{Abidi}}\ and\ \citenamefont
  {{Baldauf}}(2018)}]{AbidiBaldauf1802}%
  \BibitemOpen
  \bibfield  {author} {\bibinfo {author} {\bibfnamefont {M.~M.}\ \bibnamefont
  {{Abidi}}}\ and\ \bibinfo {author} {\bibfnamefont {T.}~\bibnamefont
  {{Baldauf}}},\ }\href {\doibase 10.1088/1475-7516/2018/07/029} {\bibfield
  {journal} {\bibinfo  {journal} {\jcap}\ }\textbf {\bibinfo {volume} {7}},\
  \bibinfo {eid} {029} (\bibinfo {year} {2018})},\ \Eprint
  {http://arxiv.org/abs/1802.07622} {arXiv:1802.07622} \BibitemShut {NoStop}%
\bibitem [{\citenamefont {{Dekel}}\ and\ \citenamefont
  {{Lahav}}(1999)}]{1999ApJ...520...24D}%
  \BibitemOpen
  \bibfield  {author} {\bibinfo {author} {\bibfnamefont {A.}~\bibnamefont
  {{Dekel}}}\ and\ \bibinfo {author} {\bibfnamefont {O.}~\bibnamefont
  {{Lahav}}},\ }\href {\doibase 10.1086/307428} {\bibfield  {journal} {\bibinfo
   {journal} {\apj}\ }\textbf {\bibinfo {volume} {520}},\ \bibinfo {pages} {24}
  (\bibinfo {year} {1999})},\ \Eprint {http://arxiv.org/abs/astro-ph/9806193}
  {astro-ph/9806193} \BibitemShut {NoStop}%
\bibitem [{\citenamefont {{Taruya}}\ and\ \citenamefont
  {{Soda}}(1999)}]{1999ApJ...522...46T}%
  \BibitemOpen
  \bibfield  {author} {\bibinfo {author} {\bibfnamefont {A.}~\bibnamefont
  {{Taruya}}}\ and\ \bibinfo {author} {\bibfnamefont {J.}~\bibnamefont
  {{Soda}}},\ }\href {\doibase 10.1086/307612} {\bibfield  {journal} {\bibinfo
  {journal} {\apj}\ }\textbf {\bibinfo {volume} {522}},\ \bibinfo {pages} {46}
  (\bibinfo {year} {1999})},\ \Eprint {http://arxiv.org/abs/astro-ph/9809204}
  {astro-ph/9809204} \BibitemShut {NoStop}%
\bibitem [{\citenamefont {{Matsubara}}(1999)}]{1999ApJ...525..543M}%
  \BibitemOpen
  \bibfield  {author} {\bibinfo {author} {\bibfnamefont {T.}~\bibnamefont
  {{Matsubara}}},\ }\href {\doibase 10.1086/307931} {\bibfield  {journal}
  {\bibinfo  {journal} {\apj}\ }\textbf {\bibinfo {volume} {525}},\ \bibinfo
  {pages} {543} (\bibinfo {year} {1999})},\ \Eprint
  {http://arxiv.org/abs/astro-ph/9906029} {astro-ph/9906029} \BibitemShut
  {NoStop}%
\bibitem [{\citenamefont {{Tegmark}}\ and\ \citenamefont
  {{Bromley}}(1999)}]{1999ApJ...518L..69T}%
  \BibitemOpen
  \bibfield  {author} {\bibinfo {author} {\bibfnamefont {M.}~\bibnamefont
  {{Tegmark}}}\ and\ \bibinfo {author} {\bibfnamefont {B.~C.}\ \bibnamefont
  {{Bromley}}},\ }\href {\doibase 10.1086/312068} {\bibfield  {journal}
  {\bibinfo  {journal} {\apjl}\ }\textbf {\bibinfo {volume} {518}},\ \bibinfo
  {pages} {L69} (\bibinfo {year} {1999})},\ \Eprint
  {http://arxiv.org/abs/astro-ph/9809324} {astro-ph/9809324} \BibitemShut
  {NoStop}%
\bibitem [{\citenamefont {Blanton}(2000)}]{Blanton2000}%
  \BibitemOpen
  \bibfield  {author} {\bibinfo {author} {\bibfnamefont {M.}~\bibnamefont
  {Blanton}},\ }\href {http://stacks.iop.org/0004-637X/544/i=1/a=63} {\bibfield
   {journal} {\bibinfo  {journal} {The Astrophysical Journal}\ }\textbf
  {\bibinfo {volume} {544}},\ \bibinfo {pages} {63} (\bibinfo {year}
  {2000})}\BibitemShut {NoStop}%
\bibitem [{\citenamefont {Blanton}\ \emph {et~al.}(2000)\citenamefont
  {Blanton}, \citenamefont {Cen}, \citenamefont {Ostriker}, \citenamefont
  {Strauss},\ and\ \citenamefont {Tegmark}}]{BlantonEtAl2000}%
  \BibitemOpen
  \bibfield  {author} {\bibinfo {author} {\bibfnamefont {M.}~\bibnamefont
  {Blanton}}, \bibinfo {author} {\bibfnamefont {R.}~\bibnamefont {Cen}},
  \bibinfo {author} {\bibfnamefont {J.~P.}\ \bibnamefont {Ostriker}}, \bibinfo
  {author} {\bibfnamefont {M.~A.}\ \bibnamefont {Strauss}}, \ and\ \bibinfo
  {author} {\bibfnamefont {M.}~\bibnamefont {Tegmark}},\ }\href
  {http://stacks.iop.org/0004-637X/531/i=1/a=1} {\bibfield  {journal} {\bibinfo
   {journal} {The Astrophysical Journal}\ }\textbf {\bibinfo {volume} {531}},\
  \bibinfo {pages} {1} (\bibinfo {year} {2000})}\BibitemShut {NoStop}%
\bibitem [{\citenamefont {{Somerville}}\ \emph {et~al.}(2001)\citenamefont
  {{Somerville}}, \citenamefont {{Lemson}}, \citenamefont {{Sigad}},
  \citenamefont {{Dekel}}, \citenamefont {{Kauffmann}},\ and\ \citenamefont
  {{White}}}]{2001MNRAS.320..289S}%
  \BibitemOpen
  \bibfield  {author} {\bibinfo {author} {\bibfnamefont {R.~S.}\ \bibnamefont
  {{Somerville}}}, \bibinfo {author} {\bibfnamefont {G.}~\bibnamefont
  {{Lemson}}}, \bibinfo {author} {\bibfnamefont {Y.}~\bibnamefont {{Sigad}}},
  \bibinfo {author} {\bibfnamefont {A.}~\bibnamefont {{Dekel}}}, \bibinfo
  {author} {\bibfnamefont {G.}~\bibnamefont {{Kauffmann}}}, \ and\ \bibinfo
  {author} {\bibfnamefont {S.~D.~M.}\ \bibnamefont {{White}}},\ }\href
  {\doibase 10.1046/j.1365-8711.2001.03894.x} {\bibfield  {journal} {\bibinfo
  {journal} {\mnras}\ }\textbf {\bibinfo {volume} {320}},\ \bibinfo {pages}
  {289} (\bibinfo {year} {2001})},\ \Eprint
  {http://arxiv.org/abs/astro-ph/9912073} {astro-ph/9912073} \BibitemShut
  {NoStop}%
\bibitem [{\citenamefont {Yoshikawa}\ \emph {et~al.}(2001)\citenamefont
  {Yoshikawa}, \citenamefont {Taruya}, \citenamefont {Jing},\ and\
  \citenamefont {Suto}}]{Yoshikawa2001}%
  \BibitemOpen
  \bibfield  {author} {\bibinfo {author} {\bibfnamefont {K.}~\bibnamefont
  {Yoshikawa}}, \bibinfo {author} {\bibfnamefont {A.}~\bibnamefont {Taruya}},
  \bibinfo {author} {\bibfnamefont {Y.~P.}\ \bibnamefont {Jing}}, \ and\
  \bibinfo {author} {\bibfnamefont {Y.}~\bibnamefont {Suto}},\ }\href
  {http://stacks.iop.org/0004-637X/558/i=2/a=520} {\bibfield  {journal}
  {\bibinfo  {journal} {The Astrophysical Journal}\ }\textbf {\bibinfo {volume}
  {558}},\ \bibinfo {pages} {520} (\bibinfo {year} {2001})}\BibitemShut
  {NoStop}%
\bibitem [{\citenamefont {{Jullo}}\ \emph {et~al.}(2012)\citenamefont
  {{Jullo}}, \citenamefont {{Rhodes}}, \citenamefont {{Kiessling}},
  \citenamefont {{Taylor}}, \citenamefont {{Massey}}, \citenamefont {{Berge}},
  \citenamefont {{Schimd}}, \citenamefont {{Kneib}},\ and\ \citenamefont
  {{Scoville}}}]{2012ApJ...750...37J}%
  \BibitemOpen
  \bibfield  {author} {\bibinfo {author} {\bibfnamefont {E.}~\bibnamefont
  {{Jullo}}}, \bibinfo {author} {\bibfnamefont {J.}~\bibnamefont {{Rhodes}}},
  \bibinfo {author} {\bibfnamefont {A.}~\bibnamefont {{Kiessling}}}, \bibinfo
  {author} {\bibfnamefont {J.~E.}\ \bibnamefont {{Taylor}}}, \bibinfo {author}
  {\bibfnamefont {R.}~\bibnamefont {{Massey}}}, \bibinfo {author}
  {\bibfnamefont {J.}~\bibnamefont {{Berge}}}, \bibinfo {author} {\bibfnamefont
  {C.}~\bibnamefont {{Schimd}}}, \bibinfo {author} {\bibfnamefont {J.-P.}\
  \bibnamefont {{Kneib}}}, \ and\ \bibinfo {author} {\bibfnamefont
  {N.}~\bibnamefont {{Scoville}}},\ }\href {\doibase
  10.1088/0004-637X/750/1/37} {\bibfield  {journal} {\bibinfo  {journal}
  {\apj}\ }\textbf {\bibinfo {volume} {750}},\ \bibinfo {eid} {37} (\bibinfo
  {year} {2012})},\ \Eprint {http://arxiv.org/abs/1202.6491} {arXiv:1202.6491}
  \BibitemShut {NoStop}%
\bibitem [{\citenamefont {{Seljak}}\ and\ \citenamefont
  {{Warren}}(2004)}]{SeljakWarren2004}%
  \BibitemOpen
  \bibfield  {author} {\bibinfo {author} {\bibfnamefont {U.}~\bibnamefont
  {{Seljak}}}\ and\ \bibinfo {author} {\bibfnamefont {M.~S.}\ \bibnamefont
  {{Warren}}},\ }\href {\doibase 10.1111/j.1365-2966.2004.08297.x} {\bibfield
  {journal} {\bibinfo  {journal} {\mnras}\ }\textbf {\bibinfo {volume} {355}},\
  \bibinfo {pages} {129} (\bibinfo {year} {2004})},\ \Eprint
  {http://arxiv.org/abs/astro-ph/0403698} {astro-ph/0403698} \BibitemShut
  {NoStop}%
\bibitem [{\citenamefont {{Bonoli}}\ and\ \citenamefont
  {{Pen}}(2009)}]{BonoliPen0810}%
  \BibitemOpen
  \bibfield  {author} {\bibinfo {author} {\bibfnamefont {S.}~\bibnamefont
  {{Bonoli}}}\ and\ \bibinfo {author} {\bibfnamefont {U.~L.}\ \bibnamefont
  {{Pen}}},\ }\href {\doibase 10.1111/j.1365-2966.2009.14829.x} {\bibfield
  {journal} {\bibinfo  {journal} {\mnras}\ }\textbf {\bibinfo {volume} {396}},\
  \bibinfo {pages} {1610} (\bibinfo {year} {2009})},\ \Eprint
  {http://arxiv.org/abs/0810.0273} {arXiv:0810.0273} \BibitemShut {NoStop}%
\bibitem [{\citenamefont {{Seljak}}\ \emph {et~al.}(2009)\citenamefont
  {{Seljak}}, \citenamefont {{Hamaus}},\ and\ \citenamefont
  {{Desjacques}}}]{Seljak0904}%
  \BibitemOpen
  \bibfield  {author} {\bibinfo {author} {\bibfnamefont {U.}~\bibnamefont
  {{Seljak}}}, \bibinfo {author} {\bibfnamefont {N.}~\bibnamefont {{Hamaus}}},
  \ and\ \bibinfo {author} {\bibfnamefont {V.}~\bibnamefont {{Desjacques}}},\
  }\href {\doibase 10.1103/PhysRevLett.103.091303} {\bibfield  {journal}
  {\bibinfo  {journal} {Physical Review Letters}\ }\textbf {\bibinfo {volume}
  {103}},\ \bibinfo {eid} {091303} (\bibinfo {year} {2009})},\ \Eprint
  {http://arxiv.org/abs/0904.2963} {arXiv:0904.2963 [astro-ph.CO]} \BibitemShut
  {NoStop}%
\bibitem [{\citenamefont {{Hamaus}}\ \emph {et~al.}(2010)\citenamefont
  {{Hamaus}}, \citenamefont {{Seljak}}, \citenamefont {{Desjacques}},
  \citenamefont {{Smith}},\ and\ \citenamefont {{Baldauf}}}]{Hamaus1004}%
  \BibitemOpen
  \bibfield  {author} {\bibinfo {author} {\bibfnamefont {N.}~\bibnamefont
  {{Hamaus}}}, \bibinfo {author} {\bibfnamefont {U.}~\bibnamefont {{Seljak}}},
  \bibinfo {author} {\bibfnamefont {V.}~\bibnamefont {{Desjacques}}}, \bibinfo
  {author} {\bibfnamefont {R.~E.}\ \bibnamefont {{Smith}}}, \ and\ \bibinfo
  {author} {\bibfnamefont {T.}~\bibnamefont {{Baldauf}}},\ }\href {\doibase
  10.1103/PhysRevD.82.043515} {\bibfield  {journal} {\bibinfo  {journal}
  {\prd}\ }\textbf {\bibinfo {volume} {82}},\ \bibinfo {eid} {043515} (\bibinfo
  {year} {2010})},\ \Eprint {http://arxiv.org/abs/1004.5377} {arXiv:1004.5377}
  \BibitemShut {NoStop}%
\bibitem [{\citenamefont {{Cai}}\ \emph {et~al.}(2011)\citenamefont {{Cai}},
  \citenamefont {{Bernstein}},\ and\ \citenamefont
  {{Sheth}}}]{CaiBernsteinSheth1007}%
  \BibitemOpen
  \bibfield  {author} {\bibinfo {author} {\bibfnamefont {Y.-C.}\ \bibnamefont
  {{Cai}}}, \bibinfo {author} {\bibfnamefont {G.}~\bibnamefont {{Bernstein}}},
  \ and\ \bibinfo {author} {\bibfnamefont {R.~K.}\ \bibnamefont {{Sheth}}},\
  }\href {\doibase 10.1111/j.1365-2966.2010.17969.x} {\bibfield  {journal}
  {\bibinfo  {journal} {\mnras}\ }\textbf {\bibinfo {volume} {412}},\ \bibinfo
  {pages} {995} (\bibinfo {year} {2011})},\ \Eprint
  {http://arxiv.org/abs/1007.3500} {arXiv:1007.3500} \BibitemShut {NoStop}%
\bibitem [{\citenamefont {{Roth}}\ and\ \citenamefont
  {{Porciani}}(2011)}]{RothPorciani2011}%
  \BibitemOpen
  \bibfield  {author} {\bibinfo {author} {\bibfnamefont {N.}~\bibnamefont
  {{Roth}}}\ and\ \bibinfo {author} {\bibfnamefont {C.}~\bibnamefont
  {{Porciani}}},\ }\href {\doibase 10.1111/j.1365-2966.2011.18768.x} {\bibfield
   {journal} {\bibinfo  {journal} {\mnras}\ }\textbf {\bibinfo {volume}
  {415}},\ \bibinfo {pages} {829} (\bibinfo {year} {2011})},\ \Eprint
  {http://arxiv.org/abs/1101.1520} {arXiv:1101.1520 [astro-ph.CO]} \BibitemShut
  {NoStop}%
\bibitem [{\citenamefont {{Baldauf}}\ \emph {et~al.}(2013)\citenamefont
  {{Baldauf}}, \citenamefont {{Seljak}}, \citenamefont {{Smith}}, \citenamefont
  {{Hamaus}},\ and\ \citenamefont {{Desjacques}}}]{Baldauf1305}%
  \BibitemOpen
  \bibfield  {author} {\bibinfo {author} {\bibfnamefont {T.}~\bibnamefont
  {{Baldauf}}}, \bibinfo {author} {\bibfnamefont {U.}~\bibnamefont {{Seljak}}},
  \bibinfo {author} {\bibfnamefont {R.~E.}\ \bibnamefont {{Smith}}}, \bibinfo
  {author} {\bibfnamefont {N.}~\bibnamefont {{Hamaus}}}, \ and\ \bibinfo
  {author} {\bibfnamefont {V.}~\bibnamefont {{Desjacques}}},\ }\href {\doibase
  10.1103/PhysRevD.88.083507} {\bibfield  {journal} {\bibinfo  {journal}
  {\prd}\ }\textbf {\bibinfo {volume} {88}},\ \bibinfo {eid} {083507} (\bibinfo
  {year} {2013})},\ \Eprint {http://arxiv.org/abs/1305.2917} {arXiv:1305.2917
  [astro-ph.CO]} \BibitemShut {NoStop}%
\bibitem [{\citenamefont {{Ginzburg}}\ \emph {et~al.}(2017)\citenamefont
  {{Ginzburg}}, \citenamefont {{Desjacques}},\ and\ \citenamefont
  {{Chan}}}]{GinzburgDesjacques1706}%
  \BibitemOpen
  \bibfield  {author} {\bibinfo {author} {\bibfnamefont {D.}~\bibnamefont
  {{Ginzburg}}}, \bibinfo {author} {\bibfnamefont {V.}~\bibnamefont
  {{Desjacques}}}, \ and\ \bibinfo {author} {\bibfnamefont {K.~C.}\
  \bibnamefont {{Chan}}},\ }\href {\doibase 10.1103/PhysRevD.96.083528}
  {\bibfield  {journal} {\bibinfo  {journal} {prd}\ }\textbf {\bibinfo {volume}
  {96}},\ \bibinfo {eid} {083528} (\bibinfo {year} {2017})},\ \Eprint
  {http://arxiv.org/abs/1706.08738} {arXiv:1706.08738} \BibitemShut {NoStop}%
\bibitem [{\citenamefont {{Jasche}}\ \emph {et~al.}(2010)\citenamefont
  {{Jasche}}, \citenamefont {{Kitaura}}, \citenamefont {{Wandelt}},\ and\
  \citenamefont {{En{\ss}lin}}}]{2010MNRAS.406...60J}%
  \BibitemOpen
  \bibfield  {author} {\bibinfo {author} {\bibfnamefont {J.}~\bibnamefont
  {{Jasche}}}, \bibinfo {author} {\bibfnamefont {F.~S.}\ \bibnamefont
  {{Kitaura}}}, \bibinfo {author} {\bibfnamefont {B.~D.}\ \bibnamefont
  {{Wandelt}}}, \ and\ \bibinfo {author} {\bibfnamefont {T.~A.}\ \bibnamefont
  {{En{\ss}lin}}},\ }\href {\doibase 10.1111/j.1365-2966.2010.16610.x}
  {\bibfield  {journal} {\bibinfo  {journal} {\mnras}\ }\textbf {\bibinfo
  {volume} {406}},\ \bibinfo {pages} {60} (\bibinfo {year} {2010})},\ \Eprint
  {http://arxiv.org/abs/0911.2493} {arXiv:0911.2493} \BibitemShut {NoStop}%
\bibitem [{\citenamefont {{Jasche}}\ and\ \citenamefont
  {{Kitaura}}(2010)}]{2010MNRAS.407...29J}%
  \BibitemOpen
  \bibfield  {author} {\bibinfo {author} {\bibfnamefont {J.}~\bibnamefont
  {{Jasche}}}\ and\ \bibinfo {author} {\bibfnamefont {F.~S.}\ \bibnamefont
  {{Kitaura}}},\ }\href {\doibase 10.1111/j.1365-2966.2010.16897.x} {\bibfield
  {journal} {\bibinfo  {journal} {\mnras}\ }\textbf {\bibinfo {volume} {407}},\
  \bibinfo {pages} {29} (\bibinfo {year} {2010})},\ \Eprint
  {http://arxiv.org/abs/0911.2496} {arXiv:0911.2496} \BibitemShut {NoStop}%
\bibitem [{\citenamefont {{Jasche}}\ and\ \citenamefont
  {{Wandelt}}(2013)}]{JascheWandelt1203}%
  \BibitemOpen
  \bibfield  {author} {\bibinfo {author} {\bibfnamefont {J.}~\bibnamefont
  {{Jasche}}}\ and\ \bibinfo {author} {\bibfnamefont {B.~D.}\ \bibnamefont
  {{Wandelt}}},\ }\href {\doibase 10.1093/mnras/stt449} {\bibfield  {journal}
  {\bibinfo  {journal} {\mnras}\ }\textbf {\bibinfo {volume} {432}},\ \bibinfo
  {pages} {894} (\bibinfo {year} {2013})},\ \Eprint
  {http://arxiv.org/abs/1203.3639} {arXiv:1203.3639} \BibitemShut {NoStop}%
\bibitem [{\citenamefont {{Kitaura}}(2013)}]{Kitaura1203}%
  \BibitemOpen
  \bibfield  {author} {\bibinfo {author} {\bibfnamefont {F.-S.}\ \bibnamefont
  {{Kitaura}}},\ }\href {\doibase 10.1093/mnrasl/sls029} {\bibfield  {journal}
  {\bibinfo  {journal} {\mnras}\ }\textbf {\bibinfo {volume} {429}},\ \bibinfo
  {pages} {L84} (\bibinfo {year} {2013})},\ \Eprint
  {http://arxiv.org/abs/1203.4184} {arXiv:1203.4184} \BibitemShut {NoStop}%
\bibitem [{\citenamefont {{Wang}}\ \emph {et~al.}(2013)\citenamefont {{Wang}},
  \citenamefont {{Mo}}, \citenamefont {{Yang}},\ and\ \citenamefont {{van den
  Bosch}}}]{2013ApJ...772...63W}%
  \BibitemOpen
  \bibfield  {author} {\bibinfo {author} {\bibfnamefont {H.}~\bibnamefont
  {{Wang}}}, \bibinfo {author} {\bibfnamefont {H.~J.}\ \bibnamefont {{Mo}}},
  \bibinfo {author} {\bibfnamefont {X.}~\bibnamefont {{Yang}}}, \ and\ \bibinfo
  {author} {\bibfnamefont {F.~C.}\ \bibnamefont {{van den Bosch}}},\ }\href
  {\doibase 10.1088/0004-637X/772/1/63} {\bibfield  {journal} {\bibinfo
  {journal} {\apj}\ }\textbf {\bibinfo {volume} {772}},\ \bibinfo {eid} {63}
  (\bibinfo {year} {2013})},\ \Eprint {http://arxiv.org/abs/1301.1348}
  {arXiv:1301.1348} \BibitemShut {NoStop}%
\bibitem [{\citenamefont {{Seljak}}\ \emph {et~al.}(2017)\citenamefont
  {{Seljak}}, \citenamefont {{Aslanyan}}, \citenamefont {{Feng}},\ and\
  \citenamefont {{Modi}}}]{Uros1706}%
  \BibitemOpen
  \bibfield  {author} {\bibinfo {author} {\bibfnamefont {U.}~\bibnamefont
  {{Seljak}}}, \bibinfo {author} {\bibfnamefont {G.}~\bibnamefont
  {{Aslanyan}}}, \bibinfo {author} {\bibfnamefont {Y.}~\bibnamefont {{Feng}}},
  \ and\ \bibinfo {author} {\bibfnamefont {C.}~\bibnamefont {{Modi}}},\ }\href
  {\doibase 10.1088/1475-7516/2017/12/009} {\bibfield  {journal} {\bibinfo
  {journal} {\jcap}\ }\textbf {\bibinfo {volume} {12}},\ \bibinfo {eid} {009}
  (\bibinfo {year} {2017})},\ \Eprint {http://arxiv.org/abs/1706.06645}
  {arXiv:1706.06645} \BibitemShut {NoStop}%
\bibitem [{\citenamefont {{Jasche}}\ and\ \citenamefont
  {{Lavaux}}(2017)}]{JascheLavaux1706}%
  \BibitemOpen
  \bibfield  {author} {\bibinfo {author} {\bibfnamefont {J.}~\bibnamefont
  {{Jasche}}}\ and\ \bibinfo {author} {\bibfnamefont {G.}~\bibnamefont
  {{Lavaux}}},\ }\href {\doibase 10.1051/0004-6361/201730909} {\bibfield
  {journal} {\bibinfo  {journal} {\aap}\ }\textbf {\bibinfo {volume} {606}},\
  \bibinfo {eid} {A37} (\bibinfo {year} {2017})},\ \Eprint
  {http://arxiv.org/abs/1706.08971} {arXiv:1706.08971} \BibitemShut {NoStop}%
\bibitem [{\citenamefont {{Modi}}\ \emph {et~al.}(2018)\citenamefont {{Modi}},
  \citenamefont {{Feng}},\ and\ \citenamefont {{Seljak}}}]{Chirag1805}%
  \BibitemOpen
  \bibfield  {author} {\bibinfo {author} {\bibfnamefont {C.}~\bibnamefont
  {{Modi}}}, \bibinfo {author} {\bibfnamefont {Y.}~\bibnamefont {{Feng}}}, \
  and\ \bibinfo {author} {\bibfnamefont {U.}~\bibnamefont {{Seljak}}},\
  }\href@noop {} {\bibfield  {journal} {\bibinfo  {journal} {ArXiv e-prints}\ }
  (\bibinfo {year} {2018})},\ \Eprint {http://arxiv.org/abs/1805.02247}
  {arXiv:1805.02247} \BibitemShut {NoStop}%
\bibitem [{\citenamefont {{Schmidt}}\ \emph {et~al.}(2018)\citenamefont
  {{Schmidt}}, \citenamefont {{Elsner}}, \citenamefont {{Jasche}},
  \citenamefont {{Nguyen}},\ and\ \citenamefont {{Lavaux}}}]{Schmidt1808}%
  \BibitemOpen
  \bibfield  {author} {\bibinfo {author} {\bibfnamefont {F.}~\bibnamefont
  {{Schmidt}}}, \bibinfo {author} {\bibfnamefont {F.}~\bibnamefont {{Elsner}}},
  \bibinfo {author} {\bibfnamefont {J.}~\bibnamefont {{Jasche}}}, \bibinfo
  {author} {\bibfnamefont {N.~M.}\ \bibnamefont {{Nguyen}}}, \ and\ \bibinfo
  {author} {\bibfnamefont {G.}~\bibnamefont {{Lavaux}}},\ }\href@noop {}
  {\bibfield  {journal} {\bibinfo  {journal} {ArXiv e-prints}\ } (\bibinfo
  {year} {2018})},\ \Eprint {http://arxiv.org/abs/1808.02002}
  {arXiv:1808.02002} \BibitemShut {NoStop}%
\bibitem [{\citenamefont {{Eisenstein}}\ \emph {et~al.}(2007)\citenamefont
  {{Eisenstein}}, \citenamefont {{Seo}}, \citenamefont {{Sirko}},\ and\
  \citenamefont {{Spergel}}}]{EisensteinRec}%
  \BibitemOpen
  \bibfield  {author} {\bibinfo {author} {\bibfnamefont {D.~J.}\ \bibnamefont
  {{Eisenstein}}}, \bibinfo {author} {\bibfnamefont {H.-J.}\ \bibnamefont
  {{Seo}}}, \bibinfo {author} {\bibfnamefont {E.}~\bibnamefont {{Sirko}}}, \
  and\ \bibinfo {author} {\bibfnamefont {D.~N.}\ \bibnamefont {{Spergel}}},\
  }\href {\doibase 10.1086/518712} {\bibfield  {journal} {\bibinfo  {journal}
  {\apj}\ }\textbf {\bibinfo {volume} {664}},\ \bibinfo {pages} {675} (\bibinfo
  {year} {2007})},\ \Eprint {http://arxiv.org/abs/astro-ph/0604362}
  {astro-ph/0604362} \BibitemShut {NoStop}%
\bibitem [{\citenamefont {{Tassev}}\ and\ \citenamefont
  {{Zaldarriaga}}(2012)}]{TassevRec}%
  \BibitemOpen
  \bibfield  {author} {\bibinfo {author} {\bibfnamefont {S.}~\bibnamefont
  {{Tassev}}}\ and\ \bibinfo {author} {\bibfnamefont {M.}~\bibnamefont
  {{Zaldarriaga}}},\ }\href {\doibase 10.1088/1475-7516/2012/10/006} {\bibfield
   {journal} {\bibinfo  {journal} {\jcap}\ }\textbf {\bibinfo {volume} {10}},\
  \bibinfo {eid} {006} (\bibinfo {year} {2012})},\ \Eprint
  {http://arxiv.org/abs/1203.6066} {arXiv:1203.6066} \BibitemShut {NoStop}%
\bibitem [{\citenamefont {{Schmittfull}}\ \emph {et~al.}(2015)\citenamefont
  {{Schmittfull}}, \citenamefont {{Feng}}, \citenamefont {{Beutler}},
  \citenamefont {{Sherwin}},\ and\ \citenamefont {{Chu}}}]{Marcel1508}%
  \BibitemOpen
  \bibfield  {author} {\bibinfo {author} {\bibfnamefont {M.}~\bibnamefont
  {{Schmittfull}}}, \bibinfo {author} {\bibfnamefont {Y.}~\bibnamefont
  {{Feng}}}, \bibinfo {author} {\bibfnamefont {F.}~\bibnamefont {{Beutler}}},
  \bibinfo {author} {\bibfnamefont {B.}~\bibnamefont {{Sherwin}}}, \ and\
  \bibinfo {author} {\bibfnamefont {M.~Y.}\ \bibnamefont {{Chu}}},\ }\href
  {\doibase 10.1103/PhysRevD.92.123522} {\bibfield  {journal} {\bibinfo
  {journal} {\prd}\ }\textbf {\bibinfo {volume} {92}},\ \bibinfo {eid} {123522}
  (\bibinfo {year} {2015})},\ \Eprint {http://arxiv.org/abs/1508.06972}
  {arXiv:1508.06972} \BibitemShut {NoStop}%
\bibitem [{\citenamefont {{Keselman}}\ and\ \citenamefont
  {{Nusser}}(2017)}]{KeselmanNusser2017}%
  \BibitemOpen
  \bibfield  {author} {\bibinfo {author} {\bibfnamefont {J.~A.}\ \bibnamefont
  {{Keselman}}}\ and\ \bibinfo {author} {\bibfnamefont {A.}~\bibnamefont
  {{Nusser}}},\ }\href {\doibase 10.1093/mnras/stx152} {\bibfield  {journal}
  {\bibinfo  {journal} {\mnras}\ }\textbf {\bibinfo {volume} {467}},\ \bibinfo
  {pages} {1915} (\bibinfo {year} {2017})},\ \Eprint
  {http://arxiv.org/abs/1609.03576} {arXiv:1609.03576} \BibitemShut {NoStop}%
\bibitem [{\citenamefont {{Zhu}}\ \emph {et~al.}(2016)\citenamefont {{Zhu}},
  \citenamefont {{Yu}}, \citenamefont {{Pen}}, \citenamefont {{Chen}},\ and\
  \citenamefont {{Yu}}}]{IsoRecZhu1611ThreeD}%
  \BibitemOpen
  \bibfield  {author} {\bibinfo {author} {\bibfnamefont {H.-M.}\ \bibnamefont
  {{Zhu}}}, \bibinfo {author} {\bibfnamefont {Y.}~\bibnamefont {{Yu}}},
  \bibinfo {author} {\bibfnamefont {U.-L.}\ \bibnamefont {{Pen}}}, \bibinfo
  {author} {\bibfnamefont {X.}~\bibnamefont {{Chen}}}, \ and\ \bibinfo {author}
  {\bibfnamefont {H.-R.}\ \bibnamefont {{Yu}}},\ }\href@noop {} {\bibfield
  {journal} {\bibinfo  {journal} {ArXiv e-prints}\ } (\bibinfo {year}
  {2016})},\ \Eprint {http://arxiv.org/abs/1611.09638} {arXiv:1611.09638}
  \BibitemShut {NoStop}%
\bibitem [{\citenamefont {{Yu}}\ \emph {et~al.}(2017)\citenamefont {{Yu}},
  \citenamefont {{Zhu}},\ and\ \citenamefont {{Pen}}}]{IsoRecYu1703Halos}%
  \BibitemOpen
  \bibfield  {author} {\bibinfo {author} {\bibfnamefont {Y.}~\bibnamefont
  {{Yu}}}, \bibinfo {author} {\bibfnamefont {H.-M.}\ \bibnamefont {{Zhu}}}, \
  and\ \bibinfo {author} {\bibfnamefont {U.-L.}\ \bibnamefont {{Pen}}},\
  }\href@noop {} {\bibfield  {journal} {\bibinfo  {journal} {ArXiv e-prints}\ }
  (\bibinfo {year} {2017})},\ \Eprint {http://arxiv.org/abs/1703.08301}
  {arXiv:1703.08301} \BibitemShut {NoStop}%
\bibitem [{\citenamefont {{Wang}}\ \emph {et~al.}(2017)\citenamefont {{Wang}},
  \citenamefont {{Yu}}, \citenamefont {{Zhu}}, \citenamefont {{Yu}},
  \citenamefont {{Pan}},\ and\ \citenamefont {{Pen}}}]{IsoRecWang1703BAO}%
  \BibitemOpen
  \bibfield  {author} {\bibinfo {author} {\bibfnamefont {X.}~\bibnamefont
  {{Wang}}}, \bibinfo {author} {\bibfnamefont {H.-R.}\ \bibnamefont {{Yu}}},
  \bibinfo {author} {\bibfnamefont {H.-M.}\ \bibnamefont {{Zhu}}}, \bibinfo
  {author} {\bibfnamefont {Y.}~\bibnamefont {{Yu}}}, \bibinfo {author}
  {\bibfnamefont {Q.}~\bibnamefont {{Pan}}}, \ and\ \bibinfo {author}
  {\bibfnamefont {U.-L.}\ \bibnamefont {{Pen}}},\ }\href@noop {} {\bibfield
  {journal} {\bibinfo  {journal} {ArXiv e-prints}\ } (\bibinfo {year}
  {2017})},\ \Eprint {http://arxiv.org/abs/1703.09742} {arXiv:1703.09742}
  \BibitemShut {NoStop}%
\bibitem [{\citenamefont {{Schmittfull}}\ \emph {et~al.}(2017)\citenamefont
  {{Schmittfull}}, \citenamefont {{Baldauf}},\ and\ \citenamefont
  {{Zaldarriaga}}}]{Marcel1704}%
  \BibitemOpen
  \bibfield  {author} {\bibinfo {author} {\bibfnamefont {M.}~\bibnamefont
  {{Schmittfull}}}, \bibinfo {author} {\bibfnamefont {T.}~\bibnamefont
  {{Baldauf}}}, \ and\ \bibinfo {author} {\bibfnamefont {M.}~\bibnamefont
  {{Zaldarriaga}}},\ }\href {\doibase 10.1103/PhysRevD.96.023505} {\bibfield
  {journal} {\bibinfo  {journal} {\prd}\ }\textbf {\bibinfo {volume} {96}},\
  \bibinfo {eid} {023505} (\bibinfo {year} {2017})},\ \Eprint
  {http://arxiv.org/abs/1704.06634} {arXiv:1704.06634} \BibitemShut {NoStop}%
\bibitem [{\citenamefont {{Shi}}\ \emph {et~al.}(2018)\citenamefont {{Shi}},
  \citenamefont {{Cautun}},\ and\ \citenamefont {{Li}}}]{2018PhRvD..97b3505S}%
  \BibitemOpen
  \bibfield  {author} {\bibinfo {author} {\bibfnamefont {Y.}~\bibnamefont
  {{Shi}}}, \bibinfo {author} {\bibfnamefont {M.}~\bibnamefont {{Cautun}}}, \
  and\ \bibinfo {author} {\bibfnamefont {B.}~\bibnamefont {{Li}}},\ }\href
  {\doibase 10.1103/PhysRevD.97.023505} {\bibfield  {journal} {\bibinfo
  {journal} {\prd}\ }\textbf {\bibinfo {volume} {97}},\ \bibinfo {eid} {023505}
  (\bibinfo {year} {2018})},\ \Eprint {http://arxiv.org/abs/1709.06350}
  {arXiv:1709.06350} \BibitemShut {NoStop}%
\bibitem [{\citenamefont {{Hada}}\ and\ \citenamefont
  {{Eisenstein}}(2018)}]{HadaEisenstein1804}%
  \BibitemOpen
  \bibfield  {author} {\bibinfo {author} {\bibfnamefont {R.}~\bibnamefont
  {{Hada}}}\ and\ \bibinfo {author} {\bibfnamefont {D.~J.}\ \bibnamefont
  {{Eisenstein}}},\ }\href {\doibase 10.1093/mnras/sty1203} {\bibfield
  {journal} {\bibinfo  {journal} {\mnras}\ }\textbf {\bibinfo {volume} {478}},\
  \bibinfo {pages} {1866} (\bibinfo {year} {2018})},\ \Eprint
  {http://arxiv.org/abs/1804.04738} {arXiv:1804.04738} \BibitemShut {NoStop}%
\bibitem [{\citenamefont {{Birkin}}\ \emph {et~al.}(2018)\citenamefont
  {{Birkin}}, \citenamefont {{Li}}, \citenamefont {{Cautun}},\ and\
  \citenamefont {{Shi}}}]{2018arXiv180908135B}%
  \BibitemOpen
  \bibfield  {author} {\bibinfo {author} {\bibfnamefont {J.}~\bibnamefont
  {{Birkin}}}, \bibinfo {author} {\bibfnamefont {B.}~\bibnamefont {{Li}}},
  \bibinfo {author} {\bibfnamefont {M.}~\bibnamefont {{Cautun}}}, \ and\
  \bibinfo {author} {\bibfnamefont {Y.}~\bibnamefont {{Shi}}},\ }\href@noop {}
  {\bibfield  {journal} {\bibinfo  {journal} {ArXiv e-prints}\ } (\bibinfo
  {year} {2018})},\ \Eprint {http://arxiv.org/abs/1809.08135}
  {arXiv:1809.08135} \BibitemShut {NoStop}%
\bibitem [{\citenamefont {{Sarpa}}\ \emph {et~al.}(2018)\citenamefont
  {{Sarpa}}, \citenamefont {{Schimd}}, \citenamefont {{Branchini}},\ and\
  \citenamefont {{Matarrese}}}]{2018arXiv180910738S}%
  \BibitemOpen
  \bibfield  {author} {\bibinfo {author} {\bibfnamefont {E.}~\bibnamefont
  {{Sarpa}}}, \bibinfo {author} {\bibfnamefont {C.}~\bibnamefont {{Schimd}}},
  \bibinfo {author} {\bibfnamefont {E.}~\bibnamefont {{Branchini}}}, \ and\
  \bibinfo {author} {\bibfnamefont {S.}~\bibnamefont {{Matarrese}}},\
  }\href@noop {} {\bibfield  {journal} {\bibinfo  {journal} {ArXiv e-prints}\ }
  (\bibinfo {year} {2018})},\ \Eprint {http://arxiv.org/abs/1809.10738}
  {arXiv:1809.10738} \BibitemShut {NoStop}%
\bibitem [{\citenamefont {{Casas-Miranda}}\ \emph {et~al.}(2002)\citenamefont
  {{Casas-Miranda}}, \citenamefont {{Mo}}, \citenamefont {{Sheth}},\ and\
  \citenamefont {{Boerner}}}]{CasasMiranda2002}%
  \BibitemOpen
  \bibfield  {author} {\bibinfo {author} {\bibfnamefont {R.}~\bibnamefont
  {{Casas-Miranda}}}, \bibinfo {author} {\bibfnamefont {H.~J.}\ \bibnamefont
  {{Mo}}}, \bibinfo {author} {\bibfnamefont {R.~K.}\ \bibnamefont {{Sheth}}}, \
  and\ \bibinfo {author} {\bibfnamefont {G.}~\bibnamefont {{Boerner}}},\ }\href
  {\doibase 10.1046/j.1365-8711.2002.05378.x} {\bibfield  {journal} {\bibinfo
  {journal} {\mnras}\ }\textbf {\bibinfo {volume} {333}},\ \bibinfo {pages}
  {730} (\bibinfo {year} {2002})},\ \Eprint
  {http://arxiv.org/abs/astro-ph/0105008} {astro-ph/0105008} \BibitemShut
  {NoStop}%
\bibitem [{\citenamefont {{Manera}}\ and\ \citenamefont
  {{Gazta{\~n}aga}}(2011)}]{2011MNRAS.415..383M}%
  \BibitemOpen
  \bibfield  {author} {\bibinfo {author} {\bibfnamefont {M.}~\bibnamefont
  {{Manera}}}\ and\ \bibinfo {author} {\bibfnamefont {E.}~\bibnamefont
  {{Gazta{\~n}aga}}},\ }\href {\doibase 10.1111/j.1365-2966.2011.18705.x}
  {\bibfield  {journal} {\bibinfo  {journal} {\mnras}\ }\textbf {\bibinfo
  {volume} {415}},\ \bibinfo {pages} {383} (\bibinfo {year} {2011})},\ \Eprint
  {http://arxiv.org/abs/0912.0446} {arXiv:0912.0446} \BibitemShut {NoStop}%
\bibitem [{\citenamefont {{Perko}}\ \emph {et~al.}(2016)\citenamefont
  {{Perko}}, \citenamefont {{Senatore}}, \citenamefont {{Jennings}},\ and\
  \citenamefont {{Wechsler}}}]{Perko1610}%
  \BibitemOpen
  \bibfield  {author} {\bibinfo {author} {\bibfnamefont {A.}~\bibnamefont
  {{Perko}}}, \bibinfo {author} {\bibfnamefont {L.}~\bibnamefont {{Senatore}}},
  \bibinfo {author} {\bibfnamefont {E.}~\bibnamefont {{Jennings}}}, \ and\
  \bibinfo {author} {\bibfnamefont {R.~H.}\ \bibnamefont {{Wechsler}}},\
  }\href@noop {} {\bibfield  {journal} {\bibinfo  {journal} {ArXiv e-prints}\ }
  (\bibinfo {year} {2016})},\ \Eprint {http://arxiv.org/abs/1610.09321}
  {arXiv:1610.09321} \BibitemShut {NoStop}%
\bibitem [{\citenamefont {{Matsubara}}(2008)}]{Matsubara0807}%
  \BibitemOpen
  \bibfield  {author} {\bibinfo {author} {\bibfnamefont {T.}~\bibnamefont
  {{Matsubara}}},\ }\href {\doibase 10.1103/PhysRevD.78.083519} {\bibfield
  {journal} {\bibinfo  {journal} {\prd}\ }\textbf {\bibinfo {volume} {78}},\
  \bibinfo {eid} {083519} (\bibinfo {year} {2008})},\ \Eprint
  {http://arxiv.org/abs/0807.1733} {arXiv:0807.1733} \BibitemShut {NoStop}%
\bibitem [{\citenamefont {{McDonald}}\ and\ \citenamefont
  {{Roy}}(2009)}]{McDonaldRoy0902}%
  \BibitemOpen
  \bibfield  {author} {\bibinfo {author} {\bibfnamefont {P.}~\bibnamefont
  {{McDonald}}}\ and\ \bibinfo {author} {\bibfnamefont {A.}~\bibnamefont
  {{Roy}}},\ }\href {\doibase 10.1088/1475-7516/2009/08/020} {\bibfield
  {journal} {\bibinfo  {journal} {\jcap}\ }\textbf {\bibinfo {volume} {8}},\
  \bibinfo {eid} {020} (\bibinfo {year} {2009})},\ \Eprint
  {http://arxiv.org/abs/0902.0991} {arXiv:0902.0991 [astro-ph.CO]} \BibitemShut
  {NoStop}%
\bibitem [{\citenamefont {{Matsubara}}(2011)}]{Matsubara1102}%
  \BibitemOpen
  \bibfield  {author} {\bibinfo {author} {\bibfnamefont {T.}~\bibnamefont
  {{Matsubara}}},\ }\href {\doibase 10.1103/PhysRevD.83.083518} {\bibfield
  {journal} {\bibinfo  {journal} {\prd}\ }\textbf {\bibinfo {volume} {83}},\
  \bibinfo {eid} {083518} (\bibinfo {year} {2011})},\ \Eprint
  {http://arxiv.org/abs/1102.4619} {arXiv:1102.4619 [astro-ph.CO]} \BibitemShut
  {NoStop}%
\bibitem [{\citenamefont {{Sheth}}\ \emph {et~al.}(2013)\citenamefont
  {{Sheth}}, \citenamefont {{Chan}},\ and\ \citenamefont
  {{Scoccimarro}}}]{Sheth1207}%
  \BibitemOpen
  \bibfield  {author} {\bibinfo {author} {\bibfnamefont {R.~K.}\ \bibnamefont
  {{Sheth}}}, \bibinfo {author} {\bibfnamefont {K.~C.}\ \bibnamefont {{Chan}}},
  \ and\ \bibinfo {author} {\bibfnamefont {R.}~\bibnamefont {{Scoccimarro}}},\
  }\href {\doibase 10.1103/PhysRevD.87.083002} {\bibfield  {journal} {\bibinfo
  {journal} {\prd}\ }\textbf {\bibinfo {volume} {87}},\ \bibinfo {eid} {083002}
  (\bibinfo {year} {2013})},\ \Eprint {http://arxiv.org/abs/1207.7117}
  {arXiv:1207.7117 [astro-ph.CO]} \BibitemShut {NoStop}%
\bibitem [{\citenamefont {{Carlson}}\ \emph {et~al.}(2013)\citenamefont
  {{Carlson}}, \citenamefont {{Reid}},\ and\ \citenamefont
  {{White}}}]{Carlson1209}%
  \BibitemOpen
  \bibfield  {author} {\bibinfo {author} {\bibfnamefont {J.}~\bibnamefont
  {{Carlson}}}, \bibinfo {author} {\bibfnamefont {B.}~\bibnamefont {{Reid}}}, \
  and\ \bibinfo {author} {\bibfnamefont {M.}~\bibnamefont {{White}}},\ }\href
  {\doibase 10.1093/mnras/sts457} {\bibfield  {journal} {\bibinfo  {journal}
  {\mnras}\ }\textbf {\bibinfo {volume} {429}},\ \bibinfo {pages} {1674}
  (\bibinfo {year} {2013})},\ \Eprint {http://arxiv.org/abs/1209.0780}
  {arXiv:1209.0780} \BibitemShut {NoStop}%
\bibitem [{\citenamefont {{Matsubara}}(2014)}]{Matsubara1304}%
  \BibitemOpen
  \bibfield  {author} {\bibinfo {author} {\bibfnamefont {T.}~\bibnamefont
  {{Matsubara}}},\ }\href {\doibase 10.1103/PhysRevD.90.043537} {\bibfield
  {journal} {\bibinfo  {journal} {\prd}\ }\textbf {\bibinfo {volume} {90}},\
  \bibinfo {eid} {043537} (\bibinfo {year} {2014})},\ \Eprint
  {http://arxiv.org/abs/1304.4226} {arXiv:1304.4226 [astro-ph.CO]} \BibitemShut
  {NoStop}%
\bibitem [{\citenamefont {{Biagetti}}\ \emph {et~al.}(2014)\citenamefont
  {{Biagetti}}, \citenamefont {{Chan}}, \citenamefont {{Desjacques}},\ and\
  \citenamefont {{Paranjape}}}]{Biagetti1310}%
  \BibitemOpen
  \bibfield  {author} {\bibinfo {author} {\bibfnamefont {M.}~\bibnamefont
  {{Biagetti}}}, \bibinfo {author} {\bibfnamefont {K.~C.}\ \bibnamefont
  {{Chan}}}, \bibinfo {author} {\bibfnamefont {V.}~\bibnamefont
  {{Desjacques}}}, \ and\ \bibinfo {author} {\bibfnamefont {A.}~\bibnamefont
  {{Paranjape}}},\ }\href {\doibase 10.1093/mnras/stu680} {\bibfield  {journal}
  {\bibinfo  {journal} {\mnras}\ }\textbf {\bibinfo {volume} {441}},\ \bibinfo
  {pages} {1457} (\bibinfo {year} {2014})},\ \Eprint
  {http://arxiv.org/abs/1310.1401} {arXiv:1310.1401} \BibitemShut {NoStop}%
\bibitem [{\citenamefont {{Assassi}}\ \emph {et~al.}(2014)\citenamefont
  {{Assassi}}, \citenamefont {{Baumann}}, \citenamefont {{Green}},\ and\
  \citenamefont {{Zaldarriaga}}}]{Valentin1402}%
  \BibitemOpen
  \bibfield  {author} {\bibinfo {author} {\bibfnamefont {V.}~\bibnamefont
  {{Assassi}}}, \bibinfo {author} {\bibfnamefont {D.}~\bibnamefont
  {{Baumann}}}, \bibinfo {author} {\bibfnamefont {D.}~\bibnamefont {{Green}}},
  \ and\ \bibinfo {author} {\bibfnamefont {M.}~\bibnamefont {{Zaldarriaga}}},\
  }\href {\doibase 10.1088/1475-7516/2014/08/056} {\bibfield  {journal}
  {\bibinfo  {journal} {\jcap}\ }\textbf {\bibinfo {volume} {8}},\ \bibinfo
  {eid} {056} (\bibinfo {year} {2014})},\ \Eprint
  {http://arxiv.org/abs/1402.5916} {arXiv:1402.5916} \BibitemShut {NoStop}%
\bibitem [{\citenamefont {{Senatore}}(2015)}]{Leonardo1406}%
  \BibitemOpen
  \bibfield  {author} {\bibinfo {author} {\bibfnamefont {L.}~\bibnamefont
  {{Senatore}}},\ }\href {\doibase 10.1088/1475-7516/2015/11/007} {\bibfield
  {journal} {\bibinfo  {journal} {\jcap}\ }\textbf {\bibinfo {volume} {11}},\
  \bibinfo {eid} {007} (\bibinfo {year} {2015})},\ \Eprint
  {http://arxiv.org/abs/1406.7843} {arXiv:1406.7843} \BibitemShut {NoStop}%
\bibitem [{\citenamefont {{Mirbabayi}}\ \emph {et~al.}(2015)\citenamefont
  {{Mirbabayi}}, \citenamefont {{Schmidt}},\ and\ \citenamefont
  {{Zaldarriaga}}}]{MSZ1412}%
  \BibitemOpen
  \bibfield  {author} {\bibinfo {author} {\bibfnamefont {M.}~\bibnamefont
  {{Mirbabayi}}}, \bibinfo {author} {\bibfnamefont {F.}~\bibnamefont
  {{Schmidt}}}, \ and\ \bibinfo {author} {\bibfnamefont {M.}~\bibnamefont
  {{Zaldarriaga}}},\ }\href {\doibase 10.1088/1475-7516/2015/07/030} {\bibfield
   {journal} {\bibinfo  {journal} {\jcap}\ }\textbf {\bibinfo {volume} {7}},\
  \bibinfo {eid} {030} (\bibinfo {year} {2015})},\ \Eprint
  {http://arxiv.org/abs/1412.5169} {arXiv:1412.5169} \BibitemShut {NoStop}%
\bibitem [{\citenamefont {{Matsubara}}\ and\ \citenamefont
  {{Desjacques}}(2016)}]{Matsubara1604}%
  \BibitemOpen
  \bibfield  {author} {\bibinfo {author} {\bibfnamefont {T.}~\bibnamefont
  {{Matsubara}}}\ and\ \bibinfo {author} {\bibfnamefont {V.}~\bibnamefont
  {{Desjacques}}},\ }\href {\doibase 10.1103/PhysRevD.93.123522} {\bibfield
  {journal} {\bibinfo  {journal} {\prd}\ }\textbf {\bibinfo {volume} {93}},\
  \bibinfo {eid} {123522} (\bibinfo {year} {2016})},\ \Eprint
  {http://arxiv.org/abs/1604.06579} {arXiv:1604.06579 [astro-ph.CO]}
  \BibitemShut {NoStop}%
\bibitem [{\citenamefont {{Vlah}}\ \emph {et~al.}(2016)\citenamefont {{Vlah}},
  \citenamefont {{Castorina}},\ and\ \citenamefont {{White}}}]{Vlah1609}%
  \BibitemOpen
  \bibfield  {author} {\bibinfo {author} {\bibfnamefont {Z.}~\bibnamefont
  {{Vlah}}}, \bibinfo {author} {\bibfnamefont {E.}~\bibnamefont {{Castorina}}},
  \ and\ \bibinfo {author} {\bibfnamefont {M.}~\bibnamefont {{White}}},\ }\href
  {\doibase 10.1088/1475-7516/2016/12/007} {\bibfield  {journal} {\bibinfo
  {journal} {\jcap}\ }\textbf {\bibinfo {volume} {12}},\ \bibinfo {eid} {007}
  (\bibinfo {year} {2016})},\ \Eprint {http://arxiv.org/abs/1609.02908}
  {arXiv:1609.02908} \BibitemShut {NoStop}%
\bibitem [{\citenamefont {Peloso}\ and\ \citenamefont
  {Pietroni}(2013)}]{Peloso:2013zw}%
  \BibitemOpen
  \bibfield  {author} {\bibinfo {author} {\bibfnamefont {M.}~\bibnamefont
  {Peloso}}\ and\ \bibinfo {author} {\bibfnamefont {M.}~\bibnamefont
  {Pietroni}},\ }\href {\doibase 10.1088/1475-7516/2013/05/031} {\bibfield
  {journal} {\bibinfo  {journal} {JCAP}\ }\textbf {\bibinfo {volume} {1305}},\
  \bibinfo {pages} {031} (\bibinfo {year} {2013})},\ \Eprint
  {http://arxiv.org/abs/1302.0223} {arXiv:1302.0223 [astro-ph.CO]} \BibitemShut
  {NoStop}%
\bibitem [{\citenamefont {Kehagias}\ and\ \citenamefont
  {Riotto}(2013)}]{Kehagias:2013yd}%
  \BibitemOpen
  \bibfield  {author} {\bibinfo {author} {\bibfnamefont {A.}~\bibnamefont
  {Kehagias}}\ and\ \bibinfo {author} {\bibfnamefont {A.}~\bibnamefont
  {Riotto}},\ }\href {\doibase 10.1016/j.nuclphysb.2013.05.009} {\bibfield
  {journal} {\bibinfo  {journal} {Nucl. Phys.}\ }\textbf {\bibinfo {volume}
  {B873}},\ \bibinfo {pages} {514} (\bibinfo {year} {2013})},\ \Eprint
  {http://arxiv.org/abs/1302.0130} {arXiv:1302.0130 [astro-ph.CO]} \BibitemShut
  {NoStop}%
\bibitem [{\citenamefont {Creminelli}\ \emph {et~al.}(2013)\citenamefont
  {Creminelli}, \citenamefont {Nore\~na}, \citenamefont {Simonovi\'c},\ and\
  \citenamefont {Vernizzi}}]{Creminelli:2013mca}%
  \BibitemOpen
  \bibfield  {author} {\bibinfo {author} {\bibfnamefont {P.}~\bibnamefont
  {Creminelli}}, \bibinfo {author} {\bibfnamefont {J.}~\bibnamefont
  {Nore\~na}}, \bibinfo {author} {\bibfnamefont {M.}~\bibnamefont
  {Simonovi\'c}}, \ and\ \bibinfo {author} {\bibfnamefont {F.}~\bibnamefont
  {Vernizzi}},\ }\href {\doibase 10.1088/1475-7516/2013/12/025} {\bibfield
  {journal} {\bibinfo  {journal} {JCAP}\ }\textbf {\bibinfo {volume} {1312}},\
  \bibinfo {pages} {025} (\bibinfo {year} {2013})},\ \Eprint
  {http://arxiv.org/abs/1309.3557} {arXiv:1309.3557 [astro-ph.CO]} \BibitemShut
  {NoStop}%
\bibitem [{\citenamefont {Creminelli}\ \emph {et~al.}(2014)\citenamefont
  {Creminelli}, \citenamefont {Gleyzes}, \citenamefont {Simonovi\'c},\ and\
  \citenamefont {Vernizzi}}]{Creminelli:2013poa}%
  \BibitemOpen
  \bibfield  {author} {\bibinfo {author} {\bibfnamefont {P.}~\bibnamefont
  {Creminelli}}, \bibinfo {author} {\bibfnamefont {J.}~\bibnamefont {Gleyzes}},
  \bibinfo {author} {\bibfnamefont {M.}~\bibnamefont {Simonovi\'c}}, \ and\
  \bibinfo {author} {\bibfnamefont {F.}~\bibnamefont {Vernizzi}},\ }\href
  {\doibase 10.1088/1475-7516/2014/02/051} {\bibfield  {journal} {\bibinfo
  {journal} {JCAP}\ }\textbf {\bibinfo {volume} {1402}},\ \bibinfo {pages}
  {051} (\bibinfo {year} {2014})},\ \Eprint {http://arxiv.org/abs/1311.0290}
  {arXiv:1311.0290 [astro-ph.CO]} \BibitemShut {NoStop}%
\bibitem [{\citenamefont {Senatore}\ and\ \citenamefont
  {Zaldarriaga}(2015)}]{Senatore:2014via}%
  \BibitemOpen
  \bibfield  {author} {\bibinfo {author} {\bibfnamefont {L.}~\bibnamefont
  {Senatore}}\ and\ \bibinfo {author} {\bibfnamefont {M.}~\bibnamefont
  {Zaldarriaga}},\ }\href {\doibase 10.1088/1475-7516/2015/02/013} {\bibfield
  {journal} {\bibinfo  {journal} {JCAP}\ }\textbf {\bibinfo {volume} {1502}},\
  \bibinfo {pages} {013} (\bibinfo {year} {2015})},\ \Eprint
  {http://arxiv.org/abs/1404.5954} {arXiv:1404.5954 [astro-ph.CO]} \BibitemShut
  {NoStop}%
\bibitem [{\citenamefont {Baldauf}\ \emph {et~al.}(2015)\citenamefont
  {Baldauf}, \citenamefont {Mirbabayi}, \citenamefont {Simonovi\'c},\ and\
  \citenamefont {Zaldarriaga}}]{Baldauf:2015xfa}%
  \BibitemOpen
  \bibfield  {author} {\bibinfo {author} {\bibfnamefont {T.}~\bibnamefont
  {Baldauf}}, \bibinfo {author} {\bibfnamefont {M.}~\bibnamefont {Mirbabayi}},
  \bibinfo {author} {\bibfnamefont {M.}~\bibnamefont {Simonovi\'c}}, \ and\
  \bibinfo {author} {\bibfnamefont {M.}~\bibnamefont {Zaldarriaga}},\ }\href
  {\doibase 10.1103/PhysRevD.92.043514} {\bibfield  {journal} {\bibinfo
  {journal} {Phys. Rev.}\ }\textbf {\bibinfo {volume} {D92}},\ \bibinfo {pages}
  {043514} (\bibinfo {year} {2015})},\ \Eprint
  {http://arxiv.org/abs/1504.04366} {arXiv:1504.04366 [astro-ph.CO]}
  \BibitemShut {NoStop}%
\bibitem [{\citenamefont {Vlah}\ \emph {et~al.}(2016)\citenamefont {Vlah},
  \citenamefont {Seljak}, \citenamefont {Chu},\ and\ \citenamefont
  {Feng}}]{Vlah:2015zda}%
  \BibitemOpen
  \bibfield  {author} {\bibinfo {author} {\bibfnamefont {Z.}~\bibnamefont
  {Vlah}}, \bibinfo {author} {\bibfnamefont {U.}~\bibnamefont {Seljak}},
  \bibinfo {author} {\bibfnamefont {M.~Y.}\ \bibnamefont {Chu}}, \ and\
  \bibinfo {author} {\bibfnamefont {Y.}~\bibnamefont {Feng}},\ }\href {\doibase
  10.1088/1475-7516/2016/03/057} {\bibfield  {journal} {\bibinfo  {journal}
  {JCAP}\ }\textbf {\bibinfo {volume} {1603}},\ \bibinfo {pages} {057}
  (\bibinfo {year} {2016})},\ \Eprint {http://arxiv.org/abs/1509.02120}
  {arXiv:1509.02120 [astro-ph.CO]} \BibitemShut {NoStop}%
\bibitem [{\citenamefont {Blas}\ \emph {et~al.}(2016)\citenamefont {Blas},
  \citenamefont {Garny}, \citenamefont {Ivanov},\ and\ \citenamefont
  {Sibiryakov}}]{Blas:2016sfa}%
  \BibitemOpen
  \bibfield  {author} {\bibinfo {author} {\bibfnamefont {D.}~\bibnamefont
  {Blas}}, \bibinfo {author} {\bibfnamefont {M.}~\bibnamefont {Garny}},
  \bibinfo {author} {\bibfnamefont {M.~M.}\ \bibnamefont {Ivanov}}, \ and\
  \bibinfo {author} {\bibfnamefont {S.}~\bibnamefont {Sibiryakov}},\ }\href
  {\doibase 10.1088/1475-7516/2016/07/028} {\bibfield  {journal} {\bibinfo
  {journal} {JCAP}\ }\textbf {\bibinfo {volume} {1607}},\ \bibinfo {pages}
  {028} (\bibinfo {year} {2016})},\ \Eprint {http://arxiv.org/abs/1605.02149}
  {arXiv:1605.02149 [astro-ph.CO]} \BibitemShut {NoStop}%
\bibitem [{\citenamefont {Senatore}\ and\ \citenamefont
  {Trevisan}(2018)}]{Senatore:2017pbn}%
  \BibitemOpen
  \bibfield  {author} {\bibinfo {author} {\bibfnamefont {L.}~\bibnamefont
  {Senatore}}\ and\ \bibinfo {author} {\bibfnamefont {G.}~\bibnamefont
  {Trevisan}},\ }\href {\doibase 10.1088/1475-7516/2018/05/019} {\bibfield
  {journal} {\bibinfo  {journal} {JCAP}\ }\textbf {\bibinfo {volume} {1805}},\
  \bibinfo {pages} {019} (\bibinfo {year} {2018})},\ \Eprint
  {http://arxiv.org/abs/1710.02178} {arXiv:1710.02178 [astro-ph.CO]}
  \BibitemShut {NoStop}%
\bibitem [{\citenamefont {{Baldauf}}\ \emph
  {et~al.}(2016{\natexlab{a}})\citenamefont {{Baldauf}}, \citenamefont
  {{Schaan}},\ and\ \citenamefont {{Zaldarriaga}}}]{TobiasDMDisplacement1505}%
  \BibitemOpen
  \bibfield  {author} {\bibinfo {author} {\bibfnamefont {T.}~\bibnamefont
  {{Baldauf}}}, \bibinfo {author} {\bibfnamefont {E.}~\bibnamefont {{Schaan}}},
  \ and\ \bibinfo {author} {\bibfnamefont {M.}~\bibnamefont {{Zaldarriaga}}},\
  }\href {\doibase 10.1088/1475-7516/2016/03/017} {\bibfield  {journal}
  {\bibinfo  {journal} {\jcap}\ }\textbf {\bibinfo {volume} {3}},\ \bibinfo
  {eid} {017} (\bibinfo {year} {2016}{\natexlab{a}})},\ \Eprint
  {http://arxiv.org/abs/1505.07098} {arXiv:1505.07098} \BibitemShut {NoStop}%
\bibitem [{\citenamefont {{Baldauf}}\ \emph
  {et~al.}(2016{\natexlab{b}})\citenamefont {{Baldauf}}, \citenamefont
  {{Schaan}},\ and\ \citenamefont {{Zaldarriaga}}}]{TobiasDMDensity1507}%
  \BibitemOpen
  \bibfield  {author} {\bibinfo {author} {\bibfnamefont {T.}~\bibnamefont
  {{Baldauf}}}, \bibinfo {author} {\bibfnamefont {E.}~\bibnamefont {{Schaan}}},
  \ and\ \bibinfo {author} {\bibfnamefont {M.}~\bibnamefont {{Zaldarriaga}}},\
  }\href {\doibase 10.1088/1475-7516/2016/03/007} {\bibfield  {journal}
  {\bibinfo  {journal} {\jcap}\ }\textbf {\bibinfo {volume} {3}},\ \bibinfo
  {eid} {007} (\bibinfo {year} {2016}{\natexlab{b}})},\ \Eprint
  {http://arxiv.org/abs/1507.02255} {arXiv:1507.02255} \BibitemShut {NoStop}%
\bibitem [{\citenamefont {{Zel'dovich}}(1970)}]{Zeldovich1970}%
  \BibitemOpen
  \bibfield  {author} {\bibinfo {author} {\bibfnamefont {Y.~B.}\ \bibnamefont
  {{Zel'dovich}}},\ }\href@noop {} {\bibfield  {journal} {\bibinfo  {journal}
  {\aap}\ }\textbf {\bibinfo {volume} {5}},\ \bibinfo {pages} {84} (\bibinfo
  {year} {1970})}\BibitemShut {NoStop}%
\bibitem [{\citenamefont {Vlah}\ \emph {et~al.}(2015)\citenamefont {Vlah},
  \citenamefont {White},\ and\ \citenamefont {Aviles}}]{Vlah:2015sea}%
  \BibitemOpen
  \bibfield  {author} {\bibinfo {author} {\bibfnamefont {Z.}~\bibnamefont
  {Vlah}}, \bibinfo {author} {\bibfnamefont {M.}~\bibnamefont {White}}, \ and\
  \bibinfo {author} {\bibfnamefont {A.}~\bibnamefont {Aviles}},\ }\href
  {\doibase 10.1088/1475-7516/2015/09/014} {\bibfield  {journal} {\bibinfo
  {journal} {JCAP}\ }\textbf {\bibinfo {volume} {1509}},\ \bibinfo {pages}
  {014} (\bibinfo {year} {2015})},\ \Eprint {http://arxiv.org/abs/1506.05264}
  {arXiv:1506.05264 [astro-ph.CO]} \BibitemShut {NoStop}%
\bibitem [{\citenamefont {{Modi}}\ \emph
  {et~al.}(2017{\natexlab{b}})\citenamefont {{Modi}}, \citenamefont {{White}},\
  and\ \citenamefont {{Vlah}}}]{Modi1706}%
  \BibitemOpen
  \bibfield  {author} {\bibinfo {author} {\bibfnamefont {C.}~\bibnamefont
  {{Modi}}}, \bibinfo {author} {\bibfnamefont {M.}~\bibnamefont {{White}}}, \
  and\ \bibinfo {author} {\bibfnamefont {Z.}~\bibnamefont {{Vlah}}},\ }\href
  {\doibase 10.1088/1475-7516/2017/08/009} {\bibfield  {journal} {\bibinfo
  {journal} {\jcap}\ }\textbf {\bibinfo {volume} {8}},\ \bibinfo {eid} {009}
  (\bibinfo {year} {2017}{\natexlab{b}})},\ \Eprint
  {http://arxiv.org/abs/1706.03173} {arXiv:1706.03173} \BibitemShut {NoStop}%
\bibitem [{\citenamefont {{Aviles}}(2018)}]{2018arXiv180505304A}%
  \BibitemOpen
  \bibfield  {author} {\bibinfo {author} {\bibfnamefont {A.}~\bibnamefont
  {{Aviles}}},\ }\href@noop {} {\bibfield  {journal} {\bibinfo  {journal}
  {ArXiv e-prints}\ } (\bibinfo {year} {2018})},\ \Eprint
  {http://arxiv.org/abs/1805.05304} {arXiv:1805.05304} \BibitemShut {NoStop}%
\bibitem [{\citenamefont {{Lazeyras}}\ and\ \citenamefont
  {{Schmidt}}(2018)}]{LazeyrasSchmidt1712}%
  \BibitemOpen
  \bibfield  {author} {\bibinfo {author} {\bibfnamefont {T.}~\bibnamefont
  {{Lazeyras}}}\ and\ \bibinfo {author} {\bibfnamefont {F.}~\bibnamefont
  {{Schmidt}}},\ }\href {\doibase 10.1088/1475-7516/2018/09/008} {\bibfield
  {journal} {\bibinfo  {journal} {\jcap}\ }\textbf {\bibinfo {volume} {9}},\
  \bibinfo {eid} {008} (\bibinfo {year} {2018})},\ \Eprint
  {http://arxiv.org/abs/1712.07531} {arXiv:1712.07531} \BibitemShut {NoStop}%
\bibitem [{\citenamefont {Lewis}\ \emph {et~al.}(2000)\citenamefont {Lewis},
  \citenamefont {Challinor},\ and\ \citenamefont {Lasenby}}]{camb}%
  \BibitemOpen
  \bibfield  {author} {\bibinfo {author} {\bibfnamefont {A.}~\bibnamefont
  {Lewis}}, \bibinfo {author} {\bibfnamefont {A.}~\bibnamefont {Challinor}}, \
  and\ \bibinfo {author} {\bibfnamefont {A.}~\bibnamefont {Lasenby}},\
  }\href@noop {} {\bibfield  {journal} {\bibinfo  {journal} {Astrophys. J.}\
  }\textbf {\bibinfo {volume} {538}},\ \bibinfo {pages} {473} (\bibinfo {year}
  {2000})},\ \Eprint {http://arxiv.org/abs/astro-ph/9911177} {astro-ph/9911177}
  \BibitemShut {NoStop}%
\bibitem [{\citenamefont {Ade}\ \emph {et~al.}(2016)\citenamefont {Ade} \emph
  {et~al.}}]{Planck15Params}%
  \BibitemOpen
  \bibfield  {author} {\bibinfo {author} {\bibfnamefont {P.~A.~R.}\
  \bibnamefont {Ade}} \emph {et~al.} (\bibinfo {collaboration} {Planck}),\
  }\href {\doibase 10.1051/0004-6361/201525830} {\bibfield  {journal} {\bibinfo
   {journal} {Astron. Astrophys.}\ }\textbf {\bibinfo {volume} {594}},\
  \bibinfo {pages} {A13} (\bibinfo {year} {2016})},\ \Eprint
  {http://arxiv.org/abs/1502.01589} {arXiv:1502.01589 [astro-ph.CO]}
  \BibitemShut {NoStop}%
\bibitem [{MPG()}]{MPGadgetWebsite}%
  \BibitemOpen
  \href {https://github.com/MP-Gadget/MP-Gadget} {}\bibinfo {howpublished}
  {https://github.com/MP-Gadget/MP-Gadget}\BibitemShut {NoStop}%
\bibitem [{\citenamefont {Feng}\ \emph {et~al.}(2018)\citenamefont {Feng},
  \citenamefont {Bird}, \citenamefont {Anderson}, \citenamefont {Font-Ribera},\
  and\ \citenamefont {Pedersen}}]{MPGadgetDOI}%
  \BibitemOpen
  \bibfield  {author} {\bibinfo {author} {\bibfnamefont {Y.}~\bibnamefont
  {Feng}}, \bibinfo {author} {\bibfnamefont {S.}~\bibnamefont {Bird}}, \bibinfo
  {author} {\bibfnamefont {L.}~\bibnamefont {Anderson}}, \bibinfo {author}
  {\bibfnamefont {A.}~\bibnamefont {Font-Ribera}}, \ and\ \bibinfo {author}
  {\bibfnamefont {C.}~\bibnamefont {Pedersen}},\ }\href {\doibase
  10.5281/zenodo.1451799} {\  (\bibinfo {year} {2018}),\
  10.5281/zenodo.1451799}\BibitemShut {NoStop}%
\bibitem [{\citenamefont {{Hand}}\ \emph {et~al.}(2018)\citenamefont {{Hand}},
  \citenamefont {{Feng}}, \citenamefont {{Beutler}}, \citenamefont {{Li}},
  \citenamefont {{Modi}}, \citenamefont {{Seljak}},\ and\ \citenamefont
  {{Slepian}}}]{nbodykitPaper}%
  \BibitemOpen
  \bibfield  {author} {\bibinfo {author} {\bibfnamefont {N.}~\bibnamefont
  {{Hand}}}, \bibinfo {author} {\bibfnamefont {Y.}~\bibnamefont {{Feng}}},
  \bibinfo {author} {\bibfnamefont {F.}~\bibnamefont {{Beutler}}}, \bibinfo
  {author} {\bibfnamefont {Y.}~\bibnamefont {{Li}}}, \bibinfo {author}
  {\bibfnamefont {C.}~\bibnamefont {{Modi}}}, \bibinfo {author} {\bibfnamefont
  {U.}~\bibnamefont {{Seljak}}}, \ and\ \bibinfo {author} {\bibfnamefont
  {Z.}~\bibnamefont {{Slepian}}},\ }\href {\doibase 10.3847/1538-3881/aadae0}
  {\bibfield  {journal} {\bibinfo  {journal} {\aj}\ }\textbf {\bibinfo {volume}
  {156}},\ \bibinfo {eid} {160} (\bibinfo {year} {2018})},\ \Eprint
  {http://arxiv.org/abs/1712.05834} {arXiv:1712.05834 [astro-ph.IM]}
  \BibitemShut {NoStop}%
\bibitem [{nbo()}]{nbodykitWebsite}%
  \BibitemOpen
  \href {\doibase 10.5281/zenodo.1336774} {}\bibinfo {howpublished}
  {https://github.com/bccp/nbodykit}\BibitemShut {NoStop}%
\bibitem [{\citenamefont {{LSST Science Collaboration}}\ \emph
  {et~al.}(2009)\citenamefont {{LSST Science Collaboration}}, \citenamefont
  {{Abell}}, \citenamefont {{Allison}}, \citenamefont {{Anderson}},
  \citenamefont {{Andrew}}, \citenamefont {{Angel}}, \citenamefont {{Armus}},
  \citenamefont {{Arnett}}, \citenamefont {{Asztalos}}, \citenamefont
  {{Axelrod}},\ and\ \citenamefont {et~al.}}]{LSSTScienceBook}%
  \BibitemOpen
  \bibfield  {author} {\bibinfo {author} {\bibnamefont {{LSST Science
  Collaboration}}}, \bibinfo {author} {\bibfnamefont {P.~A.}\ \bibnamefont
  {{Abell}}}, \bibinfo {author} {\bibfnamefont {J.}~\bibnamefont {{Allison}}},
  \bibinfo {author} {\bibfnamefont {S.~F.}\ \bibnamefont {{Anderson}}},
  \bibinfo {author} {\bibfnamefont {J.~R.}\ \bibnamefont {{Andrew}}}, \bibinfo
  {author} {\bibfnamefont {J.~R.~P.}\ \bibnamefont {{Angel}}}, \bibinfo
  {author} {\bibfnamefont {L.}~\bibnamefont {{Armus}}}, \bibinfo {author}
  {\bibfnamefont {D.}~\bibnamefont {{Arnett}}}, \bibinfo {author}
  {\bibfnamefont {S.~J.}\ \bibnamefont {{Asztalos}}}, \bibinfo {author}
  {\bibfnamefont {T.~S.}\ \bibnamefont {{Axelrod}}}, \ and\ \bibinfo {author}
  {\bibnamefont {et~al.}},\ }\href@noop {} {\bibfield  {journal} {\bibinfo
  {journal} {ArXiv e-prints}\ } (\bibinfo {year} {2009})},\ \Eprint
  {http://arxiv.org/abs/0912.0201} {arXiv:0912.0201 [astro-ph.IM]} \BibitemShut
  {NoStop}%
\bibitem [{LSS(2017)}]{LSSTwebsite}%
  \BibitemOpen
  \href {https://www.lsst.org/} {}\bibinfo {howpublished}
  {https://www.lsst.org/} (\bibinfo {year} {2017})\BibitemShut {NoStop}%
\bibitem [{\citenamefont {{Dodelson}}\ \emph {et~al.}(2016)\citenamefont
  {{Dodelson}}, \citenamefont {{Heitmann}}, \citenamefont {{Hirata}},
  \citenamefont {{Honscheid}}, \citenamefont {{Roodman}}, \citenamefont
  {{Seljak}}, \citenamefont {{Slosar}},\ and\ \citenamefont
  {{Trodden}}}]{1604.07626}%
  \BibitemOpen
  \bibfield  {author} {\bibinfo {author} {\bibfnamefont {S.}~\bibnamefont
  {{Dodelson}}}, \bibinfo {author} {\bibfnamefont {K.}~\bibnamefont
  {{Heitmann}}}, \bibinfo {author} {\bibfnamefont {C.}~\bibnamefont
  {{Hirata}}}, \bibinfo {author} {\bibfnamefont {K.}~\bibnamefont
  {{Honscheid}}}, \bibinfo {author} {\bibfnamefont {A.}~\bibnamefont
  {{Roodman}}}, \bibinfo {author} {\bibfnamefont {U.}~\bibnamefont {{Seljak}}},
  \bibinfo {author} {\bibfnamefont {A.}~\bibnamefont {{Slosar}}}, \ and\
  \bibinfo {author} {\bibfnamefont {M.}~\bibnamefont {{Trodden}}},\ }\href@noop
  {} {\bibfield  {journal} {\bibinfo  {journal} {ArXiv e-prints}\ } (\bibinfo
  {year} {2016})},\ \Eprint {http://arxiv.org/abs/1604.07626}
  {arXiv:1604.07626} \BibitemShut {NoStop}%
\bibitem [{\citenamefont {{Dor{\'e}}}\ \emph {et~al.}(2014)\citenamefont
  {{Dor{\'e}}}, \citenamefont {{Bock}}, \citenamefont {{Ashby}}, \citenamefont
  {{Capak}}, \citenamefont {{Cooray}}, \citenamefont {{de Putter}},
  \citenamefont {{Eifler}}, \citenamefont {{Flagey}}, \citenamefont {{Gong}},
  \citenamefont {{Habib}} \emph {et~al.}}]{Spherex1412}%
  \BibitemOpen
  \bibfield  {author} {\bibinfo {author} {\bibfnamefont {O.}~\bibnamefont
  {{Dor{\'e}}}}, \bibinfo {author} {\bibfnamefont {J.}~\bibnamefont {{Bock}}},
  \bibinfo {author} {\bibfnamefont {M.}~\bibnamefont {{Ashby}}}, \bibinfo
  {author} {\bibfnamefont {P.}~\bibnamefont {{Capak}}}, \bibinfo {author}
  {\bibfnamefont {A.}~\bibnamefont {{Cooray}}}, \bibinfo {author}
  {\bibfnamefont {R.}~\bibnamefont {{de Putter}}}, \bibinfo {author}
  {\bibfnamefont {T.}~\bibnamefont {{Eifler}}}, \bibinfo {author}
  {\bibfnamefont {N.}~\bibnamefont {{Flagey}}}, \bibinfo {author}
  {\bibfnamefont {Y.}~\bibnamefont {{Gong}}}, \bibinfo {author} {\bibfnamefont
  {S.}~\bibnamefont {{Habib}}},  \emph {et~al.},\ }\href@noop {} {\bibfield
  {journal} {\bibinfo  {journal} {ArXiv e-prints}\ } (\bibinfo {year}
  {2014})},\ \Eprint {http://arxiv.org/abs/1412.4872} {arXiv:1412.4872}
  \BibitemShut {NoStop}%
\bibitem [{sph(2017)}]{spherexWebsite}%
  \BibitemOpen
  \href {http://spherex.caltech.edu/} {}\bibinfo {howpublished}
  {http://spherex.caltech.edu/} (\bibinfo {year} {2017})\BibitemShut {NoStop}%
\bibitem [{SDS(2017)}]{SDSSwebsite}%
  \BibitemOpen
  \href {http://www.sdss.org/} {}\bibinfo {howpublished} {http://www.sdss.org/}
  (\bibinfo {year} {2017})\BibitemShut {NoStop}%
\bibitem [{DES()}]{DESIFDRDoc}%
  \BibitemOpen
  \href
  {http://desi.lbl.gov/wp-content/uploads/2014/04/fdr-science-biblatex.pdf}
  {\enquote {\bibinfo {title} {Desi final design report part i:
  Science,targeting, and survey design},}\ }\bibinfo {howpublished}
  {http://desi.lbl.gov/wp-content/uploads/2014/04/fdr-science-biblatex.pdf}\BibitemShut
  {NoStop}%
\bibitem [{DES(2017)}]{DESIwebsite}%
  \BibitemOpen
  \href {http://desi.lbl.gov} {}\bibinfo {howpublished} {http://desi.lbl.gov}
  (\bibinfo {year} {2017})\BibitemShut {NoStop}%
\bibitem [{\citenamefont {{Laureijs}}\ \emph {et~al.}(2011)\citenamefont
  {{Laureijs}}, \citenamefont {{Amiaux}}, \citenamefont {{Arduini}},
  \citenamefont {{Augu{\`e}res}}, \citenamefont {{Brinchmann}}, \citenamefont
  {{Cole}}, \citenamefont {{Cropper}}, \citenamefont {{Dabin}}, \citenamefont
  {{Duvet}}, \citenamefont {{Ealet}} \emph {et~al.}}]{EuclidWhitePaper}%
  \BibitemOpen
  \bibfield  {author} {\bibinfo {author} {\bibfnamefont {R.}~\bibnamefont
  {{Laureijs}}}, \bibinfo {author} {\bibfnamefont {J.}~\bibnamefont
  {{Amiaux}}}, \bibinfo {author} {\bibfnamefont {S.}~\bibnamefont {{Arduini}}},
  \bibinfo {author} {\bibfnamefont {J.~.}\ \bibnamefont {{Augu{\`e}res}}},
  \bibinfo {author} {\bibfnamefont {J.}~\bibnamefont {{Brinchmann}}}, \bibinfo
  {author} {\bibfnamefont {R.}~\bibnamefont {{Cole}}}, \bibinfo {author}
  {\bibfnamefont {M.}~\bibnamefont {{Cropper}}}, \bibinfo {author}
  {\bibfnamefont {C.}~\bibnamefont {{Dabin}}}, \bibinfo {author} {\bibfnamefont
  {L.}~\bibnamefont {{Duvet}}}, \bibinfo {author} {\bibfnamefont
  {A.}~\bibnamefont {{Ealet}}},  \emph {et~al.},\ }\href@noop {} {\bibfield
  {journal} {\bibinfo  {journal} {ArXiv e-prints}\ } (\bibinfo {year}
  {2011})},\ \Eprint {http://arxiv.org/abs/1110.3193} {arXiv:1110.3193}
  \BibitemShut {NoStop}%
\bibitem [{euc(2017{\natexlab{a}})}]{euclidECWebsite}%
  \BibitemOpen
  \href {http://www.euclid-ec.org/} {}\bibinfo {howpublished}
  {http://www.euclid-ec.org/} (\bibinfo {year}
  {2017}{\natexlab{a}})\BibitemShut {NoStop}%
\bibitem [{euc(2017{\natexlab{b}})}]{euclidESAWebsite}%
  \BibitemOpen
  \href {http://sci.esa.int/euclid/} {}\bibinfo {howpublished}
  {http://sci.esa.int/euclid/} (\bibinfo {year}
  {2017}{\natexlab{b}})\BibitemShut {NoStop}%
\bibitem [{\citenamefont {{Taruya}}\ \emph {et~al.}(2018)\citenamefont
  {{Taruya}}, \citenamefont {{Nishimichi}},\ and\ \citenamefont
  {{Jeong}}}]{Taruya1807}%
  \BibitemOpen
  \bibfield  {author} {\bibinfo {author} {\bibfnamefont {A.}~\bibnamefont
  {{Taruya}}}, \bibinfo {author} {\bibfnamefont {T.}~\bibnamefont
  {{Nishimichi}}}, \ and\ \bibinfo {author} {\bibfnamefont {D.}~\bibnamefont
  {{Jeong}}},\ }\href@noop {} {\bibfield  {journal} {\bibinfo  {journal} {ArXiv
  e-prints}\ } (\bibinfo {year} {2018})},\ \Eprint
  {http://arxiv.org/abs/1807.04215} {arXiv:1807.04215} \BibitemShut {NoStop}%
\bibitem [{\citenamefont {{McQuinn}}\ and\ \citenamefont
  {{D'Aloisio}}(2018)}]{1806.08372}%
  \BibitemOpen
  \bibfield  {author} {\bibinfo {author} {\bibfnamefont {M.}~\bibnamefont
  {{McQuinn}}}\ and\ \bibinfo {author} {\bibfnamefont {A.}~\bibnamefont
  {{D'Aloisio}}},\ }\href@noop {} {\bibfield  {journal} {\bibinfo  {journal}
  {ArXiv e-prints}\ } (\bibinfo {year} {2018})},\ \Eprint
  {http://arxiv.org/abs/1806.08372} {arXiv:1806.08372} \BibitemShut {NoStop}%
\bibitem [{\citenamefont {{Baldauf}}\ \emph
  {et~al.}(2016{\natexlab{c}})\citenamefont {{Baldauf}}, \citenamefont
  {{Codis}}, \citenamefont {{Desjacques}},\ and\ \citenamefont
  {{Pichon}}}]{Baldauf1510}%
  \BibitemOpen
  \bibfield  {author} {\bibinfo {author} {\bibfnamefont {T.}~\bibnamefont
  {{Baldauf}}}, \bibinfo {author} {\bibfnamefont {S.}~\bibnamefont {{Codis}}},
  \bibinfo {author} {\bibfnamefont {V.}~\bibnamefont {{Desjacques}}}, \ and\
  \bibinfo {author} {\bibfnamefont {C.}~\bibnamefont {{Pichon}}},\ }\href
  {\doibase 10.1093/mnras/stv2973} {\bibfield  {journal} {\bibinfo  {journal}
  {\mnras}\ }\textbf {\bibinfo {volume} {456}},\ \bibinfo {pages} {3985}
  (\bibinfo {year} {2016}{\natexlab{c}})},\ \Eprint
  {http://arxiv.org/abs/1510.09204} {arXiv:1510.09204} \BibitemShut {NoStop}%
\bibitem [{\citenamefont {{Beutler}}\ \emph {et~al.}(2014)\citenamefont
  {{Beutler}}, \citenamefont {{Saito}}, \citenamefont {{Seo}}, \citenamefont
  {{Brinkmann}}, \citenamefont {{Dawson}}, \citenamefont {{Eisenstein}},
  \citenamefont {{Font-Ribera}}, \citenamefont {{Ho}}, \citenamefont
  {{McBride}}, \citenamefont {{Montesano}}, \citenamefont {{Percival}} \emph
  {et~al.}}]{2014MNRAS.443.1065B}%
  \BibitemOpen
  \bibfield  {author} {\bibinfo {author} {\bibfnamefont {F.}~\bibnamefont
  {{Beutler}}}, \bibinfo {author} {\bibfnamefont {S.}~\bibnamefont {{Saito}}},
  \bibinfo {author} {\bibfnamefont {H.-J.}\ \bibnamefont {{Seo}}}, \bibinfo
  {author} {\bibfnamefont {J.}~\bibnamefont {{Brinkmann}}}, \bibinfo {author}
  {\bibfnamefont {K.~S.}\ \bibnamefont {{Dawson}}}, \bibinfo {author}
  {\bibfnamefont {D.~J.}\ \bibnamefont {{Eisenstein}}}, \bibinfo {author}
  {\bibfnamefont {A.}~\bibnamefont {{Font-Ribera}}}, \bibinfo {author}
  {\bibfnamefont {S.}~\bibnamefont {{Ho}}}, \bibinfo {author} {\bibfnamefont
  {C.~K.}\ \bibnamefont {{McBride}}}, \bibinfo {author} {\bibfnamefont
  {F.}~\bibnamefont {{Montesano}}}, \bibinfo {author} {\bibfnamefont {W.~J.}\
  \bibnamefont {{Percival}}},  \emph {et~al.},\ }\href {\doibase
  10.1093/mnras/stu1051} {\bibfield  {journal} {\bibinfo  {journal} {\mnras}\
  }\textbf {\bibinfo {volume} {443}},\ \bibinfo {pages} {1065} (\bibinfo {year}
  {2014})},\ \Eprint {http://arxiv.org/abs/1312.4611} {arXiv:1312.4611}
  \BibitemShut {NoStop}%
\bibitem [{\citenamefont {{Gil-Mar{\'{\i}}n}}\ \emph
  {et~al.}(2015)\citenamefont {{Gil-Mar{\'{\i}}n}}, \citenamefont
  {{Nore{\~n}a}}, \citenamefont {{Verde}}, \citenamefont {{Percival}},
  \citenamefont {{Wagner}}, \citenamefont {{Manera}},\ and\ \citenamefont
  {{Schneider}}}]{2015MNRAS.451..539G}%
  \BibitemOpen
  \bibfield  {author} {\bibinfo {author} {\bibfnamefont {H.}~\bibnamefont
  {{Gil-Mar{\'{\i}}n}}}, \bibinfo {author} {\bibfnamefont {J.}~\bibnamefont
  {{Nore{\~n}a}}}, \bibinfo {author} {\bibfnamefont {L.}~\bibnamefont
  {{Verde}}}, \bibinfo {author} {\bibfnamefont {W.~J.}\ \bibnamefont
  {{Percival}}}, \bibinfo {author} {\bibfnamefont {C.}~\bibnamefont
  {{Wagner}}}, \bibinfo {author} {\bibfnamefont {M.}~\bibnamefont {{Manera}}},
  \ and\ \bibinfo {author} {\bibfnamefont {D.~P.}\ \bibnamefont
  {{Schneider}}},\ }\href {\doibase 10.1093/mnras/stv961} {\bibfield  {journal}
  {\bibinfo  {journal} {\mnras}\ }\textbf {\bibinfo {volume} {451}},\ \bibinfo
  {pages} {539} (\bibinfo {year} {2015})},\ \Eprint
  {http://arxiv.org/abs/1407.5668} {arXiv:1407.5668} \BibitemShut {NoStop}%
\bibitem [{\citenamefont {{Beutler}}\ \emph {et~al.}(2017)\citenamefont
  {{Beutler}}, \citenamefont {{Seo}}, \citenamefont {{Saito}}, \citenamefont
  {{Chuang}}, \citenamefont {{Cuesta}}, \citenamefont {{Eisenstein}},
  \citenamefont {{Gil-Mar{\'{\i}}n}}, \citenamefont {{Grieb}}, \citenamefont
  {{Hand}}, \citenamefont {{Kitaura}} \emph {et~al.}}]{Beutler1607}%
  \BibitemOpen
  \bibfield  {author} {\bibinfo {author} {\bibfnamefont {F.}~\bibnamefont
  {{Beutler}}}, \bibinfo {author} {\bibfnamefont {H.-J.}\ \bibnamefont
  {{Seo}}}, \bibinfo {author} {\bibfnamefont {S.}~\bibnamefont {{Saito}}},
  \bibinfo {author} {\bibfnamefont {C.-H.}\ \bibnamefont {{Chuang}}}, \bibinfo
  {author} {\bibfnamefont {A.~J.}\ \bibnamefont {{Cuesta}}}, \bibinfo {author}
  {\bibfnamefont {D.~J.}\ \bibnamefont {{Eisenstein}}}, \bibinfo {author}
  {\bibfnamefont {H.}~\bibnamefont {{Gil-Mar{\'{\i}}n}}}, \bibinfo {author}
  {\bibfnamefont {J.~N.}\ \bibnamefont {{Grieb}}}, \bibinfo {author}
  {\bibfnamefont {N.}~\bibnamefont {{Hand}}}, \bibinfo {author} {\bibfnamefont
  {F.-S.}\ \bibnamefont {{Kitaura}}},  \emph {et~al.},\ }\href {\doibase
  10.1093/mnras/stw3298} {\bibfield  {journal} {\bibinfo  {journal} {\mnras}\
  }\textbf {\bibinfo {volume} {466}},\ \bibinfo {pages} {2242} (\bibinfo {year}
  {2017})},\ \Eprint {http://arxiv.org/abs/1607.03150} {arXiv:1607.03150}
  \BibitemShut {NoStop}%
\bibitem [{\citenamefont {{S{\'a}nchez}}\ \emph {et~al.}()\citenamefont
  {{S{\'a}nchez}}, \citenamefont {{Scoccimarro}}, \citenamefont {{Crocce}},
  \citenamefont {{Grieb}}, \citenamefont {{Salazar-Albornoz}}, \citenamefont
  {{Dalla Vecchia}}, \citenamefont {{Lippich}}, \citenamefont {{Beutler}},
  \citenamefont {{Brownstein}}, \citenamefont {{Chuang}} \emph
  {et~al.}}]{2017MNRAS.464.1640S}%
  \BibitemOpen
  \bibfield  {author} {\bibinfo {author} {\bibfnamefont {A.~G.}\ \bibnamefont
  {{S{\'a}nchez}}}, \bibinfo {author} {\bibfnamefont {R.}~\bibnamefont
  {{Scoccimarro}}}, \bibinfo {author} {\bibfnamefont {M.}~\bibnamefont
  {{Crocce}}}, \bibinfo {author} {\bibfnamefont {J.~N.}\ \bibnamefont
  {{Grieb}}}, \bibinfo {author} {\bibfnamefont {S.}~\bibnamefont
  {{Salazar-Albornoz}}}, \bibinfo {author} {\bibfnamefont {C.}~\bibnamefont
  {{Dalla Vecchia}}}, \bibinfo {author} {\bibfnamefont {M.}~\bibnamefont
  {{Lippich}}}, \bibinfo {author} {\bibfnamefont {F.}~\bibnamefont
  {{Beutler}}}, \bibinfo {author} {\bibfnamefont {J.~R.}\ \bibnamefont
  {{Brownstein}}}, \bibinfo {author} {\bibfnamefont {C.-H.}\ \bibnamefont
  {{Chuang}}},  \emph {et~al.},\ }\href@noop {} {\ }\BibitemShut {NoStop}%
\bibitem [{\citenamefont {{Alam}}\ \emph {et~al.}(2017)\citenamefont {{Alam}},
  \citenamefont {{Ata}}, \citenamefont {{Bailey}}, \citenamefont {{Beutler}},
  \citenamefont {{Bizyaev}}, \citenamefont {{Blazek}}, \citenamefont
  {{Bolton}}, \citenamefont {{Brownstein}}, \citenamefont {{Burden}},
  \citenamefont {{Chuang}} \emph {et~al.}}]{2017MNRAS.470.2617A}%
  \BibitemOpen
  \bibfield  {author} {\bibinfo {author} {\bibfnamefont {S.}~\bibnamefont
  {{Alam}}}, \bibinfo {author} {\bibfnamefont {M.}~\bibnamefont {{Ata}}},
  \bibinfo {author} {\bibfnamefont {S.}~\bibnamefont {{Bailey}}}, \bibinfo
  {author} {\bibfnamefont {F.}~\bibnamefont {{Beutler}}}, \bibinfo {author}
  {\bibfnamefont {D.}~\bibnamefont {{Bizyaev}}}, \bibinfo {author}
  {\bibfnamefont {J.~A.}\ \bibnamefont {{Blazek}}}, \bibinfo {author}
  {\bibfnamefont {A.~S.}\ \bibnamefont {{Bolton}}}, \bibinfo {author}
  {\bibfnamefont {J.~R.}\ \bibnamefont {{Brownstein}}}, \bibinfo {author}
  {\bibfnamefont {A.}~\bibnamefont {{Burden}}}, \bibinfo {author}
  {\bibfnamefont {C.-H.}\ \bibnamefont {{Chuang}}},  \emph {et~al.},\ }\href
  {\doibase 10.1093/mnras/stx721} {\bibfield  {journal} {\bibinfo  {journal}
  {\mnras}\ }\textbf {\bibinfo {volume} {470}},\ \bibinfo {pages} {2617}
  (\bibinfo {year} {2017})},\ \Eprint {http://arxiv.org/abs/1607.03155}
  {arXiv:1607.03155} \BibitemShut {NoStop}%
\bibitem [{\citenamefont {{Baldauf}}\ \emph
  {et~al.}(2016{\natexlab{d}})\citenamefont {{Baldauf}}, \citenamefont
  {{Mirbabayi}}, \citenamefont {{Simonovi{\'c}}},\ and\ \citenamefont
  {{Zaldarriaga}}}]{TobiasTheoError1602}%
  \BibitemOpen
  \bibfield  {author} {\bibinfo {author} {\bibfnamefont {T.}~\bibnamefont
  {{Baldauf}}}, \bibinfo {author} {\bibfnamefont {M.}~\bibnamefont
  {{Mirbabayi}}}, \bibinfo {author} {\bibfnamefont {M.}~\bibnamefont
  {{Simonovi{\'c}}}}, \ and\ \bibinfo {author} {\bibfnamefont {M.}~\bibnamefont
  {{Zaldarriaga}}},\ }\href@noop {} {\bibfield  {journal} {\bibinfo  {journal}
  {ArXiv e-prints}\ } (\bibinfo {year} {2016}{\natexlab{d}})},\ \Eprint
  {http://arxiv.org/abs/1602.00674} {arXiv:1602.00674} \BibitemShut {NoStop}%
\bibitem [{\citenamefont {Bernardeau}\ \emph {et~al.}(2002)\citenamefont
  {Bernardeau}, \citenamefont {Colombi}, \citenamefont {Gaztanaga},\ and\
  \citenamefont {Scoccimarro}}]{Bernardeau:2001qr}%
  \BibitemOpen
  \bibfield  {author} {\bibinfo {author} {\bibfnamefont {F.}~\bibnamefont
  {Bernardeau}}, \bibinfo {author} {\bibfnamefont {S.}~\bibnamefont {Colombi}},
  \bibinfo {author} {\bibfnamefont {E.}~\bibnamefont {Gaztanaga}}, \ and\
  \bibinfo {author} {\bibfnamefont {R.}~\bibnamefont {Scoccimarro}},\ }\href
  {\doibase 10.1016/S0370-1573(02)00135-7} {\bibfield  {journal} {\bibinfo
  {journal} {Phys. Rept.}\ }\textbf {\bibinfo {volume} {367}},\ \bibinfo
  {pages} {1} (\bibinfo {year} {2002})},\ \Eprint
  {http://arxiv.org/abs/astro-ph/0112551} {arXiv:astro-ph/0112551 [astro-ph]}
  \BibitemShut {NoStop}%
\bibitem [{\citenamefont {Hamann}\ \emph {et~al.}(2010)\citenamefont {Hamann},
  \citenamefont {Hannestad}, \citenamefont {Lesgourgues}, \citenamefont
  {Rampf},\ and\ \citenamefont {Wong}}]{Hamann:2010pw}%
  \BibitemOpen
  \bibfield  {author} {\bibinfo {author} {\bibfnamefont {J.}~\bibnamefont
  {Hamann}}, \bibinfo {author} {\bibfnamefont {S.}~\bibnamefont {Hannestad}},
  \bibinfo {author} {\bibfnamefont {J.}~\bibnamefont {Lesgourgues}}, \bibinfo
  {author} {\bibfnamefont {C.}~\bibnamefont {Rampf}}, \ and\ \bibinfo {author}
  {\bibfnamefont {Y.~Y.~Y.}\ \bibnamefont {Wong}},\ }\href {\doibase
  10.1088/1475-7516/2010/07/022} {\bibfield  {journal} {\bibinfo  {journal}
  {JCAP}\ }\textbf {\bibinfo {volume} {1007}},\ \bibinfo {pages} {022}
  (\bibinfo {year} {2010})},\ \Eprint {http://arxiv.org/abs/1003.3999}
  {arXiv:1003.3999 [astro-ph.CO]} \BibitemShut {NoStop}%
\bibitem [{\citenamefont {Baumann}\ \emph {et~al.}(2018)\citenamefont
  {Baumann}, \citenamefont {Green},\ and\ \citenamefont
  {Wallisch}}]{Baumann:2017gkg}%
  \BibitemOpen
  \bibfield  {author} {\bibinfo {author} {\bibfnamefont {D.}~\bibnamefont
  {Baumann}}, \bibinfo {author} {\bibfnamefont {D.}~\bibnamefont {Green}}, \
  and\ \bibinfo {author} {\bibfnamefont {B.}~\bibnamefont {Wallisch}},\ }\href
  {\doibase 10.1088/1475-7516/2018/08/029} {\bibfield  {journal} {\bibinfo
  {journal} {JCAP}\ }\textbf {\bibinfo {volume} {1808}},\ \bibinfo {pages}
  {029} (\bibinfo {year} {2018})},\ \Eprint {http://arxiv.org/abs/1712.08067}
  {arXiv:1712.08067 [astro-ph.CO]} \BibitemShut {NoStop}%
\bibitem [{\citenamefont {{Hamaus}}\ \emph {et~al.}(2011)\citenamefont
  {{Hamaus}}, \citenamefont {{Seljak}},\ and\ \citenamefont
  {{Desjacques}}}]{Hamaus1104}%
  \BibitemOpen
  \bibfield  {author} {\bibinfo {author} {\bibfnamefont {N.}~\bibnamefont
  {{Hamaus}}}, \bibinfo {author} {\bibfnamefont {U.}~\bibnamefont {{Seljak}}},
  \ and\ \bibinfo {author} {\bibfnamefont {V.}~\bibnamefont {{Desjacques}}},\
  }\href {\doibase 10.1103/PhysRevD.84.083509} {\bibfield  {journal} {\bibinfo
  {journal} {\prd}\ }\textbf {\bibinfo {volume} {84}},\ \bibinfo {eid} {083509}
  (\bibinfo {year} {2011})},\ \Eprint {http://arxiv.org/abs/1104.2321}
  {arXiv:1104.2321 [astro-ph.CO]} \BibitemShut {NoStop}%
\bibitem [{\citenamefont {{Hamaus}}\ \emph {et~al.}(2012)\citenamefont
  {{Hamaus}}, \citenamefont {{Seljak}},\ and\ \citenamefont
  {{Desjacques}}}]{Hamaus1207}%
  \BibitemOpen
  \bibfield  {author} {\bibinfo {author} {\bibfnamefont {N.}~\bibnamefont
  {{Hamaus}}}, \bibinfo {author} {\bibfnamefont {U.}~\bibnamefont {{Seljak}}},
  \ and\ \bibinfo {author} {\bibfnamefont {V.}~\bibnamefont {{Desjacques}}},\
  }\href {\doibase 10.1103/PhysRevD.86.103513} {\bibfield  {journal} {\bibinfo
  {journal} {\prd}\ }\textbf {\bibinfo {volume} {86}},\ \bibinfo {eid} {103513}
  (\bibinfo {year} {2012})},\ \Eprint {http://arxiv.org/abs/1207.1102}
  {arXiv:1207.1102 [astro-ph.CO]} \BibitemShut {NoStop}%
\bibitem [{\citenamefont {{Schmidt}}(2016)}]{Schmidt1511}%
  \BibitemOpen
  \bibfield  {author} {\bibinfo {author} {\bibfnamefont {F.}~\bibnamefont
  {{Schmidt}}},\ }\href {\doibase 10.1103/PhysRevD.93.063512} {\bibfield
  {journal} {\bibinfo  {journal} {\prd}\ }\textbf {\bibinfo {volume} {93}},\
  \bibinfo {eid} {063512} (\bibinfo {year} {2016})},\ \Eprint
  {http://arxiv.org/abs/1511.02231} {arXiv:1511.02231} \BibitemShut {NoStop}%
\bibitem [{\citenamefont {{Wang}}\ \emph {et~al.}(2009)\citenamefont {{Wang}},
  \citenamefont {{Mo}}, \citenamefont {{Jing}}, \citenamefont {{Guo}},
  \citenamefont {{van den Bosch}},\ and\ \citenamefont
  {{Yang}}}]{2009MNRAS.394..398W}%
  \BibitemOpen
  \bibfield  {author} {\bibinfo {author} {\bibfnamefont {H.}~\bibnamefont
  {{Wang}}}, \bibinfo {author} {\bibfnamefont {H.~J.}\ \bibnamefont {{Mo}}},
  \bibinfo {author} {\bibfnamefont {Y.~P.}\ \bibnamefont {{Jing}}}, \bibinfo
  {author} {\bibfnamefont {Y.}~\bibnamefont {{Guo}}}, \bibinfo {author}
  {\bibfnamefont {F.~C.}\ \bibnamefont {{van den Bosch}}}, \ and\ \bibinfo
  {author} {\bibfnamefont {X.}~\bibnamefont {{Yang}}},\ }\href {\doibase
  10.1111/j.1365-2966.2008.14301.x} {\bibfield  {journal} {\bibinfo  {journal}
  {\mnras}\ }\textbf {\bibinfo {volume} {394}},\ \bibinfo {pages} {398}
  (\bibinfo {year} {2009})},\ \Eprint {http://arxiv.org/abs/0803.1213}
  {arXiv:0803.1213} \BibitemShut {NoStop}%
\bibitem [{\citenamefont {Zeldovich}(1965)}]{ZELDOVICH1965241}%
  \BibitemOpen
  \bibfield  {author} {\bibinfo {author} {\bibfnamefont {Y.}~\bibnamefont
  {Zeldovich}}\ }(\bibinfo  {publisher} {Elsevier},\ \bibinfo {year} {1965})\
  Chap.\ \bibinfo {chapter} {V.D;
  \url{https://doi.org/10.1016/B978-1-4831-9921-4.50011-9}}, pp.\ \bibinfo
  {pages} {241 -- 379}\BibitemShut {NoStop}%
\bibitem [{\citenamefont {{Peebles}}(1980)}]{Peebles1980}%
  \BibitemOpen
  \bibfield  {author} {\bibinfo {author} {\bibfnamefont {P.~J.~E.}\
  \bibnamefont {{Peebles}}},\ }\href@noop {} {\emph {\bibinfo {title} {The
  large-scale structure of the universe, Princeton University Press}}}\
  (\bibinfo {year} {1980})\BibitemShut {NoStop}%
\bibitem [{\citenamefont {{Goroff}}\ \emph {et~al.}(1986)\citenamefont
  {{Goroff}}, \citenamefont {{Grinstein}}, \citenamefont {{Rey}},\ and\
  \citenamefont {{Wise}}}]{Goroff86}%
  \BibitemOpen
  \bibfield  {author} {\bibinfo {author} {\bibfnamefont {M.~H.}\ \bibnamefont
  {{Goroff}}}, \bibinfo {author} {\bibfnamefont {B.}~\bibnamefont
  {{Grinstein}}}, \bibinfo {author} {\bibfnamefont {S.}~\bibnamefont {{Rey}}},
  \ and\ \bibinfo {author} {\bibfnamefont {M.~B.}\ \bibnamefont {{Wise}}},\
  }\href {\doibase 10.1086/164749} {\bibfield  {journal} {\bibinfo  {journal}
  {Astrophys. J.}\ }\textbf {\bibinfo {volume} {311}},\ \bibinfo {pages} {6}
  (\bibinfo {year} {1986})}\BibitemShut {NoStop}%
\bibitem [{\citenamefont {{Crocce}}\ and\ \citenamefont
  {{Scoccimarro}}(2008)}]{2008PhRvD..77b3533C}%
  \BibitemOpen
  \bibfield  {author} {\bibinfo {author} {\bibfnamefont {M.}~\bibnamefont
  {{Crocce}}}\ and\ \bibinfo {author} {\bibfnamefont {R.}~\bibnamefont
  {{Scoccimarro}}},\ }\href {\doibase 10.1103/PhysRevD.77.023533} {\bibfield
  {journal} {\bibinfo  {journal} {\prd}\ }\textbf {\bibinfo {volume} {77}},\
  \bibinfo {eid} {023533} (\bibinfo {year} {2008})},\ \Eprint
  {http://arxiv.org/abs/0704.2783} {arXiv:0704.2783} \BibitemShut {NoStop}%
\bibitem [{\citenamefont {{Seljak}}(2009)}]{Uros0807}%
  \BibitemOpen
  \bibfield  {author} {\bibinfo {author} {\bibfnamefont {U.}~\bibnamefont
  {{Seljak}}},\ }\href {\doibase 10.1103/PhysRevLett.102.021302} {\bibfield
  {journal} {\bibinfo  {journal} {Physical Review Letters}\ }\textbf {\bibinfo
  {volume} {102}},\ \bibinfo {eid} {021302} (\bibinfo {year} {2009})},\ \Eprint
  {http://arxiv.org/abs/0807.1770} {arXiv:0807.1770} \BibitemShut {NoStop}%
\bibitem [{\citenamefont {{Norberg}}\ \emph {et~al.}(2001)\citenamefont
  {{Norberg}}, \citenamefont {{Baugh}}, \citenamefont {{Hawkins}},
  \citenamefont {{Maddox}}, \citenamefont {{Peacock}}, \citenamefont {{Cole}},
  \citenamefont {{Frenk}}, \citenamefont {{Bland-Hawthorn}}, \citenamefont
  {{Bridges}}, \citenamefont {{Cannon}} \emph {et~al.}}]{2001MNRAS.328...64N}%
  \BibitemOpen
  \bibfield  {author} {\bibinfo {author} {\bibfnamefont {P.}~\bibnamefont
  {{Norberg}}}, \bibinfo {author} {\bibfnamefont {C.~M.}\ \bibnamefont
  {{Baugh}}}, \bibinfo {author} {\bibfnamefont {E.}~\bibnamefont {{Hawkins}}},
  \bibinfo {author} {\bibfnamefont {S.}~\bibnamefont {{Maddox}}}, \bibinfo
  {author} {\bibfnamefont {J.~A.}\ \bibnamefont {{Peacock}}}, \bibinfo {author}
  {\bibfnamefont {S.}~\bibnamefont {{Cole}}}, \bibinfo {author} {\bibfnamefont
  {C.~S.}\ \bibnamefont {{Frenk}}}, \bibinfo {author} {\bibfnamefont
  {J.}~\bibnamefont {{Bland-Hawthorn}}}, \bibinfo {author} {\bibfnamefont
  {T.}~\bibnamefont {{Bridges}}}, \bibinfo {author} {\bibfnamefont
  {R.}~\bibnamefont {{Cannon}}},  \emph {et~al.},\ }\href {\doibase
  10.1046/j.1365-8711.2001.04839.x} {\bibfield  {journal} {\bibinfo  {journal}
  {\mnras}\ }\textbf {\bibinfo {volume} {328}},\ \bibinfo {pages} {64}
  (\bibinfo {year} {2001})},\ \Eprint {http://arxiv.org/abs/astro-ph/0105500}
  {astro-ph/0105500} \BibitemShut {NoStop}%
\bibitem [{\citenamefont {{Tegmark}}\ \emph {et~al.}(2004)\citenamefont
  {{Tegmark}}, \citenamefont {{Blanton}}, \citenamefont {{Strauss}},
  \citenamefont {{Hoyle}}, \citenamefont {{Schlegel}}, \citenamefont
  {{Scoccimarro}}, \citenamefont {{Vogeley}}, \citenamefont {{Weinberg}},
  \citenamefont {{Zehavi}}, \citenamefont {{Berlind}} \emph
  {et~al.}}]{2004ApJ...606..702T}%
  \BibitemOpen
  \bibfield  {author} {\bibinfo {author} {\bibfnamefont {M.}~\bibnamefont
  {{Tegmark}}}, \bibinfo {author} {\bibfnamefont {M.~R.}\ \bibnamefont
  {{Blanton}}}, \bibinfo {author} {\bibfnamefont {M.~A.}\ \bibnamefont
  {{Strauss}}}, \bibinfo {author} {\bibfnamefont {F.}~\bibnamefont {{Hoyle}}},
  \bibinfo {author} {\bibfnamefont {D.}~\bibnamefont {{Schlegel}}}, \bibinfo
  {author} {\bibfnamefont {R.}~\bibnamefont {{Scoccimarro}}}, \bibinfo {author}
  {\bibfnamefont {M.~S.}\ \bibnamefont {{Vogeley}}}, \bibinfo {author}
  {\bibfnamefont {D.~H.}\ \bibnamefont {{Weinberg}}}, \bibinfo {author}
  {\bibfnamefont {I.}~\bibnamefont {{Zehavi}}}, \bibinfo {author}
  {\bibfnamefont {A.}~\bibnamefont {{Berlind}}},  \emph {et~al.},\ }\href
  {\doibase 10.1086/382125} {\bibfield  {journal} {\bibinfo  {journal} {\apj}\
  }\textbf {\bibinfo {volume} {606}},\ \bibinfo {pages} {702} (\bibinfo {year}
  {2004})},\ \Eprint {http://arxiv.org/abs/astro-ph/0310725} {astro-ph/0310725}
  \BibitemShut {NoStop}%
\bibitem [{\citenamefont {{Zehavi}}\ \emph {et~al.}(2005)\citenamefont
  {{Zehavi}}, \citenamefont {{Zheng}}, \citenamefont {{Weinberg}},
  \citenamefont {{Frieman}}, \citenamefont {{Berlind}}, \citenamefont
  {{Blanton}}, \citenamefont {{Scoccimarro}}, \citenamefont {{Sheth}},
  \citenamefont {{Strauss}}, \citenamefont {{Kayo}}, \citenamefont {{Suto}}
  \emph {et~al.}}]{2005ApJ...630....1Z}%
  \BibitemOpen
  \bibfield  {author} {\bibinfo {author} {\bibfnamefont {I.}~\bibnamefont
  {{Zehavi}}}, \bibinfo {author} {\bibfnamefont {Z.}~\bibnamefont {{Zheng}}},
  \bibinfo {author} {\bibfnamefont {D.~H.}\ \bibnamefont {{Weinberg}}},
  \bibinfo {author} {\bibfnamefont {J.~A.}\ \bibnamefont {{Frieman}}}, \bibinfo
  {author} {\bibfnamefont {A.~A.}\ \bibnamefont {{Berlind}}}, \bibinfo {author}
  {\bibfnamefont {M.~R.}\ \bibnamefont {{Blanton}}}, \bibinfo {author}
  {\bibfnamefont {R.}~\bibnamefont {{Scoccimarro}}}, \bibinfo {author}
  {\bibfnamefont {R.~K.}\ \bibnamefont {{Sheth}}}, \bibinfo {author}
  {\bibfnamefont {M.~A.}\ \bibnamefont {{Strauss}}}, \bibinfo {author}
  {\bibfnamefont {I.}~\bibnamefont {{Kayo}}}, \bibinfo {author} {\bibfnamefont
  {Y.}~\bibnamefont {{Suto}}},  \emph {et~al.},\ }\href {\doibase
  10.1086/431891} {\bibfield  {journal} {\bibinfo  {journal} {\apj}\ }\textbf
  {\bibinfo {volume} {630}},\ \bibinfo {pages} {1} (\bibinfo {year} {2005})},\
  \Eprint {http://arxiv.org/abs/astro-ph/0408569} {astro-ph/0408569}
  \BibitemShut {NoStop}%
\bibitem [{\citenamefont {{Guo}}\ \emph {et~al.}(2013)\citenamefont {{Guo}},
  \citenamefont {{Zehavi}}, \citenamefont {{Zheng}}, \citenamefont
  {{Weinberg}}, \citenamefont {{Berlind}}, \citenamefont {{Blanton}},
  \citenamefont {{Chen}}, \citenamefont {{Eisenstein}}, \citenamefont {{Ho}},
  \citenamefont {{Kazin}} \emph {et~al.}}]{Guo1212}%
  \BibitemOpen
  \bibfield  {author} {\bibinfo {author} {\bibfnamefont {H.}~\bibnamefont
  {{Guo}}}, \bibinfo {author} {\bibfnamefont {I.}~\bibnamefont {{Zehavi}}},
  \bibinfo {author} {\bibfnamefont {Z.}~\bibnamefont {{Zheng}}}, \bibinfo
  {author} {\bibfnamefont {D.~H.}\ \bibnamefont {{Weinberg}}}, \bibinfo
  {author} {\bibfnamefont {A.~A.}\ \bibnamefont {{Berlind}}}, \bibinfo {author}
  {\bibfnamefont {M.}~\bibnamefont {{Blanton}}}, \bibinfo {author}
  {\bibfnamefont {Y.}~\bibnamefont {{Chen}}}, \bibinfo {author} {\bibfnamefont
  {D.~J.}\ \bibnamefont {{Eisenstein}}}, \bibinfo {author} {\bibfnamefont
  {S.}~\bibnamefont {{Ho}}}, \bibinfo {author} {\bibfnamefont {E.}~\bibnamefont
  {{Kazin}}},  \emph {et~al.},\ }\href {\doibase 10.1088/0004-637X/767/2/122}
  {\bibfield  {journal} {\bibinfo  {journal} {\apj}\ }\textbf {\bibinfo
  {volume} {767}},\ \bibinfo {eid} {122} (\bibinfo {year} {2013})},\ \Eprint
  {http://arxiv.org/abs/1212.1211} {arXiv:1212.1211 [astro-ph.CO]} \BibitemShut
  {NoStop}%
\end{thebibliography}%

\label{lastpage}

\end{document}